\documentclass[a4paper,11pt]{article}
\pdfoutput=1 

\usepackage{jheppub} 
\usepackage{Definitions}
\usepackage[T1]{fontenc} 
\usepackage{amsfonts,amsmath,amssymb}
\usepackage{graphicx}
\usepackage{xcolor}
\usepackage{multirow}
\usepackage{mathtools}
\usepackage{bm}
\usepackage[vcentermath]{youngtab}
\usepackage{ytableau}
\usepackage{mathtools}
\usepackage{caption}
\usepackage{subcaption}
\newcommand{\limu}[1]{\mathrel{\mathop{\sim}\limits_{\scriptstyle{#1}}}}
\usepackage{tikz-cd}

\def\PP{\mathcal{P}}
\def\CC{\mathcal{C}}
\def\DD{\mathcal{D}}
\def\D{\Delta}

\newcommand{\PtP}{\CF_{\Delta_1}\otimes \CF_{\Delta_2}}
\newcommand{\PtD}{\CF_{\Delta}\otimes \CD_p}

\newcommand{\rrel}{\rho_{\text{rel}}}

\newcommand{\reef}[1]{(\ref{#1})}
\def\Ls{j}
\def\Aj{A_{\Ls}(q)}
\def\Bj{B_{\Ls}(q)}
\def\Real{\mathbb{R}}
\def\intg{\mathbb{Z}}
\def\Cas{C}
\newcommand{\Ket}[1]{|#1\rangle}
\newcommand{\Bra}[1]{\langle#1|}

\renewcommand{\Re}{{\rm Re \, }}

\newcommand{\CO}{{\cal O}}

\newcommand{\CD}{{\cal D}}

\newcommand{\SL}{{\text{SL}}}
\newcommand{\CK}{{\cal K}}

\newcommand{\CU}{{\cal U}}

\newcommand{\CV}{{\cal V}}

\newcommand{\CF}{{\cal F}}
\newcommand{\CP}{{\cal P}}

\newcommand{\CI}{{\cal I}}

\newcommand{\mY}{{\mathbb Y}}

\newcommand{\so}{{\mathfrak{so}}}

\newcommand{\SO}{{\text{SO}}}
\newcommand{\SU}{{\text{SU}}}
\newcommand{\CJ}{{\cal J}}

\newcommand{\Db}{\bar{\Delta}}

\usepackage{amsthm}
\newtheorem*{theorem*}{Theorem}
\newtheorem{theorem}{Theorem}
\newtheorem{remark}[theorem]{Remark}

\makeatletter
\newcommand*{\rom}[1]{\expandafter\@slowromancap\romannumeral #1@}
\makeatother
\newcommand{\ket}[1]{\bigl| #1 \bigr\rangle}

\usepackage{esint}

\DeclarePairedDelimiter\floor{\lfloor}{\rfloor}
\usepackage{dsfont}

\usepackage{color}
\definecolor{darkgreen}{rgb}{0,0.5,0}
\definecolor{darkblue}{rgb}{0,0,0.6}
\definecolor{purple}{rgb}{0.4,.2,0.7}

\numberwithin{equation}{section}
\numberwithin{figure}{section}
\numberwithin{table}{section}

\def\CQ{{\cal Q}} 
\def\CH{{\cal H}}

\def\CU{{\cal U}}

\def\CN{{\cal N}}
\def\CD{{\cal D}}
\def\tr{\,{\rm tr}\,}

\title{\boldmath 
Hilbert space of  Quantum Field Theory\\ in de Sitter spacetime}

\author[a]{Joao Penedones,}
\author[a,b]{Kamran Salehi Vaziri}
\author[c]{and Zimo Sun}

\affiliation[a]{Fields and Strings Laboratory, Institute of Physics\\ École Polytechnique Fédéral de Lausanne (EPFL)
\\ Route de la Sorge, CH-1015 Lausanne, Switzerland}
\affiliation[b]{Institute of Physics, University of Amsterdam, Amsterdam, 1098 XH, The Netherlands}
\affiliation[c]{Princeton Gravity Initiative, Princeton University, Princeton, NJ 08544, USA}

\emailAdd{joao.penedones@epfl.ch}
\emailAdd{k.salehivaziri@uva.nl}
\emailAdd{zs8479@princeton.edu}

\abstract{
We study the decomposition of the Hilbert space of quantum field theory in $(d+1)$ dimensional de Sitter spacetime into Unitary Irreducible Representations (UIRs) of its isometry group \SO$(1,d+1)$. 
Firstly, we consider multi-particle states in free theories starting from  the tensor product of single-particle UIRs.
Secondly,  we study conformal multiplets of a bulk Conformal Field Theory  with symmetry group \SO$(2,d+1)$.
Our main tools are the Harish-Chandra characters and the numerical diagonalization of the (truncated) quadratic Casimir  of \SO$(1,d+1)$. 
We introduce a continuous density 
that encodes the spectrum of irreducible representations contained in a reducible one of $\SO(1,d+1)$.
Our results are complete for $d=1$ and $d=2$. In higher dimensions, we rederive and extend several results previously known in the literature. 
Our work provides the foundation for future nonperturbative bootstrap studies of Quantum Field Theory in de Sitter spacetime.

}

\begin{document} 
\maketitle
\flushbottom

\section{Introduction and summary}
\label{sec:intro}

Our universe is under accelerated expansion. 
The best known description of the dynamics of elementary particles is Quantum Field Theory (QFT).
Therefore, it is worth understanding QFT in the most symmetric expanding universe, namely de Sitter (dS) spacetime \cite{Linde:1981mu,Guth:1980zm,Maldacena:2002vr,Arkani-Hamed:2015bza,Marolf:2010nz,Baumann:2022jpr,Arkani-Hamed:2018kmz,Goodhew:2020hob,Melville:2021lst,Sleight:2019mgd,DiPietro:2021sjt,Sleight:2021plv,Anninos:2014lwa}.
More precisely, we would like to set up a non-perturbative bootstrap approach to   QFT in  dS spacetime \cite{Hogervorst:2021uvp}.
This requires a detailed understanding of how the spacetime symmetries are realized in the Hilbert space of the theory.

In a quantum theory, symmetry generators are realized as (anti-)hermitian operators acting  on the Hilbert space.
For QFT in $(d+1)$-dimensional dS, the symmetry group is $\SO(1,d+1)$.
Therefore, the Hilbert space must decompose into Unitary Irreducible Representations (UIRs) of this non-compact group.
Such UIRs have been classified long ago in various dimensions \cite{10.2307/97833,10.2307/1969129,GelNai47,ThomasSO,NewtonSO,BSMF_1961__89__9_0,10.3792/pja/1195523378}. 
Namely, there are principal series $\CP$, complementary series $\CC$, discrete series $\CD$, exceptional series  $\CV$ (of type I) and $\CU$ (of type II),  as we partially review in section \ref{repreview}.
The question we address in this paper is: 

\vspace{.3cm}
\emph{How does the Hilbert space of a QFT in dS decompose into  UIRs of  $\SO(1,d+1)$?}
\vspace{.3cm}

\noindent Clearly, this question is too difficult to solve for a general interacting QFT.
In this paper, we consider two simple cases: free theories (in sections \ref{tensorlow} and \ref{tensorhigh})
and  Conformal Field Theories (CFT) (in section \ref{CFTsec}).

The Hilbert space of a {\bf free QFT} is a multi-particle Fock space, whose decomposition into UIRs amounts to the study of tensor products of single-particle UIRs. Remarkably, this simple group theory question has not been completely solved.
We summarize our results in tables \ref{table:tensord1}, \ref{table:tensord2} and \ref{table:tensord3}. In addition, below  we give  a brief historical review of the previous literature. 
In the present work, we rederive and extend previous results by writing the product of two Harish-Chandra characters (reviewed in section \ref{repreview}) as a sum of characters of UIRs. The sum over the principal series is actually an integral over its continuous label involving a density.  For a compact Lie group, the tensor product decomposition $R_1\otimes R_2 = \oplus_{a} n_a\, R_a$, where $R_a$ is a UIR labelled by some parameter $\lambda_a$  \footnote{$\lambda_a$ can be a multi-dimensional label and in this case, $\int^\oplus\,d\lambda$ is a multi-dimensional integral.} and $n_a$ is the multiplicity of $R_a$, is encoded in a $\delta$-function type density $\rho(\lambda) = \sum_a \, n_a \delta(\lambda-\lambda_a)$, since the decomposition of $R_1\otimes R_2$ can be formally written as a direct integral $\int^{\oplus} d\lambda\,\rho(\lambda) R_\lambda$. Our computation generalizes $\rho(\lambda)$ to the noncompact group $\SO(1, d+1)$.
In fact, the analytic structure of this density also encodes information about the complementary series. 
In addition, we check our predictions by numerically diagonalizing (a truncated version) of the quadratic Casimir of  $\SO(1,d+1)$ acting on the tensor product of two UIRs.
An example of a novel result is the decomposition of the tensor product of two exceptional series $\CV_{1,0}$  described in section \ref{EEt}. For QFT in dS, this result implies that there is a photon UIR inside the two-particle states of two different massless scalars.

The Hilbert space of a {\bf CFT} in dS is equivalent to the well understood Hilbert space of a CFT in radial quantization. Therefore, we are led to another simple group theoretical task: decompose the well known unitary conformal multiplets of $\SO(2,d+1)$ into UIRs of the subgroup $\SO(1,d+1)$.
Surprisingly, this is also an unsolved problem.
We summarize our results in table \ref{table:CFT}. 
To the best of our knowledge, these are new results building on \cite{Hogervorst:2021uvp} that studied the $d=1$ case.

As you can see from the tables, the results are complete in $d=1$  and $d=2$ but there are many open questions in $d\ge 3$.
In $d=1$, both free massive QFT and interacting CFT give rise to Hilbert spaces containing all principal and discrete series with infinite degeneracy.\footnote{The exception being chiral theories. For example, if the single-particle states correspond to $\CD^+_k$, then the full multi-particle Hilbert space will decompose into $\CD^+_q$  with finite degeneracy for each $q\ge 1$.
Similarly, a chiral CFT in dS$_2$ only gives rise to discrete series UIRs (see the last paragraph of section \ref{sec:characterCFTd1}).
} 
On the contrary, the complementary series UIRs appear in finite number (possibly zero). This is reminiscent of single-particle asymptotic states in Minkowski spacetime.
By continuity, we expect this structure of the Hilbert space to hold for generic interacting QFTs in dS$_2$.
The situation is very similar for $d=2$ with the difference that there are no discrete series and there are principal series of all spins $s\in \mathbb{Z}$.

It would be very interesting to complete this analysis in $d=3$.
There are many results known in the literature. For example, Martin in  \cite{Martin} studied the tensor product of principal series times other UIRs (including principal, complementary, exceptional and discrete).
To complete the analysis, the decomposition of the tensor product of exceptional and discrete series is needed. 
It would be interesting to revisit and extend this work using Harish-Chandra characters. 
Similarly, it would be great to complete the analysis of CFT in dS$_4$.
We leave these tasks for the  future.

\newpage

\begin{table}[h!]
\begin{center}
\def\arraystretch{1.4}%
\begin{tabular}{ |c|c|c|c|c| }
\hline
 $\otimes$ 
& $\PP_{\D_1} $ & $\CC_{\D_1}$ & $\DD_{k_1}^+$ & $\DD_{k_1}^-$\\ \hline
  $\PP_{\D_2}$  
 & $\scriptstyle \int_\D \PP_{\D}  \oplus \sum_k  \DD_k^\pm $ & 
 &  
 & 
 \\
 \hline
   $\CC_{\D_2}$   
 & $\scriptstyle \int_\D \PP_{\D}   \oplus \sum_k\DD_k^\pm$ & 
 $\scriptstyle \int_\D \PP_{\D}   \oplus \sum_k\DD_k^\pm \oplus\, \CC_{\D_1+\D_2-1}$\footnotemark  &  
 & 
 \\
 \hline
  $\DD_{k_2}^+$   
  & $\scriptstyle \int_\D \PP_{\D}   \oplus \sum_k\DD_k^+$& 
  $\scriptstyle \int_\D \PP_{\D}  \oplus \sum_k\DD_k^+$ & 
  $\scriptstyle  \sum_{k\ge k_1+k_2} \DD_k^+$& 
  \\
 \hline 
 $\DD_{k_2}^-$   
  & $\scriptstyle \int_\D \PP_{\D}   \oplus \sum_k\DD_k^-$& 
$\scriptstyle  \int_\D \PP_{\D}  \oplus \sum_k\DD_k^-$ & 
$\scriptstyle  \int_\D \PP_{\D}   \oplus 
\sum_{k=1}^{|k_1-k_2|}\DD_k^{{\rm sign}(k_1-k_2)}$
& 
$\scriptstyle  \sum_{k\ge k_1+k_2} \DD_k^-$\\
 \hline
\end{tabular}
\caption{Decomposition of the tensor products of UIRs of  $\SO(1,2)$.
These are the principal series $\CP_\Delta$ with $\D \in \frac{1}{2} +i\mathbb{R}_{\ge 0}$,
the complementary series $\CC_\Delta$ with $\frac{1}{2} <\D<1$, and the discrete series 
 $\CD_k^\pm$ with $k\in \mathbb{Z}_+$.
Since the tensor product is symmetric we did not fill in the upper triangle of the table to avoid cluttering. The sum $ \sum_k \equiv \oplus_k$ runs over $k\in \mathbb{Z}_+$ unless otherwise specified and $\int_\D$ is a sum over all principal series with $\D=\frac{1}{2} + i\lambda$ and $\lambda \in \mathbb{R}_{\ge 0}$.}  
\label{table:tensord1}
\end{center}
\end{table}
\footnotetext{A complementary series  is present in the tensor product $\CC_{\D_1} \otimes \CC_{\D_2}$ of $\SO(1,2)$ if and only if $\Delta_1+\Delta_2>\frac{3}{2}$.}

\vspace{-0.3cm}
\begin{table}[h!]
\begin{center}
\def\arraystretch{1.4}%
\begin{tabular}{ |c|c|c| }
\hline
 $\otimes$ & $\CP_{\Delta_1,m_1}$ & $\CC_{\Delta_1}$ \\ \hline
 $\PP_{\Delta_2,m_2}$ & $\scriptstyle \sum_m\int_\Delta \PP_{\Delta,m}$ & $\scriptstyle \sum_m \int_\Delta \PP_{\Delta,m}$  \\  \hline
$\CC_{\Delta_2}$  & $\scriptstyle \sum_m \int_\Delta \PP_{\Delta,m}$ & $\scriptstyle\sum_m\int_\Delta \PP_{\Delta,m} \oplus\,\CC_{\Delta_1+\Delta_2-2}$\footnotemark\\ 
\hline
\end{tabular}
\caption{Decomposition of the tensor products of UIRs of  $\SO(1,3)$.
These are the principal series $\CP_{\Delta,m}$ with dimension $\D \in 1 +i \mathbb R$ and spin $m\in \mathbb{N}$, and 
the (scalar) complementary series $\CC_\Delta$ with $1 <\D<2$. 
The sum $ \sum_m = \oplus_m$ runs over all  integers and $\Delta$ is integrated along the contour $\D=1+ i\lambda$ with $\lambda \in \mathbb{R}_{\ge 0}$.
} 
\label{table:tensord2}
\end{center}
\end{table}
\footnotetext{A complementary series  is present in the tensor product $\CC_{\D_1} \otimes\CC_{\D_2}$ of $\SO(1,3)$ if and only if $\Delta_1+\Delta_2>3$.}

\vspace{-0.4cm}

\subsection*{Brief literature  review}

For $\SO(1,2)$  (or its double covering group $\SL(2,\mathbb R)$), the decomposition of the tensor product of two principal series or complementary series representations was first carried out  by Puk\'anszky \cite{Pukan}. The full list of tensor product decompositions involving all UIRs of $\SO(1,2)$ was later solved by \cite{Repka:1978}.

The decomposition of the tensor product  of  principal series and complementary series of  $\SO(1,3)$ (or its double covering group $\SL(2,\mathbb C)$) was studied by Naimark in a series of papers \cite{naimark1959decomposition,naimark1960decomposition,naimark1961decomposition}. In particular, in the last paper \cite{naimark1961decomposition}, he found that  $\CC_{\Delta_1}\otimes\CC_{\Delta_2}$ can contain one complementary series representation $\CC_{\Delta_1+\Delta_2-2}$ when $\Delta_1+\Delta_2>3$.

For  higher dimensional $\SO(1,d+1)$, it was argued that UIRs contained in $\CP_{\Delta_1,s_1}\otimes\CP_{\Delta_2,s_2}$  should also appear in the decomposition of the regular representation of $\SO(1,d+1)$ \cite{Dobrev:1976vr,Dobrev:1977qv}. Based on Hirai's result \cite{10.3792/pja/1195522333, 1966323} on the latter problem, it is inferred that $\CP_{\Delta_1,s_1}\otimes\CP_{\Delta_2,s_2}$ contains only principal series\footnote{The principal series representations are not necessarily the $\CP_{\Delta,s}$ type in the sense that the $\SO(d)$ spin $s$ should be replaced by highest weight vectors with multiple nonzero entries.  }, and when $d$ is odd, also discrete series. 
For the special case of $\SO(1,4)$, the discrete series part of this tensor product was solved by Martin \cite{Martin}.
When $s_1=s_2=0$, the discrete series should also disappear \cite{Dobrev:1976vr,Dobrev:1977qv}. Our analysis of the $\SO(d+1)$ content in section \ref{tensorhigh} can also be used to exclude discrete series when $d$ is sufficiently large. A complete understanding of complementary series in $\CC_{\Delta_1,s_1}\otimes \CC_{\Delta_2,s_2}$ is not known. Some partial results for $s_1=s_2=0$  were derived recently in \cite{https://doi.org/10.48550/arxiv.1402.2950}.
Another open question is to identify what tensor products contain complementary series.

See also~\cite{Higuchi:1986wu,Anninos:2019oka,Joung:2006gj,Basile:2016aen,Anous:2020nxu,Sengor:2022lyv,Sengor:2019mbz} for other works that studied $\SO(1,d+1)$ representation theory motivated by QFT in de Sitter.

\begin{table}[h!]
\begin{center}
\def\arraystretch{1.4}%
\begin{tabular}{ |c|c|c|c|c| }
\hline
 $\otimes$ 
& $\PP_{\D_1,0} $ & $\CC_{\D_1,0}$ & $\CV_{1,0}$ & $\dots$\\ \hline
  $\PP_{\D_2,0}$  
 & $\scriptstyle\sum_s \int_\D \PP_{\D,s}  $ & 
 &  
 & 
 \\
 \hline
   $\CC_{\D_2,0}$   
 & $\scriptstyle \sum_s \int_\D \PP_{\D,s}$ & 
 $\scriptstyle \sum_s \int_\D \PP_{\D,s} \oplus\, \sum_{n,s}\CC_{\D_1+\D_2-d-s-2n,s}$\footnotemark  &  
 & 
 \\
 \hline
  $\CV_{1,0}$   
  & $?$& 
  $?$ & 
  $\scriptstyle  \sum_s \int_\D \PP_{\D,s} \oplus \,\CU_{1,0}$& 
  \\
 \hline 
 $\dots$   
  & $  ?$& 
$  ?$ & 
$ ?$
& 
$  ?$\\
 \hline
\end{tabular}
\caption{Decomposition of the tensor product of some UIRs of  $\SO(1,d+1)$ for $d\ge3$.
We consider  the principal series $\CP_{\Delta,s}$ with dimension $\D \in \frac{d}{2} +i\mathbb{R}_{\ge 0}$ and integer spin $s\ge 0$,
the complementary series $\CC_{\Delta,s}$ with $\frac{d}{2} <\D<d$, and the exceptional series 
 $\CV_{p,0}$ with $p\in \mathbb{N}$ and $\CU_{s,t}$ with $s\in \mathbb{N}$ and $t=0,1,\dots,s-1$, described in section \ref{repreview}.
In section \ref{sec:remarksspin}, we make some comments about tensor products of UIRs with non-zero spin but the current knowledge is very incomplete.
In addition, there are several more UIRs of  $\SO(1,d+1)$ that we do not study in this paper.
Since the tensor product is symmetric we did not fill in the upper triangle of the table to avoid cluttering. The sum $ \sum_s\equiv \oplus_s$ runs over all integer spins $s\ge 0 $ and $\int_\D$ is a sum over all principal series with $\D=\frac{d}{2} + i\lambda$ and $\lambda \in \mathbb{R}_{\ge 0}$.}  
\label{table:tensord3}
\end{center}
\end{table}
\footnotetext{For each pair $(n,s)$ of non-negative integers  such that $\D_1+\D_2-d-s-2n>\frac{d}{2}$, there is complementary series in the tensor product $\CC_{\D_1,0} \otimes \CC_{\D_2,0}$ of $\SO(1,d+1)$.}

\begin{table}[h!]
\begin{center}
\def\arraystretch{1.4}%
\begin{tabular}{ |c|c|c|c| }
\hline
\multirow{1}{*}{\small{ $\SO(2,d+1) \to \SO(1,d+1)$}}
& $\mathcal{R}_{\tilde{\D},0}$ & $\mathcal{R}_{\tilde{\D},\ell}$ & $\dots $\\ \hline
\multirow{1}{*}{ $d=1$}
 & $ \int_\D \PP_\D\oplus \CC_{1-\tilde{\Delta}}$ & 
 $\int_\D \PP_\D\oplus \sum_{k=1}^{|\ell|} \CD_k^{{\rm sign}(\ell)}$ & $-$ \\ 
 \hline
 \multirow{1}{*}{ $d=2$}
 & $ \int_\D \PP_{\D,0} \oplus \CC_{2-\tilde{\Delta}}$ & 
 $ \sum_{|m|\le \ell} \int_\D \PP_{\D,m}$  &$-$ \\ 
 \hline
\multirow{1}{*}{ $d\ge 3$ }
 & $\int_\D \PP_{\D,0}\oplus \CC_{d-\tilde{\Delta},0}$ \footnotemark & 
   $ \sum_{s =0}^ \ell \int_\D \PP_{\D,s} $   \footnotemark & $? $\\
 \hline
\end{tabular}
\caption{Decomposition of UIRs $\mathcal{R}_{\tilde{\D},\ell}$ of  $\SO(2,d+1)$ into UIRs of $\SO(2,d+1)$.
We  consider conformal multiplets  $\mathcal{R}_{\tilde{\D},\ell}$ with a symmetric traceless primary state with integer spin $\ell$ and conformal dimension $\tilde \D$.  
For $d\ge 3$ and non-zero spin the results are incomplete (see section \ref{sec:CFTspin}).  
$\int_\D$ is a sum over all principal series with $\D=\frac{d}{2} + i\lambda$ and $\lambda \in \mathbb{R}_\ge 0$. 
}  
\label{table:CFT}
\end{center}
\end{table}
\footnotetext[6]{A complementary series  is present  if and only if $\tilde{\Delta}<\frac{d}{2}$.}
\footnotetext{This is our conjecture based on the structure of the quadratic Casimir for $s=0,1$ and its numerical diagonalization (see section \ref{sec:CFTspin} for more details). In principle, our numerical tests do not exclude the presence of the exceptional series $\cV$. Nevertheless, we think these are absent (we checked this in section \ref{sec:CFTspin} when the conformal multiplet saturates the unitarity bound).}

\section{A brief review of UIRs of $\SO(1, d+1)$ }\label{repreview}
The $(d+1)$ dimensional de Sitter spacetime is a hypersurface in the ambient space $\mathbb{R}^{1,d+1}$
\begin{align}\label{embdS}
-X_0^2+X_1^2+\cdots+ X_{d+1}^2=1
\end{align}
where we take the de Sitter radius to be 1. The embedding (\ref{embdS}) manifests the  isometry group  $\SO(1,d+1)$ of $\text{dS}_{d+1}$, which is generated by $L_{AB}=-L_{BA}, 0\le A,B\le d+1$ satisfying commutation relations 
\begin{align}\label{defiso}
[L_{AB}, L_{CD}]=\eta_{BC} L_{AD}-\eta_{AC} L_{BD}+\eta_{AD} L_{BC}-\eta_{BD} L_{AC}
\end{align} 
where $\eta_{AB}=\text{diag}(-1,1,\cdots, 1)$ is the metric on $\mathbb{R}^{1,d+1}$. In a unitary representation, $L_{AB}$ are realized as anti-hermitian operators on some Hilbert space.
The isomorphism between $\so(1, d+1)$ and the $d$-dimensional Euclidean conformal algebra is realized as
\begin{align}\label{defconf}
L_{ij}=M_{ij}~, \,\,\,\,\, L_{0, d+1}=D~, \,\,\,\,\, L_{d+1, i}=\frac{1}{2}(P_i+K_i)~, \,\,\,\,\, L_{0, i}=\frac{1}{2}(P_i-K_i)
\end{align}
where $D$ is the dilatation, $P_i$ ($i=1, 2,\cdots d$) are translations, $K_i$ are special conformal transformations and $M_{ij}=-M_{ji}$ are  rotations.
The commutation relations of the conformal algebra following from (\ref{defiso}) and (\ref{defconf}) are
\begin{align}\label{confalg}
&[D, P_i]=P_i~, \,\,\,\,\, [D, K_i]=-K_i~, \,\,\,\,\, [K_i, P_j]=2\delta_{ij}D-2M_{ij}~,\nonumber\\
&[M_{ij}, P_k]=\delta_{jk} P_i-\delta_{ik}P_j~, \,\,\,\,\,[M_{ij}, K_k]=\delta_{jk} K_i-\delta_{ik}K_j~,\nonumber\\
&[M_{ij}, M_{k\ell}]=\delta_{jk} M_{i\ell}-\delta_{ik} M_{j\ell}+\delta_{i\ell} M_{jk}-\delta_{j\ell} M_{ik}~.
\end{align}
The quadratic Casimir of $\SO(1, d+1)$, which commutes with all $L_{AB}$, is chosen to be 
\begin{align}\label{generalcas}
\Cas^{\SO(1, d+1)}&=\frac{1}{2}L_{AB}L^{AB}=D(d-D)+P_i K_i+\frac{1}{2} M_{ij}^2~.
\end{align}
Here $\frac{1}{2}M_{ij}^2\equiv \frac{1}{2} M_{ij}M^{ij}$ is the quadratic Casimir of $\SO(d)$ and it is negative-definite for a unitary representation since $M_{ij}$ are anti-hermitian. For example, for a spin-$s$ representation of $\SO(d)$, it takes the value of $-s(s+d-2)$.

\subsection{Classification of UIRs}
An infinite dimensional representation of $\SO(1, d+1)$ is  fixed by a scaling dimension $\Delta\in\mathbb C$ and a highest-weight vector $\lambda$ of  $\SO(d)$. In the present paper, we only consider $\lambda=(s, 0,\cdots,0)$ labelled by a nonnegative integer $s$ \footnote{When $d=2$, $s$ can be any integers.}, in other words, we will focus on symmetric traceless tensors of $\SO(d)$.  $s$ labels the spin of a field in dS$_{d+1}$. More general $\lambda$ corresponds to fields of mixed symmetry, including form fields, spinors, tensor spinors, etc. See \cite{Basile:2016aen,A_Letsios_2021, Pethybridge_2022, https://doi.org/10.48550/arxiv.2206.09851} for recent discussions on these fields.
Given a representation labelled by $\Delta$ and $s$, the quadratic Casimir is equal to $\Delta(d-\Delta)-s(d+s-2)$. For any $d\ge 3$, there are four types of UIRs apart from the trivial representation \cite{Dobrev:1977qv,Basile:2016aen,Sun:2021thf}:
\begin{itemize}
\item \textbf{Principal series} $\CP_{\Delta, s}$: $\Delta\in\frac{d}{2}+i\mathbb R$ and $s\ge 0$. The restriction of $\CP_{\Delta, s}$ to the maximal compact subgroup $\SO(d+1)$ of $\SO(1, d+1)$ is given by 
\begin{align}\label{CPcomp}
\left. \CP_{\Delta, s}\right|_{\SO(d+1)}=\bigoplus_{n=s}^\infty \bigoplus_{m=0}^s \mY_{n,m}
\end{align}
where $\mY_{n,m}$ denotes a two-row Young diagram with $n$ boxes in the first row and $m$ boxes in the second row \footnote{When $d=3$, $m$ can be negative, corresponding to anti-self-dual tensors of $\SO(4)$. In this case, it is more appropriate to understand the  symbol $\mY_{n,m}$ as a highest-weight vector $(n,m)$.}.
\item \textbf{Complementary series} $\CC_{\Delta, s}$: $0<\Delta<d$ when $s=0$ and $1<\Delta<d-1$ when $s\ge 1$. It has the same $\SO(d+1)$ content as  $\CP_{\Delta, s}$.
\item \textbf{Type \rom{1} exceptional series } $\CV_{p,0}$: $\Delta=d+p-1$ and $s=0$ for $p\ge 1$. The $\SO( d+1)$ content of $\CV_{p, 0}$ only consists of single-row Young diagrams:
\begin{align}\label{CVcomp}
\left. \CV_{p,0}\right|_{\SO(d+1)}=\bigoplus_{n=p}^\infty  \mY_{n}~.
\end{align}
These representations can be roughly thought as analytical continuation of complementary series beyond its unitarity regime, with certain $\SO(d+1)$ UIRs removed to maintain irreducibility and unitarity. The precise meaning of this analytical continuation is clarified  in appendix \ref{matrixcomp}.
\item \textbf{Type \rom{2} exceptional series } $\CU_{s, t}$: $\Delta=d+t-1$ and $s\ge 1$ with $t=0,1,2\cdots, s-1$. The $\SO( d+1)$ content of $\CU_{s, t}$ is  
\begin{align}\label{CUcomp}
\left. \CU_{s, t}\right|_{\SO(d+1)}=\bigoplus_{n=s}^\infty \bigoplus_{m=t+1}^s \mY_{n,m}~.
\end{align}
\end{itemize}
The $[\Delta, s]$ and $[d-\Delta, s]$ representations in principal series and complementary series are actually isomorphic.  Therefore, we only consider $\Delta$ with nonnegative imaginary part in principal series, and $\Delta>\frac{d}{2}$ in complementary series.  We will also use the notation  $\CF_{\Delta,s}$ for both  principal series and complementary series. Whether $\CF_{\Delta, s}$ belongs to principal series or complementary series will be clear once we specify the scaling dimension $\Delta$. $\CF_{\Delta,s}$ describes spin-$s$ massive fields in $\text{dS}_{d+1}$ of mass $m^2=\Delta(d-\Delta)$ when $s=0$ or  $m^2=(\Delta+s-2)(d+s-2-\Delta)$ when $s\ge 1$. $\CU_{s,t}$ describes partially massless gauge fields of spin $s$ and depth $t$ \cite{Deser:1983mm,Brink:2000ag,Deser:2001pe,Deser:2001us,Deser:2001wx,Deser:2001xr,Zinoviev:2001dt,Dolan:2001ih,Hinterbichler:2016fgl}. In particular, $t=s-1$ corresponds to massless gauge fields, e.g. photon, linearized graviton, etc. $\CV_{p,0}$ is expected  to describe scalar fields of mass $m^2=(1-p)(d+p-1)$ with some shift symmetry being gauged \cite{Epstein:2014jaa, Bonifacio:2018zex, Sun:2021thf}. In particular, when $p=1$, the shift symmetry is simply $\phi(x)\to \phi(x)+c$, where $c$ is an arbitrary real number.

\begin{remark}\label{discseries}
When $d=3$, $\CU_{s, t}$ is actually the direct sum of two discrete series representations $\CU^\pm_{s,t}$, which consist of $\SO(4)$ representations $\bigoplus_{n=s}^\infty \bigoplus_{m=t+1}^s \mY_{n,\pm m}$ respectively. Since discrete series is essentially the same as exceptional series for $d=3$, we will only consider the latter in $\text{dS}_4$.  In general, discrete series (which only exists when $d$ is odd) requires an $\SO(d)$ highest-weight vector $\lambda$ with $\frac{d-1}{2}$ nonzero entries. It implies that the $\SO(d+1)$ content of a discrete series with $d\ge 5$ has at least three rows in terms of Young diagram, which is completely different from exceptional series.
\end{remark}

\begin{remark}\label{dS3corr}
When $d=2$, there are only principal series and complementary series up to isomorphism \cite{Dirac:1945cm,10.2307/97833,10.2307/1969129,GelNai47}. The complementary series representations always have $s=0$.  There exists an isomorphism between $\CF_{\Delta, s}$ and $\CF_{\bar\Delta, -s}$, where $s$ is an arbitrary integer labelling the one dimensional representation of $\SO(2)$.
A generic massive spinning field in $\text{dS}_3$ is described by $\CF_{\Delta, s}\oplus \CF_{\Delta,-s}$, which is a UIR of $\text{O}(1,3)$. Gauge fields are described by principal series representations $\CP_{1,s}\cong \CP_{1, -s}$. For example, photons in $\text{dS}_3$ correspond to $\CP_{1,1}$.
\end{remark}

When $d=1$, the infinite dimensional representations of $\SO(1,2)$ are only labelled by the scaling dimension $\Delta$. So the classification of UIRs is different:
\begin{itemize}
\item \textbf{Principal series} $\CP_{\Delta}$: $\Delta\in\frac{1}{2}+i\mathbb R$. Its restriction to $\SO(2)$ yields 
\begin{align}\label{CPcomp2}
\left. \CP_{\Delta}\right|_{\SO(2)}=\bigoplus_{n\in\mathbb Z}(n)
\end{align}
where $(n)$ denotes the (one-dimensional) spin $n$ representation of $\SO(2)$.
\item \textbf{Complementary series} $\CC_{\Delta}$: $0<\Delta<1$. It has the same $\SO(2)$ content as  $\CP_{\Delta}$.
\item \textbf{Lowest-weight discrete series $\CD^+_p$}: $\Delta=p\in\mathbb Z_{>0}$. 
\item \textbf{Highest-weight discrete series $\CD^-_p$}: $\Delta=-p\in\mathbb Z_{<0}$.  
\end{itemize}
The $\SO(2)$ spectrum of the discrete series is
\begin{align}\label{CDcomp2}
\left. \CD^+_p\right|_{\SO(2)}=\bigoplus_{n\ge p}(n) ~,\qquad \qquad
\left. \CD^-_p\right|_{\SO(2)}=\bigoplus_{n\le -p}(n)~.
\end{align}
Again, the principal series and complementary series of $\SO(1,2)$ have the $\Delta\leftrightarrow 1-\Delta$ symmetry.
We will use $\CD_p$ to denote the unitary reducible representation $\CD_p^+\oplus \CD_p^-$. A massless scalar in dS$_2$, with the constant mode being removed, is described by $\CD_1$. In particular, its right-moving modes (along the global circle of dS$_2$) correspond to $\CD^+_1$ and the left-moving modes correspond to $\CD_1^-$. In general, the $\CD^+_p$ ($\CD^-_p$) describes the right (left) movers of a scalar field with mass $m^2=p(1-p)$ in dS$_2$.

\subsection{Harish-Chandra characters}
For all the UIRs reviewed above, one can define a group character (known as Harish-Chandra character or global character):
\begin{align}
\Theta_R(g)\equiv \tr_\CH (g), \,\,\,\,\, g\in\SO(1,d+1)
\end{align}
where $\CH$ denotes the infinite dimensional Hilbert space of a given UIR $R$. Such a character  exists as a distribution on $\SO(1,d+1)$. We will take $g$ to be $e^{tD}$ 
for $\SO(1,2)$,  and $e^{t D} x_1^{J_{1}}\cdots x_r^{J_r}$ for $\SO(1,d+1)$, where $t\in\mathbb R$, $r=\floor*{\frac{d}{2}}$ is the rank of $\SO(d)$, $x_j\in \text{U}(1)$ and $J_j=i M_{2j-1,2j}$ are Cartan generators of $\SO(d)$. The $t$ dependence is always via $q\equiv e^{-|t|}$, except for the spinning principal series of $\SO(1,3)$, which will be clarified in remark \ref{jlj} below.  In the $\SO(1,2)$ case, the Harish-Chandra characters are \cite{Knapp,Sun:2021thf}
\begin{itemize}
\item \textbf{Principal and complementary series}: 
\begin{align}
\Theta_\Delta(q)=\frac{q^\Delta+q^{\bar\Delta}}{1-q}~, \,\,\,\,\, \bar\Delta\equiv 1-\Delta~.
\end{align}
\item \textbf{Highest and lowest weight Discrete series}:
\begin{align}
\Theta^\pm_p(q)=\frac{q^p}{1-q}~.
\end{align} 
\end{itemize}

For generic $\SO(1,d+1)$, the explicit expression of Harish-Chandra characters depends on the parity of $d$ \cite{Basile:2016aen,Sun:2021thf}:
\begin{itemize}
\item \textbf{Principal and complementary series}: 
\begin{align}\label{kj;}
\Theta_{\Delta,s}(q,\bm x)=\chi_{\mY_s}^{\SO(d)}(\bm x) \frac{q^\Delta+q^{\bar\Delta}}{P_d(q,\bm x)}, \,\,\,\,\, \bar\Delta\equiv d-\Delta
\end{align}
where $\chi_{\mY_s}^{\SO(d)}(\bm x)$ is the $\SO(d)$ Weyl character \footnote{In this paper, we will use $\Theta$ for Harish-Chandra characters of noncompact Lie groups and use $\chi$ for Weyl characters of compact Lie groups.}  corresponding to the spin-$s$ representation $\mY_s$ and 
\begin{align}\label{Pddef}
P_d(q, \bm x)=\prod_{i=1}^r(1-x_i q)(1-x_i^{-1}q)\times\begin{cases} 1, & d=2r\\ 1-q, & d=2r+1\end{cases}
\end{align}
The presence of both $q^\Delta$ and $q^{\bar\Delta}$ in (\ref{kj;}) manifests the $\Delta\leftrightarrow d-\Delta$ symmetry.
\item  \textbf{Type \rom{1} exceptional series $\CV_{p,0}$}: 
\begin{align}\label{CVchar}
\Theta_{\CV_{p,0}}(q,\bm x)=\frac{q^{1-p}+q^{d+p-1}}{P_d(q, \bm x)}-\chi^{\SO(d+2)}_{\mY_{p-1}}(q,\bm x)
\end{align}
The small $q$ expansion of $\Theta_{\CV_{p,0}}(q,\bm x)$ starts with $\chi_{\mY_p}^{\SO(d)}(\bm x) q$. For example, when $p=1$ and $d=3$, eq. (\ref{CVchar}) becomes
\begin{align}\label{CVcharspec}
d=3: \,\, \Theta_{\CV_{1,0}}(q,x)=\frac{2\, q^3}{(1-q)(1-x q)(1-x^{-1}q)}+\frac{\chi^{\SO(3)}_{\mY_1}(x)\,q}{(1-x q)(1-x^{-1}q)}~.
\end{align}
\item  \textbf{Type \rom{2} exceptional series $\CU_{s,t}$}:
\begin{align}\label{CUchar}
\Theta_{\CU_{s,t}}(q,\bm x)=\chi_{\mY_s}^{\SO(d)}(\bm x)\frac{q^{1-t}+q^{d+t-1}}{P_d(q, \bm x)}-\chi_{\mY_t}^{\SO(d)}(\bm x)\frac{q^{1-s}+q^{d+s-1}}{P_d(q, \bm x)}+\chi^{\SO(d+2)}_{\mY_{s-1,t}}(q,\bm x)
\end{align}
where $\chi_{\mY_{s-1,t}}^{\SO(d+2)}(q,\bm x)$ is the $\SO(d+2)$ Weyl character  corresponding to the two-row Young diagram  $\mY_{s-1,t}$. 
When $d=3$, eq. (\ref{CUchar}) reduces to 
\begin{align}\label{CU3char}
d=3: \,\,\,\,\, \Theta_{\CU_{s,t}}(q,x)= 2\left(\chi_{\mY_s}^{\SO(3)}(x)\frac{q^{t+2}}{P_3(q,x)} -\chi_{\mY_t}^{\SO(3)}(x)\frac{q^{s+2}}{P_3(q,x)} \right)~.
\end{align}
The overall factor $2$ is related to the reducibility of $\CU_{s,t}$ in the $d=3$ case, discussed in remark \ref{discseries}. When $d\ge 4$, the small $q$ expansion of $\Theta_{\CU_{s,t}}(q,\bm x)$ has a universal leading term \footnote{When $d=4$, the character $\chi_{\mY_{s,t+1}}^{\SO(d)}(\bm x) $ in eq.(\ref{CUleading}) should be understood as the sum of $\chi_{\mY_{s,t+1}}^{\SO(4)}(\bm x) $ and $\chi_{\mY_{s,-(t+1)}}^{\SO(4)}(\bm x) $. }
\begin{align}\label{CUleading}
d\ge 4: \,\,\,\,\, \Theta_{\CU_{s,t}}(q,\bm x)=\chi_{\mY_{s,t+1}}^{\SO(d)}(\bm x) q^2+\CO(q^3)~.
\end{align}
whose physical origin is explained in \cite{Sun:2020sgn}.
\end{itemize}

\begin{remark}\label{jlj}
For spinning principal series  of $\SO(1,3)$, the corresponding Harish-Chandra character is given by
\begin{align}\label{2dc}
d=2:\,\,\, \Theta_{\Delta,s}(q, x)=\frac{x^s q^\Delta+x^{-s}q^{\bar\Delta}}{(1-x q)(1-x^{-1}q)}, \,\,\,\,\, q=e^{-t}
\end{align}
which manifests the isomorphism between $\CF_{\Delta, s}$ and $\CF_{\bar\Delta, -s}$. Because $q$ is equal to $e^{-t}$ rather than $e^{-|t|}$ here, the character does not have $t\to -t$ symmetry. Instead, it satisfies $\Theta_{\Delta,s}(q^{-1}, x^{-1})=\Theta_{\Delta,s}(q, x)$.
\end{remark}

\begin{remark}
There exist more general principal series representations $\CP_{\Delta,\mY}$, labelled by  a complex scaling dimension $\Delta\in\frac{d}{2}+i\mathbb R$ and an arbitrary UIR of $\SO(d)$ corresponding to the Young diagram $\mY$. The Harish-Chandra character of $\CP_{\Delta,\mY}$ is given by 
\begin{align}\label{mixedsym}
\Theta_{\CP_{\Delta,\mY}}(q,\bm x)=\chi_{\mY}^{\SO(d)}(\bm x) \frac{q^\Delta+q^{\bar\Delta}}{P_d(q,\bm x)}~.
\end{align}
\end{remark}

\section{Tensor products in $\SO(1,2)$}\label{tensorlow}
The tensor product decomposition of $\SO(1,2)$ (or its covering group $\SL(2,\mathbb R)$) UIRs has been solved  by Puk\'anszky \cite{Pukan} and Repka \cite{Repka:1978}. For instance, the tensor product of two principal series representations contain all discrete series representations and the full principal series.
In this section, we will revisit this problem with the aid of Harish-Chandra characters, focusing on defining and computing a proper density of principal series representations in any tensor product. As a byproduct, we will also show that such a density also contains information of complementary series after a simple analytic continuation.

\subsection{Some details regarding $\SO(1,2)$ UIRs}
\label{introSO12}
A particularly useful basis of $\so(1,2)$ is given by 
\begin{align}
L_0=-\frac{i}{2}(P+K)~, \,\,\,\,\ L_\pm=-\frac{i}{2}(P-K)\mp D
\end{align}
and satisfies the commutation relations $[L_0,L_\pm]=\pm L_{\pm}, [L_-,L_+]=2L_0$. In this basis, the $\SO(1,2)$ Casimir can be expressed as
\begin{align}
\Cas^{\SO(1,2)}=L_0(1-L_0)+L_+L_-~.
\end{align}
$L_0$ is the generator of the $\SO(2)$ subgroup, corresponding to rotations along the global circle of $\text{dS}_2$. Since we only consider representations of $\SO(1,2)$, its eigenvalues are integers\footnote{The eigenvalues can be half-integers if we consider the covering group $\SL(2,\mathbb R)$ which can describe spinors in $\text{dS}_2$.}. Denote the eigenstates of $L_0$ by $|n)$. In principal or complementary series of scaling dimension $\Delta$, the label $n$ takes the value of all integers. Following the conventions in \cite{Sun:2021thf},  the action of  $\so(1,2)$ on $|n)$ is 
\begin{align}\label{Lact}
L_0|n)=n|n), \,\,\,\,\, L_\pm |n)=(n\pm \Delta)|n\pm 1)~.
\end{align}
The norm of $|n)$ compatible with eq.~\reef{Lact} is given by 
\begin{align}\label{nnorm}
&\text{Principal series}\,\, \CP_\Delta:\,\,\quad \quad \quad (n|n)=1\nonumber\\
&\text{Complementary series}\,\, \CC_\Delta:\,\,\, (n|n)=\frac{\Gamma(n+\bar\Delta)}{\Gamma(n+\Delta)}=\frac{\Gamma(-n+\bar\Delta)}{\Gamma(-n+\Delta)}~.
\end{align}
In lowest-weight discrete series, e.g. $\CD^+_p$, the label $n$ has a lower bound $p$, corresponding to a lowest-weight state $|p)$ that is annihilated by $L_-$. In highest-weight discrete series, e.g. $\CD^-_p$, the label $n$ has an upper  bound $-p$, corresponding to a highest-weight state $|-p)$ that is annihilated by $L_+$. In  both $\CD^\pm_p$, we also have the action~(\ref{Lact}), with $\Delta$ being replaced by $p$. The corresponding norm of $|n)$ is 
\begin{align}
\text{Discrete series}\,\, \CD^\pm_p: \quad (n|n)=\frac{\Gamma(\pm n+1-p)}{\Gamma(\pm  n+p)}~.
\end{align}

\subsection{$\CF_{\Delta_1}\otimes \CF_{\Delta_2}$}\label{CF1tCF2}
In general, the tensor product $\CF_{\Delta_1}\otimes \CF_{\Delta_2}$ 
contains both discrete and continuous series UIRs.
Since these two types of representations are very different in nature, we  will discuss them separately. 

\subsubsection{The discrete part}

Let $|n,m)\equiv |n)\otimes |m)$ be a basis of the tensor product of  two continuous series $\CF_{\Delta_1}\otimes\CF_{\Delta_2}$  on which the action of $\so(1,2)$ follows from eq.~\reef{Lact} 
\begin{align}\label{act1}
L_\pm |n,m)=(n\pm\Delta_1)|n\pm1,m)+(m\pm\Delta_2)|n,m\pm1), \,\,\,\,\, L_0|n,m)=(n+m)|n,m)~.
\end{align}
In order to identify a discrete series representation in $\CF_{\Delta_1}\otimes\CF_{\Delta_2}$, say $\CD^+_k$, it suffices to build the corresponding lowest weight state $|k)_k$, which is characterized by the following  first order relations (we also include the trivial representation by allowing $k$ to be 0)
\begin{align}
L_-|k)_k=0, \,\,\,\,\, L_0 |k)_k=k |k)_k~.
\end{align}
Being an eigenstate of $L_0$, the state $|k)_k$ should take the following from
\begin{align}\label{g2}
|k)_k=\sum_{n\in\mathbb Z}a_n|n,k-n)
\end{align}
where $a_n$ are coefficients to be fixed. Applying the lowest-weight condition to~\reef{g2}, we obtain a recurrence relation for $a_n$
\begin{align}
a_n(n-\Delta_1)+a_{n-1}(k-n+\bar\Delta_2)=0
\end{align}
which can be solved up to an overall normalization factor, denoted by $c$
\begin{align}\label{g2d}
a_n= c\frac{\Gamma(n+\Delta_2-k)}{\Gamma(n+\bar\Delta_1)}~.
\end{align}
With the lowest-weight state $|k)_k$ being constructed, the existence of $\CD_k^+$ in $\CF_{\Delta_1}\otimes\CF_{\Delta_2}$ is equivalent to the normalizability of $|k)_k$. It is straightforward to check that the explicit expression of $_k(k|k)_k$ is independent of whether $\CF_{\Delta_i}$ is a principal series or complementary series:
\begin{align}\label{kkp}
_k( k|k)_k=|c|^2 
\sum_{n\in\mathbb Z} \frac{\Gamma(n+\Delta_2-k)\Gamma(n+\bar\Delta_2-k)}{\Gamma(n+\Delta_1)\Gamma(n+\bar\Delta_1)}~.
\end{align}
The large $n$ behavior of  the summands in eq.~\reef{kkp} is $n^{-2k}$.
Therefore, as long as $k\in\mathbb Z_+$, the sum is convergent and hence $|k)_k$ is normalizable. When $k=0$, it is non-normalizable, which implies the absence of the trivial representation.
A similar result holds for the highest-weight representations $D^-_k$. Altogether, the tensor product $\CF_{\Delta_1}\otimes\CF_{\Delta_2}$ contains all the discrete series representations and each has multiplicity one, since the coefficients $\{a_n\}$ are unique (up to an overall phase) once we fix the normalization. The trivial representation does not appear in this tensor product.

\vspace{10pt}

\subsubsection{The continuous part}\label{sec:1dPxPCon}
In addition to the discrete series, $\CF_{\Delta_1}\otimes\CF_{\Delta_2}$ is also known to contain all principal series representations, and in some cases, one complementary series representation. We aim to identify a proper density of these representations. Presumably, this density encodes conditions for the appearance of complementary series representations, given that complementary series can be thought as an analytical continuation of principal series, at least at the level of Harish-Chandra characters.

\,

\noindent{}\textbf{Character analysis}

In the representation theory of compact groups, the Weyl character is a powerful tool to compute multiplicities because the tensor decomposition $R_1\otimes R_2=\oplus_a \left(R_a^{\oplus n_a}\right)$ of some  compact Lie group $G$ is equivalent to the character relation $\chi^G_{R_1}\chi^G_{R_2}=\sum_a n_a \chi^G_{R_a}$, where $n_a$ denotes the multiplicity of the UIR $R_a$.  In particular, it shows what representations would be present in the tensor products if their $n_a\neq 0$.  

Consider a simple example with $G=\SO(3)$, and $R_1, R_2$ being spin 1 and spin 2 representations respectively.  The multiplicity $n_\ell$ of each spin $\ell$ representation in $R_1\otimes R_2$ should satisfy 
\begin{align}\label{chch}
\chi_1(\theta)\chi_2(\theta)=\sum_{\ell\in\mathbb N} n_\ell \chi_\ell(\theta)~,\qquad \chi_\ell(\theta)=\frac{\sin(\ell+\frac{1}{2})\theta}{\sin\frac{\theta}{2}}\,,
\end{align}
where $ \chi_\ell(\theta)$ is the character of spin $\ell$ representation.
Define an inner product $(\chi_\ell, \chi_{\ell'})\equiv \int_0^{2\pi}\,d\theta \sin^2\frac{\theta}{2}\chi_\ell(\theta)\chi_{\ell'}(\theta)$, then it is straightforward to check that  $\chi_\ell$ are orthogonal to each other with respect to this inner product, i.e.~$(\chi_\ell, \chi_{\ell'})=\pi\delta_{\ell\ell'}$, which leads to an integral representation of the multiplicity $n_\ell$ 
\begin{align}\label{nleq}
n_\ell &= \frac{1}{\pi}\int_0^{2\pi}\, d\theta\, \frac{\sin(\ell+\frac{1}{2})\theta}{\sin\frac{\theta}{2}}\sin\left(\frac{3}{2}\,\theta\right)\sin\left(\frac{5}{2}\,\theta\right)\nonumber\\
&=\frac{1}{2\pi}\int_0^{2\pi}\,\left(1+2\cos\theta+\cdots+2\cos(\ell\theta)\right)(\cos\theta-\cos(4\theta))~.
\end{align}
Based on this integral representation, it is clear that $n_\ell$ equals to 1 when $\ell=1,2,3$, and vanishes otherwise. Therefore, we can conclude the following tensor product decomposition of $\SO(3)$:
\begin{align}
[3]\otimes [5]=[3]\oplus [5]\oplus [7]
\end{align}
where $[n]$ denotes the spin $\frac{n-1}{2}$ representation of $\SO(3)$.
For the tensor product of any spin $m$ and spin $n$ representations, it amounts to replacing  $\cos\theta-\cos(4\theta)$ in eq.~(\ref{nleq}) by $\cos(m-n)\theta-\cos(m+n+1)\theta$, which then leads to $n_\ell=1$ when $|m-n|\le \ell\le m+n$, and vanishes otherwise.

Now let us apply this idea to $G=\SO(1,2)$, with $R_j=\CP_{\Delta_j}, \Delta_j=\frac{1}{2}+i\mu_j$ belonging to principal series. Since $\CP_{\Delta_1}\otimes\CP_{\Delta_2}$ only includes principal series and discrete series, we expect
\begin{align}\label{khlk}
\Theta_{\Delta_1}(q)\Theta_{\Delta_2}(q)=\int_0^\infty\,d\lambda\, \CK(\lambda)\,\Theta_{\frac{1}{2}+i\lambda}(q)+\sum_{k\ge 1,\pm}\Theta^\pm_k(q)~,
\end{align}
along the lines of eq.~\reef{chch}, where $\CK(\lambda)$ can be roughly thought as a density of $\CP_{\frac{1}{2}+i\lambda}$. Plugging in the explicit expression of these characters given in section \ref{repreview} with $q=e^{-|t|}$, we find that eq.~\reef{khlk} effectively fixes the Fourier transformation of $\CK(\lambda)$
\begin{align}\label{eq:mother of decomposition}
\int_{\mathbb R}\,d\lambda \,\CK(\lambda) e^{i\lambda t}=\frac{\cos(\mu_1+\mu_2)t+\cos(\mu_1-\mu_2)t-1}{|\sinh\frac{t}{2}|}
\end{align}
where we have implicitly extended $\CK(\lambda)$ to an even function on the whole real line, which is compatible with the shadow symmetry of principal series. Unfortunately, the singularity of $\frac{\cos(\mu_1+\mu_2)t+\cos(\mu_1-\mu_2)t-1}{|\sinh\frac{t}{2}|}$ at $t=0$ implies that $\CK(\lambda)$ cannot be obtained by an inverse Fourier transformation. This is consistent with the fact, which we shall show later both analytically and numerically, that the tensor product $\CP_{\Delta_1}\otimes \CP_{\Delta_2}$ has a continuous spectrum of principal series.  In other words, given an arbitrary interval $[a,b]\subset \mathbb R_{\ge 0}$, there exist infinite number of principal series representations $\CP_{\frac{1}{2}+i\lambda}$ with $\lambda\in[a,b]$, contained in $\CP_{\Delta_1}\otimes \CP_{\Delta_2}$, and hence the density of principal series blows up \cite{Stanford_2017}. 

In order to get a finite $\CK(\lambda)$, we may consider a regularization scheme for the inverse Fourier transformation. A simple regularization scheme is to introduce a hard cutoff $t>\epsilon>0$: 
\small
\begin{align}\label{eq:epsilon reg}
\CK_\epsilon(\lambda)&=\int_\epsilon^\infty \frac{dt}{\pi} \cos (\lambda t)\frac{\cos\left(\mu_1+\mu_2\right)t +\cos\left(\mu_1-\mu_2\right)t-1}{|\sinh \frac{t}{2}|}\nonumber\\
&=- \frac{2}{\pi} \log \left(e^{\gamma_E} \, \epsilon \right) + \frac{1}{\pi} \sum_\pm\psi\left(\half\pm i\lambda\right) - \frac{1}{2\pi} \sum_{\pm,\pm, \pm}\psi\left(\half\pm i\lambda \pm i \mu_1 \pm i \mu_2\right)+\CO(\epsilon)
\end{align}
\normalsize
We define the hard-cutoff renormalized kernel as the finite part of this expression:
\be\label{eq:hard cutoff}
\CK_{\text{hc}}(\lambda)=  \frac{1}{\pi} \sum_\pm\psi\left(\half\pm i\lambda\right) 
- \frac{1}{2\pi} \sum_{\pm, \pm, \pm}\psi\left(\half\pm i\lambda \pm i \mu_1 \pm i \mu_2\right)
\ee
where the index in $\CK_{\text{hc}} $ is to label the hard-cutoff renormalization. Note that the $\mu_i$-independent term is from the discrete series contribution and the $\mu_i$-dependent term is from the left hand side of equation~\reef{khlk}.

In fact, this renormalized density should be valid more generally. To see that one thinks of \eqref{eq:mother of decomposition} as an equality between distributions.\footnote{We thank Petr Kravchuk for enlightening discussions about this issue.} In other words, it becomes a true equality  if we integrate both sides against a smooth test function,
\begin{align}\label{eq:Kintegratedf(t)}
\int dt f(t) \int_{\mathbb R}\,d\lambda \,\CK(\lambda) e^{i\lambda t}=\int dt f(t) \frac{\cos(\mu_1+\mu_2)t+\cos(\mu_1-\mu_2)t-1}{|\sinh\frac{t}{2}|}\,,
\end{align}
where   $f(t)$ vanishes at $t=0$ (and is polynomially bounded when $t\to \pm \infty$). Notice that a constant shift $\CK(\lambda) \to \CK(\lambda)+const$ drops out of this equation because $f(0)=0$.
This is the only ambiguity in the distribution $\CK(\lambda)$. Therefore, we expect that different renormalization schemes lead to 
$\CK_{\text{hc}}(\lambda) $ up to a constant shift.

Another possible regularization scheme we consider is a Pauli-Villars type regularization in the same spirit as \cite{Anninos:2020hfj}, where it is argued that the density of single particle states $\rho(\omega)$ of a scalar field $\phi$, say in $\text{dS}_2$, of scaling dimension $\Delta$, can be computed as the Fourier transformation of the Harish-Chandra character $\Theta_{\Delta}(t)=\frac{e^{-\Delta t}+e^{-\bar\Delta t}}{1-e^{-|t|}}$. Due to the singularity at $t=0$, the density of single particle states suffers from a divergence. Then \cite{Anninos:2020hfj} introduced a Pauli-Villars regularization, which effectively replaces $\Theta_{\Delta}$ by 
\begin{align}
\Theta^\Lambda_\Delta(t)\equiv \frac{e^{-\Delta t}+e^{-\bar\Delta t}}{1-e^{-|t|}}-\frac{e^{-\Delta_\Lambda t}+e^{-\bar\Delta_\Lambda t}}{1-e^{-|t|}}
\end{align}
where $\Delta_\Lambda=\frac{1}{2}+i\Lambda$ for some large scale $\Lambda$. The second character in $\Theta^\Lambda_\Delta$ corresponds to a very heavy particle $\phi_\Lambda$ with a mass $(\frac{1}{4}+\Lambda^2)^{\half}\sim\Lambda$. Then a regularized density $\rho_\Lambda(\omega)$, defined as the Fourier transformation of $\Theta^\Lambda_\Delta$ for energy $\omega$ much smaller than the cutoff scale $\Lambda$, can also be thought as  the relative density of single particle states between $\phi$ and the very heavy field $\phi_\Lambda$ at low energy. A renormalized density is identified as the finite part of  $\rho_\Lambda(\omega)$ in the large $\Lambda$ limit. In our setup, we implement the Pauli-Villars regularization by introducing two more principal series representations $\CP_{\Delta_3}$ and $\CP_{\Delta_4}$, where $\Delta_3=\frac{1}{2}+i\mu_3$ and $\Delta_4=\frac{1}{2}+i\mu_4$, with $\mu_3, \mu_4$ being large.  Then a relative density $\CK_{\text{rel}}(\lambda)$ of principal series between $\CP_{\Delta_1}\otimes\CP_{\Delta_2}$ and $\CP_{\Delta_3}\otimes\CP_{\Delta_4}$ should satisfy
\begin{align}\label{khlks1}
\Theta_{\Delta_1}(q)\Theta_{\Delta_2}(q)-\Theta_{\Delta_3}(q)\Theta_{\Delta_4}(q)=\int_0^\infty\,d\lambda\, \CK_{\rm rel}(\lambda)\,\Theta_{\frac{1}{2}+i\lambda}(q)
\end{align}
because the discrete parts of the two tensor products cancel out. In eq.~\reef{khlks1}, the relative density $\CK_{\rm rel}(\lambda)$ is well-defined and can be easily evaluated as an inverse Fourier transformation \footnote{The inverse Fourier transformation is equivalent to saying that principal series characters are $\delta$ function normalizable with respect to  the inner product $(\Theta_{\Delta_1},\Theta_{\Delta_2})\equiv \int_{0}^\infty dt\,\sinh^2\left(\frac{t}{2}\right)\Theta_{\Delta_1}(e^{-t})\Theta_{\Delta_2}(e^{-t})$, which is the reminiscence of the inner product we have introduced for $\SO(3)$ characters. However, there is an extra subtlety in the $\SO(1,2)$ case because principal series characters are not orthogonal to discrete series characters with respect to this inner product. So the character relation itself along the lines of~(\ref{khlk}) is not sufficient for deriving the multiplicities of discrete series representations.} by using eq.~\reef{CIresult}
\begin{align}\label{CKrelf}
\CK_{\rm rel}(\lambda)&=\frac{1}{2\pi}\sum_{\pm,\pm,\pm}\int_0^\infty dt\, \left(\frac{e^{-(\frac{1}{2}\pm i\mu_1\pm i\mu_2\pm i\lambda)t}}{1-e^{-t}}-\frac{e^{-(\frac{1}{2}\pm i\mu_3\pm i\mu_4\pm i\lambda)t}}{1-e^{-t}}\right)\nonumber\\
&=\frac{1}{2\pi}\sum_{\pm,\pm,\pm}\left(\psi\left(\frac{1}{2}\pm i\mu_3\pm i\mu_4\pm i\lambda\right)-\psi\left(\frac{1}{2}\pm i\mu_1\pm i\mu_2\pm i\lambda\right)\right)~.
\end{align}
Taking $\mu_3=\mu_4=\Lambda$ very large, we get for $\lambda\ll\Lambda$:
\small
\begin{align}\label{11L}
\mu_3=\mu_4=\Lambda:\,\,\,\CK_{\text{PV,1}}(\lambda)
& \limu{\Lambda \to \infty} \frac{2}{\pi}\log(2\Lambda)\!+\!\frac{1}{\pi}\sum_{\pm}\psi\left(\frac{1}{2}\pm i\lambda\right)\!-\!\frac{1}{2\pi}\sum_{\pm,\pm,\pm}\psi\left(\frac{1}{2}\!\pm \!i\mu_1\!\pm\! i\mu_2\!\pm\! i\lambda\right)
\end{align}
\normalsize
which has the same finite part as~\reef{eq:hard cutoff}. 
This is not surprising because, in the limit $\mu_3=\mu_4=\Lambda\to \infty $,  \eqref{khlks1} turns into \eqref{khlk} up to rapidly oscillating terms that integrate to zero against any smooth test function $f(t)$.

On the other hand, taking $\mu_3=2\Lambda$ and $\mu_4=\Lambda$ very large, we get instead 
\begin{align}\label{21L}
\mu_3=2\mu_4=2\Lambda: \,\,\,\,\CK_{\text{PV,2}}(\lambda)
\limu{\Lambda \to \infty} \frac{2}{\pi}\log(3\Lambda^2)-\frac{1}{2\pi}\sum_{\pm,\pm,\pm}\psi\left(\frac{1}{2}\pm i\mu_1\pm i\mu_2\pm i\lambda\right)~.
\end{align}
Comparing eq.~\reef{eq:epsilon reg} and eq.~\reef{11L} with eq.~\reef{21L}, we conclude that one needs to be careful with the choice of regularisation scheme. 
Nonetheless, the relative density  $\CK_{\rm rel}(\lambda)$ always makes sense as long as we keep $\mu_3$ and $\mu_4$ finite. Therefore,  we will mostly use the notion of relative density henceforth.

At this point, $\CK_{\rm rel}(\lambda)$ has been derived by using characters and is called a relative density simply because of intuitions from representation theory of compact Lie groups. We will justify the name ``relative density'' shortly by reconstructing   $\CK_{\rm rel}(\lambda)$ numerically as the relative density of eigenvalues of two large but finite matrices, with each eigenvalue being identified as a UIR of $\SO(1,2)$.
Before describing the numerical approach, we  digress a little bit and discuss what we can learn about complementary series in $\CC_{\Delta_1}\otimes\CC_{\Delta_2}, \Delta_j=\frac{1}{2}+\mu_j$ from the relative density $\CK_{\rm rel}(\lambda)$. Treating complementary series as a direct analytical continuation of principal series, we expect eq.~\reef{CKrelf} to hold for $\CC_{\Delta_1}\otimes\CC_{\Delta_2}$ upon the replacement $i\mu_j\to \mu_j,  j=1,2$. However, this naive guess breaks down manifestly when $\mu_1+\mu_2>\frac{1}{2}$ because $\frac{e^{-(\frac{1}{2}-\mu_1-\mu_2\pm i\lambda)t}}{1-e^{-t}}$ would blow up as $t\to \infty$ or equivalently $q\to 0$. This phenomenon signals  the appearance of a complementary series representation $\CC_{\mu_1+\mu_2}$ in eq.~\reef{khlks1}, which potentially should cancel the exponentially blowing up behavior at large $t$. To justify this argument from a different viewpoint, notice that although $\CK_{\rm rel}(\lambda)$ does not make sense as an inverse Fourier transformation when $\mu_1+\mu_2>\frac{1}{2}$, its  $\psi$ function realization clearly admits  well-defined analytical continuation in this region, i.e. 
\small
\begin{align}\label{hkljh}
\CC_{\Delta_1}\otimes\CC_{\Delta_2}:\,\,\,\CK_{\rm rel}(\lambda)=\frac{1}{2\pi}\sum_{\pm,\pm,\pm}\left(\psi\left(\frac{1}{2}\pm i\mu_3\pm i\mu_4\pm i\lambda\right)-\psi\left(\frac{1}{2}\pm \mu_1\pm \mu_2\pm i\lambda\right)\right)~.
\end{align}
\normalsize
Then it is natural to assume that $\CK_{\rm rel}(\lambda)$ given by eq.~\reef{hkljh} correctly captures the relative density of principal series between $\CC_{\Delta_1}\otimes\CC_{\Delta_2}$ and $\CP_{\Delta_3}\otimes\CP_{\Delta_4}$, no matter whether $\mu_1+\mu_2$ is larger than $\frac{1}{2}$ or not. In both cases, evaluating the integral $\int_0^\infty d\lambda\,\CK_{\rm rel}(\lambda)\Theta_{\frac{1}{2}+i\lambda}$ using the general formula~\reef{CJthired}, we find 
\small
\begin{align}\label{CKTint}
\int_0^\infty d\lambda\,\CK_{\rm rel}(\lambda)\Theta_{\frac{1}{2}+i\lambda}(q)=\begin{cases}\Theta_{\Delta_1}(q)\Theta_{\Delta_2}(q)-\Theta_{\Delta_3}(q)\Theta_{\Delta_4}(q),&0\le\mu_1+\mu_2<\frac{1}{2} \\ \Theta_{\Delta_1}(q)\Theta_{\Delta_2}(q)-\Theta_{\Delta_3}(q)\Theta_{\Delta_4}(q)-\Theta_{\mu_1+\mu_2}(q), &\frac{1}{2}<\mu_1+\mu_2<1\end{cases}
\end{align}
\normalsize
which implies that in the tensor product $\CC_{\Delta_1}\otimes\CC_{\Delta_2}$, when $\mu_1+\mu_2$ crosses the value $\frac{1}{2}$ from below, a complementary series representation of $\Delta=\mu_1+\mu_2$ appears. Altogether, based on the fact that $\CP_{\Delta_3}\otimes \CP_{\Delta_4}$ does not contain any complementary series representations, we can conclude that $\CC_{\Delta_1}\otimes \CC_{\Delta_2}$ contains one complementary series representation, i.e. $\CC_{\mu_1+\mu_2}$ when $\frac{1}{2}<\mu_1+\mu_2<1$, and zero such representation otherwise. This result is consistent with \cite{Pukan,Repka:1978} and will be confirmed numerically. At the special point $\mu_1+\mu_2=\frac{1}{2}$, the integral $\int_0^\infty d\lambda\,\CK_{\rm rel}(\lambda)\Theta_{\frac{1}{2}+i\lambda}$ can be computed by using eq. (\ref{a=0J})
\small
\begin{align}
\mu_1+\mu_2=\frac{1}{2}: \,\,\,\, \int_0^\infty d\lambda\,\CK_{\rm rel}(\lambda)\Theta_{\frac{1}{2}+i\lambda}(q)=\Theta_{\Delta_1}(q)\Theta_{\Delta_2}(q)-\Theta_{\Delta_3}(q)\Theta_{\Delta_4}(q)-\frac{1}{2}\Theta_{\frac{1}{2}}(q).
\end{align}
\normalsize
where $\Theta_{\frac{1}{2}}(q)$ is the character of $\CP_{\frac{1}{2}}$. It suggests that in this case there is a $\delta$ function at $\lambda=0$ on top of the the smooth distribution $\CK_{\rm rel}(\lambda)$.

\begin{remark}\label{polecrossing}
From a contour integral point of view, the extra term $\Theta_{\mu_1+\mu_2}$ in eq.~\reef{CKTint} appears because certain poles of $\psi\left(\frac{1}{2}\!-\!\mu_1\!-\!\mu_2\!\pm\! i\lambda\right)$ cross the integration contour along the real line in the $\lambda$ plane, when we vary the value of $\mu_1+\mu_2$ from 0 and 1. We will see lots of times in this paper that whenever such pole crossing happens some complementary series representations pop out. So the pole crossing phenomenon can be thought as the smoking gun of complementary series.
\end{remark}

\vspace{10pt}

\noindent{}\textbf{Numerical check}

Apart from characters, Casimir operators are also  useful tools to study tensor product decomposition. More generally, 
identifying $G$-irreps in a reducible representation $R$  of $G$ is equivalent to diagonalizing  Casimirs of $G$. For example, let $G=\SO(3)$ and $R$ be the tensor product to two spin $\frac{1}{2}$ representations.\footnote{Strictly speaking, spin $\frac{1}{2}$ are representations of $\SU(2)$ and not of $\SO(3)$. However, the tensor product of two spin $\frac{1}{2}$ representations is a (reducible) representation of $\SO(3)$.}
  The matrix form of $\SO(3)$ quadratic Casimir in the four dimensional Hilbert space of $R$ is given by 
\begin{align}
\Cas^{\SO(3)}=\left(
\begin{array}{cccc}
 2 & 0 & 0 & 0 \\
 0 & 1 & 1 & 0 \\
 0 & 1 & 1 & 0 \\
 0 & 0 & 0 & 2 \\
\end{array}
\right)
\end{align}
Diagonalizing this matrix yields an eigenvalue 0 which corresponds to a trivial representation, and a triply degenerated eigenvalue 2 which corresponds to a spin 1 representation. Altogether, the diagonalization procedure implies the  tensor product decomposition $[2]\otimes [2]=[1]\oplus [3]$. Actually, the computation can be simplified knowing that only representations of integer spin can  appear in the tensor product $[2]\otimes [2]$.
Since such representations always have a one dimensional  $L_z=0$ eigenspace, it suffices to diagonalize $\Cas^{\SO(3)}$ in the $L_z=0$ subspace of $[2]\otimes[2]$, which is spanned by $|\pm \frac{1}{2},\mp\frac{1}{2}\rangle$. The matrix form of $\Cas^{\SO(3)}$ restricted to this subspace is given by 
\begin{align}
\left.\Cas^{\SO(3)}\right|_{L_z=0}=
\left(
\begin{array}{cc}
 1 & 1 \\
 1 & 1 \\
\end{array}
\right)
\end{align}
Its eigenvalues are $0$ and $2$, corresponding to the irreps $[1]$ and $[3]$ respectively.

Similarly, in order to identify the continuous families of representations in the case of $G=\SO(1,2)$ and $R=\CF_{\Delta_1}\otimes \CF_{\Delta_2}$, we can simply  diagonalize the Casimir $\Cas^{\SO(1,2)}$ in the $L_0=0$  sector, because $\CF_\Delta$ always has a one dimensional eigenspace of  $L_0=0$ while the discrete series representations do not have any $L_0=0$ state.

 Let $\CH_0$ be the $L_0=0$ subspace of $\CF_{\Delta_1}\otimes \CF_{\Delta_2}$, and a (non-normalized) basis of $\CH_0$ is given by $|\psi_n)\equiv |n,-n), n\in \mathbb Z$. The norm of $|\psi_n)$ follows from eq.~\reef{nnorm}
\begin{align}
(\psi_n|\psi_n)=\begin{cases} 1, & \CF_{\Delta_1}=\CP_{\Delta_1}, \CF_{\Delta_2}=\CP_{\Delta_2} \\ 
\frac{\Gamma(\bar\Delta_2-n)}{\Gamma(\Delta_2-n)}, & \CF_{\Delta_1}=\CP_{\Delta_1}, \CF_{\Delta_2}=\CC_{\Delta_2} \\ 
\frac{\Gamma(\bar\Delta_1+n)}{\Gamma(\Delta_1+n)}\frac{\Gamma(\bar\Delta_2-n)}{\Gamma(\Delta_2-n)}, & \CF_{\Delta_1}=\CC_{\Delta_1}, \CF_{\Delta_2}=\CC_{\Delta_2} \\ 
\end{cases}
\end{align}
The action of $\Cas^{\SO(1,2)}$  on $|\psi_n)$ takes the form
\small
\begin{align}
\Cas^{\SO(1,2)}|\psi_n)=(2n^2+\Delta_1\bar\Delta_1+\Delta_2\bar\Delta_2)|\psi_n)-(n-\Delta_1)(n-\Delta_2)|\psi_{n-1})-(n+\Delta_1)(n+\Delta_2)|\psi_{n+1})
\end{align}
\normalsize
which commutes with a  $\mathbb Z_2$ action $\mathcal T: |\psi_n)\to |\psi_{-n})$. So the Hilbert space $\CH_0$ splits into eigenspaces of $\mathcal T$, denoted by $\CH_\pm$. The $\mathcal T= +1$ subspace $\CH_+$ is spanned by $|\psi^+_n)\equiv \frac{1}{2}\left(|\psi_n)+|\psi_{-n})\right)$ with $n\ge 0$, 
and the $\mathcal T= -1$ subspace $\CH_-$ is spanned by $|\psi^-_m)\equiv  \frac{1}{2}\left(|\psi_m)-|\psi_{-m})\right)$ with  $m\ge 1$. 

Define matrix elements of the $\SO(1,2)$ Casimir in $\CH_\pm$ as 
\begin{align}
\CQ^{(\pm)}_{mn}\equiv \frac{(\psi^\pm_m|\Cas^{\SO(1,2)}|\psi^\pm_n)}{\sqrt{(\psi^\pm_m|\psi^\pm_m)(\psi^\pm_n|\psi^\pm_n)}}~.
\end{align}
$\CQ^{(\pm)}$ are sparse matrices, whose  nonzero entries are given by 
\begin{align}\label{QpmPC}
&\CQ^{(+)}_{nn}=2n^2+\Delta_1\bar\Delta_1+\Delta_2\bar\Delta_2,\,\,\,\,\, \CQ^{(+)}_{n+1,n}=\left(\CQ^{(+)}_{n,n+1}\right)^*=\begin{cases} \sqrt{2}\beta_0,\,\,\,\, &n=0\\ \beta_n,\,\,\,\, & n\ge 1\end{cases}\nonumber\\
&\CQ^{(-)}_{nn}=2n^2+\Delta_1\bar\Delta_1+\Delta_2\bar\Delta_2,\,\,\,\, \CQ^{(-)}_{n+1,n}=\left(\CQ^{(-)}_{n,n+1}\right)^*=\beta_n,\,\,\,\, n\ge 1
\end{align}
where 
\begin{align}\label{betacases}
\beta_n=-\begin{cases}(n+\Delta_1)(n+\Delta_2), &\CP_{\Delta_1}\otimes\CP_{\Delta_2}\\ 
(n+\Delta_1)\sqrt{(n+\Delta_2)(n+\bar\Delta_2)}, &\CP_{\Delta_1}\otimes\CC_{\Delta_2}\\ 
\sqrt{(n+\Delta_1)(n+\bar\Delta_1)(n+\Delta_2)(n+\bar\Delta_2)}, &\CC_{\Delta_1}\otimes\CC_{\Delta_2}\\\end{cases}
\end{align}
Diagonalizing $\CQ^{(\pm)}$ is equivalent to solving their spectrum. First, we can show that both $\CQ^{(\pm)}$ have a continuous spectrum on $\left[\frac{1}{4},\infty \right)$. Since our argument only relies on the asymptotic behavior of $\CQ^{(\pm)}$, we will   use $\CQ^{(+)}$ to illustrate the idea. For any $q_\lambda=\frac{1}{4}+\lambda^2\in \left[\frac{1}{4},\infty\right)$, let $v_n(\lambda)$ be the corresponding eigenstate, satisfying 
\begin{align}\label{CQrec}
\CQ^{(+)}_{n,n-1}v_{n-1}(\lambda)+\CQ^{(+)}_{n,n}v_{n}(\lambda)+\CQ^{(+)}_{n,n+1}v_{n+1}(\lambda)=q_\lambda v_n(\lambda)~.
\end{align}
Such a $v_n(\lambda)$ is uniquely determined up to an overall normalization. For large $n$, plug the ans\"atz $v_n(\lambda)\sim\frac{1}{n^\alpha}\left(1+\frac{a_1}{n}+\frac{a_2}{n^2}+\cdots\right)$ into the recurrence relation~\reef{CQrec}, and we find $\alpha$ to be
\begin{align}\label{eq:alphapm}
\alpha_\pm(\lambda)=\begin{cases}\bar\Delta_1+\bar\Delta_2-\frac{1}{2}\pm i\lambda, &\CP_{\Delta_1}\otimes\CP_{\Delta_2}\\ 
\bar\Delta_1\pm i\lambda, &\CP_{\Delta_1}\otimes\CC_{\Delta_2}\\ 
\frac{1}{2}\pm i\lambda, &\CC_{\Delta_1}\otimes\CC_{\Delta_2}\\\end{cases}
\end{align}
Altogether, the leading large $n$ asymptotic behavior of $v_n(\lambda)$ is 
\begin{align}\label{jk'kl}
v_n(\lambda)=\frac{R_+}{n^{\alpha_+(\lambda)}}\left(1+\CO(n^{-1})\right)+\frac{R_-}{n^{\alpha_-(\lambda)}}\left(1+\CO(n^{-1})\right)
\end{align}
where $R_\pm$ are two constants that cannot be fixed by the asymptotic analysis. The asymptotic behavior in eq.~\reef{jk'kl} implies that the eigenvectors $v_n(\lambda)$ are $\delta$-function normalizable. The reason is as follows 
\small
\begin{align}
\left(v(\lambda_1), v(\lambda_2)\right)&\sim\sum_n \frac{R_+^* R_+}{n^{1-i(\lambda_1-\lambda_2)}}+\frac{R_-^* R_-}{n^{1+i(\lambda_1-\lambda_2)}}+\frac{R_+^* R_-}{n^{1-i(\lambda_1+\lambda_2)}}+\frac{R_-^* R_+}{n^{1+i(\lambda_1+\lambda_2)}}\nonumber\\
&\sim\int\, ds\left(R_+^* R_+ e^{i(\lambda_1\!-\!\lambda_2)s}\!+\!R_-^* R_- e^{-i(\lambda_1\!-\!\lambda_2)s}\!+\!R_+^* R_- e^{i(\lambda_1\!+\!\lambda_2)s}\!+\!R_-^* R_+ e^{-i(\lambda_1\!+\!\lambda_2)s}\right)\nonumber\\
&\sim 2\pi (R_+^* R_+ +R_-^* R_- )\delta(\lambda_1-\lambda_2)
\end{align}
\normalsize
where in the last line we have put $\delta(\lambda_1+\lambda_2)=0$ because  $\lambda$ is assumed to be nonnegative. Therefore, we conclude that $\CF_{\Delta_1}\otimes \CF_{\Delta_2}$ contains $\SO(1,2)$ principal series representation $\CP_{\frac{1}{2}+i\lambda}$ for all $\lambda \ge 0$.

 Eigenvectors of $\CQ^{(\pm)}$ with eigenvalue $\frac{1}{4}-\lambda^2$, which corresponds to a complementary series $\CC_{\frac{1}{2}+\lambda}$, take the same form as~\reef{jk'kl}, with the rotation $i\lambda\to \lambda$ in $\alpha_\pm(\lambda)$. Then  for these eigenvectors,  normalizability is equivalent to the vanishing of $R_-$. However, checking whether $R_-=0$ requires more than asymptotic analysis. The (non)existence problem of eigenvalues below $\frac{1}{4} $ has been solved by Puk\'anszky \cite{Pukan} by studying the resolvent $R_\pm(z)=\frac{1}{\CQ^{(\pm)}-z}$, which encodes the full spectrum of $\CQ^{(\pm)}$.  Instead of reviewing his lengthy and technical analysis, we will propose an efficient  numerical method to extract the full spectrum of $\CQ^{(\pm)}$, which as we shall see, admits a straightforward generalization to all spectrum problems in this paper. More importantly, the numerical method provides an  intuitive and physical picture to understand why  $\CK_{\text{rel}}(\lambda)$ in eq.~\reef{CKrelf} and eq.~\reef{hkljh} can be viewed as a relative density of principal series.

\begin{figure}[]
     \centering
     \begin{subfigure}[t]{0.45\textwidth}
         \centering
         \includegraphics[width=\textwidth]{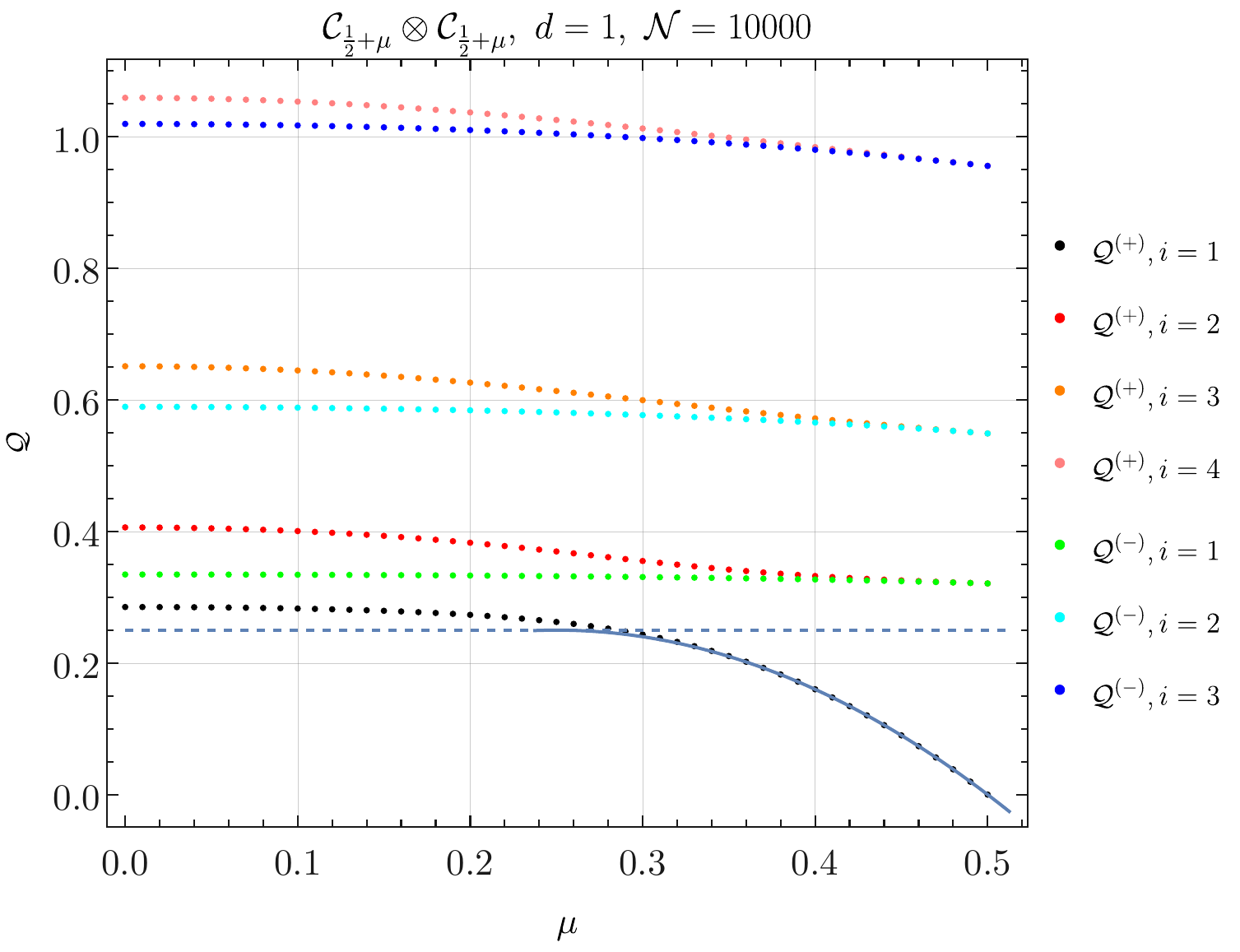}
             \end{subfigure}
     \hfill
     \begin{subfigure}[t]{0.45\textwidth}
         \centering
         \includegraphics[width=\textwidth]{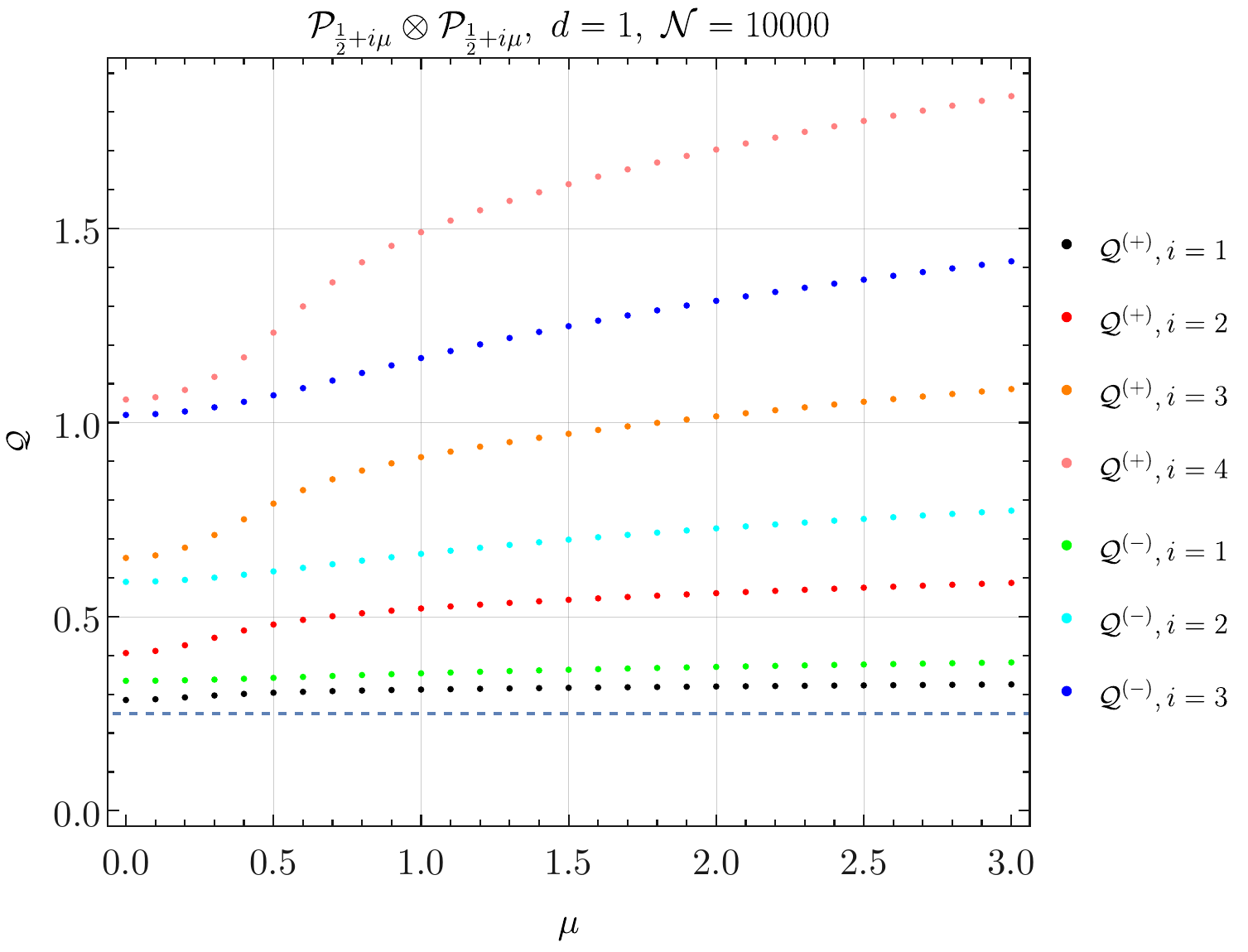}
     \end{subfigure}
             \caption{The first a few low-lying eigenvalues of truncated $\CQ^{(\pm)}$ in $\CF_\Delta\otimes\CF_{\Delta}$. Left: Tensor product of two complementary series representations $\CC_{\frac{1}{2}+\mu}\otimes\CC_{\frac{1}{2}+\mu}$ for different $\mu$. The solid curve is $\CQ=2\mu(1-2\mu)$, the $\SO(1,2)$ Casimir corresponding to  $\CC_{2\mu}$. Right: Tensor product of two principal series representations $\CP_{\frac{1}{2}+i\mu}\otimes\CP_{\frac{1}{2}+i\mu}$ for different $\mu$. The dashed line in both left and right panels is $\CQ=\frac{1}{4}$, separating the regions corresponding to principal series and complementary series.}
                  \label{fig:CxCd1}
     \end{figure}

In order to perform a numerical analysis of $\CQ^{(\pm)}$, we  cut off the Hilbert space $\CH_0$ at $-\CN\le n\le \CN$. From a $\text{dS}_2$ picture, it is equivalent to cutting off the angular momentum (along the circle in global coordinates) of the two particles corresponding to $\CF_{\Delta_1}$ and $\CF_{\Delta_2}$ respectively.
In this truncated Hilbert space $\CH_0^{(\CN)}$,  $\CQ^{(+)}$ becomes an $(\CN+1)\times (\CN+1)$ matrix and $\CQ^{(-)}$ becomes an $\CN\times\CN$ matrix. 
 Taking  $\CQ^{(+)}$ as an example, it has $\CN+1$ eigenvalues, denoted by $q_0\le q_1\le\cdots\le q_\CN$.
Any $q_a$ smaller than $\frac{1}{4}$ at large $\CN$ corresponds to  a complementary series representation and $q_a\ge \frac{1}{4}$ belongs to principal series.  As discussed above, in the case of $\CC_{\Delta_1}\otimes \CC_{\Delta_2}$ with $\mu_1+\mu_2>\half$,  there is a single complementary series representation with dimension $\D=\mu_1+\mu_2$, which has  $\SO(1,2)$ Casimir $(\mu_1+\mu_2)(1-(\mu_1+\mu_2))$. This can be seen in the left panel of fig.~\reef{fig:CxCd1}, where we plot the first four eigenvalues of  $\CQ^{(+)}$ and the first three eigenvalues of  $\CQ^{(-)}$ in the truncated tensor product $\CC_{\frac{1}{2}+\mu}\otimes \CC_{\frac{1}{2}+\mu}$ as a function of $\mu$. The cutoff $\CN$ is chosen to be $5000$. $\CQ^{(-)}$ is always larger than $\frac{1}{4}$ and $\CQ^{(+)}>\frac{1}{4}$  for $0<\mu\lesssim\frac{1}{4}$. When $\mu>\frac{1}{4}$, the smallest eigenvalue $\CQ^{(+)}_{\rm min}$ of $\CQ^{(+)}$  becomes smaller than $\frac{1}{4}$ and lies on the quadratic curve $\CQ=2\mu(1-2\mu)$, which implies the existence of a complementary series representation $\CC_{2\mu}$ in this regime. We also illustrate the convergence of $\CQ^{(+)}_{\rm min}(\CN)$ to the expected value $\CQ^{(+)}_{\rm min}(\infty)=(\mu_1+\mu_2)(1-(\mu_1+\mu_2))$ in the $\CN \to \infty$ limit  in the left and middle panels of fig.~\reef{fig:CxCd1Precision}.  
\begin{figure}[]
     \centering
     \begin{subfigure}[t]{0.3\textwidth}
         \centering
         \includegraphics[width=\textwidth, height=1.5in]{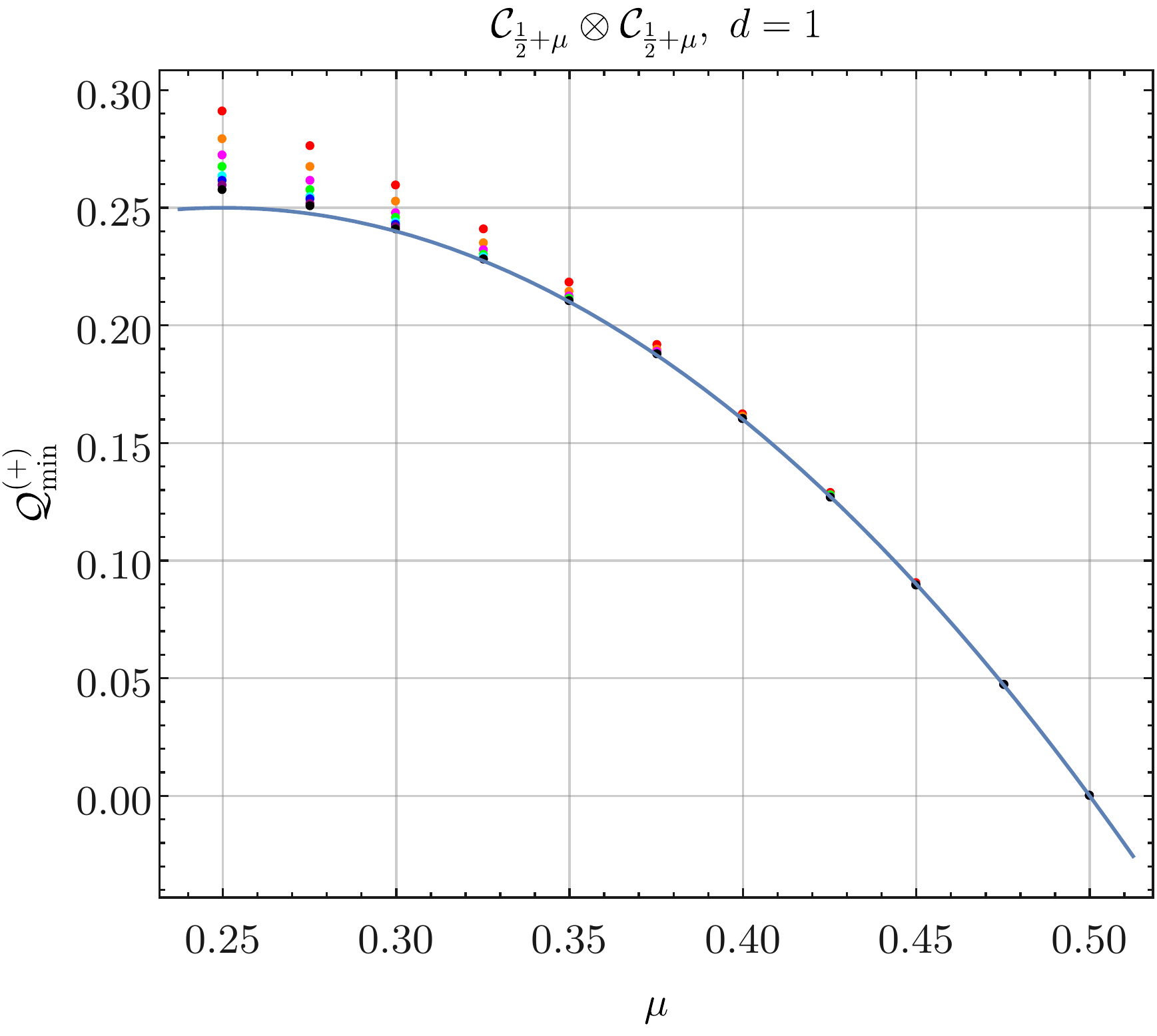}
             \end{subfigure}
     \hfill
     \begin{subfigure}[t]{0.3\textwidth}
         \centering
         \includegraphics[width=\textwidth, height=1.5in]{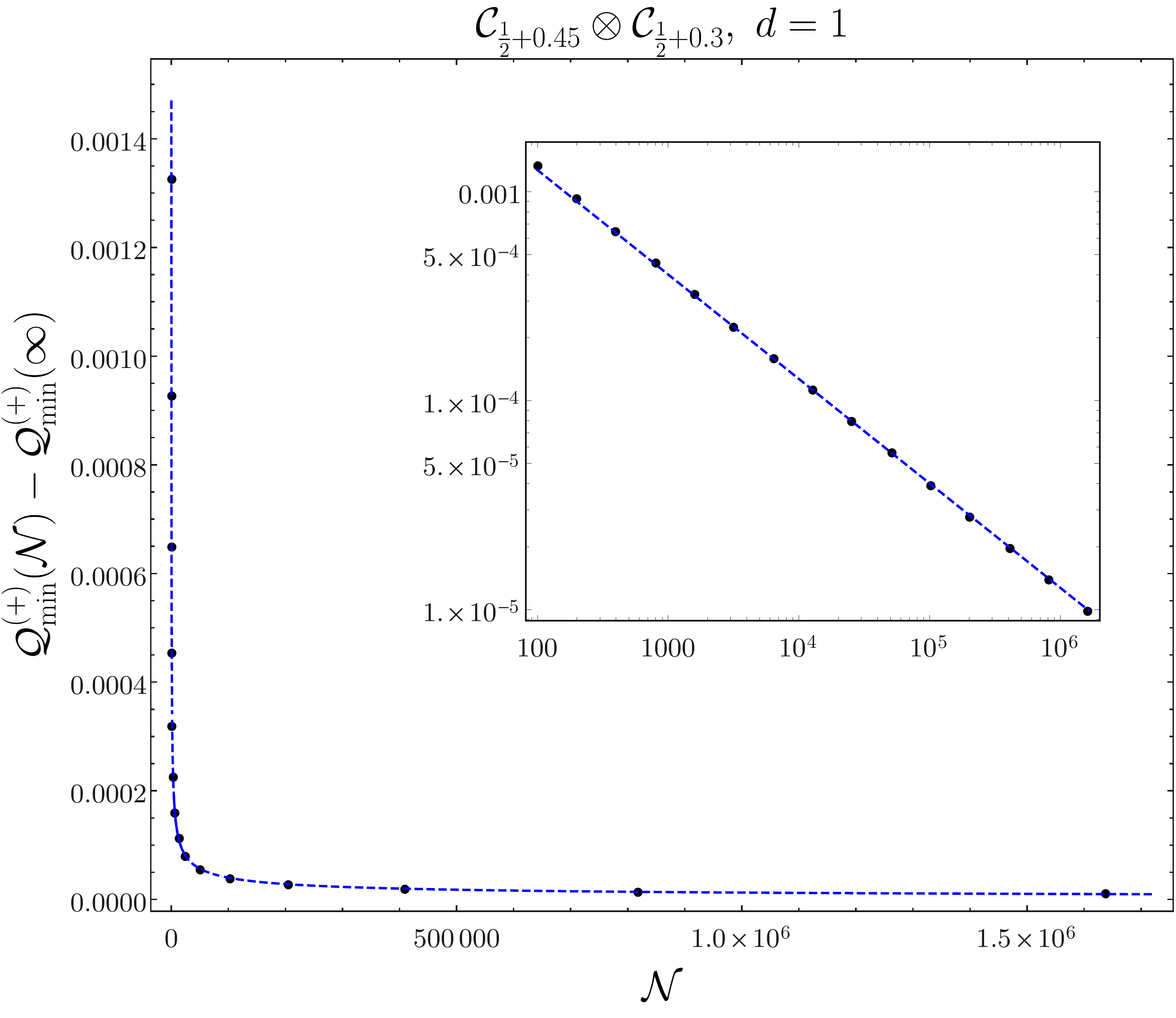}
     \end{subfigure}
          \hfill
     \begin{subfigure}[t]{0.3\textwidth}
         \centering
         \includegraphics[width=\textwidth, height=1.5in]{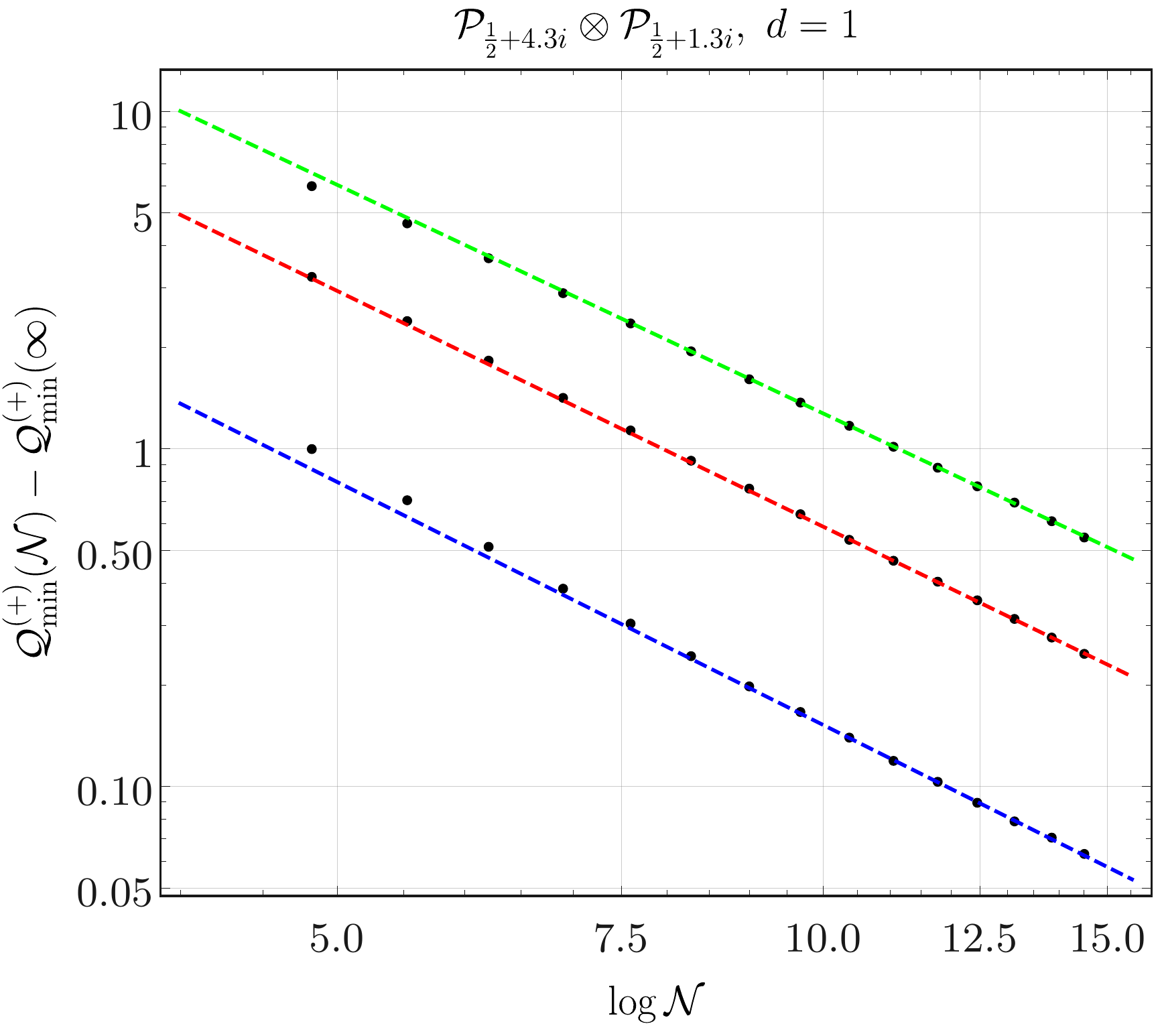}
     \end{subfigure}
             \caption{The convergence of the minimal eigenvalue of $\CQ^{(+)}_{\rm min}$. Left: $\CQ^{(+)}_{\rm min}$ in the tensor product $\CC_{\frac{1}{2}+\mu}\otimes \CC_{\frac{1}{2}+\mu}$ for different $\mu$ and different $\CN$. The solid line is $2\mu(1-2\mu)$. Middle: The convergence to $\CQ^{(+)}_{\rm min}(\infty)=\frac{3}{15}$ for $\CC_{\frac{1}{2}+0.45}\otimes \CC_{\frac{1}{2}+0.3}$ with the inset of a loglog-plot. The dashed line is the fit of the curve $a/\sqrt{ \CN}$ to the data. Right: The convergence of first three eigenvalues of $\CQ$ to $\frac{1}{4}$ for $\CP_{\frac{1}{2}+4.3i}\otimes \CP_{\frac{1}{2}+1.3i}$, illustrated in a log-log plot with $\log(\CN)$ as the x-axis.  The dashed lines are fits of type $a \log(\CN)^b $, showing that the gap would close eventually in large $\CN$ limit.}
                  \label{fig:CxCd1Precision}
     \end{figure}
Empirically, we find that there is a power low convergence of $\CQ^{(+)}_{\text{min}}(\CN)$ to the exact value given by 
\be
\label{eq:precision CxC1d}
\CQ^{(+)}_{\text{min}}(\CN) - \CQ^{(+)}_{\text{min}}(\infty) \; \; \limu{\CN\to\infty} \; \; \frac{1}{\CN^{2(\mu_1+\mu_2)-1}}~.
\ee

In the case of $\CP_{\Delta_1}\otimes \CP_{\Delta_2}$, all eigenvalues of $\CQ^{(\pm)}$ are larger than $\frac{1}{4}$, e.g. the right panel of  fig.~\reef{fig:CxCd1}. So each eigenvalue $q_n$ in the spectrum of, say $\CH_+$, can be parameterized by a positive number $\lambda_n=\sqrt{q_n-\frac{1}{4}}$, which effectively corresponds to a principal series representation of scaling dimension $\Delta_n=\frac{1}{2}+i\lambda_n$. A coarse-grained density of principal series in truncated $\CH_+$ can then be defined as the inverse spacing of $\lambda_n$
\begin{align}\label{brdef}
\bar\rho^+_\CN(\lambda_n)\equiv \frac{2}{\lambda_{n+1}-\lambda_{n-1}}~.
\end{align}
Similarly, we define a coarse-grained density $\bar\rho^-_\CN(\lambda_n)$ of principal series in truncated $\CH_-$. Adding up $\bar\rho_\CN^\pm$ as  interpolated functions yields the full coarse-grained density $\bar\rho_\CN(\lambda)$ of principal series in the tensor product $\CP_{\Delta_1}\otimes \CP_{\Delta_2}$. This coarse-grained density suffers from divergence in the $\CN\to \infty$ limit because we have shown above that $\{\lambda_n\}$  converge to a continuous spectrum in this limit. In the same spirit as the character analysis, we remove this divergence by considering a relative coarse-grained density (dropping the label $\CN$ whenever we have a relative density)
\be\label{eq:rhorel}
\rrel= \bar{\rho}^{\CP_{\Delta_1}\otimes \CP_{\Delta_2}}_{\CN} - \bar{\rho}^{\CP_{\Delta_3}\otimes \CP_{\Delta_4}}_{\CN} 
\ee

\begin{figure}[t!]
     \centering
     \begin{subfigure}[t]{0.43\textwidth}
         \centering
         \includegraphics[width=\textwidth]{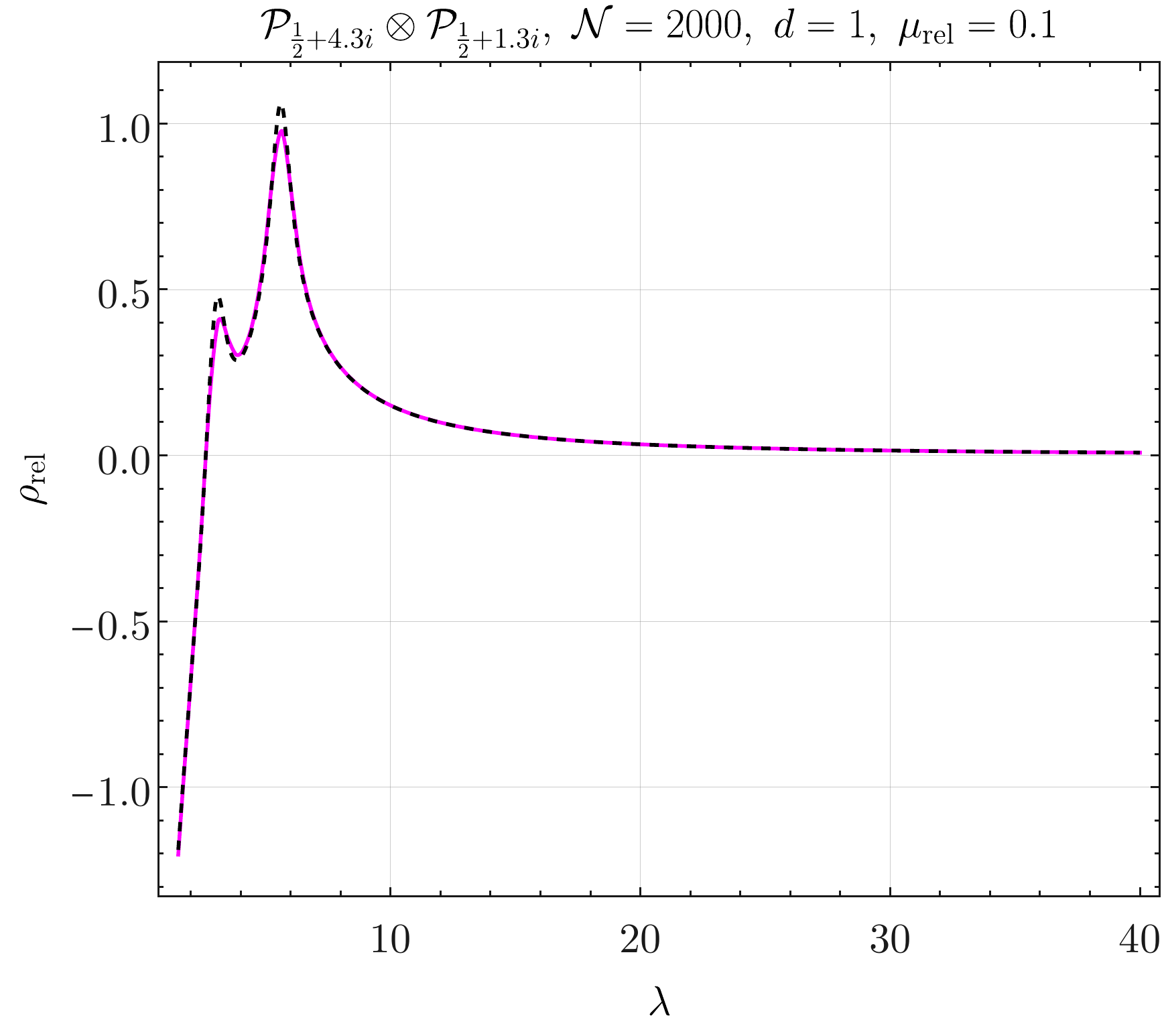}
         \label{rrelzoomout}
     \end{subfigure}
     \hfill
     \begin{subfigure}[t]{0.5\textwidth}
         \centering
         \includegraphics[width=\textwidth]{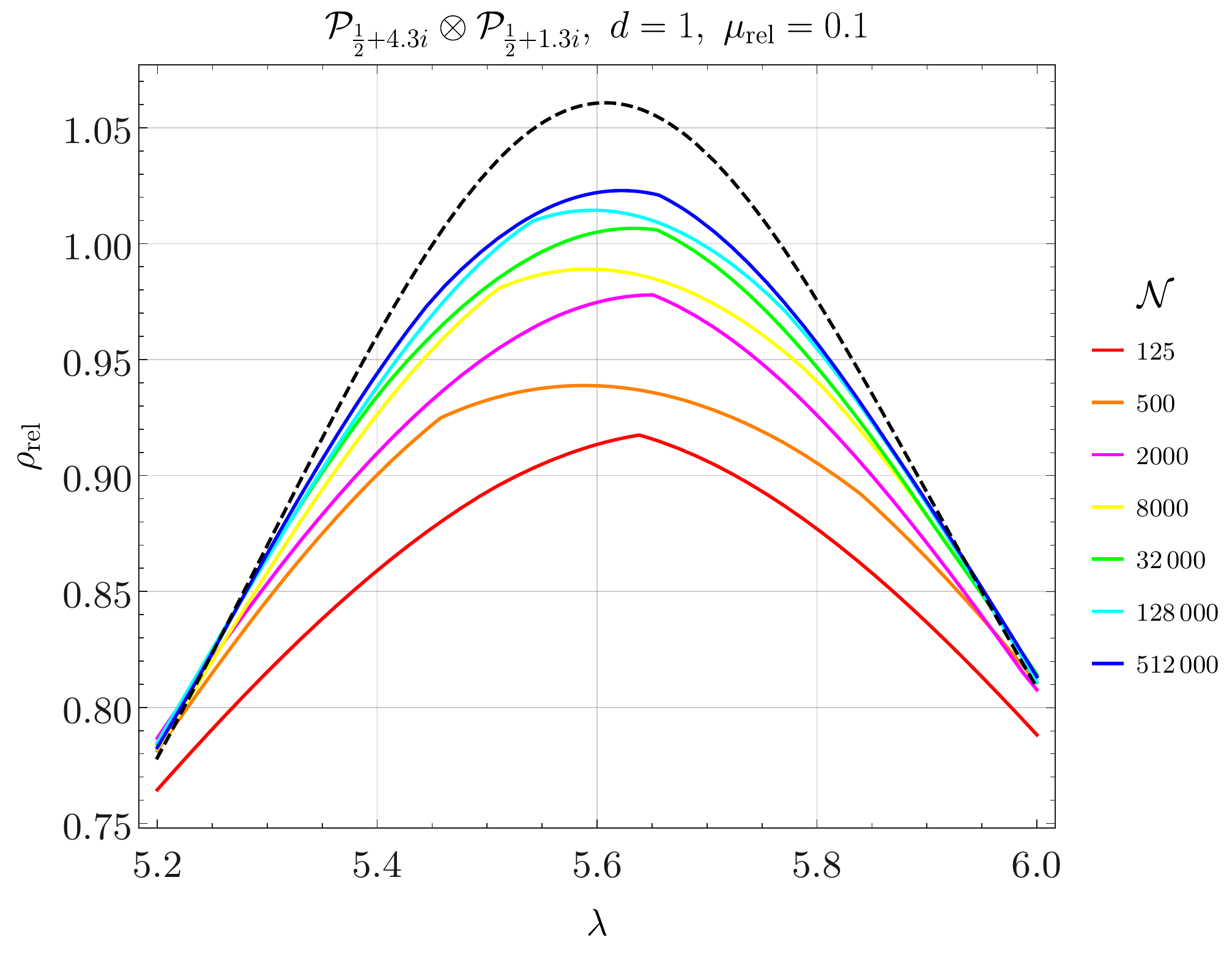}
         \label{fig:rrel1dZoom}
     \end{subfigure}
     \caption{ Relative density of principal series between $\CP_{\frac{1}{2}+4.3i}\otimes\CP_{\frac{1}{2}+1.3i}$ and $\CP_{\frac{1}{2}+0.1i}\otimes\CP_{\frac{1}{2}+0.1i}$.
     Left: The dashed black line corresponds to $\CK_{\text{rel}}$, given by eq. (\ref{CKrelf}), and the pink line is $\rho_\text{rel}$ for $\CN=2000$.
      The right panel zooms into the region around the peak at $\lambda=5.6$ and shows how $\rho_{\text{rel}}$ approaches $\CK_{\text{rel}}$ by increasing $\CN$.} \label{fig:rrel1d}
\end{figure}

\begin{figure}
     \centering
     \begin{subfigure}[t]{0.45\textwidth}
         \centering
                   \includegraphics[width=\textwidth]{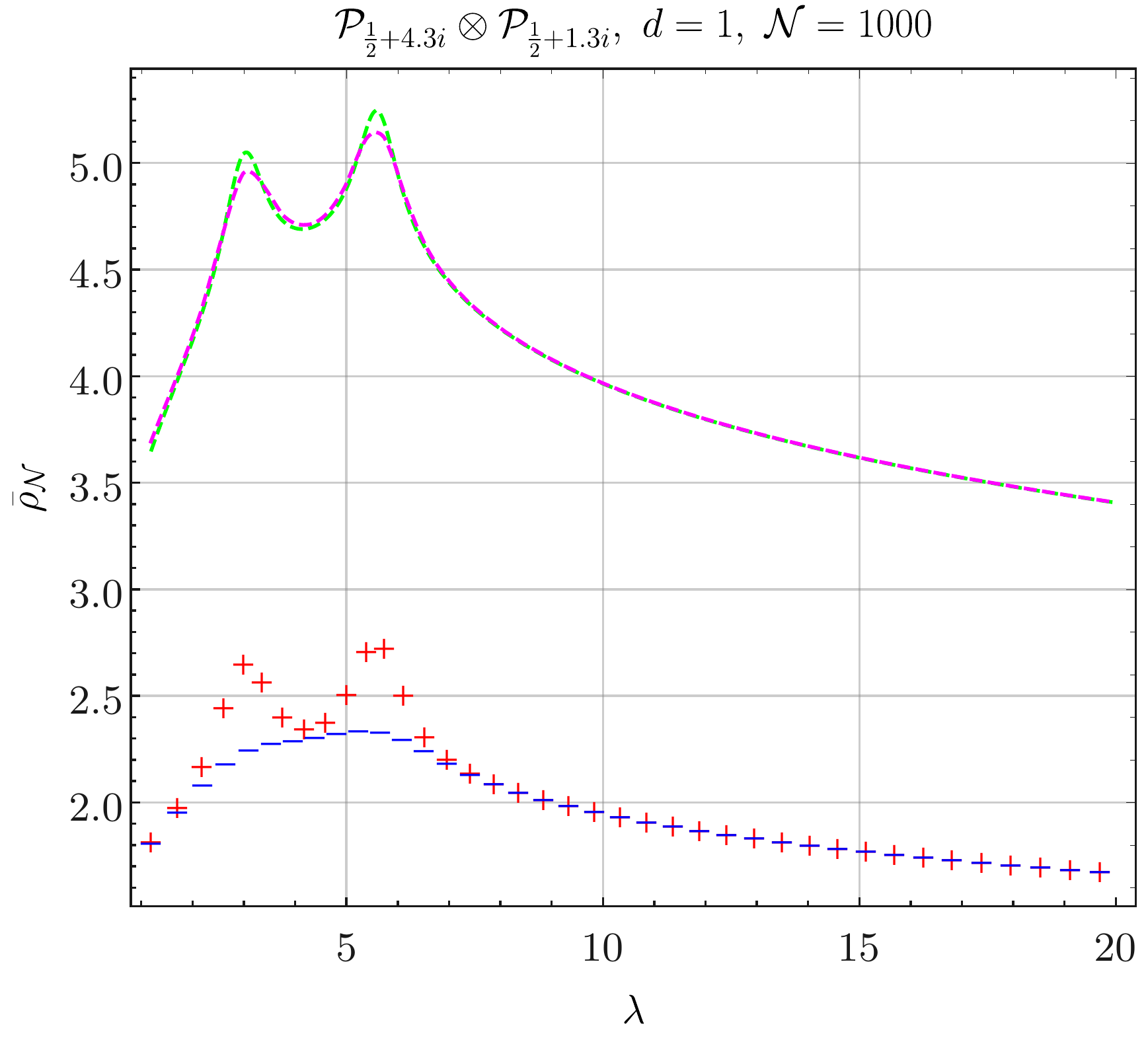}
         \label{rbarpm}
     \end{subfigure}
     \hfill
     \begin{subfigure}[t]{0.45\textwidth}
         \centering
                   \includegraphics[width=\textwidth]{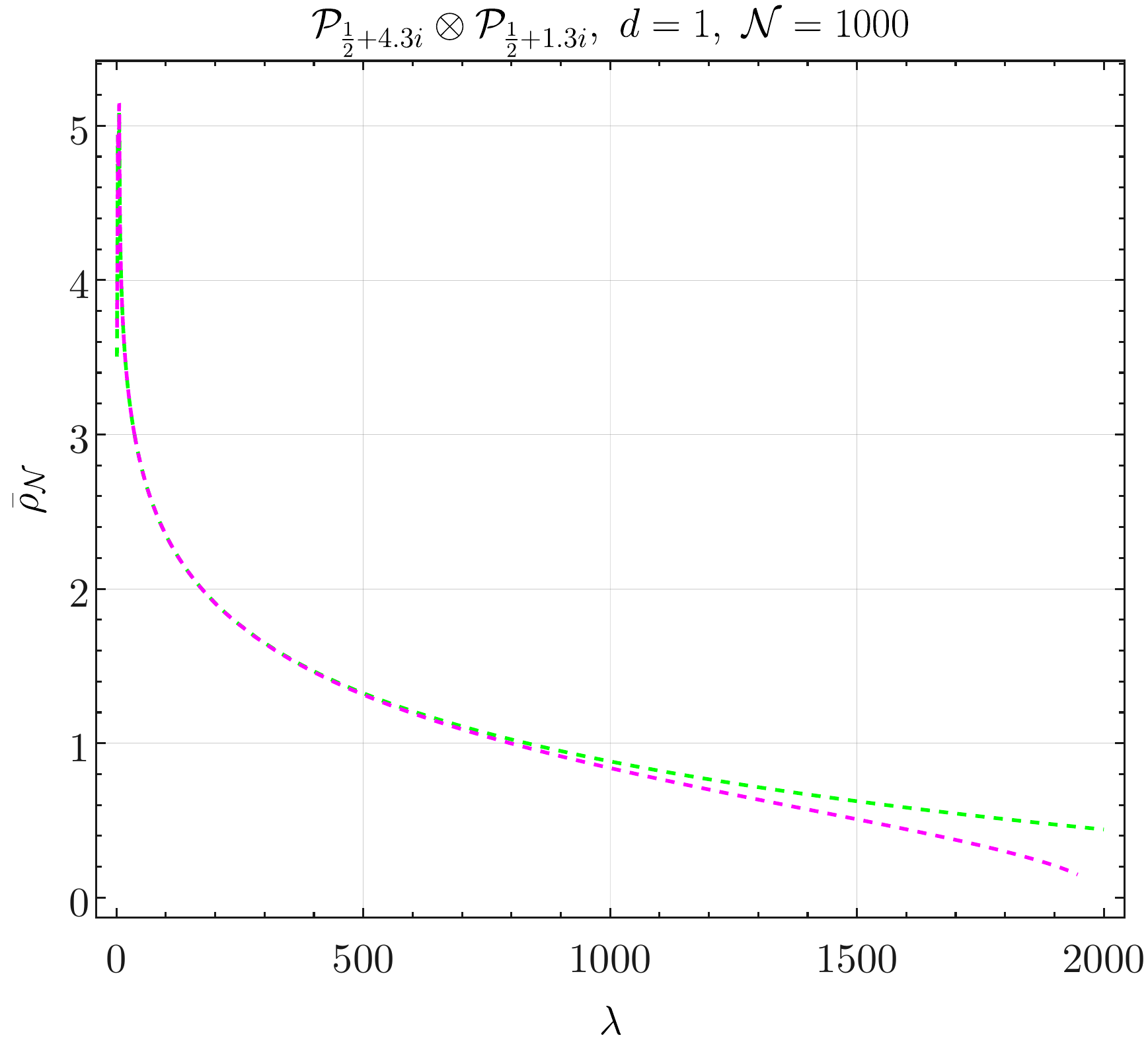}
         \label{fig:rbarLargelambda}
     \end{subfigure}
     \caption{Density of principal series in the tensor product $\CP_{\frac{1}{2}+4.3i}\otimes\CP_{\frac{1}{2}+1.3i}$. Left: The red and blue dots show the coarse-grained density $\bar\rho_\CN^+$ and $\bar\rho_\CN^-$ respectively, with the cutoff $\CN$ being 1000. The pink dashed line is $\bar\rho_\CN$, derived from adding the interpolations of  $\bar\rho_\CN^+$ and $\bar\rho_\CN^-$. The green dashed line is $\CK_{\epsilon}$ in ~\reef{eq:epsilon reg}, with $\epsilon=1.4\times 10^{-4}$ . The right panel shows a larger range of $\lambda$. The matching between $\bar\rho_\CN$ and $\CK_{\epsilon}$ starts to fail when $\lambda$ is roughly the same order as the cut-off scale $\CN$. 
     }\label{fig:rhobar1d}
\end{figure}

As a very nontrivial check, we find  that the  relative coarse-grained density $\rho_{\text{rel}}$ defined in eq. (\ref{eq:rhorel})
matches perfectly with $\CK_{\text{rel}}$ given by eq. (\ref{CKrelf}) when $\lambda$ is much smaller than the cut-off $\CN$. In fig. (\ref{fig:rrel1d}) we show this remarkable matching for the case of $\Delta_1=\frac{1}{2}+4.3 i, \Delta_2=\frac{1}{2}+1.3 i$ and $\Delta_3=\Delta_4=\frac{1}{2}+0.1 i$. By construction of $\rho_{\text{rel}}$, the matching confirms the meaning of $\CK_{\text{rel}}$, which is derived purely from characters,  as a (relative) density of principal series.

It is worth mentioning the observation that the coarse-grained $\bar{\rho}_\CN$ (which is regularized by $\CN$) for a fixed large $\CN$ has a good agreement with $\CK_{\epsilon}$ (which is regularized by $\epsilon$) defined in equation~\reef{eq:epsilon reg} by choosing the regularization $\epsilon$ such that $\bar{\rho}_\CN$ and $\CK_{\epsilon}$ match at some scale $\lambda_*$, in the spirit of renormalization. 
In fig.~\reef{fig:rhobar1d}, we show the match between $\bar{\rho}_\CN$ with $\CN=1000$ and  $\CK_{\epsilon}$ with $\epsilon=1.4\times 10^{-4}$ for $\CP_{\frac{1}{2}+4.3 i}\otimes \CP_{\frac{1}{2}+1.3 i}$. As we will discuss in appendix~\ref{sec:L0s}, this phenomenon disappears  for $\bar{\rho}_\CN$ in $L_0 \neq0$ sectors.

\subsection{$\CF_{\Delta}\otimes \CD_p$}
As we have seen in subsection \ref{CF1tCF2} that highest-weight and lowest-weight discrete series representations always appear in pairs in $\CF_{\Delta_1}\otimes \CF_{\Delta_2}$, we are mainly interested in  the tensor product of $\CF_\Delta$ and a reducible representation $\CD_p=\CD^+_p\oplus \CD^-_p$ in this subsection.
\vspace{5pt}
\subsubsection{The discrete part}

Let $|n,m)$ be a basis of $\CF_\Delta\otimes \CD^+_p$, where $n\in\mathbb Z$ and $m\ge p$.
First, we want to show that this tensor product  contains exactly one copy of each $\CD^+_k, k\ge 1$. As in the previous case, it amounts to constructing the lowest-weight state $|k)_k$ of each $\CD^+_k$. Since such a state has eigenvalue $k$ with respect to $L_0$, it can be expressed as a linear combination of all $|k-n,n)$
\begin{align}
|k)_k=\sum_{n\ge p}a_n \, |k-n,n)~.
\end{align}
 The condition $L_-|k)_k=0$ yields the following recurrence relation
 \begin{align}\label{reca1}
 (n-p)a_n+(k-n+\bar\Delta)a_{n-1}=0
 \end{align}
 whose solution is given by $a_n=\frac{\Gamma(n+\Delta-k)}{\Gamma(n+1-p)}$ up to an  overall normalization constant $c$. Then the norm of $|k)_k$ becomes 
 \begin{align}
_k(k|k)_k=|c|^2\sum_{n\ge p}\frac{\Gamma(n+\Delta-k)\Gamma(n+\bar\Delta-k)}{\Gamma(n+1-p)\Gamma(n+p)}\sim |c|^2 \sum_n \,n^{-2k}~.
 \end{align}
The sum is convergent for $k=1,2,\cdots$ and hence all $\CD^+_k$ belongs to $\CF_\Delta\otimes \CD_p^+$. The degeneracy of each $\CD^+_k$ follows from the uniqueness of the solution to eq.~\reef{reca1}.

The same strategy can be used to exclude all highest-weight discrete series representations $\CD^-_\ell$. For example, let $|-\ell)_\ell$ be the highest-weight state of $\CD^-_\ell$ and it has to take the following form, if exists
\begin{align}
|-\ell)_\ell=\sum_{n\ge p}b_n \, |-\ell-n,n)~.
\end{align}
 The condition $L_+|-\ell)_\ell=0$ leads to a recurrence relation
 \begin{align}\label{reca2}
 (\Delta-n-\ell)b_n+(p+n-1)b_{n-1}=0, \,\,\,\,\, n\ge p+1
 \end{align}
and an initial condition  $b_p=0$, which further imply that all $b_n$ are actually vanishing. Therefore, $\CD^-_\ell$ does not belong to  $\CF_\Delta\otimes \CD_p^+$. 

Similarly, one can show  that $\CF_\Delta\otimes \CD_p^-$ contains exactly one copy of each highest-weight discrete series representations and does not contain any lowest-weight ones. Altogether, the discrete part of $\CF_{\Delta}\otimes \CD_p$ is the same as that of $\CF_{\Delta_1}\otimes \CF_{\Delta_2}$, namely
\begin{align}\label{eq:Discrete of PtD}
\CF_{\Delta}\otimes \CD_p\supseteq \bigoplus_{k\ge 1}\left(\CD^+_k\oplus \CD^-_k\right)~.
\end{align}

\subsubsection{The continuous part}\label{sec:cont FtD}

\noindent{}\textbf{Character analysis}

Using the tensor product of two principal series representations $\CP_{\Delta_3}\otimes\CP_{\Delta_4}$ to regularize $\CF_\Delta\otimes \CD_p$, the discrete series part cancel out and then the relative density of principal series should satisfy
\begin{align}
\Theta_{\Delta}(q)\Theta_{p}(q)-\Theta_{\Delta_3}(q)\Theta_{\Delta_4}(q)=\int_0^\infty\,d\lambda\, \CK_{\text{rel}}(\lambda)\Theta_{\Delta_\lambda}(q)
\end{align}
where we have used the fact that such a tensor product does not contain any complementary series representations \cite{Repka:1978}. This is  consistent with the explicit expression of  $\CK_{\text{rel}}(\lambda)$:
\begin{align}\label{eq:charPtD}
 \CK_{\text{rel}}(\lambda)=&-\frac{1}{\pi}\sum_{\pm}\left(\psi\left(\Delta+p-\frac{1}{2}\pm i\lambda\right)+\psi\left(\bar\Delta+p-\frac{1}{2}\pm i\lambda\right)\right)\nonumber\\
 &+\frac{1}{2\pi}\sum_{\pm,\pm,\pm}\psi\left(\frac{1}{2}\pm i\mu_3\pm i\mu_4\pm i\lambda\right)
\end{align}
because it does not admit any pole crossing along the lines of remark \ref{polecrossing}.

\vspace{5pt}

\noindent{}\textbf{Numerical check}

To identify the continuous families of states in $\CF_\Delta\otimes \CD_p^\pm$, it is sufficient to diagonalize the $\SO(1,2)$ Casimir in the $L_0=0$ subspace $\CH_0$ of $\CF_\Delta\otimes \CD_p^\pm$. For example, a basis of $\CH_0$ in $\CF_\Delta\otimes \CD_p^+$ is  $|\psi_n)=|-n,n),~n\ge p$. The non-vanishing matrix elements of $\Cas^{\SO(1,2)}$ with respect to this basis are found to be 
\begin{align}\label{eq:CQ PD}
\CQ_{n,n}=2n^2+\Delta\bar\Delta+p(1-p), \,\,\,\,\, \CQ_{n+1,n}=\left(\CQ_{n,n+1}\right)^*=-\sqrt{(n+p)(n+1-p)}\,\gamma_n
\end{align}
where 
\begin{align}
\gamma_n=\begin{cases} n+\Delta, &\CF_\Delta=\CP_\Delta\\ \sqrt{(n+\Delta)(n+\bar\Delta)}, &\CF_\Delta=\CC_\Delta\end{cases}
\end{align}
In $\CF_\Delta\otimes \CD_p^-$, we get exactly the same matrix representation for $\Cas^{\SO(1,2)}$. Similar to the discussion for the case of $\PtP$, we are going to diagonalize the $\CQ$ matrix to derive the spectrum. Let us mention that with a similar large $n$ argument that led to recursion relation~\reef{CQrec} -- where complementary series can be replaced by discrete series with $\D=p$, one can show that $\PtD$ contains all the $\SO(1,2)$ principal series: $\CP_{\frac{1}{2}+i\lambda}$ with $\lambda \ge 0$.

\begin{figure}[t]
     \centering
     \begin{subfigure}[t]{0.45\textwidth}
         \centering
         \includegraphics[width=\textwidth]{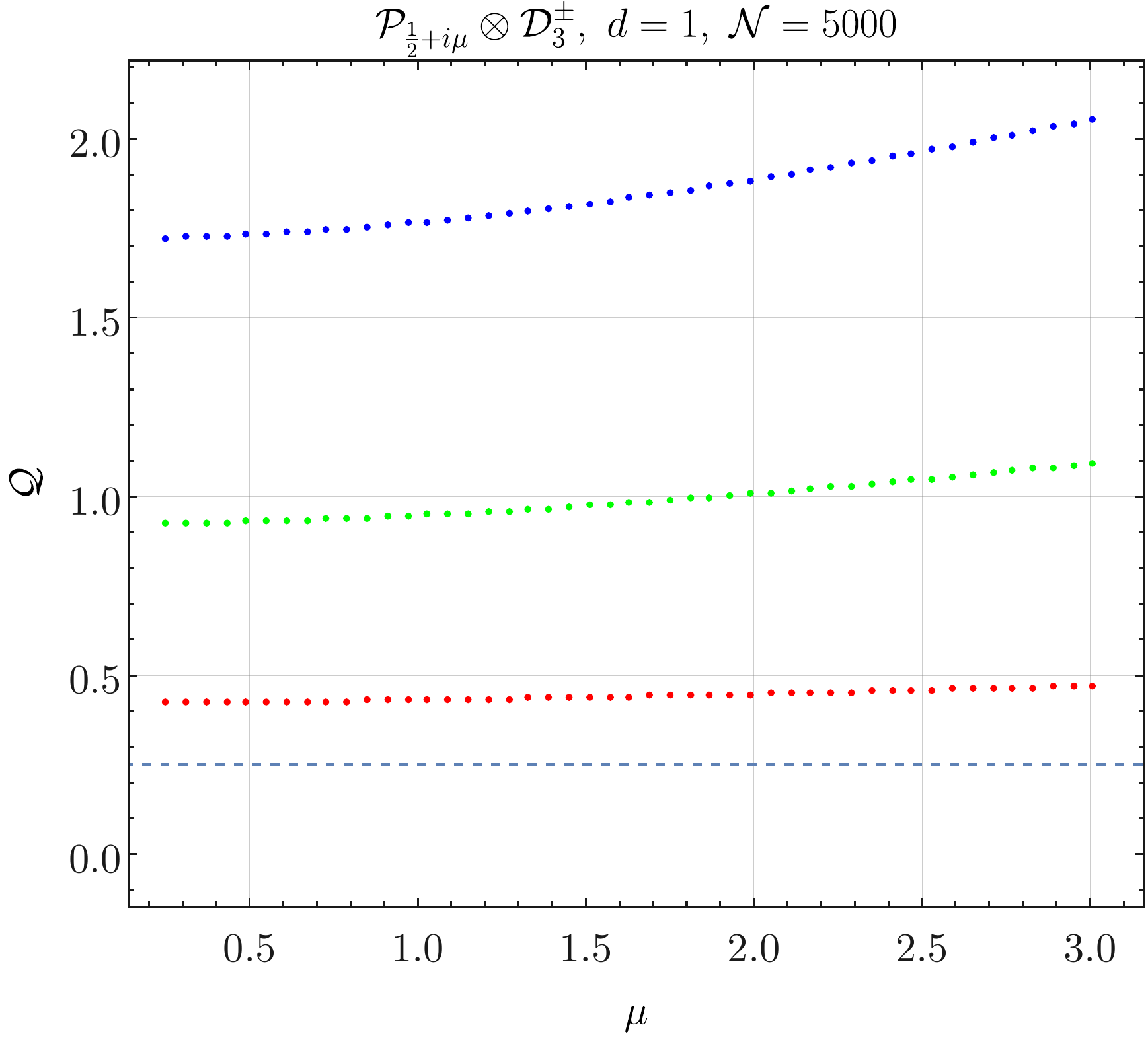}
         \label{fig:PxDd1}
     \end{subfigure}
     \hfill
     \begin{subfigure}[t]{0.45\textwidth}
         \centering
         \includegraphics[width=\textwidth]{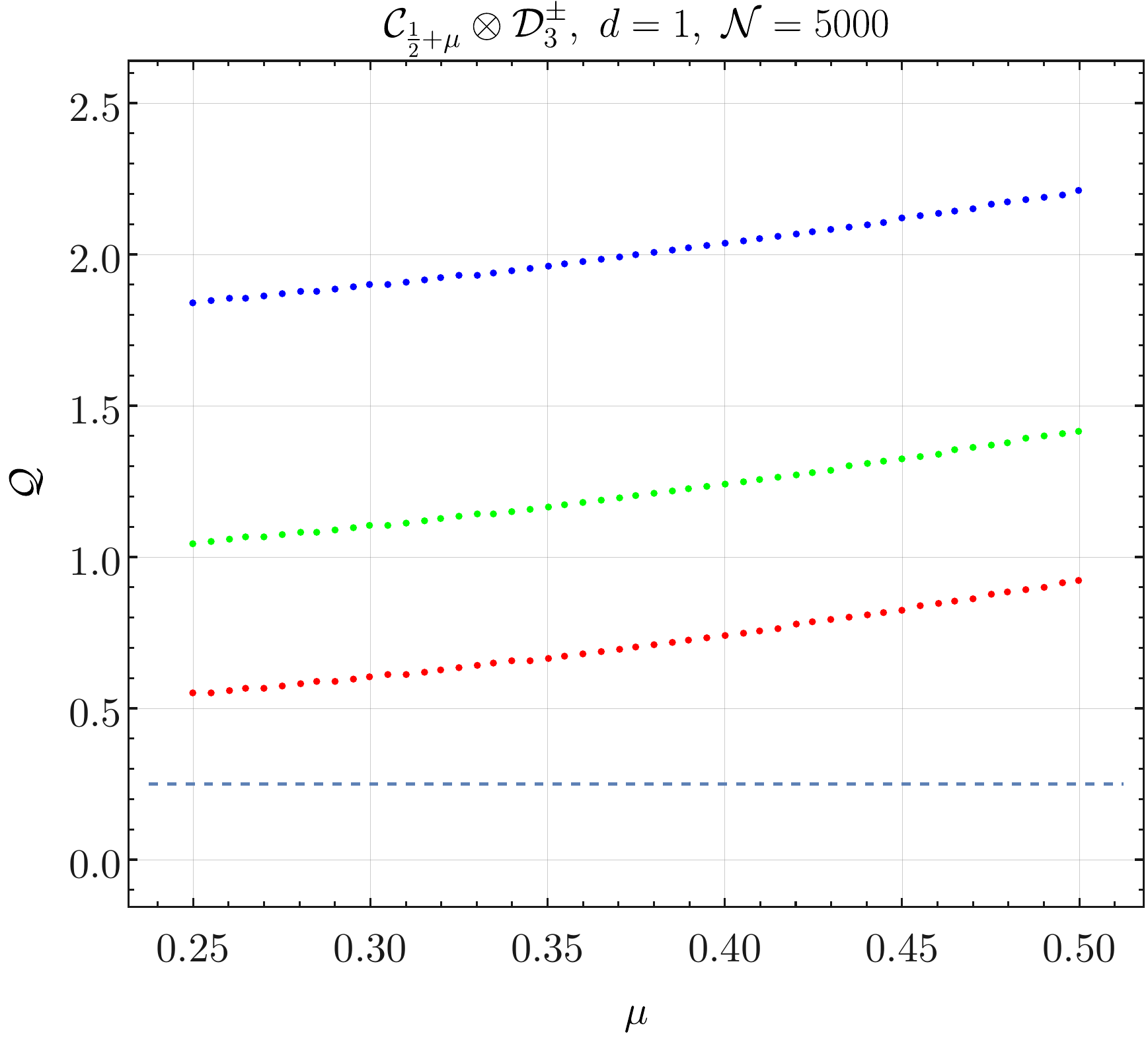}
         \label{fig:CxDd1}
     \end{subfigure}
     \caption{The low-lying eigenvalues of $\CQ$ for $\CF_\Delta\otimes\CD^\pm_3$. $\CF_{\Delta}=\CP_{\frac{1}{2}+i\mu}$ is a principal series representation in the left panel and $\CF_{\Delta}=\CC_{\frac{1}{2}+\mu}$ is a complementary series series in the right panel. The cutoff is 5000 for both.}{\label{fig:PCxDd1}}
\end{figure}

In fig. (\ref{fig:PCxDd1}), we present the numerical result of diagonalizing the $\CQ$ matrix. In $L_0=0$ sector, we do not expect to see the discrete series -- while the $L_0\neq 0$ sectors include discrete series following~\reef{eq:Discrete of PtD} -- see appendix~\ref{sec:L0s} for the discussion on $L_0\neq 0$ sectors.  As discussed in~\reef{eq:charPtD}, in the $\CC_\D \otimes \CD_p$ tensor product decomposition, no complementary series appears, in agreement with fig. (\ref{fig:PCxDd1}).      

\begin{figure}[t]
     \centering
     \begin{subfigure}[t]{0.43\textwidth}
         \centering
         \includegraphics[width=\textwidth]{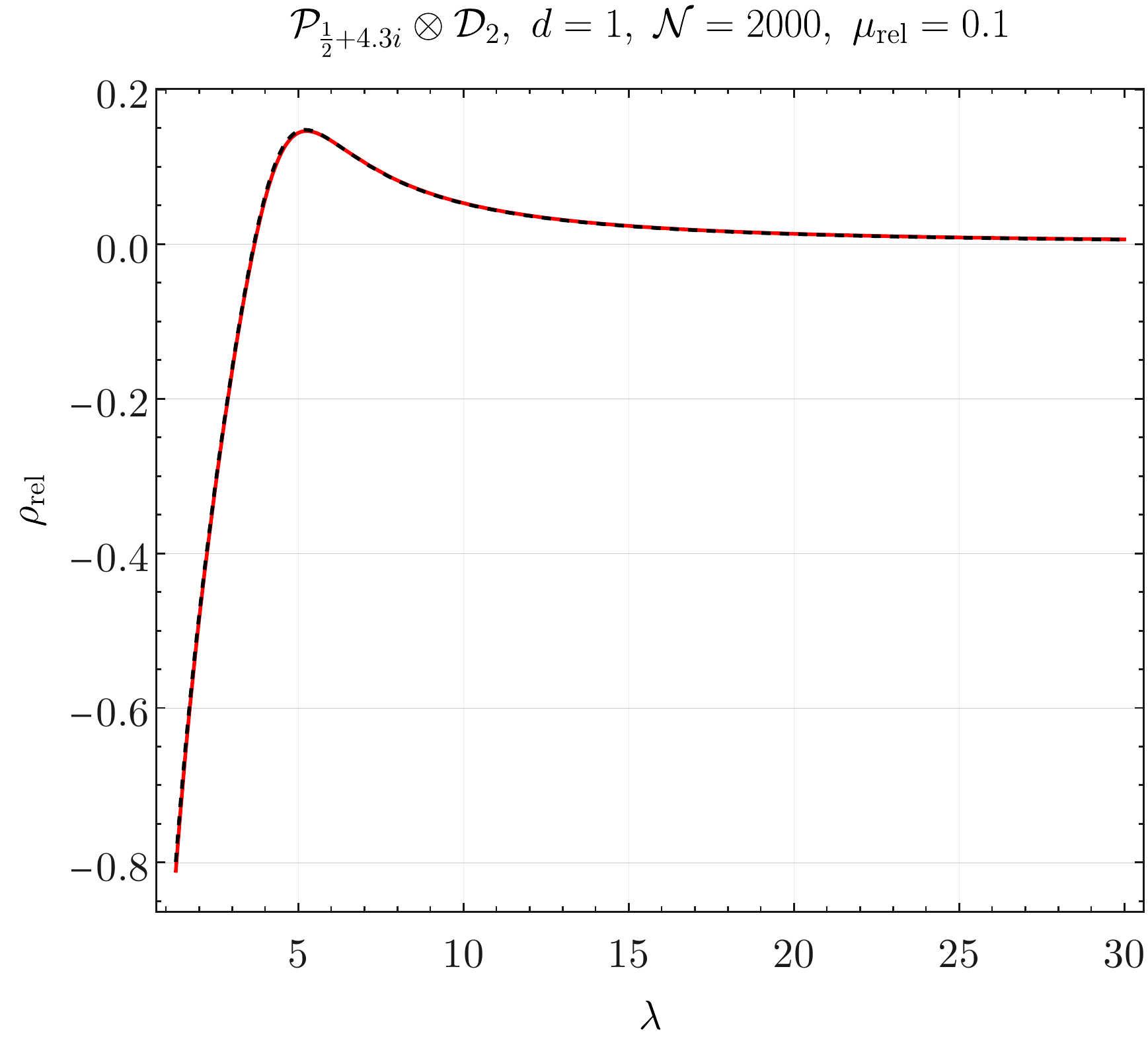}
         \label{fig:rrel1dPxDsmall}
     \end{subfigure}
     \hfill
     \begin{subfigure}[t]{0.51\textwidth}
         \centering
         \includegraphics[width=\textwidth]{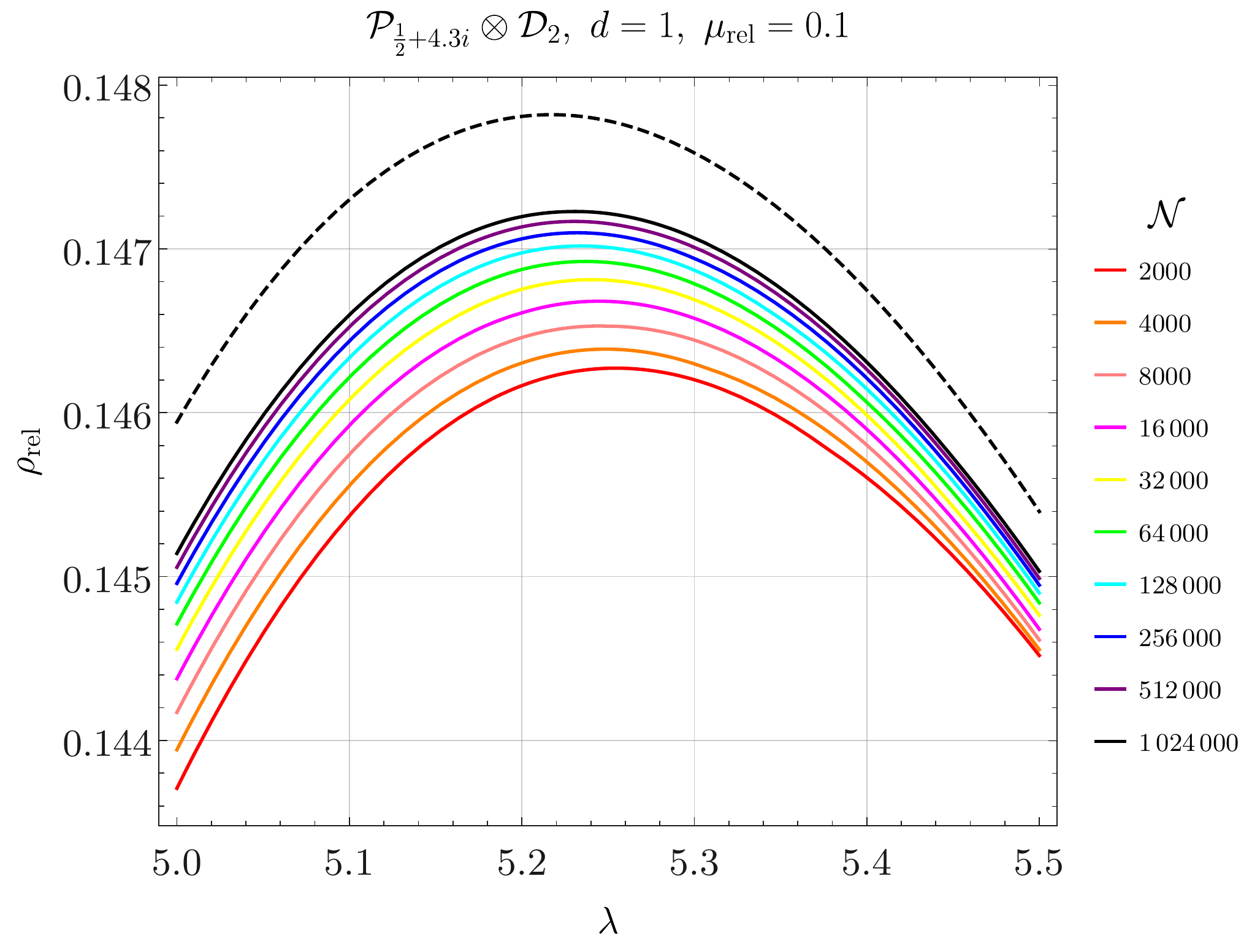}
         \label{fig:rrelZoom}
     \end{subfigure}
     \caption{Relative density of principal series between $\CP_{\frac{1}{2}+4.3 i}\otimes\CD_2$ and $\CP_{\frac{1}{2}+0.1 i}\otimes\CP_{\frac{1}{2}+0.1 i}$. In the left panel, the red line is  $\rho_\text{rel}$ and the dashed black line is $\CK_{\text{rel}}$, given by eq. (\ref{eq:charPtD}) with $\mu_3=\mu_4=\mu_{\rm rel}$.  The right panel zooms into the local maximum and shows how $\rho_{\text{rel}}$ approaches $\CK_{\text{rel}}$ by increasing $\CN$.}{\label{fig:rrel1dPxD}}
\end{figure}

Analogous to section~\ref{sec:1dPxPCon}, one can define a density of states for $\PtD$ by diagonalizing the matrix $\CQ$ and parametrize the eigenvalues $q_n>\frac{1}{4}$ corresponding to principal series with $\lambda_n=\sqrt{q_n-\frac{1}{4}}$ . Strictly speaking, the spectral density derived from $\CQ$ in eq.~\reef{eq:CQ PD} corresponds to $\CF_\Delta\otimes \CD_p^+$ or $\CF_\Delta\otimes \CD_p^-$ separately. Therefore, we define
\begin{align}\label{eq:rbarPD}
\bar\rho_\CN^{\CF_\Delta\otimes \CD_p}(\lambda_n)\equiv 2\times \frac{2}{\lambda_{n+1}-\lambda_{n-1}}~
\end{align}
where the factor of two is to add up the equal contributions of $\CF_\Delta\otimes \CD_p^+$ and $\CF_\Delta\otimes \CD_p^-$.
In fig. (\ref{fig:rrel1dPxD}), we present the results of $\rrel$ defined similar to~\reef{eq:rhorel} as
\be
\rrel\equiv \bar{\rho}^{\CF_\Delta\otimes \CD_{p}}_{\CN} - \bar{\rho}^{\CP_{\half+i\mu_{\text{rel}}}\otimes \CP_{\half+i\mu
_\text{rel}}}_{\CN}~.
\ee
There is a perfect agreement between $\rrel$ in large $\CN$ and $\CK_{\text{rel}}$ found in~\reef{eq:charPtD}.

\subsection{$\CD_{p_1}\otimes \CD_{p_2}$}

The tensor product $\CD_{p_1}\otimes \CD_{p_2}$ consists of four parts
\begin{align}
\CD_{p_1}\otimes \CD_{p_2}=\left(\CD_{p_1}^+\otimes \CD_{p_2}^+\right)\oplus \left(\CD_{p_1}^-\otimes \CD_{p_2}^-\right)\oplus \left(\CD_{p_1}^+\otimes \CD_{p_2}^-\right)\oplus \left(\CD_{p_1}^-\otimes \CD_{p_2}^+\right)
\end{align}
where  $\CD_{p_1}^+\otimes \CD_{p_2}^+$  and $\CD_{p_1}^-\otimes \CD_{p_2}^-$ are trivial in the sense that they do not have a continuous part. Taking $\CD_{p_1}^+\otimes \CD_{p_2}^+$  as an example, the reason is that in the full Hilbert space, $L_0$ has a positive minimal eigenvalue $p_1+p_2$. In addition, it is an easy task to write down the full decomposition
\begin{align}
\CD_{p_1}^+\otimes \CD_{p_2}^+=\bigoplus_{k\ge p_1+p_2}\CD^+_{k},\,\,\,\,\, \CD_{p_1}^-\otimes \CD_{p_2}^-=\bigoplus_{k\ge p_1+p_2}\CD^-_{k}
\end{align}
Therefore we will focus on the tensor product of a lowest-weight discrete series representation  and a highest-weight discrete series representation, e.g. $\CD_{p_1}^+ \otimes \CD^-_{p_2}$, which certainly contains the continuous families of states (because there exist $L_0=0$ states).

\vspace{5pt}

\subsubsection{The discrete part}

The existence of a discrete series representation $\CD_k^+$ in the tensor product $\CD_{p_1}^+ \otimes \CD^-_{p_2}$ correspond to a normalizable $L_0=k$ state annihilated by $L_-$. Assuming $p_1\ge p_2$, the most general form of such a state is 
\begin{align}
|k)_k=\sum_{n\ge p_2}a_n |n+k,-n)
\end{align}
where the coefficients $a_n$ should be zero if $n+k\le p_1-1$. The lowest-weight condition $L_-|k)_k=0$ yields
\begin{align}
0&=a_{p_2}(k+p_2-p_1)|k+p_2-1,-p_2)\nonumber\\
&+\sum_{n\ge p_2}(a_{n+1}(k+n+1-p_1)-a_n(n+p_2))|k+n,-n-1)
\end{align}
which is equivalent to an initial condition and a recurrence relation
\begin{align}
a_{p_2}(k+p_2-p_1)=0, \,\,\,\,\, a_{n+1}(k+n+1-p_1)-a_n(n+p_2)=0~.
\end{align}
When $k\ge p_1-p_2+1$, all $a_n$ are identically vanishing. When $k\le p_1-p_2$, all $a_n$ with $p_2\le n\le p_1-k-1$ are zero and  the rest $a_n$ are given by  
\begin{align}
a_n=c\frac{\Gamma(n+p_2)}{\Gamma(k+n+1-p_1)}, \,\,\,\,\, n\ge p_1-k
\end{align}
where $c$ is an arbitrary normalization constant. The norm of $|k)_k$ becomes
\begin{align}
_k(k|k)_k=|c|^2 \sum_{n\ge p_1-k}\frac{\Gamma(n+p_2)\Gamma(n+1-p_2)}{\Gamma(n+k+1-p_1)\Gamma(n+k+p_1)}\sim |c|^2\sum_n n^{-2k}
\end{align}
which is convergent for $k\in\mathbb Z_+$.  Therefore $\{\CD^+_1,\cdots \CD^+_{p_1-p_2}\}$ belong to $\CD^+_{p_1}\otimes \CD^-_{p_2}$ when $p_1\ge p_2$.

For $\CD^-_k$, the highest-weight state $|-k)_k$ should take the following form
\begin{align}
|-k)_k=\sum_{n\ge p_1} b_n |n,-n-k)~.
\end{align}
The highest-weight condition $L_+|-k)_k=0$ yields 
\begin{align}\label{bnrec}
b_{n+1}=\frac{n+p_1}{n+k+1-p_2} b_n, \,\,\,\,\, b_{p_1}=0~.
\end{align}
Since $p_1\ge p_2$ only solution of eq.~\reef{bnrec} is $b_n\equiv 0$ for any $n\ge p_1$ and hence there does not  exist any $\CD_k^-$.

Similarly, we can show that when $p_1\le p_2$, the discrete part of $\CD^+_{p_1}\otimes \CD^-_{p_2}$ only consists of highest weight discrete series representations $\{\CD^-_1,\cdots \CD^-_{p_2-p_1}\}$. Altogether, the discrete part of $\CD^+_{p_1}\otimes \CD^-_{p_2}$ for any $p_1$ and $p_2$ can be summarized as 
\begin{align}
\CD^+_{p_1}\otimes \CD^-_{p_2}\supseteq\begin{cases}  \bigoplus_{k=1}^{p_1-p_2}\CD^+_k, & p_1>p_2\\
\varnothing, &p_1=p_2\\ \bigoplus_{k=1}^{p_2-p_1}\CD^-_k, & p_1<p_2\end{cases}
\end{align}
which further implies that 
\begin{align}
\left(\CD_{p_1}^+\otimes \CD_{p_2}^-\right) \oplus \left(\CD_{p_1}^-\otimes \CD_{p_2}^+\right)\supseteq \bigoplus_{1\le k\le |p_1-p_2|}\left( \CD^+_k\oplus\CD^-_k\right)
\end{align}
and 
\begin{align}\label{discDD}
\CD_{p_1}\otimes \CD_{p_2}\supseteq \bigoplus_{1\le k\le |p_1-p_2|,\,\, k\ge p_1+p_2}  \CD_k ~.
\end{align}

\subsubsection{The continuous part}

\noindent{}\textbf{Character analysis}

Use $\CP_{\Delta_3}\otimes \CP_{\Delta_4}$ to regularize the tensor product of $\CD_{p_1}\otimes \CD_{p_2}$. According to eq.~\reef{discDD}, the discrete series parts of  $\CD_{p_1}\otimes \CD_{p_2}$ and $\CP_{\Delta_3}\otimes \CP_{\Delta_4}$ have a finite mismatch. Taking into account this mismatch, the relative density of principal series should satisfy
\begin{align}
\Theta_{p_1}(q)\Theta_{p_2}(q)-\Theta_{\Delta_3}(q)\Theta_{\Delta_4}(q)=\int_0^\infty\,d\lambda\, \CK_{\text{rel}}(\lambda) \Theta_{\Delta_\lambda}(q)-\sum_{|p_1\!- p_2|\! < k < p_1\!+ p_2  }\Theta_k(q)
\end{align}
where $\Theta_k(q)=\frac{2\, q^k}{1-q}$ is the character of $\CD^+_k\oplus \CD^-_k$. Solving this equation yields 
\begin{align}\label{eq:KrelDxD}
\CK_{\text{rel}}(\lambda)& =\frac{1}{2\pi}\sum_{\pm,\pm,\pm}\psi\left(\frac{1}{2}\pm i\mu_3\pm i\mu_4\pm i\lambda\right)-\frac{2}{\pi}\sum_\pm\psi\left(p_1+p_2-\frac{1}{2}\pm i\lambda\right)\nonumber\\
&+\sum_{|p_1\!- p_2|< k < p_1\!+ p_2 }\frac{2}{\pi}\frac{k-\frac{1}{2}}{\left(k-\frac{1}{2}\right)^2+\lambda^2}~.
\end{align}
Since $p_1+p_2$ is always larger than $\frac{1}{2}$, there cannot be pole crossing in $\CK_{\text{rel}}(\lambda)$ and hence there is no complementary series in $\CD_{p_1}\otimes \CD_{p_2}$.

\vspace{5pt}

\noindent{}\textbf{Numerical check}

     \begin{figure}[t]
         \centering
         \includegraphics[width=0.6\textwidth]{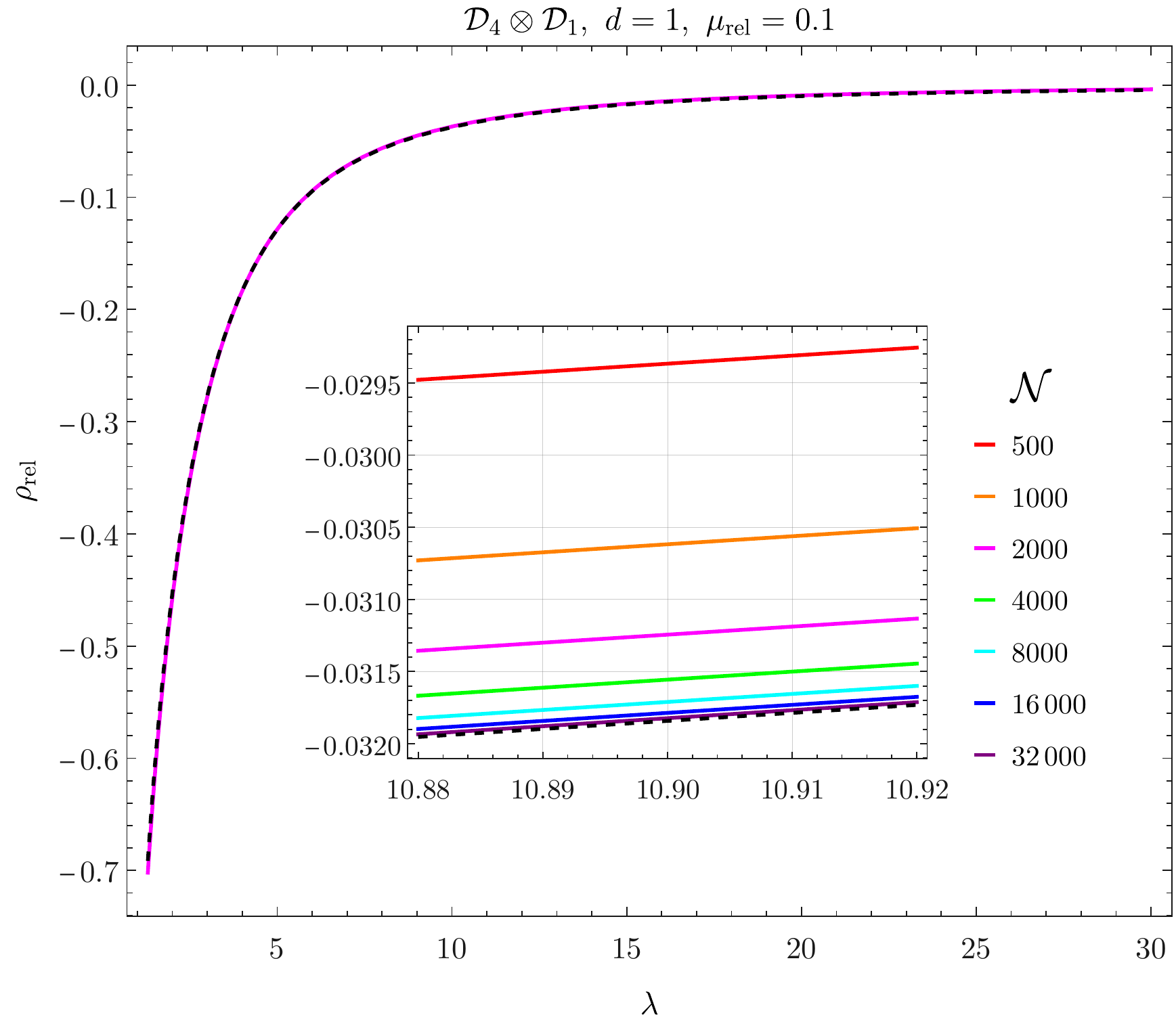}
         \caption{Demonstration of agreement between $\rho_\text{rel}$ (pink line) and $\CK_\text{rel}$ (black dashed line) for the tensor product $\CD_4\otimes\CD_1$ with cutoff $\CN=2000$. Inset plot shows the convergence of $\rho_\text{rel}$ to $\CK_\text{rel}$ (black dashed line) in large $\CN$. }{\label{fig:DxD1d}}
\end{figure}

For $\CD^+_{p_1}\otimes \CD^{-}_{p_2}$ and $\CD^+_{p_2}\otimes \CD^{-}_{p_1}$, the matrix elements of $\Cas^{\SO(1,2)}$ in the $L_0=0$ subspace take the same form
\begin{align}\label{dji}
 &\CQ_{nn}=2n^2+p_1(1-p_1)+p_2(1-p_2)\nonumber\\
&\CQ_{n+1,n}=\CQ_{n,n+1}=-\sqrt{(n+p_1)(n+p_2)(n+1-p_1)(n+1-p_2)}
\end{align}
where $n\ge \max(p_1,p_2)$. $L_0=0$ does not have access to discrete series but similar to how it was shown in section~\ref{sec:1dPxPCon} and section~\ref{sec:cont FtD}, one can use eq. (\ref{dji}) to see that all the principal series representations appear in the tensor product decomposition. 
 By introducing a cut-off $n\le \CN$, we truncate the infinite dimensional matrix $\CQ$, which leads to a coarse-grained density  $\bar\rho^{\CD_{p_1}\otimes \CD_{p_2}}_{\CN}$ of principal series.
 In fig.~\reef{fig:DxD1d}, similar to previous sections, we find a perfect agreement between spectral kernel from character analysis in~\reef{eq:KrelDxD} and the adopted version of~\reef{eq:rhorel}:
\begin{align}
\rrel\equiv \bar{\rho}^{\CD_{p_1}\otimes \CD_{p_2}}_{\CN} - \bar{\rho}^{\CP_{\half+i\mu_{\text{rel}}}\otimes \CP_{\half+i\mu
_\text{rel}}}_{\CN}.
\end{align}


\subsection{Identical particles}
\subsubsection{Two-particle Hilbert space}
When $\Delta_1=\Delta_2=\Delta$, the symmetrized tensor product $\CF_\Delta\odot\CF_\Delta\subset \CF_\Delta\otimes\CF_\Delta$ describes the two-particle Hilbert space of a scalar field with mass $m=\sqrt{\Delta(1-\Delta)}$. Since  $\CF_\Delta\odot\CF_\Delta$ is the permutation invariant subsector of $\CF_\Delta\otimes\CF_\Delta$, the $\SO(1,2)$ invariant subspaces in $\CF_\Delta\odot\CF_\Delta$ can be obtained by imposing permutation symmetry on the decomposition of $\CF_\Delta\otimes\CF_\Delta$. For the discrete part of $\CF_\Delta\otimes\CF_\Delta$, as shown in subsection \ref{CF1tCF2}, each $\CD^+_k$ is generated by a lowest-weight state $|k)_k$, c.f. eq.~(\ref{g2}) and eq.~(\ref{g2d}):
\begin{align}
\label{Dkstates}
|k)_k=\sum_n a_n |n,k-n), \,\,\,\,\, a_n=c \frac{\Gamma(n+\Delta-k)}{\Gamma(n+\bar\Delta)}~.
\end{align}
Noticing  $a_{k-n}=(-)^k a_n$, we find that $|k)_k$ and the whole representation $\CD_k^+$ has a parity $(-)^k$ under permutation. Therefore only $\CD_k^+$ with even $k$ can exist in $\CF_\Delta\odot\CF_\Delta$. The same conclusion also holds for the the lowest-weight discrete series. Altogether, the discrete part of $\CF_\Delta\odot\CF_\Delta$ consists of $\CD_2^\pm, \CD_4^\pm, \CD_6^\pm,\cdots$

For the continuous part of $\CF_\Delta\otimes\CF_\Delta$, the $\mathbb Z_2$ action $\mathcal T$ that maps $|\psi_n)$ to $|\psi_{-n})$ coincides with the permutation operator since $|\psi_n)=|n,-n)$ by definition. Then the eigenspaces $\CH_\pm\subset \CH_0$ of $\mathcal T$ have eigenvalue $\pm 1$   under permutation respectively. Therefore, the principal series and complementary series contained in $\CF_\Delta\odot\CF_\Delta$ correspond to the spectrum of $\SO(1,2)$ Casimir restricted to the permutation invariant subspace $\CH_+$.

To understand the continuous part from character side, we need the following well-established fact in representation theory of compact Lie groups. Let $R$ be an irrep of $G$, then the character of $R\odot R$ is given by\footnote{It is actually very easy to prove this relation. Let $|n\rangle$ be an normalized and orthogonal basis of $R$. One can easily check that $\Pi_+=\frac{1}{2}\sum_{n,m}|n,m\rangle\langle n,m|+|n,m\rangle\langle m,n|$ is a projection operator onto the permutation invariant subspace of $R\otimes R$. Therefore evaluating the character $\chi_{R\odot R}(g)$ is equivalent to computing the trace of $g \Pi_+$ in  $R\otimes R$, i.e. 
\begin{align}
\chi_{R\odot R}(g)&=\tr(g\Pi_+)=\frac{1}{2}\sum_{n,m}\left(\langle n,m |g|n,m\rangle +\langle n,m |g|m,n\rangle\right)\nonumber\\
&=\frac{1}{2}\left(\sum_n \langle n |g|n\rangle\right)^2+\frac{1}{2}\sum_{n,m}\langle n|g|m\rangle \langle m|g|n\rangle
\end{align}
where the first term is exactly $\frac{1}{2}\chi_R(g)^2$. In the second term, using $\sum_m |m\rangle\langle m|=1_R$, we are left with trace of $g^2$, which is nothing but the character $\chi_R(g^2)$.}
\begin{align}\label{charofRsR}
\chi_{R\odot R}(g)=\frac{1}{2}\left(\chi_R(g^2)+\chi_R(g)^2\right)~.
\end{align}
For $G=\SO(1,2)$ and $R=\CF_\Delta$, eq.~(\ref{charofRsR}) means 
\begin{align}\label{hujhi}
\Theta_{\CF_\Delta\odot\CF_\Delta}(q)=\frac{1}{2}\frac{q^{2\Delta}+q^{2\bar\Delta}+2 q}{(1-q)^2}+\frac{1}{2} \frac{q^{2\Delta}+q^{2\bar\Delta}}{1-q^2}~.
\end{align}
For $\Delta_k=\frac{1}{2}+i\mu_k, k=1,2$, let $\CK_{\text{rel}}(\lambda)$ be the relative density of principal series between $\CP_{\Delta_1}\odot\CP_{\Delta_1}$ and $\CP_{\Delta_2}\odot\CP_{\Delta_2}$, and then it should satisfy 
\small
\begin{align}
\int_0^\infty\, d\lambda\,\CK_{\text{rel}}(\lambda) \Theta_{\frac{1}{2}+i\lambda}(q)
=\frac{1}{2}\left(\frac{q^{2\Delta_1}+q^{2\bar\Delta_1}-q^{2\Delta_2}-q^{2\bar\Delta_2}}{(1-q)^2}+\frac{q^{2\Delta_1}+q^{2\bar\Delta_1}-q^{2\Delta_2}-q^{2\bar\Delta_2}}{1-q^2}\right)
\end{align}
\normalsize
which yields
\begin{align}\label{ypiup}
\CK_{\text{rel}}(\lambda)&=\frac{1}{4\pi}\sum_{\pm\pm}\left(\psi\left(\frac{1}{2}\pm 2i\mu_2\pm i\lambda\right)-\psi\left(\frac{1}{2}\pm 2i \mu_1\pm i\lambda\right)\right)\nonumber\\
&+\frac{1}{4}\sum_\pm \left(\frac{1}{\cosh(\pi(\lambda\pm 2\mu_1))}-\frac{1}{\cosh(\pi(\lambda\pm 2\mu_2))}\right)~.
\end{align}

As explained in \cite{Sun:2021thf},  the single particle states of a free massive scalar field in dS$_2$ can be identified with the principal series representation $\CP_{\Delta}$ if the mass squared $m^2 = \Delta(1-\Delta) >\frac{1}{4}$.
The states  $|n)$, introduced in section \ref{introSO12}, form an eigenbasis of $L_0$ which is the $\SO(2)$ generator of global  dS$_2$.
The discussion above, implies that the two-particle Hilbert space of the same scalar field contains the discrete series representations $\CD_2^\pm, \CD_4^\pm, \CD_6^\pm,\cdots$.
For example, $\CD_k^+$ contains the normalizable lowest-weight state $|k)_k$, given in \eqref{Dkstates}, that is annihilated by the generator $L_-$. One can think of such UIRs as purely right moving in dS$_2$.

\begin{figure}[t]
     \centering
     \begin{subfigure}[t]{0.43\textwidth}
         \centering
         \includegraphics[width=\textwidth]{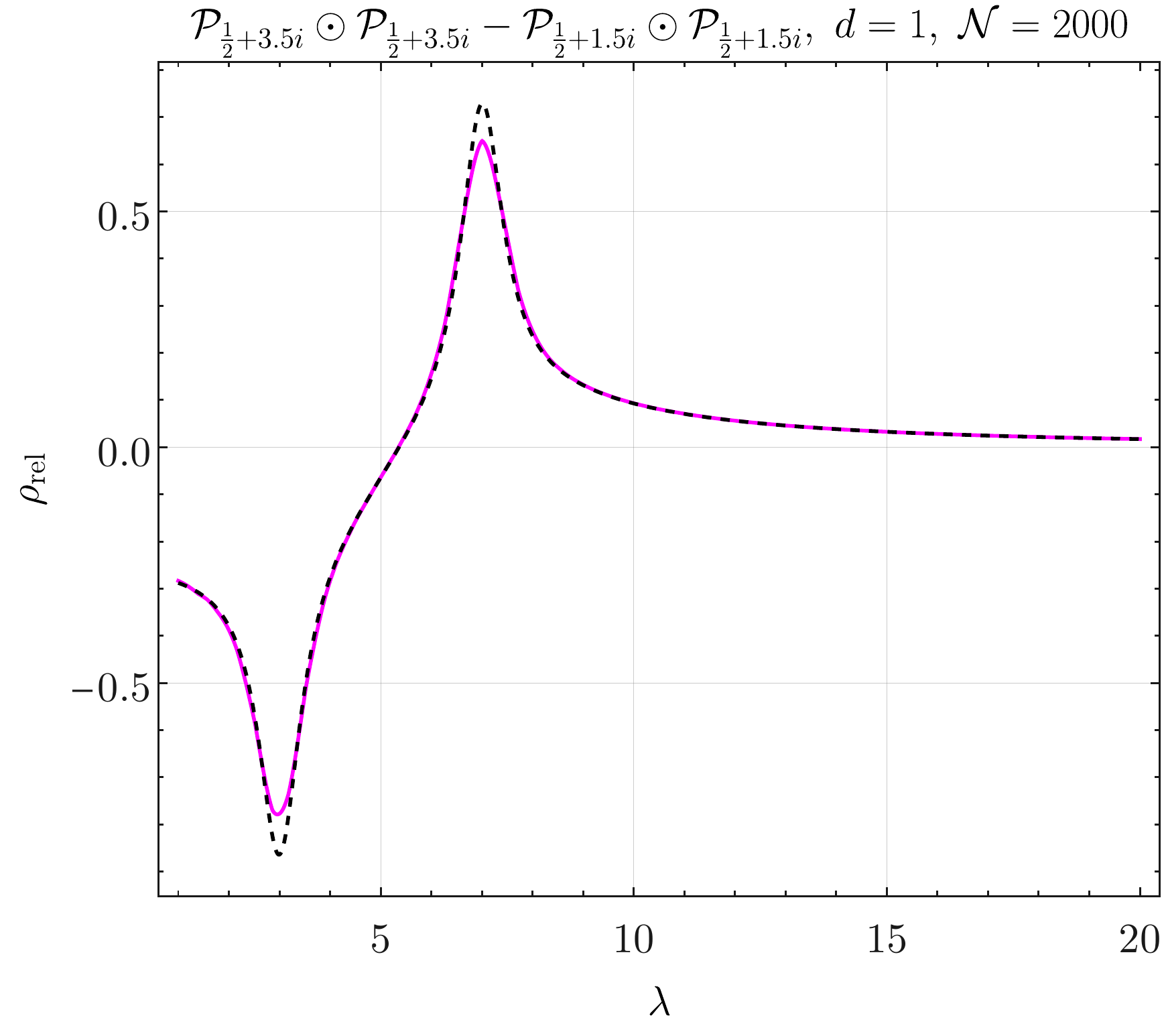}
         \label{rrel1dIdent}
     \end{subfigure}
     \hfill
     \begin{subfigure}[t]{0.5\textwidth}
         \centering
         \includegraphics[width=\textwidth]{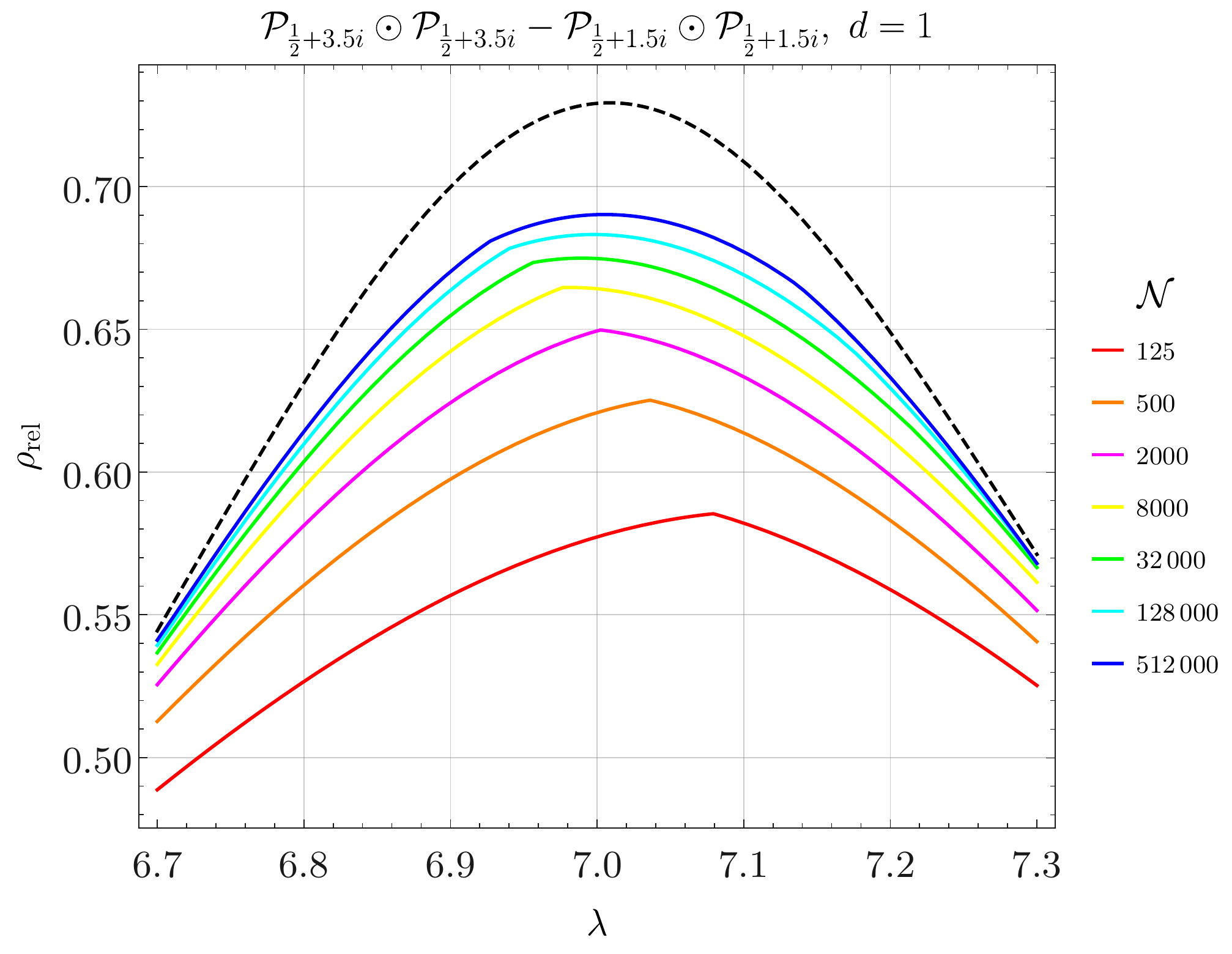}
               \label{fig:rrel1dZoomIdent}
     \end{subfigure}
     \caption{Relative density of principal series between $\CP_{\frac{1}{2}+3.5 i}\odot\CP_{\frac{1}{2}+3.5 i}$ and $\CP_{\frac{1}{2}+1.5 i}\odot\CP_{\frac{1}{2}+1.5 i}$. In the left panel, the pink line is  $\rho_\text{rel}$ and black dashed line is $\CK_{\text{rel}}$, given by eq. (\ref{ypiup}).  The right panel zooms into the local maximum and shows how $\rho_{\text{rel}}$ approaches $\CK_{\text{rel}}$ (black dashed line) by increasing $\CN$.}\label{fig:Identical1d}
\end{figure}

Similar to section~\ref{sec:1dPxPCon}, for a numerical check of the analysis above, we need to construct the matrix representation of the Casimir operator restricted to the $L_0=0$ sector of $\CF_\Delta\odot\CF_\Delta$. However, it does not require any extra work since the $L_0=0$ sector of the symmetrized tensor $\CF_\Delta\odot\CF_\Delta$ is nothing but the subspace $\CH_+$ of the usual tensor product $\CF_\Delta\otimes \CF_\Delta$. So the Casimir is given by $\CQ^{(+)}$ of $\CF_\Delta\otimes \CF_\Delta$, c.f. eq. (\ref{QpmPC}). This simple observation has two immediate applications. First, $\CF_{\Delta}\odot\CF_{\Delta}$ and $\CF_{\Delta}\otimes\CF_{\Delta}$ contain the same complementary series representation, if exists, because  we have checked numerically  $\CQ^{(-)}>\frac{1}{4}$. Second, the coarse-grained density of principal series in (truncated) $\CF_{\Delta}\odot\CF_{\Delta}$  is the same as $\bar\rho_{\CN}^+(\lambda)$ for (truncated) $\CF_\Delta\otimes\CF_\Delta$, c.f. eq. (\ref{brdef}). With the coarse-grained density known, we further define a relative density $\rho_{\rm rel}$ between $\CF_{\Delta_1}\odot\CF_{\Delta_1}$ and $\CF_{\Delta_2}\odot\CF_{\Delta_2}$. A comparison of $\rho_{\rm rel}$ and  $\CK_{\text{rel}}$ (c.f. eq. (\ref{ypiup})) is shown in fig. (\ref{fig:Identical1d}).

\subsubsection{Counting complementary series in multi-particle Hilbert space}
In section \ref{CF1tCF2}, we have shown that the tensor product $\CC_{\frac{1}{2}+\mu_1}\otimes\CC_{\frac{1}{2}+\mu_2}$ contains exactly one complementary series representation of scaling dimension $\mu_1+\mu_2$ when $\mu_1+\mu_2>\frac{1}{2}$. This result also holds for symmetrized tensor product in the sense that $\CC_{2\mu}\in \CC_{\frac{1}{2}+\mu}\odot\CC_{\frac{1}{2}+\mu}$  when $2\mu>\frac{1}{2}$. In the QFT language, it means that given a light scalar\footnote{ We use the word ``light'' for a field whose single-particle Hilbert space is described by a complementary series representation. } $\phi$ of scaling dimension $\Delta=\frac{1}{2}+\mu$ or equivalently mass $m^2=\frac{1}{4}-\mu^2$, its two-particle Hilbert space contains an invariant subspace corresponding to a  light scalar of mass $m^2=2\mu(1-2\mu)$ when $\mu>\frac{1}{4}$. It is then natural to ask if there is any light scalar in any $M$-particle Hilbert space of $\phi$, and if yes, what is the corresponding mass. Group theoretically, the $M$-particle Hilbert space is described by the symmetrized  tensor product $\CC_\Delta^{\odot M}$, whose character $\Theta_{\CC^{\odot M}_{\Delta}}(q)$ can be extracted from the following generating function
\begin{align}
\mathbb P(q,  t)\equiv \exp \left(\sum_{k\ge 1} \Theta_{\CC_\Delta}(q^k) \,\frac{t^k}{k}\right)=\sum_{M\ge 0}\Theta_{\CC^{\odot M}_{\Delta}}(q) \, t^M
\end{align}
In particular, it is easy to check for $M=2$ we recover eq. (\ref{hujhi}).

As discussed in section \ref{sec:1dPxPCon}, the first several terms in the  small $q$ expansion of $\Theta_{\CC^{\odot M}_{\Delta}}(q)$ contain all the information regarding invariant subspaces of $\CC^{\odot M}_{\Delta}$ that correspond to complement series. The generating function $\mathbb P(q,  t)$ provides a simple way to derive such expansions as follows. 
In the exponent of $\mathbb P(q, t)$, for each fixed $k$, we expand  $\Theta_{\CC_\Delta}(q^k)$ as a series $(q^{k\Delta}+q^{k\bar\Delta})\sum_{n\ge 0} q^{n k}$. Switching the order of the two infinite sums, then the sum over $k$ yields two logarithmic functions, which allows us to rewrite $\mathbb P(q,  t)$ as 
\begin{align}
\mathbb P(q,  t)&=\prod_{n\ge 0}\frac{1}{1-t \,q^{\Delta+n}}\frac{1}{1-t \,q^{\bar\Delta+n}}=\sum_{\bm k, \bm \ell}\, q^{\alpha (\bm k, \bm \ell)}t^{\sum_{n} (k_n+\ell_n)}
\end{align}
where $\mu=\Delta-\frac{1}{2}$ and 
\begin{align}
\alpha(\bm k, \bm \ell)=\sum_{n\ge 0}\left(n+\frac{1}{2}\right)(k_n+\ell_n)+\mu \sum_{n\ge 0}(k_n-\ell_n)
\end{align}
Fixing the particle number $\sum_{n}(k_n+\ell_n)=M$, we have 
\begin{align}
\alpha(\bm k, \bm \ell)=\Delta M+\sum_{n\ge 0} n(k_n+\ell_n)-2\mu\sum_{n\ge 0}\ell_n
\end{align}
Noticing that $\sum_{n\ge 0} n(k_n+\ell_n)\ge 0$ and $\sum_{n\ge 0}\ell_n\le \sum_{n\ge 0}(k_n+\ell_n)=M$, where the two ``='' hold if and only if $\bm k=0$ and $\bm \ell=(M, 0,0,\cdots)$, we find the minimal value of  $\alpha(\bm k, \bm \ell)$ to be $\bar\Delta M$. To obtain the second minimal value of $\alpha(\bm k, \bm \ell)$, which corresponds to $\ell_0\le M-1$, we rewrite it as
\begin{align}
\alpha(\bm k, \bm \ell)=\Delta M-2\mu \ell_0+\sum_{n\ge 1}( n (k_n+\ell_n)-2\mu \ell _n)\ge \Delta M-2\mu \ell_0 +\sum_{n\ge 1} n k_n
\end{align}
where we have used $2\mu<1\le n$ in the sum over $n$. Then it is easy to see $\alpha(\bm k, \bm \ell)\ge M\bar\Delta+2\mu$ when $\ell_0\le M-1$. The equality holds for $\bm k=(1,0,0,\cdots)$ and $\bm\ell=(M-1, 0,0,\cdots)$. Altogether, we obtain the leading and subleading terms in the small $q$ expansion of $\Theta_{\CC^{\odot M}_{\Delta}}(q)$
\begin{align}
\Theta_{\CC^{\odot M}_{\Delta}}(q)= q^{M\bar\Delta}+ q^{M\bar\Delta+2\mu}+\text{higher order terms in}\,\, q
\end{align}
When $M\bar\Delta<\frac{1}{2}$, which corresponds to $\mu>\frac{M-1}{2M}$, there has to be one  copy of $\CC_{1-M\bar\Delta}$ in the symmetrized tensor product $\CC^{\odot M}_{\Delta}$ to cancel $q^{M\bar\Delta}$ since the latter  cannot be reproduced by an integral over principal series characters. On the other hand, because of  $M\bar\Delta+2\mu=(M-2)\bar\Delta+1\ge 1$ for $M\ge 2$, there should not be any extra complementary series representations. It is worth mentioning that the same results also hold for the tensor product $\CC_\Delta^{\otimes M}$ which can be easily checked by using the rule of $\CC_{\Delta_1}\otimes  \CC_{\Delta_2}$. A simple reason for the coincidence from the character viewpoint is that the first two terms in the small $q$ expansion of  $\Theta_{\CC^{\otimes M}_{\Delta}}(q)$ are $q^{M\bar\Delta}+M q^{M\bar\Delta+2\mu}$. 
Altogether, in the multi-particle Hilbert space of $\phi$, there exist $n_{\rm max}\equiv \left \lfloor{\frac{1}{1-2\mu}}\right \rfloor$ complementary series representations, with scaling dimensions $1-\bar\Delta, 1-2\bar\Delta, \cdots, 1-n_{\rm max} \bar\Delta $.

\section{Tensor products in $\SO(1, d+1)$}
\label{tensorhigh}
In this section, we generalize the character and numerical methods developed in section \ref{tensorlow} to study tensor product decomposition of higher dimensional $\SO(1,d+1)$, focusing on principal and complementary series of spin 0.
Let us first consider the tensor product of two scalar principal series representations  $\CP_{\Delta_1}\otimes \CP_{\Delta_2}$ with $\Delta_k=\frac{d}{2}+i\mu_k$. Each $\CP_{\Delta_k}$ consists of all the single-row representations of $\SO(d+1)$. Using the following decomposition rule of orthogonal groups
\begin{align}\label{mell}
\mY_{n}\otimes \mY_\ell=\bigoplus_{a=0}^{\text{min}(n,\ell)}\bigoplus_{b=0}^{\text{min}(n,\ell)-a}\mY_{n+\ell-2a-b,b}~,
\end{align}
we can immediately conclude that the $\SO(d+1)$ content of $\CP_{\Delta_1}\otimes \CP_{\Delta_2}$  has at most two rows in the Young diagram language. This condition excludes a large class of $\SO(1, d+1)$ UIRs, including discrete series when $d\ge 5$, because discrete series representations should contain $\SO(d+1)$ content that has $\frac{d+1}{2}\ge 3$ rows.

Thus, $\CP_{\Delta_1}\otimes \CP_{\Delta_2}$ only contains the four types of UIRs which are reviewed in section \ref{repreview}. 
Among these UIRs, the exceptional series should be thought as the analogue of  discrete series in the $d=1$ case, since they can be characterized by certain linear relations (in terms of $\so(1,d+1)$ generators)\footnote{When $d=1$, these linear relations simply refer to the highest-weight or lowest-weight conditions, and for higher $d$, they are discussed in  appendix \ref{Noex}.} while the two continuous series can only be identified by the quadratic operator $\Cas^{\SO(1,d+1)}$. In appendix \ref{Noex}, we show that the exceptional series  cannot be present in this tensor product by directly checking the linear relations. Further assuming the absence of complementary series, which is actually proved  by \cite{Dobrev:1976vr} and will also be confirmed numerically later, the decomposition of  $\CP_{\Delta_1}\otimes \CP_{\Delta_2}$ can be written {\it schematically} as 
\begin{align}\label{PPsch2}
\CP_{\Delta_1}\otimes \CP_{\Delta_2}=\sum_{s\ge 0}\int_0^\infty d\lambda\, \CK^{(s)}(\lambda)\, \CP_{\Delta_\lambda, s}, \,\,\,\,\,\Delta_\lambda=\frac{d}{2}+i\lambda
\end{align}
where $\CK^{(s)}(\lambda)$ is {\it formally} the density of spin $s$ principal series. Proceeding with eq.~(\ref{PPsch2}) and replacing each UIR by the corresponding Harish-Chandra character, we would end up with a divergent $\CK^{(s)}(\lambda)$. Following the strategy used in section \ref{tensorlow}, we regularize $\CK^{(s)}(\lambda)$ by introducing two more principal series representation $\CP_{\Delta_3}$ and $\CP_{\Delta_4}$ as a reference, which then yields
\begin{align}\label{PPSch2char}
\Theta_{\Delta_1}(q,\bm x)\Theta_{\Delta_2}(q,\bm x)-\Theta_{\Delta_3}(q,\bm x)\Theta_{\Delta_4}(q,\bm x)=\sum_{s\ge 0}\int_0^\infty d\lambda\, \CK^{(s)}_{\rm rel}(\lambda)\, \Theta_{\lambda,s}(q,\bm x)
\end{align}
where $\Theta_{\lambda,s}$ means  the character of $\CP_{\Delta_\lambda,s}$. The character equation (\ref{PPSch2char}) implies that $\CK^{(s)}_{\rm rel}(\lambda)$ should be understood as the relative density of spin $s$ principal series between the two tensor products $\CP_{\Delta_1}\otimes \CP_{\Delta_2}$ and $\CP_{\Delta_3}\otimes \CP_{\Delta_4}$. 

\subsection{The relative density}
By extending $\CK^{(s)}_{\rm rel}(\lambda)$ to an even function of $\lambda$ on the real line and using the explicit expression of Harish-Chandra characters, c.f. eq.~(\ref{kj;}), we rewrite (\ref{PPSch2char}) as
\small
\begin{align}\label{Chareq2}
\sum_{s\ge 0}\chi^{\SO(d)}_{\mY_s}(\bm x)\int_{\mathbb R}d\lambda\,\CK^{(s)}_{\rm rel}(\lambda)e^{i\lambda t}=\frac{4\,e^{-\frac{d}{2}|t|}}{P_d(e^{-|t|}, \bm x)}\left(\cos(\mu_1 t)\cos(\mu_2 t)-\cos(\mu_3 t)\cos(\mu_4 t)\right)
\end{align}
\normalsize
where we have made the substitution $q=e^{-|t|}$. To obtain an integral equation for each $\CK^{(s)}_{\rm rel}(\lambda)$, we need to expand the R.H.S of (\ref{Chareq2}) into an infinite sum of $\chi^{\SO(d)}_{\mY_s}(\bm x)$. Such an expansion exists thanks to the following relation 
\begin{align}\label{techguess0}
\sum_{s=0}^\infty \, \chi^{\SO(d)}_{\mY_s} (\bm x)\, q^s=\frac{1-q^2}{P_d(q,\bm x)}
\end{align}
which is proved in appendix \ref{sumW} for any dimension $d\ge 3$.
Plugging eq.~(\ref{techguess0}) into eq.~(\ref{Chareq2}) yields
\begin{align}\label{CKsint}
\int_{\mathbb R}d\lambda\,\CK^{(s)}_{\rm rel}(\lambda)e^{i\lambda t}=2 \, e^{-(\frac{d}{2}+s-1)|t|}\, \frac{\cos(\mu_1 t)\cos(\mu_2 t)-\cos(\mu_3 t)\cos(\mu_4 t)}{|\sinh(t)|}~.
\end{align}
Then $\CK^{(s)}_{\rm rel}(\lambda)$ can be computed as an inverse Fourier transform using eq.~(\ref{CIresult})
\begin{align}\label{CKsfinal}
\CK^{(s)}_{\rm rel}(\lambda)=\frac{1}{4\pi}\sum_{\pm,\pm,\pm}\left[\psi\left(\frac{\frac{d}{2}+s\pm i\lambda\pm i\mu_3\pm i\mu_4}{2}\right)-\psi\left(\frac{\frac{d}{2}+s\pm i\lambda\pm i\mu_1\pm i\mu_2}{2}\right)\right]
\end{align}
where $\sum_{\pm,\pm,\pm}$ means summing over the 8 sign combinations. Altogether, there are 16 $\psi$-functions in $\CK^{(s)}_{\rm rel}(\lambda)$.

\subsection{Truncated model of the relative density}
To ascertain it makes sense to identify $\CK^{(s)}_{\rm rel}(\lambda)$ as a relative density,  we would like to compare this to a model with discretized spectrum of eigenvalues. In particular, we will focus on the $s=0$ case, which corresponds to scalar principal series.

As reviewed in section \ref{repreview}, an exclusive feature of any $\CF_{\Delta, \ell}$ with $\ell=0$ is that it contains the trivial representation of $\SO(d+1)$. Therefore we should be able to extract all information about $\CP_{\Delta_\lambda}$ in $\CP_{\Delta_1}\otimes\CP_{\Delta_2}$ by only studying the $\SO(d+1)$ singlet subspace of the latter, denoted by $\CH_0$.
$\SO(d+1)$ singlets can only appear in the tensor product of two $\mY_n$ subspaces, one in $\CP_{\Delta_1}$ and the other one in $\CP_{\Delta_2}$. This structure yields a natural way to define a truncated model $\CH^{(\CN)}_0$ of $\CH_0$ by simply cutting off the $\SO(d+1)$ spins in each $\CP_{\Delta_i}$, e.g. $0\le n\le \CN$. More explicit, let $\left\{|n,\Delta_i \rangle_{a_1\cdots a_n}, a_j=1,\cdots, d+1\right\}$ be a  basis of the  $\mY_n$ subspace in $\CP_{\Delta_i}$, where $(a_1\cdots a_n)$ are symmetric and traceless. The normalization of $|n,\Delta_i \rangle_{a_1\cdots a_n}$ is specified by eq.~(\ref{expinner}). We can then construct a normalized and orthonormal basis of  the $(\CN+1)$ dimensional Hilbert space $\CH^{(\CN)}_0$ as 
\begin{align}
|\psi_n\rangle=\frac{1}{\sqrt{D^{d+1}_n}}|n,\Delta_1\rangle_{a_1\cdots a_n}|n,\Delta_2\rangle_{a_1\cdots a_n}, \quad 0\le n\le \CN
\end{align}
where $D^{d+1}_{n}=\text{dim}_{\SO(d+1)}(\mY_n)$.  Defining $\mathcal Q_{nm}\equiv \langle \psi_n|\CC_2^{\SO(1, d+1)}|\psi_m\rangle$,  the nonvanishing entries of $\mathcal Q$  are given by eq.~(\ref{PPoff})
\begin{align}\label{CQ12}
&\mathcal Q_{nn}=\Delta_{1}\bar\Delta_{1}+\Delta_{2}\bar\Delta_{2}+2n(n+d-1)\nonumber\\
&\mathcal Q_{n+1,n}=\mathcal Q^{*}_{n,n+1}=-\sqrt{\frac{(n+1)(n+d-1)}{(n+\frac{d-1}{2})(n+\frac{d+1}{2})}}(\Delta_{1}+n)(\Delta_{2}+n)~.
\end{align}
For large $\CN$, each eigenvalue $q_{n}$ of $\CQ$ that is larger than $\frac{d^2}{4}$ corresponds to a scalar principal series representation with  $\Delta=\frac{d}{2}+i\lambda_n$, where $\lambda_n=\sqrt{q_n-\frac{d^2}{4}}$. A coarse-grained density of scalar principal series representations in the truncated model defined by $\CQ$ is given by  the inverse spacing of $\{\lambda_n\}$
\begin{align}\label{brdef1}
\bar\rho_\CN(\lambda_n)\equiv \frac{2}{\lambda_{n+1}-\lambda_{n-1}}
\end{align}
 This coarse-grained density blows up when $\CN\to \infty$ because the matrix $\mathcal Q$ has a continuous spectrum in $\left[\frac{d^2}{4},\infty\right)$ in the large $\CN$ limit. Let $a_n(\lambda)$ be an eigenvector of $\mathcal Q$ with eigenvalue $\frac{d^2}{4}+\lambda^2, \lambda\ge 0$, i.e. 
 \begin{align}
 \mathcal Q_{n,n-1} a_{n-1}(\lambda)+\mathcal Q_{n,n}a_{n}(\lambda)+ \mathcal Q_{n,n+1} a_{n+1}(\lambda)=\left(\frac{d^2}{4}+\lambda^2\right) a_n(\lambda)~.
 \end{align}
 This recursion relation leads to the following asymptotic behavior of $a_n(\lambda)$
 \begin{align}\label{eigenasy}
 a_n(\lambda)=R_+ \frac{n^{i(\mu_1+\mu_2+\lambda)}}{\sqrt{n}}\left(1+\CO(1/n)\right)+R_- \frac{ n^{i(\mu_1+\mu_2-\lambda)}}{\sqrt{n}}(\left(1+\CO(1/n)\right)
 \end{align}
 where the coefficients $R_\pm$ cannot be determined from an asymptotic analysis of the recursion relation. The asymptotic behavior~\reef{eigenasy} implies that the eigenvectors of $\mathcal Q$  are $\delta$-function normalizable in the $\CN\to \infty$ limit.
\begin{figure}[t!]
     \centering
     \begin{subfigure}[t]{0.45\textwidth}
         \centering
         \includegraphics[width=\textwidth]{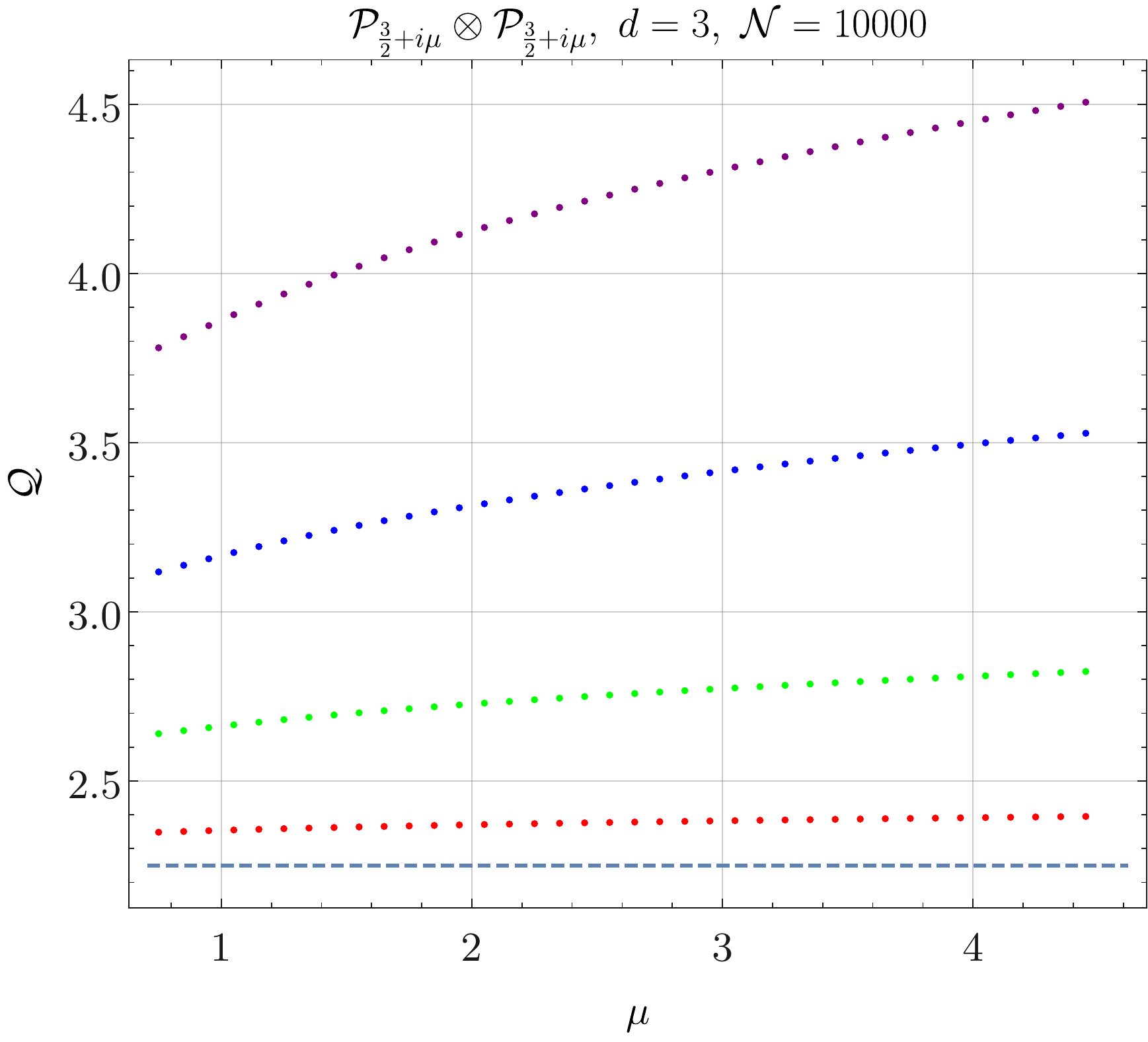}
         \label{fig:hdPxPspectrum}
     \end{subfigure}
     \hfill
     \begin{subfigure}[t]{0.46\textwidth}
         \centering
         \includegraphics[width=\textwidth]{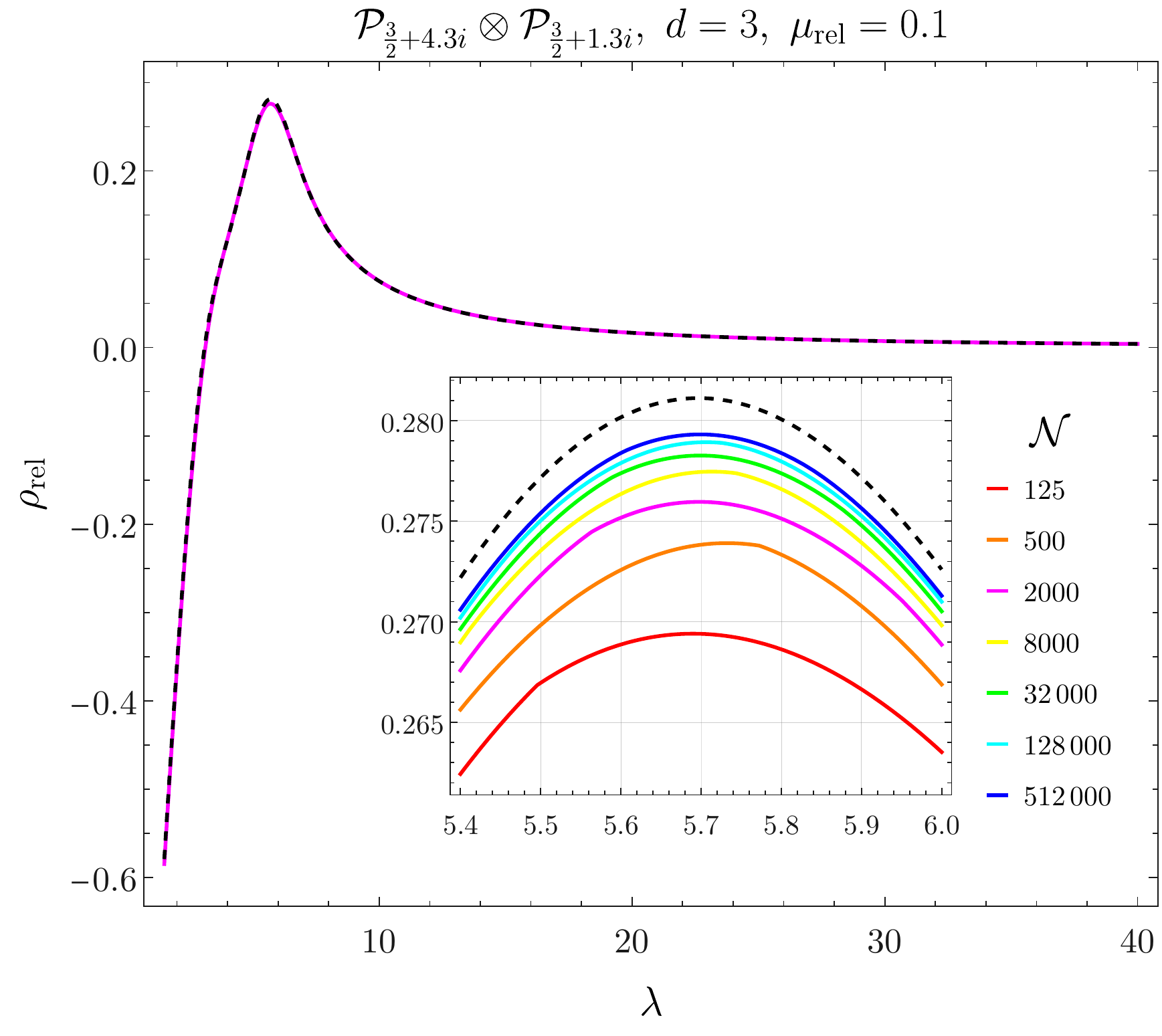}
         \label{fig:rrelhdPxP}
     \end{subfigure}
     \caption{Tensor product of two scalar principal series representations of $\SO(1,4)$.  Left: Low-lying eigenstates of the matrix $\CQ$ of the tensor product $\CP_{\frac{3}{2}+i\mu} \otimes  \CP_{\frac{3}{2}+i\mu}$ for various $\mu$. The cutoff is $\CN=10^4$.
    Right: Comparison of the relative coarse-grained  density $\rho_{\text{rel}}(\lambda)$ (pink solid line) given by eq. (\ref{uoiu}) to  $\CK^{(0)}_{\text{rel}}(\lambda)$ (black dashed line) given by eq. (\ref{CKsfinal}) for $\D_1={\frac{3}{2}+4.3i}, \D_2 ={\frac{3}{2}+1.3i}, \D_3=\D_4= \frac{3}{2}+i\mu_\text{rel}=\frac{3}{2}+0.1i$, and $\CN=2000$. The inset plot, zooming into the range $5.4<\lambda<6$, illustrates how $\rho_{\rm rel}$ approaches  $\CK^{(0)}_{\text{rel}}(\lambda)$ as we increase $\CN$.}{\label{fig:hdPxP}}
\end{figure}

 The difference of two coarse-grained densities, on the other hand, has a finite large $\CN$ limit. For example, given  principal series $\CP_{\Delta_3}$ and $\CP_{\Delta_4}$, we can construct another  $(\CN+1)\times(\CN+1)$ matrix like eq.~\reef{CQ12} and extract a coarse-grained density  $\bar{\rho}^{\CP_{\Delta_3}\otimes \CP_{\Delta_4}}_{\CN} $ from it.  Define a relative coarse-grained density as 
 \begin{align}\label{uoiu}
\rrel\equiv \bar{\rho}^{\CP_{\Delta_1}\otimes \CP_{\Delta_2}}_{\CN} - \bar{\rho}^{\CP_{\Delta_3}\otimes \CP_{\Delta_4}}_{\CN}~.
 \end{align}
 In fig. (\ref{fig:hdPxP}), we numerically show that  $\rrel(\lambda)$ approaches $\CK^{(0)}_{\rm rel}(\lambda)$, c.f. eq.~\reef{CKsfinal}, as $\CN\to\infty$. 

 Similar to the observation we had in section~\ref{sec:1dPxPCon}, we can define a  Pauli-Villars regularization of $\CK^{(0)}(\lambda)$ by taking $\mu_3=\mu_4=\Lambda$ large in $\CK^{(0)}_{\rm rel}(\lambda)$
 \small
  \be\label{eq:CKepsilonhd}
\CK_{\rm PV,1}^{(0)}(\lambda)= \frac{1}{\pi}\log\Lambda+\frac{1}{2\pi}\sum_{\pm}\psi\left(\frac{\frac{d}{2}\pm i\lambda}{2}\right)- \frac{1}{4\pi}\sum_{\pm,\pm,\pm}\psi\left(\frac{\frac{d}{2}\pm i\lambda\pm i\mu_1\pm i\mu_2}{2}\right)+\CO(\Lambda^{-1})~.
 \ee
 \normalsize
We notice that for each fixed $\CN$, $\CK_{\rm PV,1}^{(0)}(\lambda)$ matches the coarse-grained density $\bar\rho_\CN(\lambda)$ with a properly chosen $\Lambda$. In fig.~\reef{fig:rhoNPxPhd},  we show the remarkable match for the case of $\CP_{\frac{3}{2}+4.3i}\otimes\CP_{\frac{3}{2}+1.3 i}$. In the coarse-grained density the cutoff is $\CN=2000$, and in the Pauli-Villars regularization the cutoff is $\Lambda\approx4000$.

 \begin{figure}[t]
     \centering
         \includegraphics[width=0.6\textwidth]{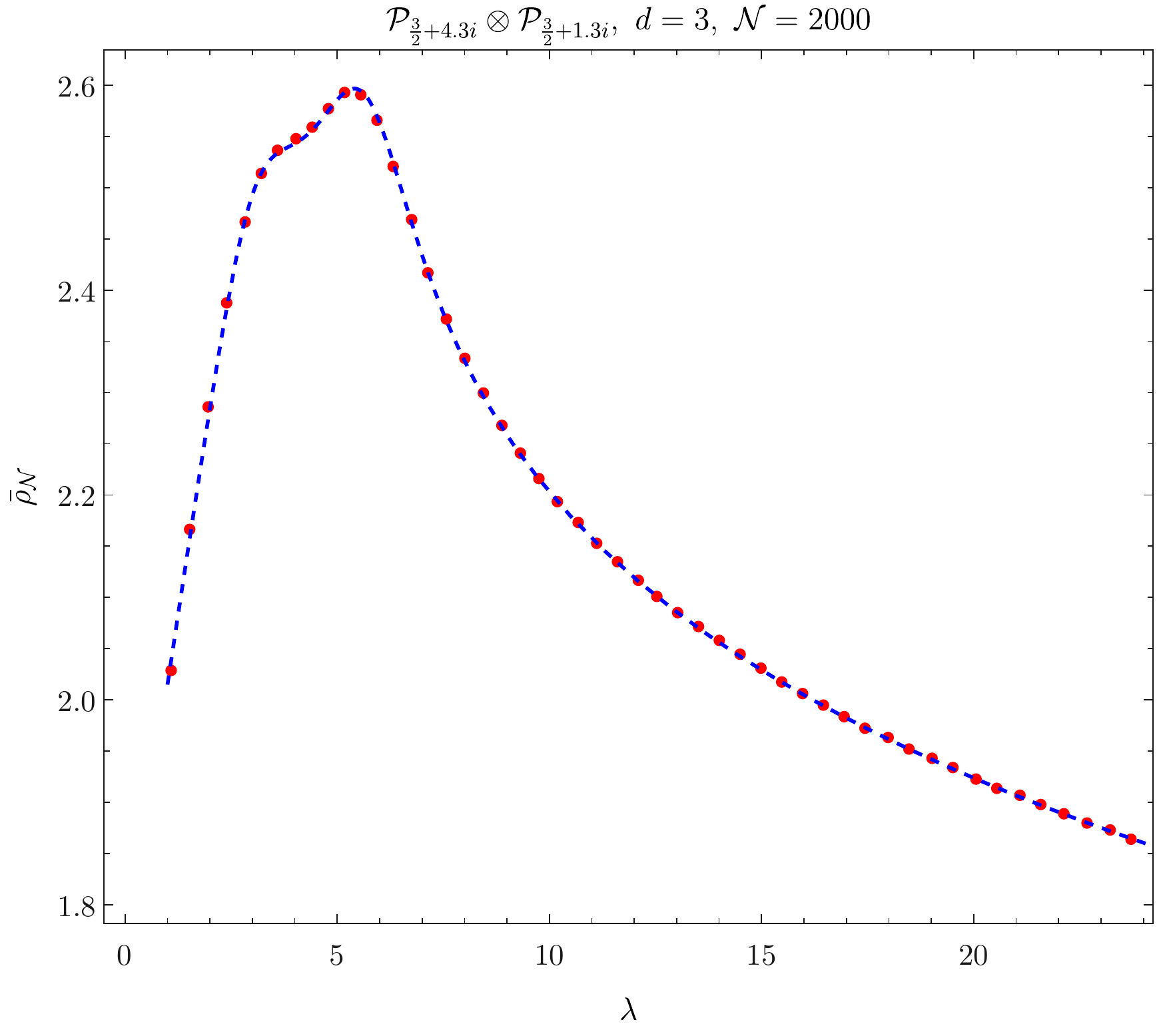}
         \caption{Plot of the course-grained spectral density $\bar{\rho}_\CN$ (red dots) and its perfect match with $\CK^{(s=0)}_{\rm PV,1}(\lambda)$ (blue dashed line) defined in eq.~\reef{eq:CKepsilonhd} with $\Lambda\approx 4000$.} 
         \label{fig:rhoNPxPhd}
     \end{figure}

\subsection{Analytical continuation to complementary series}\label{anac}

\begin{figure}[t]
     \centering
     \begin{subfigure}[t]{0.4\textwidth}
         \centering
         \includegraphics[width=\textwidth]{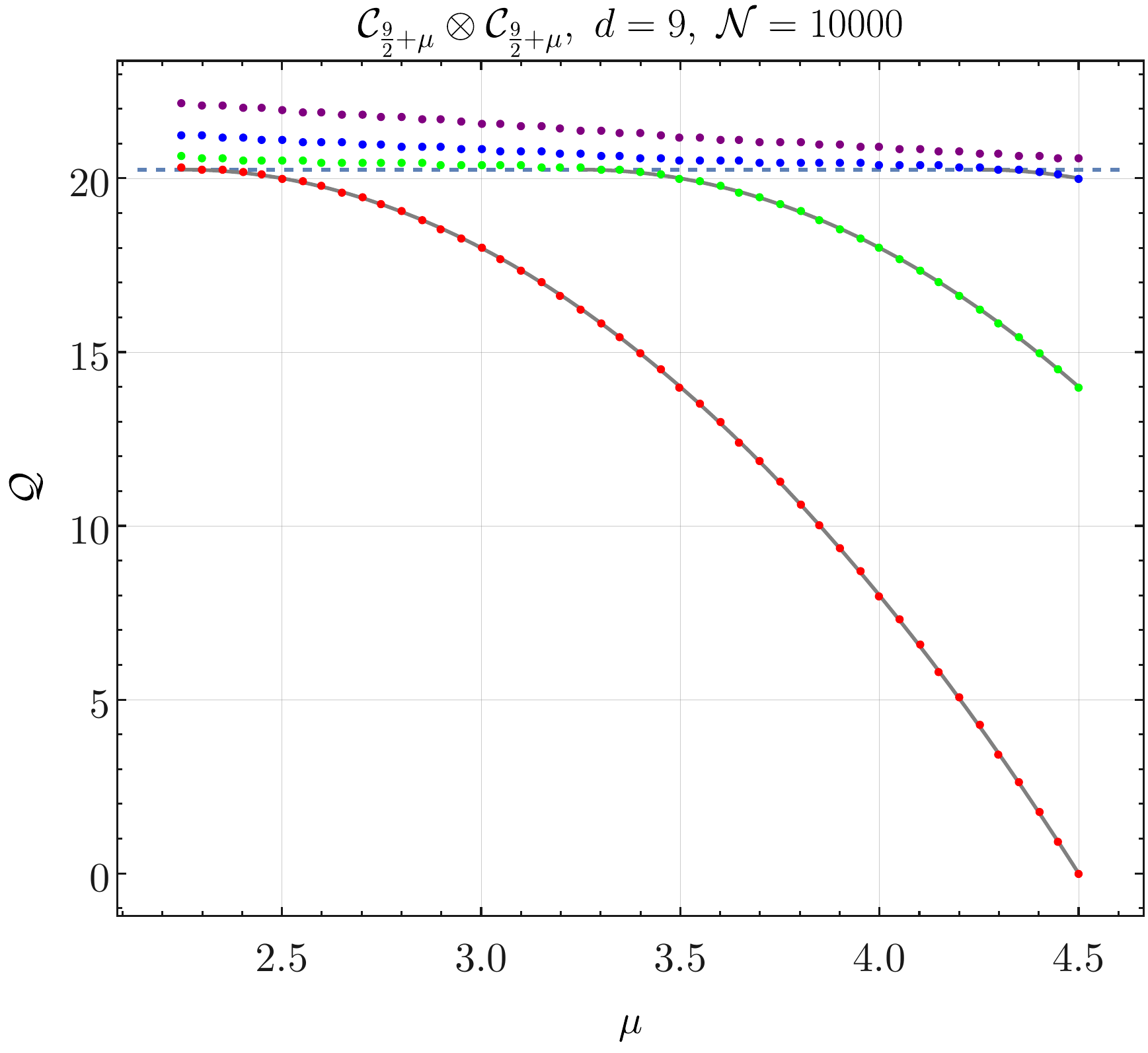}
         \label{fig:rrelCxChd}
     \end{subfigure}
     \hfill
     \begin{subfigure}[t]{0.45\textwidth}
         \centering
         \includegraphics[width=\textwidth, height=2.2in]{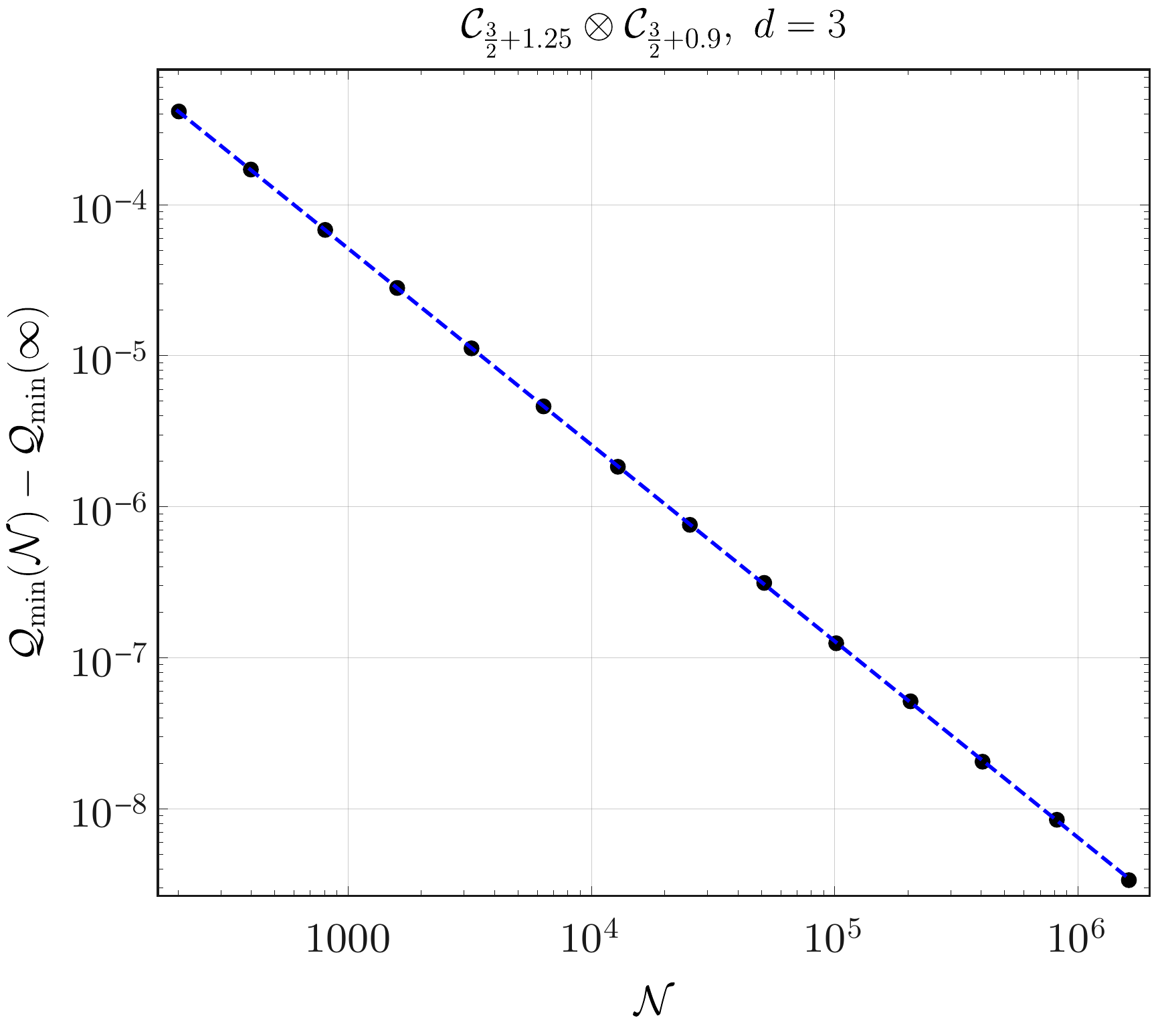}
         \label{fig:ConvCxChd}
     \end{subfigure}
     \caption{Tensor product of two complementary series representations. Left: The first four eigenvalues of $\CQ$ for the tensor product $\CC_{\frac{9}{2}+\mu}\otimes\CC_{\frac{9}{2}+\mu}$ of $\SO(1,10)$ with the cutoff being $\CN=10^4$. The dashed line at $\CQ=\frac{81}{4}$ is the critical line between principal series and complementary series. The three solid lines are $\CQ=(2\mu-2n)(9+2n-2\mu), n\in\{0,1,2\}$, corresponding to the $\SO(1,10)$ Casimir of $\CC_{2\mu-2n}, n\in\{0,1,2\}$. Right: The convergence of the minimal eigenvalue of $\CQ$ to its $\CN\to\infty$ limit for the tensor product $\CC_{\frac{3}{2}+1.25}\otimes\CC_{\frac{3}{2}+0.9}$ of $\SO(1,4)$.}{\label{fig:rhoCxChd}}
\end{figure}

While studying the tensor product of a principal series and a complementary series, e.g. $\CP_{\Delta_1}\otimes \CC_{\Delta_2}$, the whole procedure of solving $\CK^{(s)}_{\rm rel}(\lambda)$ above works exactly the same way up to the analytical continuation $i\mu_2\to \mu_2$. However, this simple analytical continuation prescription is not always valid in the case of $\CC_{\Delta_1}\otimes \CC_{\Delta_2}$ with $\Delta_1=\frac{d}{2}+\mu_1$ and $\Delta_2=\frac{d}{2}+\mu_2$, because as we have shown in appendix \ref{psifour}, the integral (\ref{CKsint}) does not hold after the replacement $i\mu_1\to \mu_1, i\mu_2\to \mu_2$ if $\frac{d}{2}+s-\mu_1-\mu_2<0$. For example, assume $\frac{d}{2}+s+2N_s<\mu_1+\mu_2<\frac{d}{2}+s+2(N_s+1)$ for some nonnegative integer $N_s$ and then eq.~\reef{CJthired} implies the following modification of eq.~\reef{CKsint}
\begin{align}\label{CKsintcont}
\int_{\mathbb R}d\lambda\,\CK^{(s)}_{\rm rel}(\lambda)e^{i\lambda t}=&2 \, e^{-(\frac{d}{2}+s-1)|t|}\, \frac{\cosh(\mu_1 t)\cosh(\mu_2 t)-\cos(\mu_3 t)\cos(\mu_4 t)}{|\sinh(t)|}\nonumber\\
&-2\sum_{n=0}^{N_s}\cosh\left[\left(\frac{d}{2}+s-\mu_1-\mu_2+2n\right)t\right]
\end{align}
where 
\begin{align}\label{CKscont}
\CK^{(s)}_{\rm rel}(\lambda)=\frac{1}{4\pi}\sum_{\pm,\pm,\pm}\left[\psi\left(\frac{\frac{d}{2}+s\pm i\lambda\pm i\mu_3\pm i\mu_4}{2}\right)-\psi\left(\frac{\frac{d}{2}+s\pm i\lambda\pm \mu_1\pm \mu_2}{2}\right)\right]
\end{align}
is a direct analytical continuation of eq.~(\ref{CKsfinal}).
 Because of the extra terms in (\ref{CKsintcont}) , the character equation (\ref{PPSch2char}) should change accordingly as (suppressing the arguments of characters)
\begin{align}\label{CCchar}
\Theta_{\Delta_1}\Theta_{\Delta_2}-\Theta_{\Delta_3}\Theta_{\Delta_4}&=\sum_{s\ge 0}\int_0^\infty d\lambda\, \CK^{(s)}_{\rm rel}(\lambda)\, \Theta_{\lambda,s}+\sum_{s}\sum_{n=0}^{N_s} \Theta_{\mu_1+\mu_2-s-2n}
\end{align}
where the sum over $s$ is finite because $s<\mu_1+\mu_2-\frac{d}{2}<\frac{d}{2}$. Altogether, the character equation (\ref{CCchar}) implies that in the tensor product of complementary series representations $\CC_{\Delta_1}\otimes\CC_{\Delta_2}$, we get only principal series when $\mu_1+\mu_2\le \frac{d}{2}$ \cite{Dobrev:1976vr}, and we can also get additionally a finite number of complementary series representations when $\mu_1+\mu_2>\frac{d}{2}$. The scaling dimensions of these complementary series representations for each fix spin $s$ are 
\begin{align}\label{muss}
\mu_1+\mu_2-s, \,\,\,\,\mu_1+\mu_2-s-2,\,\,\,\, \mu_1+\mu_2-s-4,\cdots, \mu_1+\mu_2-s-2N_s
\end{align}
where $N_s$ satisfies
\begin{align}\label{murange}
\frac{d}{2}+s+2N_s<\mu_1+\mu_2<\frac{d}{2}+s+2\left(N_s+1\right)~.
\end{align}

When $d=2$, $\mu_1+\mu_2$ is bounded by $2$ and hence $s=N_s=0$ in eq.~\reef{murange}. In other words, there can be at most one complementary series representation, which has spin 0 and scaling dimension $\mu_1+\mu_2$, if exists.
This result was first derived in \cite{naimark1961decomposition} by Naimark. In higher $d$, the $s=0$ version of eq.~\reef{muss} was proved in \cite{https://doi.org/10.48550/arxiv.1402.2950} by directly constructing the intertwining map between $\CC_{\Delta_1}\otimes\CC_{\Delta_2}$ and $\CC_{\mu_1+\mu_2-2n}$.  It would be interesting to generalize the intertwining map to incorporate complementary series with nonzero $s$.

\begin{figure}[t]
     \centering
         \includegraphics[width=0.6\textwidth]{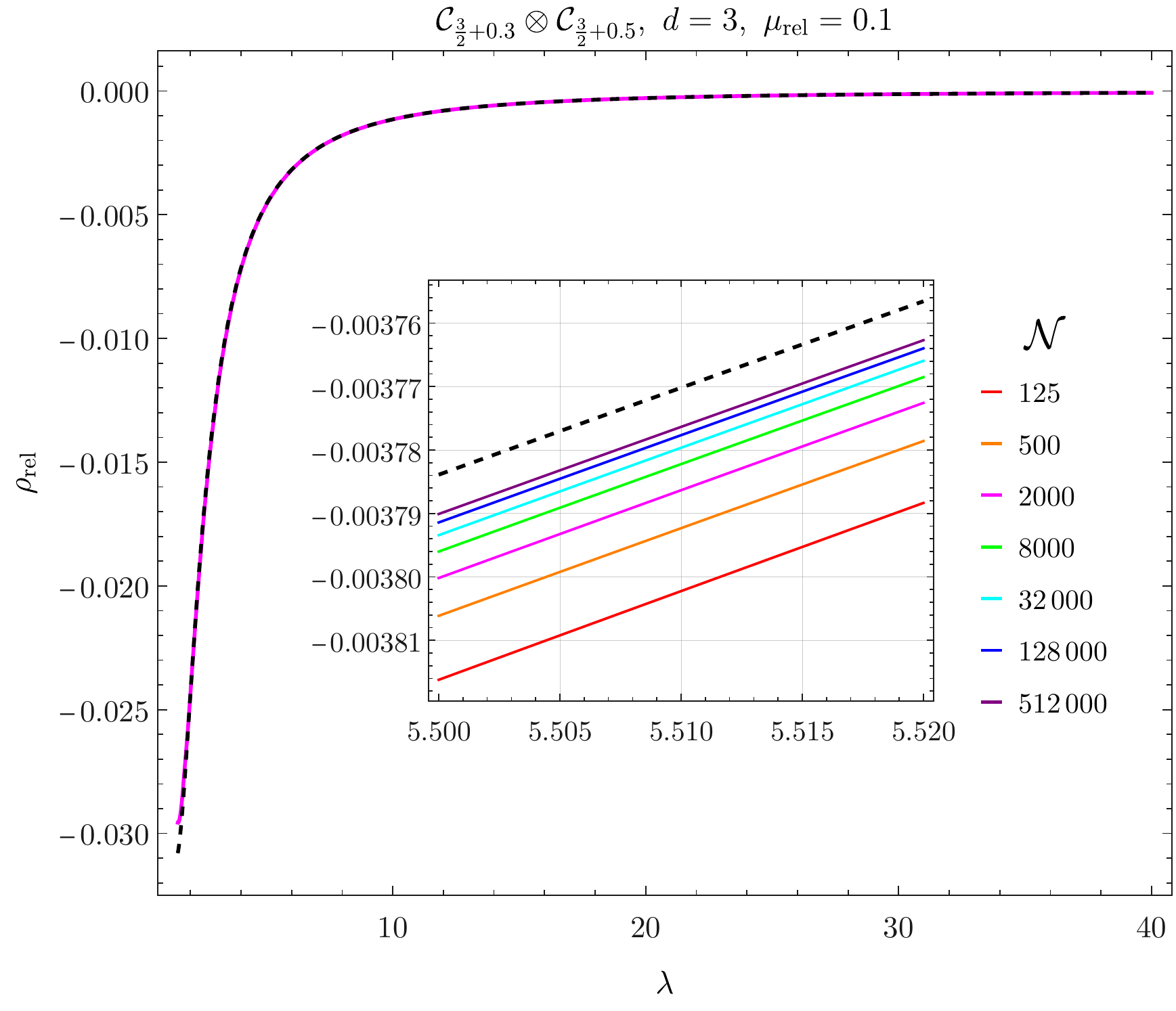}
         \caption{Comparison of the relative coarse-grained  density $\rho_{\text{rel}}(\lambda)$ (pink solid line) given by eq. (\ref{ccpp}) to  $\CK^{(0)}_{\text{rel}}(\lambda)$ (black dashed line) given by eq. (\ref{CKscont}) for $\mu_1=0.3, \mu_2=0.5, \mu_3=\mu_4= \mu_\text{rel}=0.1$, and $\CN=2000$. The inset plot, zooming into the range $5.500<\lambda<5.520$, illustrates how $\rho_{\rm rel}$ approaches  $\CK^{(0)}_{\text{rel}}(\lambda)$ (black dashed line)  as we increase $\CN$.}
         \label{fig:rrelCxChd}
     \end{figure}

The truncated model $\CQ$ described in the previous subsection can also be used to test these complementary series representations numerically. The nonvanishing matrix elements of $\CQ$ in the  $\CC_{\Delta_1}\otimes\CC_{\Delta_2}$ case are given by eq.~(\ref{CCoff})
\begin{align}\label{CQ12p}
&\mathcal Q_{nn}=\Delta_{1}\bar\Delta_{1}+\Delta_{2}\bar\Delta_{2}+2n(n+d-1)\nonumber\\
&\mathcal Q_{n+1,n}=\mathcal Q_{n,n+1}=-\sqrt{\frac{(n+1)(n+d-1)(\Delta_{1}+n)(\bar\Delta_{1}+n)(\Delta_{2}+n)(\bar\Delta_{2}+n)}{(n+\frac{d-1}{2})(n+\frac{d+1}{2})}}~.
\end{align}
where $0\le n\le \CN$.
We plot the first four eigenvalues of $\CQ$ in fig.(\ref{fig:rhoCxChd}), with $d=9, \Delta_1=\Delta_2=\frac{9}{2}+\mu$ and $\CN=10^4$. The smallest eigenvalue enters the complementary series regime around $\mu=\frac{d}{4}=2.25$ and lies on the line $2\mu(9-2\mu)$, corresponding to $\CC_{2\mu}$. The second eigenvalue starts to be smaller than $\frac{d^2}{4}=20.25$ around $\mu=\frac{d}{4}+1=3.25$ and then lies on the line $(2\mu-2)(11-2\mu)$, corresponding to $\CC_{2\mu-2}$. The third eigenvalue becomes the Casimir of $\CC_{2\mu-4}$ roughly in the region $4.25<\mu<4.5$, since it lies on the line  $(2\mu-4)(13-2\mu)$. The fourth eigenvalue is always larger than $\frac{d^2}{4}=20.25$  and hence cannot belong to complementary series. These numerical results agree perfectly with eq.~(\ref{muss}) and eq.~(\ref{murange}) for $s=0$, which are derived from characters.

Apart from these discrete eigenvalues corresponding to complementary series, the matrix $\mathcal Q$ defined by (\ref{CQ12p}) also has a continuous spectrum in  $[\frac{d^2}{4},\infty)$ when $\CN\to \infty$. The reason is that  eigenvectors of $\mathcal Q$ satisfying $\mathcal Q_{nm}a_m(\lambda)=(\frac{d^2}{4}+\lambda^2)a_n(\lambda)$, have the asymptotic behavior 
 \begin{align}\label{eigenasy1}
 a_n(\lambda)=R_+ \frac{n^{i\lambda}}{\sqrt{n}}\left(1+\CO(1/n)\right)+R_- \frac{ n^{-i\lambda}}{\sqrt{n}}(\left(1+\CO(1/n)\right)
 \end{align}
which implies that eigenvectors of different $\lambda$ are $\delta$-function normalizable. The continuous spectrum corresponds to scalar principal series. We can define a coarse-grained density $\bar\rho_\CN(\lambda)$ of these representations, along the lines of eq. (\ref{brdef1}), by numerically diagonalizing the truncated matrix $\CQ$. We also define a relative coarse-grained density
\begin{align}\label{ccpp}
\rho_{\rm rel}(\lambda)=\bar\rho_\CN^{\CC_{\Delta_1}\otimes\CC_{\Delta_2}}-\bar\rho_{\CN}^{\CP_{\Delta_3}\otimes\CP_{\Delta_4}}
\end{align}
that has a finite large $\CN$ limit. In fig. (\ref{fig:rrelCxChd}), we illustrate the agreement between $\rho_{\rm rel}(\lambda)$ and $\CK^{(0)}_{\rm rel}(\lambda)$
in eq. (\ref{CKscont}), using the tensor products $\CC_{\frac{3}{2}+0.3}\otimes \CC_{\frac{3}{2}+0.5}$ and $\CP_{\frac{3}{2}+0.1 i}\otimes \CP_{\frac{3}{2}+0.1 i}$.

\subsection{Photon from massless scalars}\label{EEt}
We have shown that the tensor product $\CF_{\Delta_1}\otimes\CF_{\Delta_2}$ only contains UIRs in the continuous families, which describe  massive fields in dS$_{d+1}$. Given the similarity between scalar complementary series and type \rom{1} exceptional series, which is mentioned in section \ref{repreview} and described in more detail in appendix \ref{matrixcomp}, one would naively expect the  same result for $\CV_{p_1,0}\otimes\CV_{p_2,0}$. However, we will show that this naive expectation is not true by using $\CV_{1,0}\otimes\CV_{1,0}$ as an example. In particular, the tensor product $\CV_{1,0}\otimes\CV_{1,0}$ contains a type \rom{2} exceptional series representation $\CU_{1, 0}$. In field theory language, it means that the two-particle  Hilbert space of two {\it different} massless particles admits an invariant subspace corresponding to a photon. To make the whole discussion very explicit, we will focus on the case of dS$_{4}$. The same holds for any higher dimensional dS space. 

Comparing $\CV_{1,0}$ and representations in scalar complementary series, the main difference is the absence of $\SO(4)$ singlets in $\CV_{1,0}$.
Fortunately, this property does not affect almost any  conclusion in appendix \ref{Noex}, except for $\CU_{1,0}=\CU^+_{1,0}\oplus\CU^-_{1,0}$. To see what appends to $\CU_{1,0}$, let us first recall that it is characterized by a normalizable state $|\chi\rangle_{a,b}=-|\chi\rangle_{b,a}$\footnote{$\CU^\pm_{1,0}$ corresponds to imposing (anti-)self-dual condition on this state, i.e. $\frac{1}{2}\epsilon_{abcd}|\chi)_{c,d}=\pm |\chi)_{a,b}$.} that satisfies $L_{0b}|\chi\rangle_{a,b}=0$ because  $\CU_{1,0}$ does not contain spin 1 representation of $\SO(4)$. In $\CV_{1,0}\otimes\CV_{1,0}$, such a state has to be a linear combination of\
\begin{align}\label{yio}
|\chi_n)_{a,b}=|n\rangle_{aa_2\cdots a_n}|n\rangle_{ba_2\cdots a_n}-(a\leftrightarrow b), \,\,\,\,\, n\ge 1
\end{align}
where $|n\rangle_{a_1\cdots a_n}$ is a basis of the $\mY_n$ component in $\CV_{1,0}$. Let $|\chi)_{a,b}=\sum_{n\ge 1} c_n |\chi_n)_{a,b}$. In appendix \ref{Noex}, we show that the condition  $L_{0b}|\chi\rangle_{a,b}=0$ leads to a recurrence relation of all $c_n$, c.f. eq.~(\ref{crec2}),
\begin{align}\label{crec3}
c_n \alpha_n+c_{n+1}\beta_{n+1}\frac{n+3}{n+1}=0, \,\,\,\,\, n\ge 1
\end{align}
where $\alpha_n$ and $\beta_n$ are given by eq.~(\ref{abP1}), which, for $\Delta=d=3, \bar\Delta=0$, reduces to
\begin{align}
\alpha_n=\sqrt{\frac{n(n+1)(n+3)}{2(n+2)}},\,\,\,\,\, \beta_n=-\sqrt{\frac{(n-1)n(n+2)}{2(n+1)}}~.
\end{align}
Besides the recurrence relation (\ref{crec3}), the requirement  $L_{0b}|\chi\rangle_{a,b}=0$ actually also implies an initial  relation $c_1\beta_1=0$. For principal series or complementary series, $c_1\beta_1=0$ is equivalent to $c_1=0$ because $\beta_1\not=0$, which then leads to the vanishing of all $c_n$ due to eq.~(\ref{crec3}). In the $\CV_{1,0}$ case, the initial relation does not impose any constraint on $c_1$ because $\beta_1=0$. A general solution of all $c_n$ is 
\begin{align}
c_n=\frac{A}{(n+1)(n+2)}
\end{align}
where $A$ is an arbitrary constant. Using the inner product of $|\chi_n)_{a,b}$, c.f. eq.~\reef{cc3}, we find that $|\chi\rangle_{a,b}$ is normalizable 
\begin{align}
_{a,b}\langle \chi|\chi\rangle_{a,b}=2|A|^2\sum_{n\ge 1}\frac{1}{n(n+2)}<\infty
\end{align}
and hence $\CU_{1,0}$ belongs to $\CV_{1,0}\otimes\CV_{1,0}$. Altogether, we can conclude that $\CV_{1,0}\otimes\CV_{1,0}$ contains exactly one exceptional series representation $\CU_{1,0}$. We also want to mention that  the same conclusion does not hold for the symmetrized tensor product $\CV_{1,0}\odot\CV_{1,0}$ since the state $|\chi_n)_{a,b}$ in eq. (\ref{yio}) is manifestly anti-symmetric under permutation. So it is impossible to identify a photon like state in the two-particle Hilbert space of a single massless scalar field.

\begin{figure}[t!]
     \centering
     \hspace*{\fill}%
     \begin{subfigure}[t]{0.25\textwidth}
         \centering
         \includegraphics[width=\textwidth]{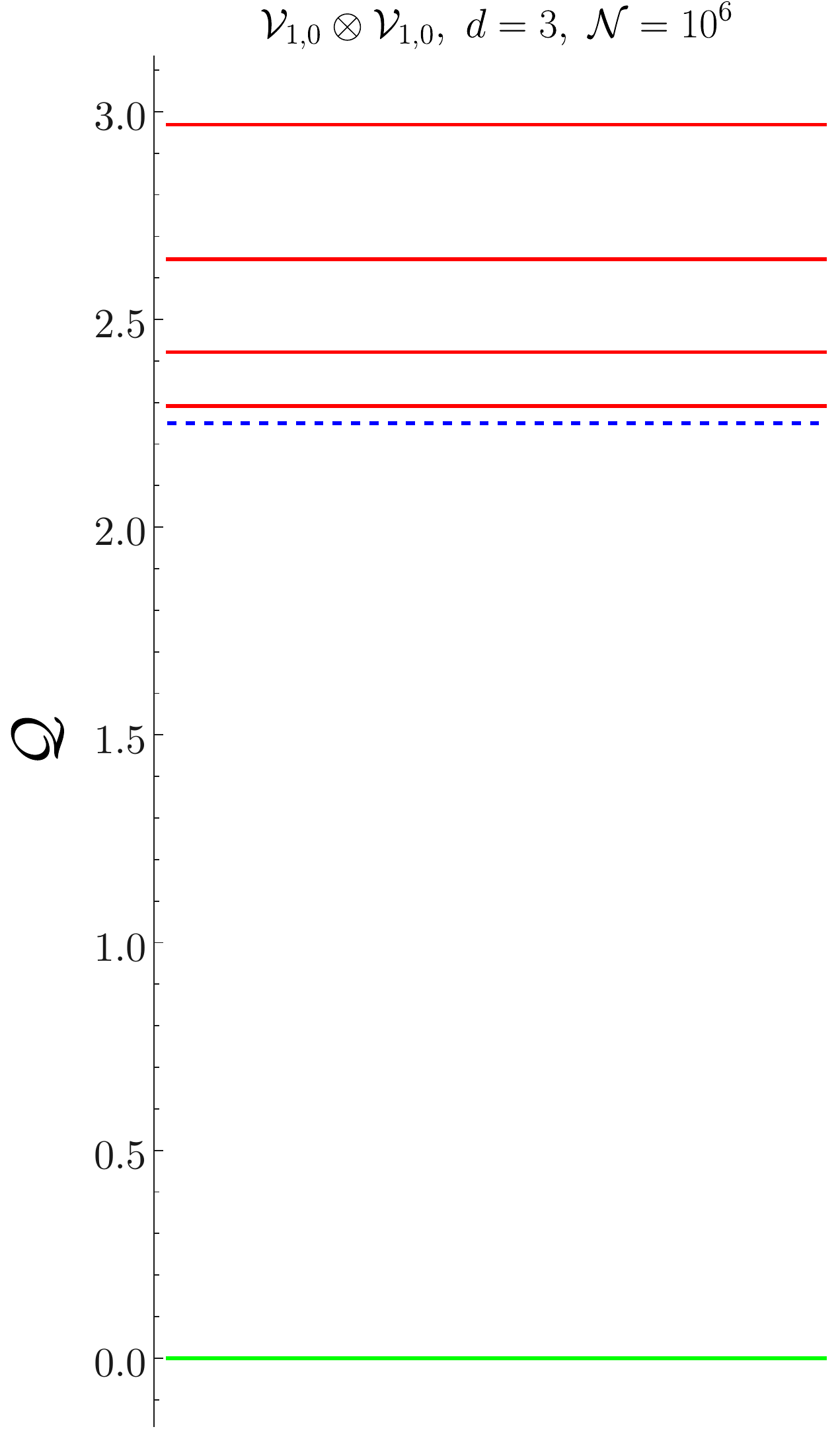}
         \caption{}
         \label{fig:CasVxVhd}
     \end{subfigure}
     \hfill
     \begin{subfigure}[t]{0.46\textwidth}
         \centering
         \includegraphics[width=\textwidth]{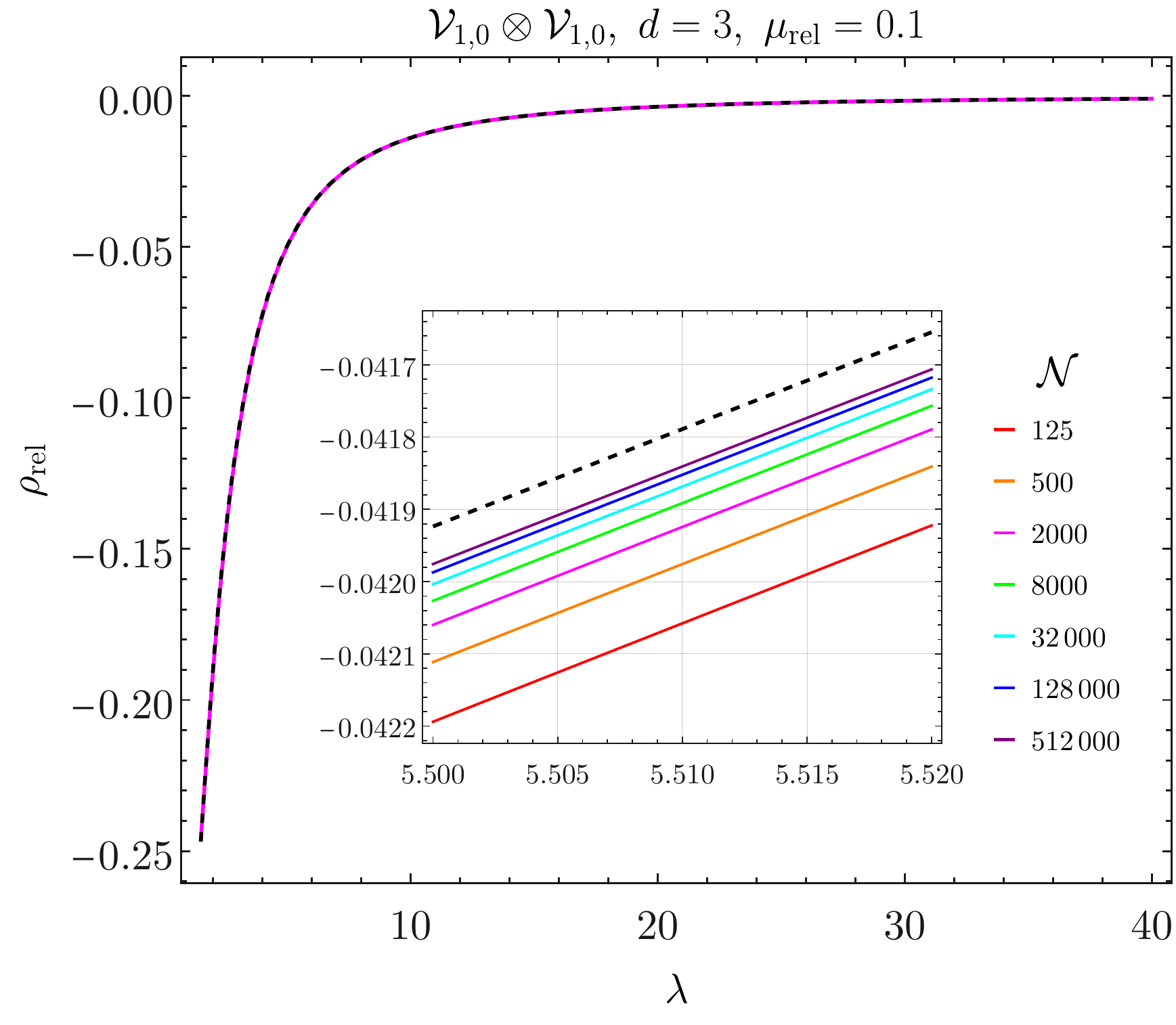}
         \caption{}
         \label{fig:rrelVxVhd}
     \end{subfigure}
     \hspace*{\fill}%
     \caption{The tensor product $\CV_{1,0}\otimes\CV_{1,0}$ of $\SO(1,4)$. Left: Some low-lying eigenvalues of $\CQ$, given by eq.~(\ref{CV10ma}). The dashed line is $\CQ=\frac{9}{4}$. Right: Comparison of the relative coarse-grained density $\rho_{\rm rel}(\lambda)$ given by eq. (\ref{hkhk})  to $\CK_{\rm rel}^{(0)}(\lambda)$ given by eq.~(\ref{jhiue}) with $\mu_3=\mu_4=0.1$. The inset plot zooms into $5.50\le \lambda\le 5.52$, showing how $\rho_{\rm rel}(\lambda)$ approaches $\CK_{\rm rel}^{(0)}(\lambda)$ (black dashed line) by increasing $\CN$.}{\label{fig:VxVhd}}
\end{figure}

Next, we move to study the continuous part of $\CV_{1,0}\otimes\CV_{1,0}$, assuming that it does not support complementary series, which will be checked numerically later. Using $\CP_{\Delta_3}\otimes\CP_{\Delta_4}$ as a regulator,  the relative density of principal series $\CK_{\rm rel}^{(s)}(\lambda)$ between $\CV_{1,0}\otimes\CV_{1,0}$ and $\CP_{\Delta_3}\otimes\CP_{\Delta_4}$ is defined as 
\begin{align}\label{defVVd}
\Theta_{\CV_{1,0}}(q,x)^2-\Theta_{\Delta_3}(q,x)\Theta_{\Delta_4}(q,x)=\sum_{s\ge 0}\int_0^\infty\, d\lambda\,\CK_{\rm rel}^{(s)}(\lambda)\Theta_{\Delta_\lambda}(q, x)+\Theta_{\CU_{1,0}}(q, x)
\end{align}
where $\Theta_{\CV_{1,0}}(q,x)$ is given by eq.~(\ref{CVcharspec}),  and $\Theta_{\CU_{1,0}}(q,x)$ is given by eq.~(\ref{CU3char}) with $s=1, t=0$.
As we have seen  in previous sections, in order to compute $\CK_{\rm rel}^{(s)}(\lambda)$, we need to expand $\Theta_{\CV_{1,0}}(q,x)^2 P_3(q,x)$ in terms of $\SO(3)$ character. In this discussion, we will focus on the $s=0$ case (the rest of densities can be derived similarly), and hence it suffices to know the constant term in this expansion. Using eq.~(\ref{CVcharspec}) and eq.~(\ref{techguess}), we find the constant term to be $\frac{4\,q^6}{1-q^2}+\frac{q^2(1+3q^3)}{1+q}$. Plugging it into eq.~(\ref{defVVd}) yields
\small
\begin{align}\label{hlkh}
\int_{\mathbb R}\, d\lambda\,\CK_{\rm rel}^{(0)}(\lambda) e^{i\lambda t}=\left(\frac{4\,e^{-\frac{9}{2}|t|}}{1-e^{-2 |t|}}-\sum_{\pm,\pm}\frac{e^{-(\frac{3}{2}\pm i\mu_3\pm i\mu_4)|t|}}{1-e^{-2 |t|}}\right)+\frac{e^{-\frac{1}{2}|t|}+3\,e^{-\frac{7}{2}|t|}}{1+e^{- |t|}}+2\, e^{-\frac{3}{2}|t|}
\end{align}
\normalsize
where the last term comes from $\Theta_{\CU_{1,0}}(q, x)$. Then  $\CK_{\rm rel}^{(0)}(\lambda)$ can be computed as an inverse Fourier transformation of the R.H.S of eq.(\ref{hlkh}), term by term. For the first two terms, the inverse Fourier transformation follows from eq.~(\ref{CIresult}). For the third term, one can use the integral formula, $\int_0^\infty dt \frac{e^{-z t}}{1+e^{-t}}=\frac{1}{2}\psi(\frac{1+z}{2})-\frac{1}{2}\psi(\frac{z}{2})$, where $\Re z>0$. For the last term, it is an elementary integral. Altogether, we obtain
\begin{align}\label{jhiue}
\CK_{\rm rel}^{(0)}(\lambda) &= \frac{1}{4\pi}\sum_{\pm,\pm,\pm}\psi\left(\frac{\frac{3}{2}\pm i\lambda\pm i\mu_3\pm i\mu_4}{2}\right)-\frac{1}{\pi}\sum_{\pm}\psi\left(\frac{\frac{9}{2}\pm i\lambda}{2}\right)\nonumber\\
&+\frac{3}{2\pi} \frac{1}{\lambda^2+\frac{1}{4}}-\frac{3}{2\pi} \frac{1}{\lambda^2+\frac{9}{4}}+\frac{15}{2\pi} \frac{1}{\lambda^2+\frac{25}{4}}-\frac{2}{e^{\pi\lambda}+e^{-\pi\lambda}}~.
\end{align}
which can also be recovered by taking $d=3, s=0$ and  $\mu_1=\mu_2=\frac{3}{2}$ in eq. (\ref{CKscont}). This is not a coincidence but a result of the fact that $\CV_{1,0}$ of any $\SO(1, d+1)$ is the boundary point of complementary series up to a trivial representation of $\SO(d+1)$. From the numerical side, this is also obvious because the  $\SO(1,4)$ Casimir in the $\SO(4)$ singlet subspace of $\CV_{1,0}\otimes\CV_{1,0}$ can be derived from that of  $\CC_{\Delta_1}\otimes\CC_{\Delta_2}$ case, c.f. eq.~(\ref{CQ12p}), by taking $\Delta_1=\Delta_2=3$ and  cutting off $n$ at 1, i.e.
\begin{align}\label{CV10ma}
\CQ_{n,n}=2n(n+2), \,\,\,\,\, \CQ_{n+1,n}=\CQ_{n,n+1}=-n(n+3), \,\,\,\,\, n\ge 1~.
\end{align}

For numerical diagonalization, we impose a hard cut-off $n\le \CN$, and then $\CQ$ becomes an $\CN\times\CN$ matrix. In the left panel of fig. (\ref{fig:VxVhd}), we plot the first four eigenvalues of $\CQ$ with the cut-off $\CN$ being $10^6$. They are all larger that $\frac{9}{4}$, implying the absence of complementary series representations in $\CV_{1,0}\otimes\CV_{1,0}$. Since the truncated $\CQ$ is larger than $\frac{9}{4}$, its eigenvalues induce a coarse-grained density of principal series representations $\bar\rho_\CN(\lambda)$, which diverges in the $\CN\to\infty $ limit. We remove the divergence by introducing $\CP_{\Delta_3}\otimes\CP_{\Delta_4}$ and considering a relative density 
\begin{align}\label{hkhk}
\rho_{\rm rel}(\lambda)=\bar\rho_\CN^{\CV_{1,0}\otimes\CV_{1,0}}(\lambda)-\bar\rho_\CN^{\CP_{\Delta_3}\otimes\CP_{\Delta_4}}(\lambda)
\end{align}
In the right panel of fig.~\ref{fig:VxVhd}, we show a match between eq.~(\ref{hkhk}) and eq.~(\ref{jhiue}) with $\Delta_3=\Delta_4=\frac{3}{2}+0.1 i$.

\subsection{Some remarks on nonzero spin}
\label{sec:remarksspin}
For the tensor product of two spinning principal series representations, e.g. $\CP_{\Delta_1,s_1}\otimes \CP_{\Delta_2,s_2}$, it has been proved by \cite{Dobrev:1977qv} that there can only be principal series and discrete series. Here principal series means a larger class of representations $\CP_{\Delta, \mY}$ that are labelled by a scaling dimension $\Delta\in\frac{d}{2}+i\mathbb  R_{\ge 0}$ and a Young diagram $\mY$.  The $\CP_{\Delta, s}$ representation is a special case of $\CP_{\Delta, \mY}$ when $\mY$ has only one row, i.e. $\mY=\mY_s$. The character of $\CP_{\Delta, \mY}$ is given by eq.~(\ref{mixedsym}). 

When $d$ is even, $\SO(1,d+1)$ does not admit discrete series, and hence $\CP_{\Delta_1,s_1}\otimes \CP_{\Delta_2,s_2}$ is decomposed to principal series only. Based on this fact, we are going to discuss what kinds of principal series representations can appear in this tensor product by using  character analysis. Since regularization is not relevant in this analysis, we just use the  character relation {\it formally}
\begin{align}\label{formalchar}
\Theta_{\Delta_1, s_1}(q,\bm x)\Theta_{\Delta_2, s_2}(q,\bm x)=\sum_{\mY}\int_0^\infty\, d\lambda\, \CK_{\mY}(\lambda) \Theta_{\Delta_\lambda, \mY}(q,\bm x)~.
\end{align}
For $\Delta_1=\frac{d}{2}+i\mu_1, \Delta_2=\frac{d}{2}+i\mu_2$ and $q=e^{-|t|}$, the eq.~(\ref{formalchar}) is equivalent to 
\begin{align}\label{fcl}
\sum_{\mY}\chi_\mY^{\SO(d)}(\bm x)\int_0^\infty\, d\lambda\, \CK_{\mY}(\lambda)e^{i\lambda t}=4\frac{\chi_{\mY_{s_1}}^{\SO(d)}(\bm x)\chi_{\mY_{s_2}}^{\SO(d)}(\bm x)}{P_d(e^{-|t|},\bm x)}e^{-\frac{d}{2}|t|}\cos(\mu_1 t)\cos(\mu_2 t)~.
\end{align}
Then we need to expand the $\bm x$ dependent part on the R.H.S of eq.~(\ref{fcl}) as a series  summing over all $\SO(d)$ characters. It involves two steps. 
First, using eq.~(\ref{techguess0}), we obtain
\begin{align}\label{ssde}
\frac{\chi_{\mY_{s_1}}^{\SO(d)}(\bm x)\chi_{\mY_{s_2}}^{\SO(d)}(\bm x)}{P_d(q,\bm x)}=\sum_{s\ge 0} \chi_{\mY_{s_1}}^{\SO(d)}(\bm x)\chi_{\mY_{s_2}}^{\SO(d)}(\bm x) \chi_{\mY_{s}}^{\SO(d)}(\bm x)\frac{q^s}{1-q^2}~.
\end{align}
Next, we express the product of  three $\SO(d)$ characters $ \chi_{\mY_{s_1}}^{\SO(d)}(\bm x)\chi_{\mY_{s_2}}^{\SO(d)}(\bm x) \chi_{\mY_{s}}^{\SO(d)}(\bm x)$ as a finite sum of $\SO(d)$ characters, which is equivalent to the tensor product decomposition of $\mY_{s_1}\otimes \mY_{s_2}\otimes \mY_{s}$ as $\SO(d)$ representations. Thus, we can conclude that a principal series representation $\CP_{\Delta,\mY}$  appears in $\CP_{\Delta_1,s_1}\otimes \CP_{\Delta_2,s_2}$ if and only if $\mY$ belongs to $\mY_{s_1}\otimes \mY_{s_2}\otimes \mY_{s}$ for some $s\ge 0$. For example, when $s_1=1$ and $s_2=0$, the tensor products $\mY_1\otimes \mY_0\otimes \mY_{s}=\mY_1\otimes  \mY_{s}$ yield all single row representations and all $\mY_{n,1}, n\ge 1$. So the tensor product of a spin 1 and a spin 0 principal series representations contains all $\CP_{\Delta, s}\, (s\ge 0)$ and all $\CP_{\Delta, \mY_{s,1}}\, (s\ge 1)$. For $s_1=s_2=1$, using the decomposition rule
\begin{align}
\mY_1\otimes\mY_1\otimes \mY_s&=\mY_1\otimes \left(\mY_{s-1}\oplus \mY_{s+1}\oplus \mY_{s,1}\right)\nonumber\\
&=\mY_{s-2}\oplus 3\mY_{s}\oplus\mY_{s+2}\oplus 2\left(\mY_{s+1,1}\oplus \mY_{s-1,1}\right)\oplus \mY_{s,2}\oplus\mY_{s,1,1}~,
\end{align}
one can easily find that the tensor product of two spin 1 principal series representations contains all $\CP_{\Delta, s}\, (s\ge 0)$, $\CP_{\Delta, \mY_{s,1}}\, (s\ge 1)$, $\CP_{\Delta, \mY_{s,2}}\, (s\ge 2)$ and $\CP_{\Delta, \mY_{s,1,1}}\, (s\ge 1)$. Although, we start with even $d$, we believe this result is valid for both even and odd $d$.

In \cite{Dobrev:1977qv}, the possible $\mY$ were found by a much more sophisticated method, that is independent of the parity of $d$. The result can be described as follows. Let $\rho$ be a UIR of $\SO(d)$ contained in $\mY_{s_1}\otimes\mY_{s_2}$ and $\sigma$ be a UIR of $\SO(d-1)$ contained in the restriction of $\rho$ to $\SO(d-1)$. A principal series representation $\CP_{\Delta,\mY}$ belongs to $\CP_{\Delta_1,s_1}\otimes \CP_{\Delta_2,s_2}$ if and only if  some $\sigma$ obtained in this way belongs to the restriction of $\mY$ to $\SO(d-1)$.  
As shown in \cite{Kravchuk:2016qvl}, this constraint on $\mY$ is equivalent to saying that $\mY$ belongs to $\mY_{s_1}\otimes\mY_{s_2}\otimes \mY_{s}$ for some $s\ge 0$. To see this, we only need a simple fact that given two irreducible representations  $\mY$ and $\mY'$ of $\SO(d)$, the tensor product $\mY\otimes\mY'$ contains the trivial representation if and only if $\mY=\mY'$. Then identifying an irreducible component $\mY$ in some reducible representation $R$ amounts to checking if $\bullet\in\mY\otimes R$, where $\bullet$ denotes the trivial representation. Based on this fact, we can immediately see that the following equivalence
\begin{align}
\label{identitityTP}
\mY \subset \mY_{s_1}\otimes\mY_{s_2}\otimes \sum_{s\ge 0} \mY_{s}  \quad\Leftrightarrow \quad
\exists \,s\ge 0 : \mY_s \subset \mY  \otimes \mY_{s_1}\otimes\mY_{s_2} \,.
\end{align}
as  both sides mean $\bullet\in\mY_{s_1}\otimes\mY_{s_2}\otimes\mY\otimes\mY_s$ for some $s$.
Then, reduce the last formula from $\SO(d)$ to $\SO(d-1)$.
Since $\mY_s $ are the only irreducible representations of $\SO(d)$ that give rise to $\SO(d-1)$ singlets, we conclude that \eqref{identitityTP} is equivalent to the statement that there is at least one $\SO(d-1)$ singlet inside  the $\SO(d)$ tensor product $\mY  \otimes \mY_{s_1}\otimes\mY_{s_2} $. But this is equivalent to the statement that  
there is at least one  irreducible representation $\sigma$ of $\SO(d-1)$ that appears both inside 
$\mY$ and   $\mY_{s_1}\otimes\mY_{s_2} $.

With the possible $\SO(d)$ structures $\mY$ known, we can in principle compute the relative density of each $\CP_{\Delta_\lambda,\mY}$ between $\CP_{\Delta_1,s_1}\otimes\CP_{\Delta_2,s_2}$ and $\CP_{\Delta_3,s_1}\otimes\CP_{\Delta_4,s_2}$. For example, in the case of $s_1=1, s_2=0$, the relative densities of $\CP_{\Delta_\lambda,\mY_{s}}$ \, ($s\ge 0$) and $\CP_{\Delta_\lambda,\mY_{s,1}}$ \, ($s\ge 1$) should satisfy (suppressing the arguments of characters)
\begin{align}\label{1010}
\Theta_{\Delta_1, 1}\Theta_{\Delta_2}-\Theta_{\Delta_3, 1}\Theta_{\Delta_4}=\sum_{s\ge 0}\int_0^\infty\, d\lambda\, \CK_{\rm rel}^{(s)}(\lambda) \Theta_{\Delta_\lambda, s}+\sum_{s\ge 1}\int_0^\infty\, d\lambda\, \CK_{\rm rel}^{(s,1)}(\lambda) \Theta_{\Delta_\lambda, \mY_{s,1}}~.
\end{align}
As we have seen in eq.~(\ref{ssde}), the difference compared to the $s_1=s_2=0$ case comes mainly from the following expansion
\begin{align}\label{10exp}
\frac{\chi^{\SO(d)}_{\mY_1}(\bm x)}{P_d (q,\bm x)}&=\sum_{s\ge 0}\chi^{\SO(d)}_{\mY_1}(\bm x)\chi^{\SO(d)}_{\mY_s}(\bm x)\frac{q^s}{1-q^2}\nonumber\\
&=\frac{\chi^{\SO(d)}_{\mY_1}(\bm x)}{1-q^2}+\sum_{s\ge 1}\left(\chi^{\SO(d)}_{\mY_{s-1}}(\bm x)+\chi^{\SO(d)}_{\mY_{s+1}}(\bm x)+\chi^{\SO(d)}_{\mY_{s,1}}(\bm x)\right)\frac{q^s}{1-q^2}\nonumber\\
&=\frac{q}{1-q^2}+\sum_{s\ge 1}\chi^{\SO(d)}_{\mY_{s}}(\bm x)\frac{q^{s-1}+q^{s+1}}{1-q^2}+\sum_{s\ge 1}\chi^{\SO(d)}_{\mY_{s,1}}(\bm x)\frac{q^{s}}{1-q^2}~.
\end{align}
Plugging (\ref{10exp}) into the character relation (\ref{1010}), we obtain
\small
\begin{align}\label{Krs}
\CK_{\rm rel}^{(s)}(\lambda)=\frac{1}{4\pi}\begin{cases}\sum_{\pm,\pm,\pm}\left[\psi\left(\frac{\frac{d}{2}+1\pm i\lambda\pm i\mu_3\pm i\mu_4}{2}\right)-\psi\left(\frac{\frac{d}{2}+1\pm i\lambda\pm i\mu_1\pm i\mu_2}{2}\right)\right], &s=0\\
\sum_{\pm,\pm,\pm,\pm}\left[\psi\left(\frac{\frac{d}{2}+s\pm 1\pm i\lambda\pm i\mu_3\pm i\mu_4}{2}\right)-\psi\left(\frac{\frac{d}{2}+s\pm 1\pm i\lambda\pm i\mu_1\pm i\mu_2}{2}\right)\right], & s\ge 1
\end{cases}
\end{align}
\normalsize
and 
\small
\begin{align}\label{Krs1}
\CK_{\rm rel}^{(s,1)}(\lambda)=\frac{1}{4\pi}\sum_{\pm,\pm,\pm}\left[\psi\left(\frac{\frac{d}{2}+s\pm i\lambda\pm i\mu_3\pm i\mu_4}{2}\right)-\psi\left(\frac{\frac{d}{2}+s\pm i\lambda\pm i\mu_1\pm i\mu_2}{2}\right)\right], \,\,\,\, s\ge 1~.
\end{align}
\normalsize
The analytical continuation $i\mu_1\to \mu_1, i\mu_2\to \mu_2$ in eq.~(\ref{Krs}) and eq.~(\ref{Krs1}) leads to relative density of principal series between $\CC_{\Delta_1,1}\otimes \CC_{\Delta_2}$ and $\CP_{\Delta_3,1}\otimes\CP_{\Delta_4}$, where $\Delta_1=\frac{d}{2}+\mu_1$ and $\Delta=\frac{d}{2}+\mu_2$. After the analytical continuation, $\mu_1\in(0,\frac{d}{2}-1)$ and $\mu_2\in(0,\frac{d}{2})$. There can be pole crossing when we vary $\mu_1+\mu_2$ if $d$ is large enough. For example, when $d=4$, pole crossing happens in $\psi(\frac{2-\mu_1-\mu_2\pm i\lambda}{2})$, which is contained in $\CK^{(1)}_{\rm rel}$. It implies that when $\mu_1+\mu_2>2$, besides principal series representations, there also exists a spin 1 complementary series representation $\CC_{\mu_1+\mu_2, 1}$. A complete analysis of complementary series for higher dimension is very similar to subsection \ref{anac}.

For odd $d$, discrete series representations can appear in $\CF_{\Delta_1,s_1}\otimes\CF_{\Delta_2,s_2}$, given that one of the spins is nonzero. The full spectrum of discrete series representations in such a tensor product is not known for a generic odd $d$. When $d\ge 9$, we can exclude discrete series representations in $\CF_{\Delta_1,s_1}\otimes\CF_{\Delta_2,s_2}$, by simply analyzing its $\SO(d+1)$ components. On the one hand, a discrete series of $\SO(1,d+1)$ should contain $\SO(d+1)$ UIRs corresponding to Young diagrams with $\frac{d+1}{2}\ge 5$ rows. On the other hand, the $\SO(d+1)$ content of both $\CF_{\Delta_1,s_1}$ and $\CF_{\Delta_2, s_2}$ has at most two rows and hence their tensor product yields $\SO(d+1)$ UIRs that have at most four rows. When $d=3$, the problem is solved by Martin \cite{Martin2, Martin} if at least one of the $\CF_{\Delta_i,s_i}$ belongs to principal series. For example,  according to Martin, $\CP_{\Delta_1,1}\otimes\CP_{\Delta_2}$ contains all
$\CU^\pm_{s,0}, s\ge 1$, with multiplicity one for each\footnote{Martin used Dixmier's notation $\pi^{\pm}_{p,q}$ \cite{Dix} to denote discrete series representations, where $p\ge 1$ and $q=1,2,\cdots p$. In our notation,  $\pi^{\pm}_{p,q}$ should be identified as $\CU^\pm_{p, q-1}$.}. In particular, $\CU^+_{1,0}\oplus \CU^-_{1,0} $ is isomorphic to the single-particle Hilbert space of a photon in $\text{dS}_{d+1}$.

\subsubsection{The  case of \SO$(1,3)$}

As an explicit example, let's decompose the tensor product $\CP_{\Delta, s_1}\otimes \CP_{\Delta_2, s_2}$ of  $\SO(1,3)$. Due to the isomorphism $\CP_{\Delta, s}\cong \CP_{\bar\Delta, -s}$, we will always consider nonnegative spins, and let $\Delta$ run along the whole line $1+i\mathbb R$ when $s\ge 1$ and the half line $1+i\mathbb R_{\ge 0}$ when $s=0$.  Without loss of generality, we also assume $s_1\ge s_2$. Formally,  the following character equation is expected to hold as a manifestation of the tensor product decomposition
\begin{align}\label{2dchareq}
\Theta_{\Delta_1, s_1}(q, x)\Theta_{\Delta_2, s_2}(q, x)=\int_{0}^\infty \, \CK^{(0)}(\lambda)\Theta_{1+i\lambda}(q, x)
+\sum_{s\ge 1}\int_{\mathbb R}\, \CK^{(s)}(\lambda)\Theta_{1+i\lambda, s}(q, x)
\end{align}
where $\Delta_1=1+i\mu_1, \Delta_2=1+i\mu_2$, and $\CK^{(s)}(\lambda)$ is understood as the density of spin $s$ principal series. Using eq. (\ref{2dc}) for $\Theta_{\Delta, s}(q, x)$, we find that (\ref{2dchareq}) is equivalent to 
\small
\begin{align}\label{hhhhh}
\frac{q(x^{s_1}q^{i\mu_1}+x^{-s_1}q^{-i\mu_1})(x^{s_2}q^{i\mu_2}+x^{-s_2}q^{-i\mu_2})}{(1-x q)(1-x^{-1}q)}&=\sum_{s\ge 1}\int_{\mathbb R} d\lambda \,\CK^{(s)}(\lambda)\left(x^s\, q^{i\lambda}+x^{-s}\, q^{-i\lambda}\right)\nonumber\\
&+\int_{0}^\infty d\lambda \,\CK^{(0)}(\lambda)\left( q^{i\lambda}+ q^{-i\lambda}\right)
\end{align}
\normalsize
Denote the L.H.S of eq. (\ref{hhhhh}) by $\Phi$, the R.H.S imposes a very nontrivial constraint, namely $\Phi(q, x)=\Phi(q^{-1}, x^{-1})$, or equivalently $\Phi(t, \theta)=\Phi(-t,-\theta)$ if we make the substitution $q=e^{-t}$ and $x=e^{i\theta}$. It easy to check that $\Phi(t, \theta)=\frac{2\cos(\mu_1t-s_1\theta)\cos(\mu_2t-s_2\theta)}{\cosh t-\cos \theta}$ satisfies this condition.

To proceed, we need to match the coefficients of any $x^{\pm s}$ on both sides of eq. (\ref{hhhhh}). It requires the  Fourier expansion of $(1-x q)^{-1}(1-x^{-1}q)^{-1}$, which depends on the sign of $1-q$:
\begin{align}
\frac{1}{(1-x q)(1-x^{-1}q)}=\sum_{n\in\mathbb Z}\, x^n\times \begin{cases} \frac{q^{|n|}}{1-q^2}& 0<q<1\\   \frac{q^{-|n|}}{q^2-1} & q>1\end{cases}
\end{align}
Because of the pole at $q=1$ for each Fourier coefficient, $\CK^{(s)}(\lambda)$ defined in (\ref{2dchareq}) is divergent. In order to regularize it, we introduce another tensor product  $\CP_{\Delta_3}\otimes \CP_{\Delta_4}$, and compute a relative density $\CK_{\rm rel}^{(s)}(\lambda)$ of principal series between between  $\CP_{\Delta, s_1}\otimes \CP_{\Delta_2, s_2}$ and $\CP_{\Delta_3}\otimes \CP_{\Delta_4}$. By matching the coefficient of $x^0$, we find 
\small
\begin{align}\label{uoiui}
\CK_{\rm rel}^{(0)}(\lambda)&=-\frac{1}{4\pi}\sum_{\pm,\pm}\left(\psi\left(\frac{s_1+s_2+1\pm i(\mu_1+\mu_2)\pm i\lambda}{2}\right)+\psi\left(\frac{s_1-s_2+1\pm i(\mu_1-\mu_2)\pm i\lambda}{2}\right)\right)\nonumber\\
&+\frac{1}{4\pi}\sum_{\pm,\pm,\pm}\psi\left(\frac{1\pm i\mu_3\pm i\mu_4\pm i\lambda}{2}\right)
\end{align}
\normalsize
When $s_1=s_2=0$, (\ref{uoiui}) reduces to eq. (\ref{CKsfinal}) with $d=2, s=0$. Pole crossing does not happen if we move $\mu_1, \mu_2$ along the real axis, or analytically continue one of the two $\CP_{\Delta_i, s_i}$ to  a complementary series representation. It suggests the absence of complementary series in any $\CP\otimes\CP$ or $\CP\otimes\CC$ of $\SO(1,3)$. The $\CC\otimes\CC$ case has been studied in section \ref{anac}. Similarly, by matching $x^s$, we obtain 
\small
\begin{align}\label{uoiui}
\CK_{\rm rel}^{(s)}(\lambda)&=-\frac{1}{4\pi}\sum_{\pm}\left(\psi\left(\frac{|s\!\pm\!(s_1\!+\!s_2)|\!+\!1\!\mp \!i(\mu_1\!+\!\mu_2)\!-\! i\lambda}{2}\right)\!-\!\psi\left(\frac{|s\!\pm\!(s_1\!-\!s_2)|\!+\!1\!\mp\! i(\mu_1\!-\!\mu_2)\!-\! i\lambda}{2}\right)\right)\nonumber\\
&-\frac{1}{4\pi}\sum_{\pm}\left(\psi\left(\frac{|s\!\pm\!(s_1\!+\!s_2)|\!+\!1\!\pm \!i(\mu_1\!+\!\mu_2)\!+\! i\lambda}{2}\right)\!-\!\psi\left(\frac{|s\!\pm\!(s_1\!-\!s_2)|\!+\!1\!\pm\! i(\mu_1\!-\!\mu_2)\!+\! i\lambda}{2}\right)\right)\nonumber\\
&+\frac{1}{4\pi}\sum_{\pm,\pm,\pm}\psi\left(\frac{s+1\pm i\mu_3\pm i\mu_4\pm i\lambda}{2}\right).
\end{align}
\normalsize

\section{Conformal field theory in de Sitter}\label{CFTsec}
Since dS spacetime is conformally equivalent to   flat spacetime, the full Hilbert space $\CH_{\text{CFT}}$ of a CFT in $\text{dS}_{d+1}$ is a direct sum of unitary and irreducible primary representations of the Lorentzian conformal group $\SO(2,d+1)$. (A short review of $\SO(2,d+1)$ and its representations can be found in appendix \ref{SO2dreview}).
Then understanding the representation structure of  $\CH_{\text{CFT}}$  under the dS isometry group  $\SO(1,d+1)$ amounts to  decomposing primary representations of $\SO(2,d+1)$ into UIRs of $\SO(1,d+1)$.

In representation theory of compact groups,  the Weyl character is a powerful tool to decompose a UIR $R$ of $G$ to UIRs of its subgroup $H$. For example, let $G=\SO(4), H=\SO(3)$ and $R=\tiny\yng(2,1)$. The $\SO(4)$ Weyl character associated to $\tiny\yng(2,1)$ is 
 \begin{align}\label{SO4char}
\chi^{\SO(4)}_{\mbox{\scriptsize\tiny\yng(2,1)}}(x, y)=\frac{(x+y) (x y+1) \left(x^2 y^2+1\right)}{x^2 y^2}~.
\end{align}
 Since $G$ and $H$ have only one common Cartan generator (denoted by $J_3$), we take $y=1$ in eq.~(\ref{SO4char}), which effectively computes $\tr \, x^{J_3}$ in the representation $\tiny\yng(2,1)$, and then the character becomes 
 \begin{align}
\chi^{\SO(4)}_{\tiny\yng(2,1)}(x)=\left(x^2+x+1+x^{-1}+x^{-2}\right)+\left(x+1+x^{-1}\right)
\end{align}
which is manifestly the sum of $\SO(3)$ characters corresponding to spin 1 and spin 2 representations respectively.
 Altogether, we obtain the following simple branching rule
\begin{align}
\left.\yng(2,1)\right|_{\SO(3)}=\yng(1)\oplus\yng(2)
\end{align}
More generally, let $k$ be  the dimension of the Cartan subalgebra of $H$. Then the branching rule $R|_H=\oplus_i R_i^{\oplus n_i}$ is equivalent to the character relation
\begin{align}
\chi^G_R(x_1,x_2,\cdots, x_k)=\sum_{i} n_i \,\chi^H_{R_i}(x_1,x_2,\cdots, x_k)
\end{align}
where $n_i$ denotes the degeneracy of each UIR $R_i$ of $H$.

In this section, we will generalize the character method above to the noncompact pair  $(G, H)=(\SO(2,d+1), \SO(1,d+1))$. When $d=1$, we take $R$ to be any unitary  primary representation of $\SO(2,2)$, and when $d\ge 2$, we will focus on unitary {\it scalar} primary representations of $\SO(2,d+1)$. Some partial results regarding the spinning primary representations will also be mentioned.

\subsection{$\SO(2,d+1)$ group characters and noncompact generators}\label{kjlkll'}
Given a primary representation of $\SO(2,d+1)$, the usual character considered in CFT is defined as the trace of $e^{-\beta H}$ decorated by $\SO(d+1)$ Cartan generators, where $\beta>0$ for the convergence of trace. Rigorously speaking, such a character (which will be referred to as a CFT character henceforth) is {\it not} a group character because $e^{-\beta H}$ does not belong to the group $\SO(2,d+1)$. For a scalar primary representation $\mathcal R_\Delta$, the CFT character is given by \cite{Basile:2016aen}
\begin{align}\label{CFTchar}
\Theta^{\text{CFT}_{d+1}}_{\mathcal R_\Delta}(\beta, \bm y)=\frac{e^{-\Delta\beta}}{P_{d+1}(e^{-\beta}, \bm y)}
\end{align}
where $\bm y=(y_1,y_2,\cdots, y_{r'}), r'=\floor*{\frac{d+1}{2}}$ are fugacities of $\SO(d+1)$ rotations. For example, when $d=0$ the CFT character is simply $\frac{e^{-\Delta\beta}}{1-e^{-\beta}}$, and when $d=1$ it is $\frac{e^{-\Delta\beta}}{(1-y e^{-\beta})(1-y^{-1} e^{-\beta})}$.

Unfortunately, this well-known CFT character cannot be used to study the reduction from $\SO(2,d+1)$ to $\SO(1,d+1)$ in that its definition involves  the $\SO(2)$ generator $H$ which is not accessible to the $\SO(1,d+1)$ subgroup. Instead, we need the $\SO(2,d+1)$ character  defined with respect to the $\SO(1,d+1)$ Cartan generators, namely \footnote{We assume $0<q<1$ without loss of generality.} 
\begin{align}\label{ewr}
\Theta^{\SO(2,d+1)}_{\mathcal R_\Delta}(q,\bm x)\equiv \tr_{\mathcal R_{\Delta}} \left(q^D x_1^{J_1}\cdots x_r^{J_r}\right)
\end{align}
where $J_1, J_2,\cdots, J_r$ are the Cartan generators of $\SO(d)$, with $r=\floor*{\frac{d}{2}}$. Since $q^D x_1^{J_1}\cdots x_r^{J_r}$ is a group element of $\SO(1,d+1)$ when $x_j\in\rm U(1)$, we will call $\Theta^{\SO(2,d+1)}_{\mathcal R_\Delta}(q,\bm x)$ an $\SO(2,d+1)$ group character. It exists in the distributional sense because the trace in eq.(\ref{ewr}) is, roughly speaking, an oscillating sum. For $\SO(2,1)$, the trace is 
computed exactly in \cite{Sun:2020ame}.
In particular, the $\SO(2,1)$ group character of  $\mathcal R_\Delta$ is found to be $\frac{q^\Delta}{1-q}$,  very similar to its CFT counterpart. This result will be the main building block for us to systematically  compute the $\SO(2,d+1)$ group character of $\mathcal R_\Delta$  for any $d$. The strategy is to decompose a primary representation $\mathcal R_\Delta$ of $\SO(2,d+1)$ into primary representations of $\SO(2,1)$ while keeping track of the $\SO(d)$ spin of each $\SO(2,1)$ piece in the decomposition.
In other words, we need the reduction of $\mathcal R_\Delta$ from $\SO(2,d+1)$ to $\SO(2,1)\times\SO(d)$. The CFT character allows us to solve such a reduction very easily. When $d=2r$ is even, we have $P_{d+1}(e^{-\beta}, \bm y)= (1-e^{-\beta})P_d(e^{-\beta}, \bm y)$, and using eq. (\ref{techguess}) for $P_d(e^{-\beta}, \bm y)^{-1}$ we obtain
\begin{align}\label{2dd1}
\Theta^{\text{CFT}_{d+1}}_{\mathcal R_\Delta}(\beta, \bm y)
&=\frac{e^{-\Delta\beta}}{1-e^{-\beta}}\, \sum_{s\ge 0}\chi^{\SO(d)}_{\mY_s}(\bm y)\frac{e^{-s\beta}}{1-e^{-2\beta}}=\sum_{n, s\ge 0}\Theta^{\text{CFT}_{1}}_{\mathcal R_{\Delta+2n+s}}(\beta)\chi^{\SO(d)}_{\mY_s}(\bm y)
\end{align}
where  $(1-e^{-2\beta})^{-1}$ is expanded into a Taylor series of $e^{-2\beta}$, and each summand for fixed $n$ and $s$ is expressed as  the product of a $\text{CFT}_1$ character and an $\SO(d)$ character. It implies the following decomposition
\begin{align}\label{2dddec}
\left.\mathcal R_\Delta\right|^{\SO(2,d+1)}_{\SO(2,1)\times \SO(d)}=\bigoplus_{n, s\ge 0}\left[\mathcal R_{\Delta+2n+s}\right]_{\SO(2,1)}\otimes \left[\mY_s\right]_{\SO(d)}
\end{align}
Applying the $\SO(2,d+1)$ character to this decomposition yields
\begin{align}\label{2dd}
\Theta^{\SO(2,d+1)}_{\mathcal R_\Delta}(q,\bm x)=\sum_{n,s\ge 0} \Theta^{\SO(2,1)}_{\mathcal R_{\Delta+2n+s}}(q)\chi^{\SO(d)}_{\mY_s}(\bm x)
\end{align}
Since $\SO(2,1)$ group characters and $\text{CFT}_1$ characters take the same form upon the replacement $q\to e^{-\beta}$, eq. (\ref{2dd1}) and eq. (\ref{2dd}) guarantee that this identification also holds for any even $d$, namely
\begin{align}\label{jio}
\text{Even}\,\, d: \,\,\,\,\Theta^{\SO(2, d+1)}_{\mathcal R_\Delta}(q, \bm x)=\frac{q^\Delta}{P_{d+1}(q, \bm x)}
\end{align}
When $d=2r+1$ is odd,  $\SO(d+1)$ has one more Cartan generator than $\SO(d)$. Therefore, in order to derive the reduction from $\SO(2,d+1)$ to $\SO(2,1)\times\SO(d)$, we should set $y_{r+1}=1$ in eq. (\ref{CFTchar}). Let $\hat{\bm y}$ be the first $r$ components of $\bm y$. Since $P_{d+1}(e^{-\beta}, \bm y)=\prod_{i=1}^{r+1}(1-y_i e^{-\beta})(1-y_i^{-1} e^{-\beta})$, taking $y_{r+1}$ yields
\begin{align}
P_{d+1}(e^{-\beta}, \hat{\bm y}, 1)=\left(1-e^{-\beta}\right)^2\prod_{i=1}^{r}(1-y_i e^{-\beta})(1-y_i^{-1} e^{-\beta})=(1-e^{-\beta})P_d(e^{-\beta}, \hat{\bm y})
\end{align}
Applying eq. (\ref{techguess}) for $P_d(e^{-\beta}, \hat{\bm y})^{-1}$ we obtain
\begin{align}\label{2dd2}
\Theta^{\text{CFT}_{d+1}}_{\mathcal R_\Delta}(\beta, \hat{\bm y}, 1)
=\sum_{n, s\ge 0}\Theta^{\text{CFT}_{1}}_{\mathcal R_{\Delta+2n+s}}(\beta)\chi^{\SO(d)}_{\mY_s}(\hat{\bm y})
\end{align}
which  leads to the decomposition (\ref{2dddec}), and 
\begin{align}\label{odddk}
\text{Odd}\,\, d:\,\,\,\,\, \Theta^{\SO(2, d+1)}_{\mathcal R_\Delta}(q, \bm x)=\frac{q^\Delta}{P_{d+1}(q, \bm x, 1)}
\end{align}
Altogether, eq. (\ref{jio}) and eq. (\ref{odddk}) can be uniformly expressed as 
\begin{align}\label{alldl}
\Theta^{\SO(2, d+1)}_{\mathcal R_\Delta}(q, \bm x)=\frac{q^\Delta}{(1-q)P_{d}(q, \bm x)}
\end{align}
Similarly, for a spinning primary representation $\mathcal R_{\Delta,\ell}$, one can show that the corresponding $\SO(2,d+1)$ character is given by 
\begin{align}\label{dlc}
\Theta^{\SO(2, d+1)}_{\mathcal R_{\Delta,\ell}}(q, \bm x)=\frac{q^\Delta}{(1-q)P_{d}(q, \bm x)}\times\begin{cases} \chi^{\SO(d+1)}_{\mY_\ell}(\bm x), & d=2r\\  \chi^{\SO(d+1)}_{\mY_\ell}(\bm x, 1), & d=2r+1\end{cases}
\end{align}

\subsection{From $\SO(2,2)$ to $\SO(1,2)$ }\label{sec:SO22 to SO12}
For $d=1$, the restriction of any primary representation of  $\SO(2,2)$ to $\SO(2,1)$ is partially solved in \cite{Hogervorst:2021uvp}. It is found that an $\SO(2,2)$ primary representation $\mathcal R_{\Delta,\ell}$ of energy $\Delta$ and spin $\ell$, satisfying the unitarity bound $\Delta\ge |\ell|$, contains  $\SO(1,2)$ principal series $\CP_{\frac{1}{2}+i\lambda}$ for all $\lambda$. In addition, there is a complementary series representation $\CC_{1-\Delta}$ when $\Delta<\frac{1}{2}$ and  $\ell=0$, and $|\ell|$ discrete series representations $\CD^{\text{sign}(\ell)}_{1}, \CD^{\text{sign}(\ell)}_{2},\cdots, \CD^{\text{sign}(\ell)}_{|\ell|}$ when $\ell\not=0$. However, a properly defined density of the principal series is still missing. We will solve this problem by using Harish-Chandra characters.

\subsubsection{Character analysis}
\label{sec:characterCFTd1}
Using a straightforward generalization of the derivation in subsection \ref{kjlkll'}, we find that the $\SO(2,2)$ group character of the primary representation $\mathcal R_{\Delta,\ell}$ is given by
\begin{align}\label{hlkhh}
\Theta^{\SO(2,2)}_{\mathcal R_{\Delta,\ell}}(q)=\frac{q^\Delta}{(1-q)^2}, \,\,\,\,\, 0<q<1~.
\end{align}
This character is insensitive to the spin $\ell$ because we have only turned on one Cartan generator while $\SO(2,2)$ has two commuting generators, one counting the energy and the other counting the spin. In the spinning case, i.e. $\Delta>|\ell|>0$, we expect  the following formal expansion 
\begin{align}\label{eq:theta SO22}
\Theta^{\SO(2,2)}_{\mathcal R_{\Delta,\ell}}(q)=\int_0^\infty \, d\lambda\, \CK(\lambda;\ell) \frac{q^{\Delta_\lambda}+q^{\bar\Delta_\lambda}}{1-q}+\sum_{k=1}^{|\ell|}\frac{q^{k}}{1-q}
\end{align}
where $\CK(\lambda;\ell)$ is a formal ``density'' of $\SO(1,2)$ principal series. As we have seen many times in previous sections, $\CK(\lambda;\ell)$ defined this way is actually divergent because $\Theta^{\SO(2,2)}_{\mathcal R_{\Delta,\ell}}(q)$ has a double pole at $q=1$ while $\SO(1,2)$ characters only have a single pole at $q=1$. To regularize the divergence in $ \CK(\lambda;\ell)$, we introduce another spinless primary representation $\mathcal R_{\Delta'}$ with $\Delta'>\frac{1}{2}$ such that it does not contain any complementary series representation of $\SO(1,2)$ and define a relative density $\CK_{\rm rel}(\lambda;\ell)$ as 
\begin{align}\label{22rel}
\Theta^{\SO(2,2)}_{\mathcal R_{\Delta,\ell}}(q)-\Theta^{\SO(2,2)}_{\mathcal R_{\Delta'}}(q)=\int_0^\infty \, d\lambda\, \CK_{\rm rel}(\lambda;\ell)\Theta_{\Delta_\lambda}(q)+\sum_{k=1}^{|\ell|}\frac{q^{k}}{1-q}\,.
\end{align}
From eq.~(\ref{22rel}), we can easily solve the relative density using (\ref{CIresult})
\small
\begin{align}\label{CKrel90}
\CK_{\rm rel}(\lambda;\ell)&=\frac{1}{\pi}\int_0^\infty\,dt\,\left(\frac{e^{-(\Delta-\frac{1}{2})t}-e^{-(\Delta'-\frac{1}{2})t}}{1-e^{-t}}-\sum_{k=1}^{|\ell|}e^{-(k-\frac{1}{2})t}\right)
\cos(\lambda t)\nonumber\\
&=\frac{1}{2\pi}\sum_{\pm}\left[\psi\left(\Delta'-\frac{1}{2}\pm i\lambda \right)-\psi\left(\Delta-\frac{1}{2}\pm i\lambda\right) \right]-\frac{1}{\pi}\sum_{k=1}^{|\ell|}\frac{k-\frac{1}{2}}{(k-\frac{1}{2})^2+\lambda^2}
\end{align}
\normalsize
where the first term is independent of $\ell$ and the second term is independent of $\Delta$. Equation (\ref{22rel}) also holds for  $\ell=0$ as long as $\Delta\ge\frac{1}{2}$, since $\mathcal R_{\Delta}$ does not contain complementary series of $\SO(1,2)$ in this regime, and the corresponding $\CK_{\rm rel}(\lambda;0)$ gives the relative density of principal series between $\mathcal R_\Delta$ and $\mathcal R_{\Delta'}$. Taking $\Delta<\frac{1}{2}$ for $\ell=0$, the eq.~(\ref{22rel}) should be modified as 
\begin{align}\label{22relc}
0<\Delta<\frac{1}{2}:\,\,\,\,\,\Theta^{\SO(2,2)}_{\mathcal R_{\Delta}}(q)-\Theta^{\SO(2,2)}_{\mathcal R_{\Delta'}}(q)=\int_0^\infty \, d\lambda\, \CK_{\rm  rel}(\lambda;0) \Theta_{\Delta_\lambda}(q)+\frac{q^{\Delta}+q^{1-\Delta}}{1-q} \,.
\end{align}
where we have used eq.~(\ref{CJthired}) with $n=0$ for the integral $\int_0^\infty \, d\lambda\, \CK_{\rm  rel}(\lambda;0) \Theta_{\Delta_\lambda}(q)$. The extra term $\frac{q^{\Delta}+q^{1-\Delta}}{1-q}$ is consistent with the existence of an $\SO(1,2)$ complementary series representation $\CC_{1-\Delta}$ in the primary representation $\mathcal R_\Delta$ when $0<\Delta<\frac{1}{2}$.

\begin{figure}[t]
     \centering
     \begin{subfigure}[t]{0.45\textwidth}
         \centering
          \includegraphics[width=\textwidth]{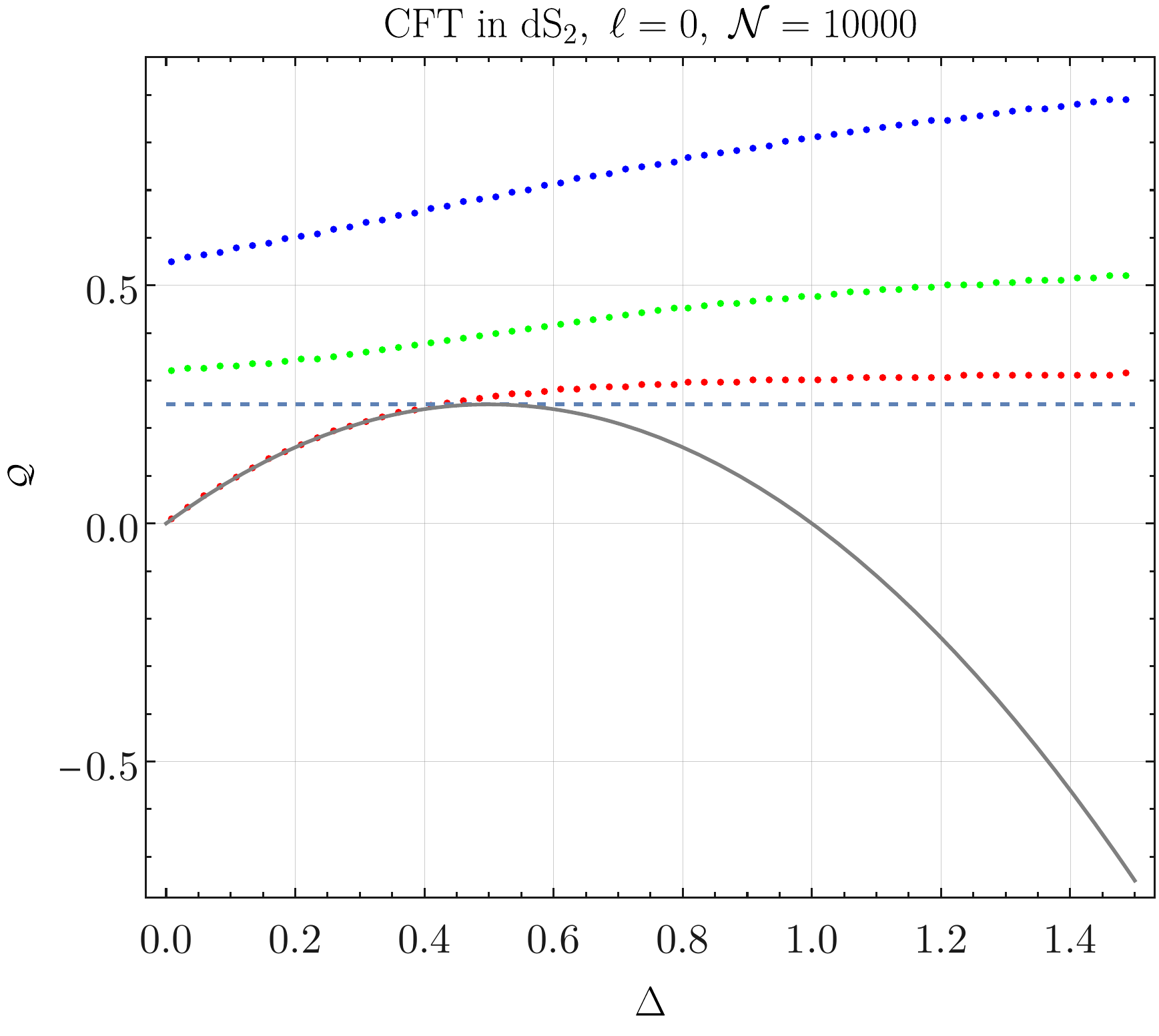}
         \label{fig:CFT1dCas}
     \end{subfigure}
     \hfill
     \begin{subfigure}[t]{0.45\textwidth}
         \centering
                 \includegraphics[width=\textwidth]{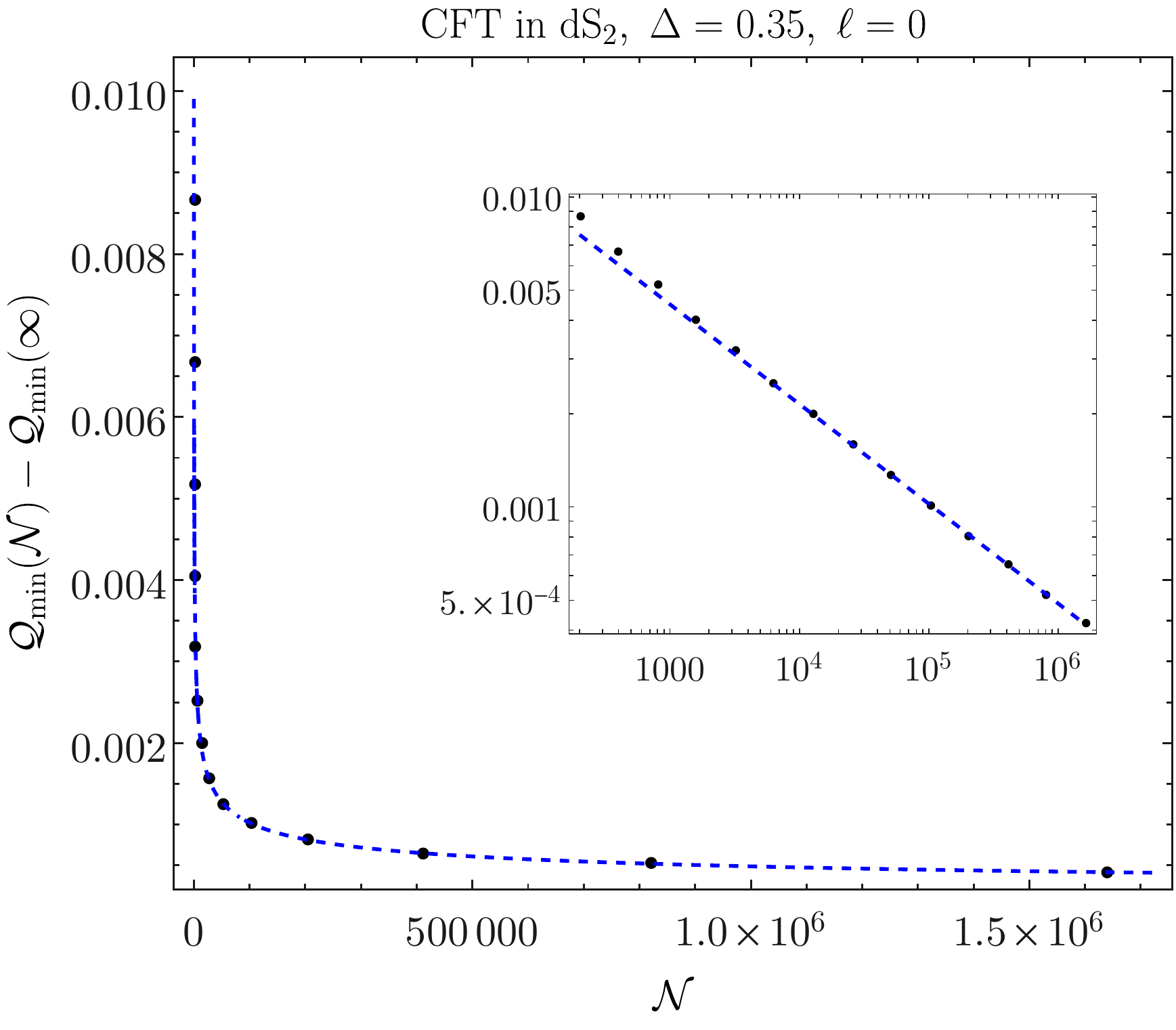}
                  \label{fig:ConvCFT1d}
     \end{subfigure}
     \caption{Low-lying eigenvalues of $\CQ$. Left: The three smallest eigenvalues for various positive $\Delta$. The dashed line is $\CQ=\frac{1}{4}$ and the gray line is $\CQ=\D(1-\D)$. Right: The convergence of the smallest eigenvalue of $\CQ$ to its large $\CN$ asymptote $\CQ_\text{min}(\infty)=\D(1-\D)$.  Inset plot is the log-log version and dashed blue line is a power law fit.
     } \label{fig:CFT1d}
\end{figure}

[\textcolor{blue}{new}]When $\Delta=|\ell|$, which corresponds to a chiral field in CFT$_2$, the primary representation $\mathcal R_{|\ell|,\ell}$ contains an infinite number of null states. More explicitly, we have a basis $\{|n,\bar n), n, \bar n\ge 0\}$ for any $\mathcal R_{\Delta,\ell}$ defined by eq. (\ref{nbn}). According to eq. (\ref{nbnbnn}), states with $\bar n\not=0$ are null when $\Delta=\ell$, and states with nonzero $n$ are null when $\Delta=-\ell$. In addition, it is straightforward to check that the physical states form the $\SO(1,2)$ discrete series representation $\CD^\pm_\ell$ when $\Delta=\pm\ell$. Altogether, the $\SO(2,2)$ UIR corresponding to a chiral field only gives a discrete series representation when restricted to $\SO(1,2)$.

\subsubsection{Numerical check}\label{sec:SO22 to SO12 Num}
The $\SO(1,2)$ Casimir restricted to the subspace spanned by scalar descendants, i.e. $\SO(2,2)$ descendants  that are also $\SO(2)$ singlets,  contains all the information about principal and complementary series representations in $\mathcal R_{\Delta,\ell}$. Denote the normalized scalar descendant states by $|n\rangle, n\ge | {\rm min} (0,\ell)|$, where each $|n\rangle$ is at level $2n+\ell$.  The matrix elements of $\Cas^{\SO(1,2)}$ with respect to the $|n\rangle$ basis have been derived in the appendix D of \cite{Hogervorst:2021uvp}. Define $\CQ_{n,m}=\langle n|\Cas^{\SO(1,2)}|m\rangle$, and the nonvanishing entries of $\CQ$ are 
\begin{align}
&\CQ_{n,n}=2n(n+\ell)+\Delta(\ell+2n+1)\nonumber\\
&\CQ_{n+1,n}=\CQ_{n,n+1}=-\sqrt{(n+1)(n+\Delta)(n+\ell+1)(n+\Delta+\ell)}~.
\end{align}
Similar to previous sections, we can show that this implies $\Cas^{\SO(1,2)}$ has continuous spectrum on principal series. As before, we may introduce a truncation cutoff $\CN$ on $n$ and diagonalize the $\CQ$  matrix to find a finite set of eigenstates $q_i$.  Any eigenvalue smaller than $\frac{1}{4}$ for large $\CN$ corresponds to a complementary series representation with $\SO(1,2)$ Casimir given by this eigenvalue. The first three eigenvalues of $\CQ$ for  scalar primary representations $\mathcal R_{\Delta}$ are plotted as a function of $\Delta$ in the left panel of fig. (\ref{fig:CFT1d}), where the cutoff is chosen to be $\CN=10000$. For $\Delta$ smaller than roughly $\frac{1}{2}$, the smallest eigenvalue of $\CQ$ is below the critical line $\CQ=\frac{9}{4}$ and lies on the curve $\Delta(1-\Delta)$, suggesting the appearance of  a single complementary series representation $\CC_{1-\Delta}$ in $\mathcal{R}_\Delta$ in the large $\CN$ limit. The convergence to the expected value $\Delta(1-\Delta)$ for the $\Delta=0.35$ case is illustrated in the right panel of  fig. (\ref{fig:CFT1d}).

We parametrize the eigenstates with $q_i>\frac{1}{4}$ corresponding to the principal series of dimensions $\D_i =\half+i\lambda_i$ with $\lambda_i=\sqrt{q_i-\frac{1}{4}}$. A coarse-grained density of states is given by the inverse of the spacing of $\{\lambda_i\}$:
\be
\bar{\rho}_\CN(\lambda_i)= \frac{2}{\lambda_{i+1}-\lambda_{i-1}}~.
\ee
Like before, we consider a relative coarse-grained density, which has a finite $\CN\to\infty$ limit
\be\label{relco1}
\rho_{\text{rel}}=\bar{\rho}_\CN^{\mathcal{R}_{\D,\ell}} - \bar{\rho}^{\mathcal{R}_{\D^\prime}}_\CN~.
\ee
Its remarkable match with the character based density $\CK_{\text{rel}}(\lambda;\ell)$
is shown in fig.\reef{fig:rhoCFT1d}.

Similar to the previous sections, we observe a match between the coarse-grained density $\bar{\rho}_\CN$ with Pauli-Villars type renormalized $\CK(\lambda,\ell)$. For more details see appendix~\ref{sec:L0s} and fig. (\ref{fig:rhoCFTs}).

\begin{figure}[t]
     \centering
          \includegraphics[width=0.6\textwidth]{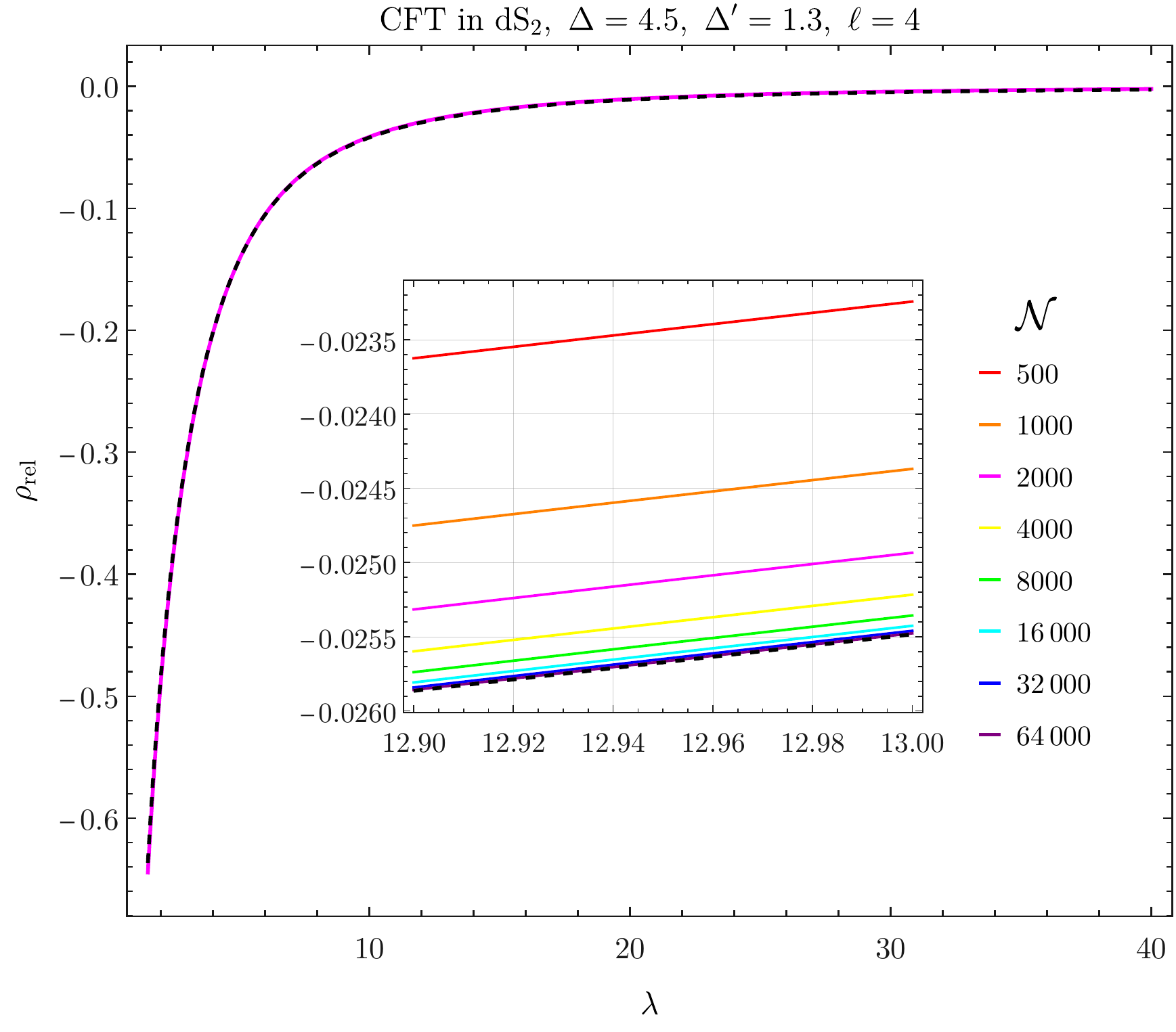}
         \caption{Comparison of the relative coarse-grained density $\rho_{\text{rel}}(\lambda)$ (pink solid line) defined in eq.~(\ref{relco1}) to  $\CK_{\text{rel}}(\lambda)$ (black dashed line) derived in eq.~(\ref{CKrel90}), with  $\ell=4,\D=4.5, \D^\prime =1.3$ and  $\CN=2000$. In the inset plot, the convergence to $\CK_{\text{rel}}(\lambda)$ (black dashed line) in the large $\CN$ limit is shown where we zoomed in to the range $12.9<\lambda<13$.}
         \label{fig:rhoCFT1d}
\end{figure}

\subsection{From SO$(2, 3)$ to SO$(1,3)$}\label{23to13}
Because SO$(1,3)$ only has principal series and complementary series, the reduction of any unitary primary representation $\mathcal R_{\Delta, \ell}$ of $\SO(2,3)$ to $\SO(1,3)$ is actually simpler than its lower dimensional and higher dimensional counterparts. In this subsection, we perform a character analysis of such a reduction. 
First, the $\SO(2,3)$ character of $\mathcal R_{\Delta, \ell}$ is given by eq. (\ref{dlc}):
\begin{align}
\Theta^{\SO(2,3)}_{\mathcal R_{\Delta,\ell}}(q, x)=\frac{\chi^{\SO(3)}_{\mY_\ell}(x)}{(1-x\,q )(1-x^{-1} q)}\frac{q^\Delta}{1-q}
\end{align}
where $\chi^{\SO(3)}_{\mY_\ell}(x)=\sum_{m=-\ell }^\ell x^m$. Due to the singularity at $q=1$, the $\SO(2,3)$ character $\Theta^{\SO(2,3)}_{\mathcal R_{\Delta,\ell}}(q, x)$ itself does not admit a well-defined decomposition into $\SO(1,3)$ characters. So we introduce another primary representation $\mathcal R_{\Delta', \ell}$ with $\Delta'> 1$. Then the difference $\Theta^{\SO(2,3)}_{\mathcal R_{\Delta,\ell}}(q, x)-\Theta^{\SO(2,3)}_{\mathcal R_{\Delta',\ell}}(q, x)$ can be expanded in terms of $\SO(1,3)$ characters. For $\Delta>1$, using the Fourier transformation
\begin{align}\label{123ed}
\frac{q^{\Delta-1}-q^{\Delta'-1}}{1-q}&=\int_{\mathbb R}\, d\lambda\, e^{i\lambda t}\CK_{\rm rel}(\lambda), \,\,\,\,\, q= e^{-|t|}\nonumber\\
\CK_{\rm rel}(\lambda)&=\frac{1}{2\pi}\sum_{\pm}\left[\psi(\Delta'-1\pm i\lambda)-\psi(\Delta-1\pm i\lambda)\right]
\end{align}
we obtain
\begin{align}\label{sdff}
\Theta^{\SO(2,3)}_{\mathcal R_{\Delta,\ell}}(q, x)-\Theta^{\SO(2,3)}_{\mathcal R_{\Delta',\ell}}(q, x)&=\int_0^\infty d\lambda\, \CK_{\rm rel}(\lambda) \Theta^{\SO(1,3)}_{1+i\lambda}(q, x)\nonumber\\
&+\sum_{m=1}^\ell \int_0^\infty d\lambda\, \CK_{\rm rel}(\lambda)\left( \Theta^{\SO(1,3)}_{1+i\lambda,m}(q, x)+ \Theta^{\SO(1,3)}_{1+i\lambda,-m}(q, x)\right)
\end{align}
where $\Theta^{\SO(1,3)}_{1+i\lambda,m}(q, x)+ \Theta^{\SO(1,3)}_{1+i\lambda,-m}(q, x)=\frac{(x^m+x^{-m})(q^{1+i\lambda}+q^{1-i\lambda})}{(1-x q)(1-x^{-1}q)}$ is the SO$(1,3)$ character of the reducible representation $\CP_{1+i\lambda, m}\oplus \CP_{1+i\lambda, -m}$, which is isomorphic to $\CP_{1-i\lambda, m}\oplus \CP_{1-i\lambda, -m}$ because of $\CP_{1+i\lambda, m}\cong\CP_{1-i\lambda, -m}$. This reducible representation describes a massive field of spin $|m|$ in dS$_3$. When $\ell=0$,  eq. (\ref{sdff}) implies that there are only scalar principal series representations in $\mathcal R_{\Delta}$ with $\Delta>1$. Decreasing $\Delta$ from $1^+$ to the unitarity bound $\frac{1}{2}$, pole crossing happens in the integral $\int_0^\infty d\lambda\, \CK_{\rm rel}(\lambda) \Theta^{\SO(1,3)}_{1+i\lambda}(q, x)$, which signals the appearance of a complementary series representation. Indeed, for $\frac{1}{2}<\Delta<1$ and $\Delta'>1$, we find 
\begin{align}
\Theta^{\SO(2,3)}_{\mathcal R_{\Delta}}(q, x)-\Theta^{\SO(2,3)}_{\mathcal R_{\Delta'}}(q, x)&=\int_0^\infty d\lambda\, \CK_{\rm rel}(\lambda) \Theta^{\SO(1,3)}_{1+i\lambda}(q, x)+\Theta^{\SO(1,3)}_{2-\Delta}(q, x)
\end{align}
Therefore, in addition to the principal series $\CP_{1+i\lambda}$, there is also a complementary series representation $\CC_{2-\Delta}$ in $\mathcal R_\Delta$ if $\Delta$ is smaller than 1. For $\ell\ge 1$, unitarity requires $\Delta\ge 1+\ell\ge 2$, so there cannot be any $\SO(1,3)$ complementary series representation in $\mathcal R_{\Delta,\ell}$. Instead, eq. (\ref{sdff}) implies that besides $\CP_{1+i\lambda}$, there are also spinning principal series $\CP_{1+i\lambda, \pm m}$ with  $m=1,2,
\cdots,\ell$ in $\mathcal R_{\Delta,\ell}$. Principal series  of different spins share the same (relative) density $\CK_{\rm rel}(\lambda)$.

\subsection{From $\SO(2,d+1)$ to $\SO(1,d+1)$: scalar primary representations}
Let $|\Delta\rangle$ be the primary state of $\mathcal R_\Delta$. A generic descendant of $|\Delta\rangle$ should be a linear combination of $L_{a_1}^+\cdots L_{a_k}^+|\Delta\rangle$, where $L^+_a$ are raising operators given by eq.~(\ref{confgen}). Treating $L_{a_1}^+\cdots L_{a_k}^+$ as a symmetrized tensor of $k$ spin $1$ representations of $\SO(d+1)$, which can be decomposed as a direct sum of $\mY_k, \mY_{k-2},\mY_{k-4}, \cdots$, then the possible irreducible $\SO(d+1)$ structures in $\mathcal R_\Delta$ are all its single-row representations. According to the review in section \ref{repreview}, the absence of two-row representations of $\SO(d+1)$ implies that the only possible $\SO(1,d+1)$ structure in $\mathcal R_\Delta$ are the spinless principal series, complementary series, and the type \rom{1} exceptional  series $\CV_{s,0}$. 

We first prove  that $\CV_{s,0}$ is not present in $\mathcal R_\Delta$. A $\CV_{s,0}$ representation in $\mathcal R_\Delta$ is characterized by a spin-$s$ state $|\phi)_{a_1\cdots a_s}$ satisfying $L_{a_10}|\phi)_{a_1\cdots a_s}=0$ because $\CV_{s,0}$ does not contain any $\mY_k$ components with $k<s$ while $L_{a_10}|\phi)_{a_1\cdots a_s}$ has the symmetry of $\mY_{s-1}$\footnote{More rigorously, $\CV_{s, 0}$ is characterized by a spin-$s$ state $|\phi)_{a_1\cdots a_s}$ such that the $\mY_{s-1}$ and $\mY_{s, 1}$ components of $L_{a}|\phi)_{a_1\cdots a_s}$ are vanishing. It is straightforward to check that a state satisfying these conditions is an eigenstate of $C^{\SO(1,d+1)}$ with the desired eigenvalue. When $|\phi)_{a_1\cdots a_s}\in\mathcal R_\Delta$, the $\mY_{s,1}$ part vanishes automatically since $\mathcal R_\Delta$ does not contain any two-row representation of $\SO(d+1)$.\label{xxx}}. In the primary representation $\mathcal R_\Delta$, any spin-$s$ state is a linear combination of $(L^+\cdot L^+)^n(L^+_{a_1}\cdots L^+_{a_s}-\text{trace})|\Delta\rangle, n\ge 0$, which in the index-free formalism can be written compactly as 
\begin{align}\label{phinsdef}
|\phi_n^s)\equiv \left(L^+\cdot L^+\right)^n \left(z\cdot L^+\right)^s|\Delta\rangle
\end{align}
where $z^a$ is an auxiliary  null vector in $\mathbb C^{d+1}$. Define $|\phi,z)=\sum_{n\ge 0} a_n |\phi^s_n)$ and then the analogue of 
$L_{a_10}|\phi)_{a_1\cdots a_s}=0$  in the index-free formalism is 
\begin{align}
L_{a0}\CD_{z^a} |\phi,z)=0, \,\,\,\,\, \CD_{z_a}=\partial_{z^a}-\frac{1}{d+2z\cdot\partial_z-1}z_a\partial_z^2~.
\end{align}
 Using eq.~(\ref{L-mns}) and (\ref{bgexplicit}), we find 
 \begin{align}
& L^+_a\CD_{z^a}|\phi_n^s)=\frac{s(d+s-2)}{d+2s-3}|\phi^{s-1}_{n+1})\nonumber\\
&L^-_a\CD_{z^a}|\phi_n^s)=\frac{4s(d+s-2)(n+s+\Delta-1)(n+s+\frac{d-1}{2})}{d+2s-3}|\phi^{s-1}_{n})~.
 \end{align}
 Since $L_{a0}=\frac{i}{2}\left(L^+_a+L^-_a\right)$, the requirement $L_{a0}\CD_{z^a} |\phi,z)=0$ leads to a recurrence relation of the coefficients $a_n$
 \begin{align}\label{huioljh}
 4 (n+s+\Delta)\left(n+s+\frac{d+1}{2}\right)a_{n+1}+a_{n}=0
 \end{align}
where $n\ge-1$ and $a_{-1}\equiv 0$. The equation corresponding to $n=-1$ holds if and only if $a_0=0$, which further requires all $a_n$ to vanish because of the recurrence relation eq.~(\ref{huioljh}). Altogether, $\mathcal R_\Delta$ does not contain any $\SO(1,d+1) $ representation belonging to the  type \rom{1} exceptional series, and hence the only possible $\SO(1,d+1)$ species is spinless principal or  complementary series representation.

\subsubsection{Character analysis}
Assuming the absence of complementary series in $\mathcal R_\Delta$ when $\Delta$ is larger than some critical value $\Delta_c$, which will be confirmed by numerics shortly, we expect the following character relation  to hold
\begin{align}\label{2drel}
\Theta^{\SO(2,d+1)}_{\mathcal R_\Delta}(q,\bm x)-\Theta^{\SO(2,d+1)}_{\mathcal R_{\Delta'}}(q,\bm x)=\int_0^\infty d\lambda\, \CK_{\rm rel}(\lambda)\Theta^{\SO(1,d+1)}_{\Delta_\lambda}(q,\bm x), \,\,\,\,\, \Delta,\Delta'>\Delta_c
\end{align}
where $\Theta^{\SO(1,d+1)}_{\Delta_\lambda}(q,\bm x)$ is the $\SO(1, d+1)$ character of principal series representation $\CP_{\frac{d}{2}+i\lambda}$, and $\CK_{\rm rel}(\lambda)$ is supposed to be  the relative density of $\CP_{\frac{d}{2}+i\lambda}$ between the two primary representations $\mathcal R_\Delta$ and $\mathcal R_{\Delta'}$. Using eq. (\ref{alldl}) and making the change of variable $q\to e^{-|t|}$, then the character relation (\ref{2drel})  is equivalent to the Fourier transform
\begin{align}\label{CKeqdd}
\frac{e^{-(\Delta-\frac{d}{2}) |t|}-e^{-(\Delta'-\frac{d}{2})|t|}}{1-e^{-|t|}}=\int_{\mathbb R} d\lambda\, \CK_{\rm rel}(\lambda)e^{i\lambda t}
\end{align}
where we have extended $ \CK_{\rm rel}(\lambda)$ to an even function on the whole real line. For $\Delta, \Delta'>\frac{d}{2}$, the solution of eq.~(\ref{CKeqdd}) is 
\begin{align}\label{MbA}
\CK_{\rm rel}(\lambda)=\frac{1}{2\pi}\sum_{\pm}\left[\psi\left(\Delta'-\frac{d}{2}\pm i\lambda\right)-\psi\left(\Delta-\frac{d}{2}\pm i\lambda\right)\right]~.
\end{align}
Since the unitarity bound on $\mathcal R_\Delta$ only requires $\Delta>\frac{d-1}{2}$, $\Delta-\frac{d}{2}$ can be negative.
In this case,  eq.~(\ref{CKeqdd}) needs a modification, which follows from the integral eq.~(\ref{CJthired}) with $n=0$
\small
\begin{align}\label{CKeqdd1}
\frac{d-1}{2}<\Delta<\frac{d}{2}:\,\,\,\, \frac{e^{-(\Delta-\frac{d}{2}) |t|}-e^{-(\Delta'-\frac{d}{2})|t|}}{1-e^{-|t|}}=\int_{\mathbb R} d\lambda\, \CK_{\rm rel}(\lambda)e^{i\lambda t}+\frac{e^{-(\Delta-\frac{d}{2})t}+e^{(\Delta-\frac{d}{2})t}}{1-e^{-|t|}}
\end{align}
\normalsize
and then the character relation  (\ref{2drel}) should change accordingly as (suppressing the arguments of characters)
\begin{align}
\Theta^{\SO(2,d+1)}_{\mathcal R_\Delta}-\Theta^{\SO(2,d+1)}_{\mathcal R_{\Delta'}}=\int_0^\infty d\lambda\, \CK_{\rm rel}(\lambda)\Theta^{\SO(1,d+1)}_{\Delta_\lambda}+\Theta_{d-\Delta}^{\SO(1,d+1)}
\end{align}
which suggests $\Delta_c=\frac{d}{2}$ and the existence of a single complementary series representation $\CC_{d-\Delta}$ in $\mathcal R_\Delta$ when $\frac{d-1}{2}<\Delta<\frac{d}{2}$. The next step is to develop a numerical scheme to test these claims.

\subsubsection{Numerical check}
 \begin{figure}[t]
     \centering
     \begin{subfigure}[t]{0.43\textwidth}
         \centering
          \includegraphics[width=\textwidth]{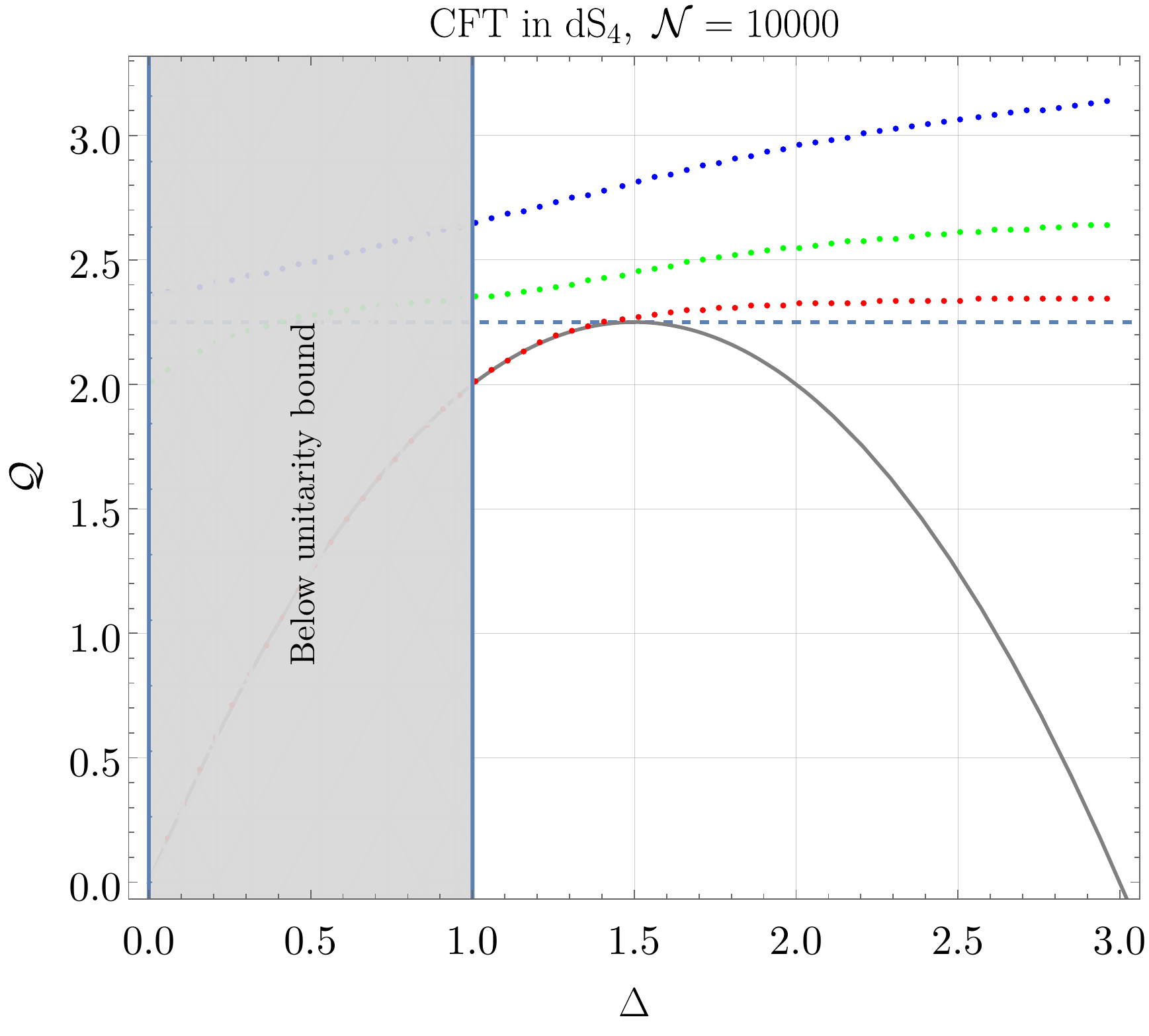}
       \label{fig:CFThdCas}
     \end{subfigure}
     \hfill
     \begin{subfigure}[t]{0.47\textwidth}
         \centering
                 \includegraphics[width=\textwidth]{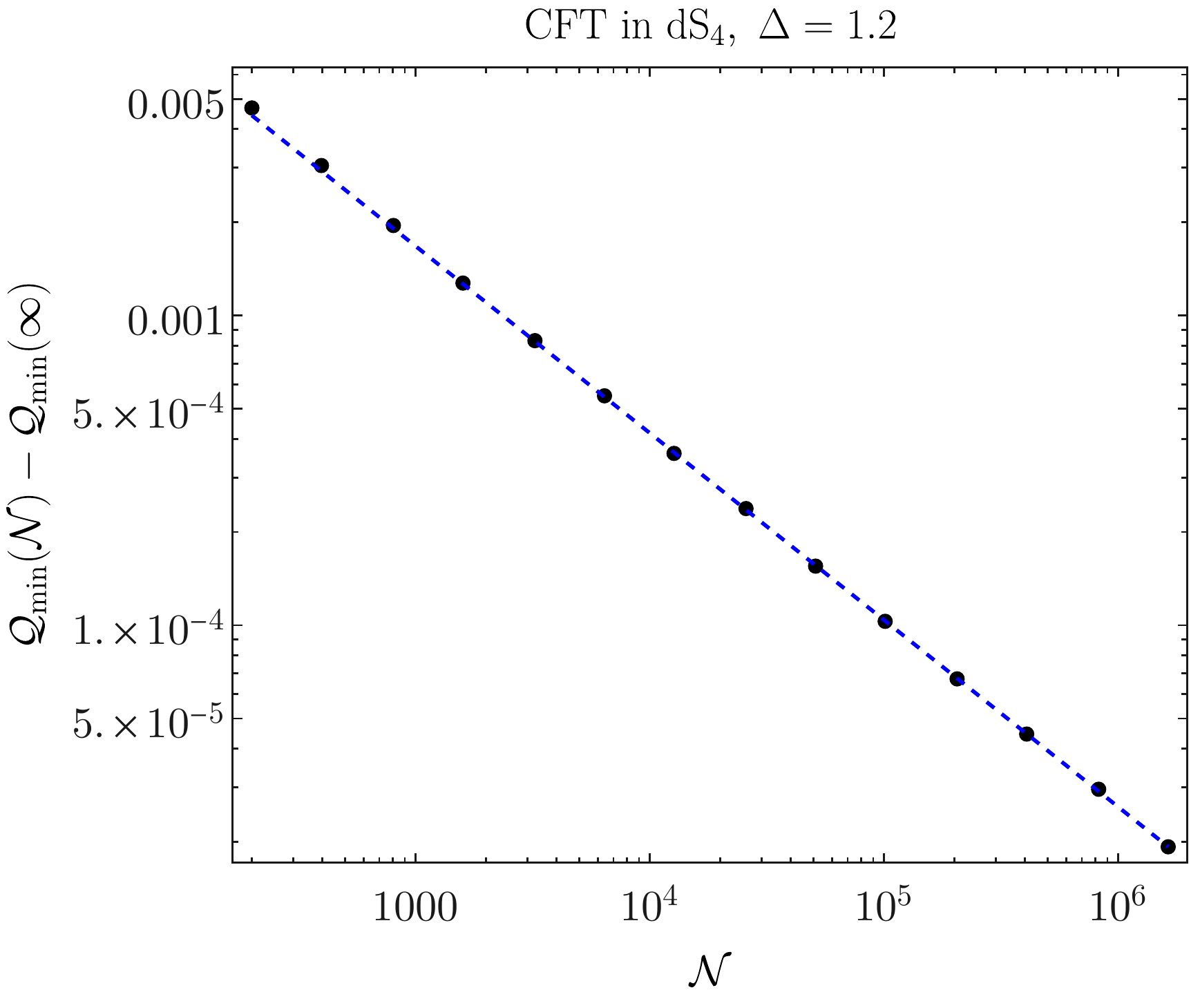}
                  \label{fig:ConvCFThd}
     \end{subfigure}
     \caption{Low-lying eigenvalues of the $\SO(1,4)$ Casimir matrix for the primary representation $\mathcal{R}_\D$ of a CFT in dS$_4$. Left: The first three eigenvalues of $\CQ$, given by eq. (\ref{yiou}), with the cut-off being $\CN=10^4$. The gray line is  $\CQ=\D(3-\D)$ and the dashed line is $\CQ=\frac{9}{4}$. Right: The convergence of $\CQ_{\rm min}(\CN)$ to its expected value $1.2\times (3-1.2)=2.16$ in the $\CN\to\infty$ limit.}{\label{fig:CFThd}}
\end{figure}

\begin{figure}[t]
     \centering
          \includegraphics[width=0.6\textwidth]{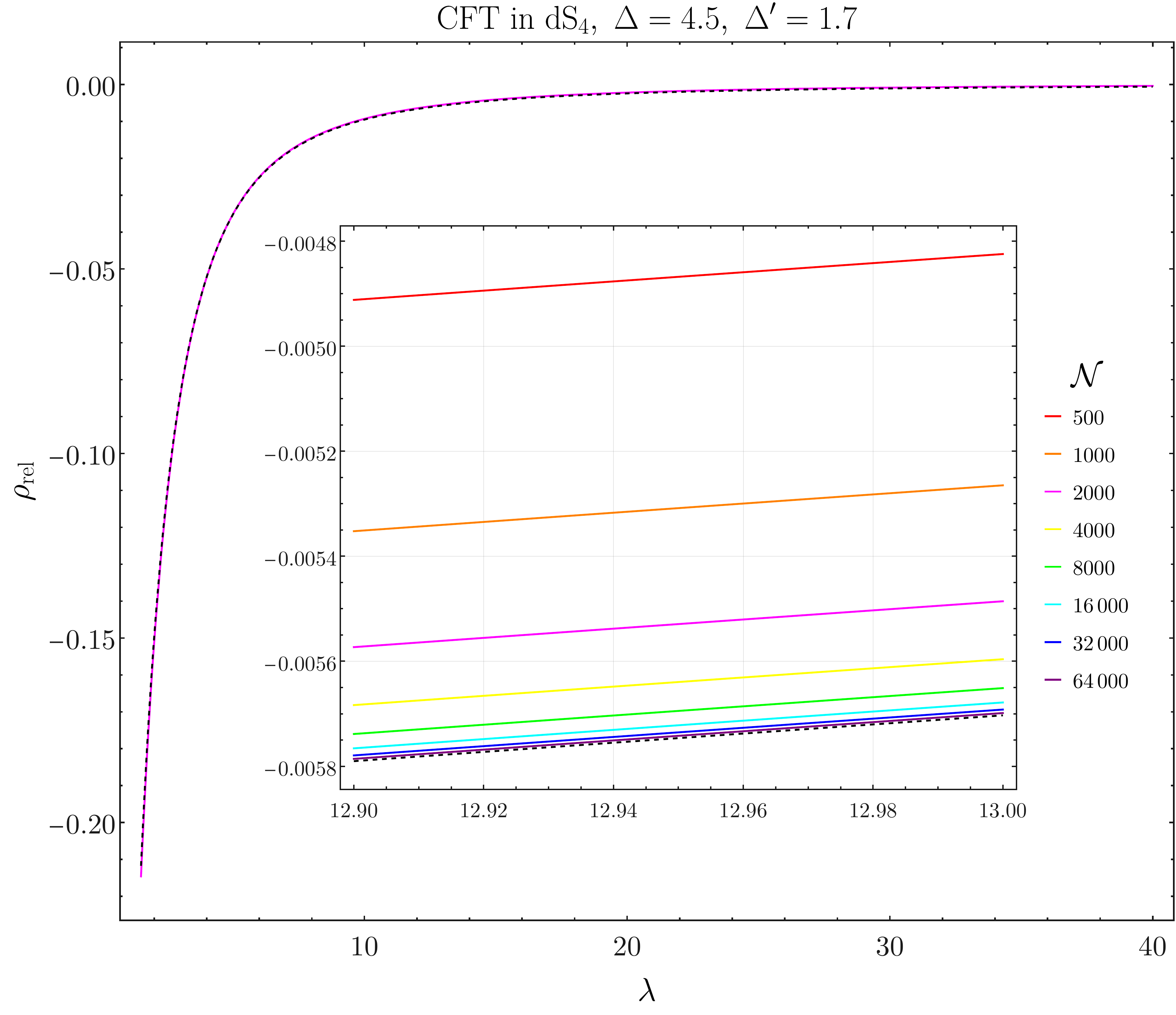}
         \caption{ Comparison of the relative coarse-grained  density $\rho_{\text{rel}}(\lambda)$ (pink solid line) to  $\CK_{\text{rel}}(\lambda)$ (black dashed line) given by eq. (\ref{MbA}) for $\D=4.5, \D^\prime =1.7$ and $\CN=2000$. The  inset plot shows the convergence to $\CK_{\text{rel}}(\lambda)$ in the large $\CN$ limit by zooming in the region $12.9<\lambda<13$. 
}         \label{fig:rhoCFThd}
     \end{figure}

To identify those continuous families of states, we need to study the spectrum of the $\SO(1,d+1)$ Casimir in the $\SO(d+1)$ singlet subspace of $\mathcal R_\Delta$, which is spanned by $|\phi_n)=\left(L^+\cdot L^+\right)^n|\Delta\rangle$. The  $\SO(1,d+1)$ Casimir acting on $|\phi_n)$ is given by  eq.~(\ref{1dCas})
\small
 \begin{align}
\Cas^{\SO(1,d+1)}|\phi_n)=\frac{1}{4}|\phi_{n+1})+\frac{1}{4}L^-\cdot L^-|\phi_n)+\left(\frac{1}{2}L^+\cdot L^-+\frac{d+1}{2}\Delta+(d+1)n\right)|\phi_n)
\end{align}
\normalsize
where both $L^-\cdot L^-|\phi_n)$ and $L^+\cdot L^-|\phi_n)$ are computed in appendix \ref{phinsprop}, c.f. eq.~(\ref{Lm2}) and 
eq.~(\ref{Lpmneigen}) respectively.
Combining all the ingredients yields 
\small
\begin{align}
\Cas^{\SO(1,d+1)}|\phi_n)&=\frac{1}{2}\left(4n(n+\Delta)+(d+1)\Delta\right)|\phi_n)+\frac{1}{4}|\phi_{n+1})\nonumber\\
&+4n\,(n+\Delta-1)\left(n+\frac{d-1}{2}\right)\left(n+\Delta-\frac{d+1}{2}\right)|\phi_{n-1})
\end{align}
\normalsize 
Define normalized states $|\phi_n\rangle\equiv \frac{1}{\sqrt{(\phi_n|\phi_n)}}|\phi_n)$, where  $(\phi_n|\phi_n)$ is computed in eq.~(\ref{phinnorm}), then the nonvanishing matrix elements of $\Cas^{\SO(1,d+1)}$ with respect to the normalized basis $|\phi_n\rangle$ are 
\begin{align}\label{yiou}
&\CQ_{nn}=2n(n+\Delta)+\frac{1}{2}(d+1)\Delta\nonumber\\
&\CQ_{n+1,n}=\CQ_{n,n+1}=\sqrt{(n+1)(n+\Delta)\left(n+\frac{d+1}{2}\right)\left(n+\Delta-\frac{d-1}{2}\right)}
\end{align}
 where $\CQ_{mn}\equiv \langle\phi_m |\Cas^{\SO(1,d+1)}|\phi_n\rangle$.
For each $q_\lambda=\frac{d^2}{4}+\lambda^2, \lambda\ge 0$ which corresponds to  a principal series representation $\CP_{\frac{d}{2}+i\lambda}$, the matrix $\CQ$ admits an eigenvector $v_n(\lambda)$ satisfying $\CQ_{nm}v_m(\lambda)=q_\lambda v_n(\lambda)$, and its asymptotic behavior at large $n$ is given by $v_n(\lambda)\sim \frac{R}{n^{\frac{1}{2}+i\lambda}}+c.c$, which implies that all $v_n(\lambda)$ are $\delta$-function normalizable. Therefore, $\Cas^{\SO(1,d+1)}$ has a continuous spectrum in the region $\left[\frac{d^2}{4}, \infty\right)$. It is also possible for the infinite dimensional matrix $\CQ$ to have eigenvalues smaller than $\frac{d^2}{4}$. To see these eigenvalues intuitively, we consider a truncated version of $\CQ_{nm}$ by imposing a hard cut-off $\CN$ on $n$ and $m$, which effectively cuts off the energy at $\Delta+2\CN$ in the primary representation $\mathcal R_\Delta$. Then we can numerically diagonalize the truncation of $\CQ$.  In fig. (\ref{fig:CFThd}), with $\CN$ chosen to be $10^4$, we plot the first three eigenvalues of $\CQ$ for primary representations $\mathcal R_\Delta$ of $\SO(2, 4)$. It shows that when $1<\Delta\lesssim\frac{3}{2}$, the smallest eigenvalue of $\CQ$ becomes smaller than $\frac{9}{4}$, lying on the line $\CQ=\Delta(3-\Delta)$ which is equal to the $\SO(1,4)$ Casimir of the complementary series representations  $\CC_{3-\Delta}$, and when $\Delta>\frac{3}{2}$ all the eigenvalues are larger than $\frac{9}{4}$. This numerical result confirms the prediction of a complementary series representation $\CC_{d-\Delta}$ in $\mathcal R_\Delta$ when $\frac{d-1}{2}<\Delta<\frac{d}{2}$. The numerical diagonalization also allows us to extract information about principal series contained in $\mathcal R_\Delta$. For each fixed $\CN$, we can define a coarse-grained density of principal series as the inverse spacing of $\lambda_n=\sqrt{q_n-\frac{d^2}{4}}$, where $q_n$ denote eigenvalues of $\CQ$ that are larger than $\frac{d^2}{4}$. Given two primary representations $\mathcal R_\Delta$ and $\mathcal R_{\Delta'}$, we obtain similarly a relative coarse-grained density $\rho_{\rm rel}$ that is finite in the large $\CN$ limit.  Fig. (\ref{fig:rhoCFThd}) confirms that  $\rho_{\rm rel}$ matches $\CK_{\rm rel}$ (c.f. eq. \eqref{MbA}), derived from charaters.
 
\subsection{From $\SO(2,d+1)$ to $\SO(1,d+1)$: Some remarks on spinning primary representations}
\label{sec:CFTspin}
Let $\mathcal{R}_{\Delta,\ell}$ be a primary representation of $\SO(2,d+1)$, built from a spin-$\ell$ primary state $|\Delta\rangle_{a_1\cdots a_\ell}$ . As the first step to constrain the possible $\SO(1,d+1)$ species in $\mathcal{R}_{\Delta,\ell}$, we list all  $\SO(d+1)$ irreducible representations in it. Descendants of $|\Delta\rangle_{a_1\cdots a_\ell}$  are linear combinations of $L^+_{b_1}\cdots L^+_{ b_k}|\Delta\rangle_{a_1\cdots a_\ell}$. Treating $L^+_{b_1}\cdots L^+_{ b_k}$ as a symmetrized tensor product of $k$ spin 1 representations of $\SO(d+1)$, which can be decomposed as a direct sum $\oplus_{m}\mY_{k-2m}$, then the possible $\SO(d+1)$ structures of descendants are encoded in the tensor products $\mY_{n}\otimes \mY_\ell, n\ge 0$, where $\mY_{\ell}$ corresponds to the primary states and $\mY_{n}$ corresponds to $(L^+_{b_1}\cdots L^+_{b_n}-\text{trace})$. The tensor product decomposition of $\mY_{n}\otimes \mY_\ell$ is given by 
eq.~(\ref{mell}):
\begin{align}
\mY_{n}\otimes \mY_\ell=\bigoplus_{a=0}^{\text{min}(n,\ell)}\bigoplus_{b=0}^{\text{min}(n,\ell)-a}\mY_{n+\ell-2a-b,b}~.
\end{align}
Some direct results of this decomposition are:
\begin{itemize}
\item Since $b\le \text{min}(n,\ell)\le \ell$,  $\mathcal{R}_{\Delta,\ell}$ does not contain any $\mY_{ss}$ with $s\ge \ell+1$. It further implies that spin $s$ UIRs of $\SO(1,d+1)$ cannot appear in $\mathcal{R}_{\Delta,\ell}$ if $s\ge \ell+1$. Therefore, the possible $\SO(1,d+1)$ UIRs in $\mathcal R_{\Delta,\ell}$ are 
\begin{align}\label{poss}
\CP_{\frac{d}{2}+i\lambda, s}, \,\,\,\,\, \CC_{\frac{d}{2}+\mu, s},\,\,\,\,\,\CU_{s,t}~,
\end{align}
for $s=0,1,\cdots, \ell$, and the type \rom{1} exceptional series.
\item The existence of some $\mY_{ss}$ requires $n+\ell-2a-b=b$, which is equivalent to $a+b=\frac{n+\ell}{2}$. On the other hand, we have $a+b\le \text{min}(n,\ell)\le\frac{n+\ell}{2}$. Therefore, $n=\ell$ and $a+b=\ell$. In other words, $\mY_{ss}$ with $0\le s\le \ell$ only appears in the tensor product $\mY_\ell\otimes \mY_{\ell}$ and appears exactly once.
Altogether, the $\mY_{ss}$-type descendants of $|\Delta\rangle_{a_1\cdots a_{\ell}}$ are of the form
\begin{align}\label{sndef}
|s,n)_{a_1\cdots a_s, b_1\cdots b_s}\equiv (L^+\cdot L^+)^n\,\hat\Pi_{ss}\left(L^+_{a_1}\cdots L^+_{a_s}L^+_{b_{s+1}}\cdots L^+_{b_\ell}|\Delta\rangle_{b_1\cdots b_\ell}\right)
\end{align}
where $n\ge 0, 0\le s\le \ell$ and $\hat\Pi_{ss}$ is a projector operator onto the $\mY_{ss}$ part. For example, when $s=1$, $\hat\Pi_{11}$ simply antisymmetrizes $a_1$ and $b_1$. For higher $s$, it antisymmtrizes the $s$ pairs of indices $[a_1, b_1], \cdots, [a_s, b_s]$, and then projects out all types of trace.
\end{itemize}

The goal for the remaining part of this section  is to gain more intuitions of the possible $\SO(1,d+1)$ UIRs contained in $\mathcal R_{\Delta,\ell}$,  at least numerically. For this purpose, we first study the spectrum  of $C^{\SO(1, d+1)}$ restricted to the subspace spanned by all $|s, n)_{a_1\cdots a_s, b_1\cdots b_s}$ for any fixed $s\in\{0,1,\cdots,\ell\}$, since it encodes information of spin $s$ principal and complementary series and $\CU_{s,t}$ in $\mathcal R_{\Delta,\ell}$. In appendix \ref{snprop}, we have derived the matrix representation of $C^{\SO(1, d+1)}$  in this subspace for $s=0$ and 1. Denote the matrix by $\mathcal Q^{(s)}$ and the nonvanishing matrix elements are given by eq. (\ref{needs}):
\begin{align}\label{hlkjss}
&\mathcal Q^{(s)}_{nn}=2n(n+\Delta+\ell)+\frac{(d+2\ell+1)\Delta-(d-1)\ell}{2}+C^{\SO(d)}(\mY_s)\nonumber\\
&\mathcal Q^{(s)}_{n+1, n}=\mathcal Q^{(s)}_{n, n+1}=\sqrt{(n+1)(n+\Delta+\ell)\left(n+\Delta-\frac{d-1}{2}\right)\left(n+\ell+\frac{d+1}{2}\right)}
\end{align}
where the diagonal entries of $\mathcal Q^{(s)}$ hold for {\it any} $s$ and the off-diagonal ones are {\it only} computed for $s=0$ and $1$.
Surprisingly, we get the same off-diagonal entries for $s=0$ and 1. Assuming that $\CQ^{(s)}_{n,n+1}=\CQ^{(s)}_{n+1,n}$ are given by eq. (\ref{hlkjss}) for any $s$, then $\mathcal Q^{(s)}=\mathcal Q^{(0)}+C^{\SO(d)}(\mY_s)$. This relation has many interesting implications. First, using the asymptotic behavior of $\CQ^{(0)}$ for 
large $n$, one can show that it has a continuous spectrum on $[\frac{d^2}{4}, \infty)$ and hence $\mathcal R_{\Delta,\ell}$ contains all the scalar principal series. Second, noticing that $C^{\SO(1,d+1)}(\CF_{\frac{d}{2}+i\lambda,s})=\frac{d^2}{4}+\lambda^2+C^{\SO(d)}(\mY_s)$, we can also immediately conclude the existence of all spin $s$ principal series with $s\in\{0,1,\cdots,\ell\}$. Third, the existence of $\CU_{s, t}$ is equivalent to  the existence of the eigenvalue $(1-t)(d+t-1)+C^{\SO(d)}(\mY_s)$ of $\CQ^{(s)}$, which is also equivalent to  the existence of the eigenvalue $(1-t)(d+t-1)$ of $\CQ^{(0)}$. This eigenvalue corresponds to $\CF_{d+t-1}$ of $\SO(1, d+1)$ which is nonunitary unless $t=0$ \footnote{Rigorously speaking, $\CV_{t, 0}$ has the same Casimir  $(1-t)(d+t-1)$. However, $\CQ^{(0)}$ is defined in the $\SO(d+1)$ singlet subspace and $\CV_{t, 0}$ does not contain any $\SO(d+1)$ singlet. So $\CV_{t, 0}$ can be easily excluded.}. Since we start with a unitary CFT in dS$_{d+1}$, we do not expect any nonunitary representations of $\SO(1,d+1)$. To test the $t=0$ one and other complementary series numerically, we can truncate $\CQ^{(0)}$ by a hard cut-off $n\le \CN$ and perform a diagonalization.  In fig. \reef{fig:CFThdl}, we show low-lying eigenvalues of the truncated $\CQ^{(0)}$ for various primary representation $\mathcal R_{\Delta, \ell}$ in various dimensions. The absence of any eigenvalue below $\frac{d^2}{4}$ seems to exclude complementary series and type \rom{2} exceptional series in $\mathcal R_{\Delta, \ell}$  with $\ell\ge 1$.

\begin{figure}[]
     \centering
     \begin{subfigure}[t]{0.3\textwidth}
         \centering
         \includegraphics[width=\textwidth]{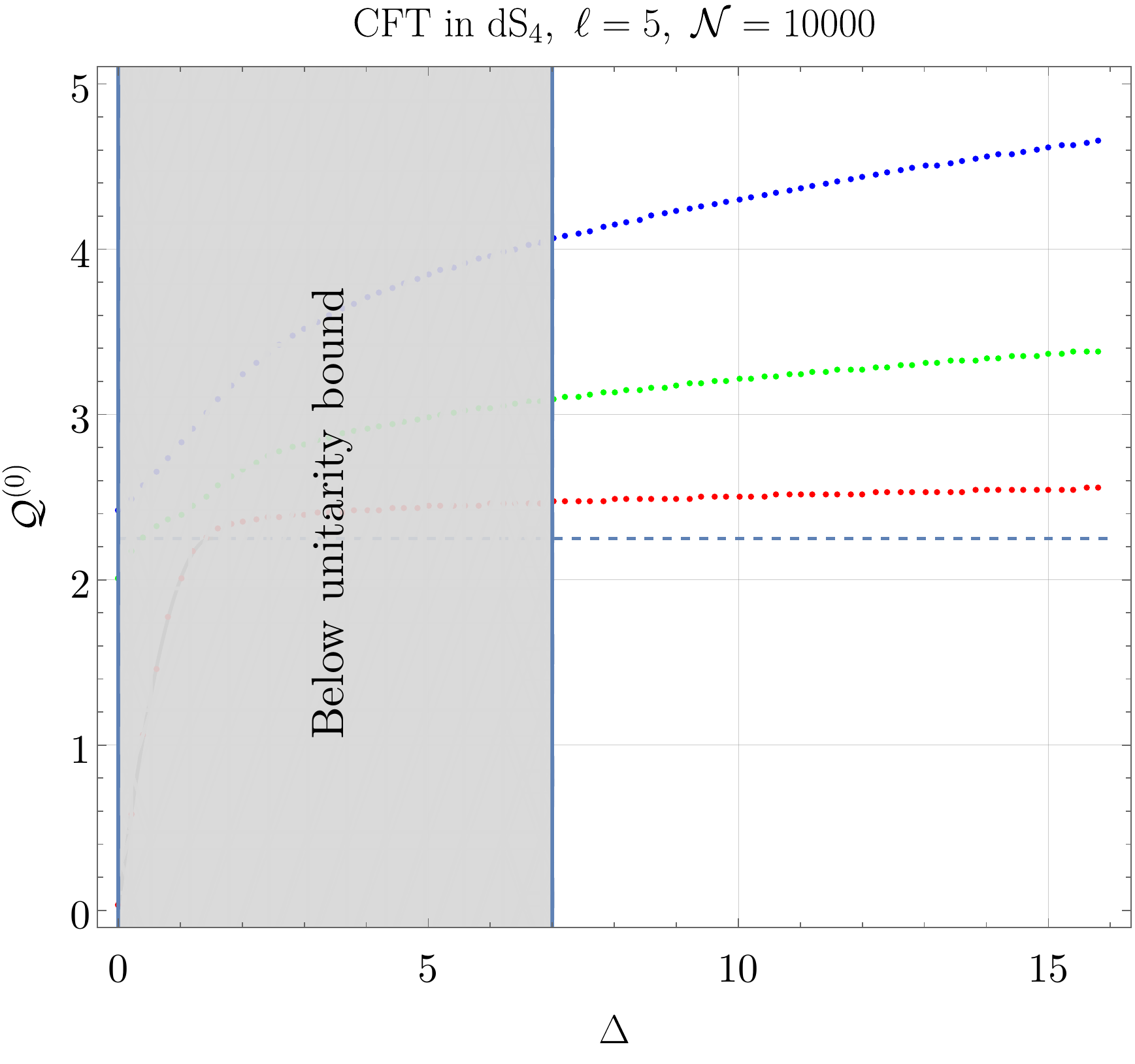}
             \end{subfigure}
     \hfill
     \begin{subfigure}[t]{0.3\textwidth}
         \centering
         \includegraphics[width=\textwidth]{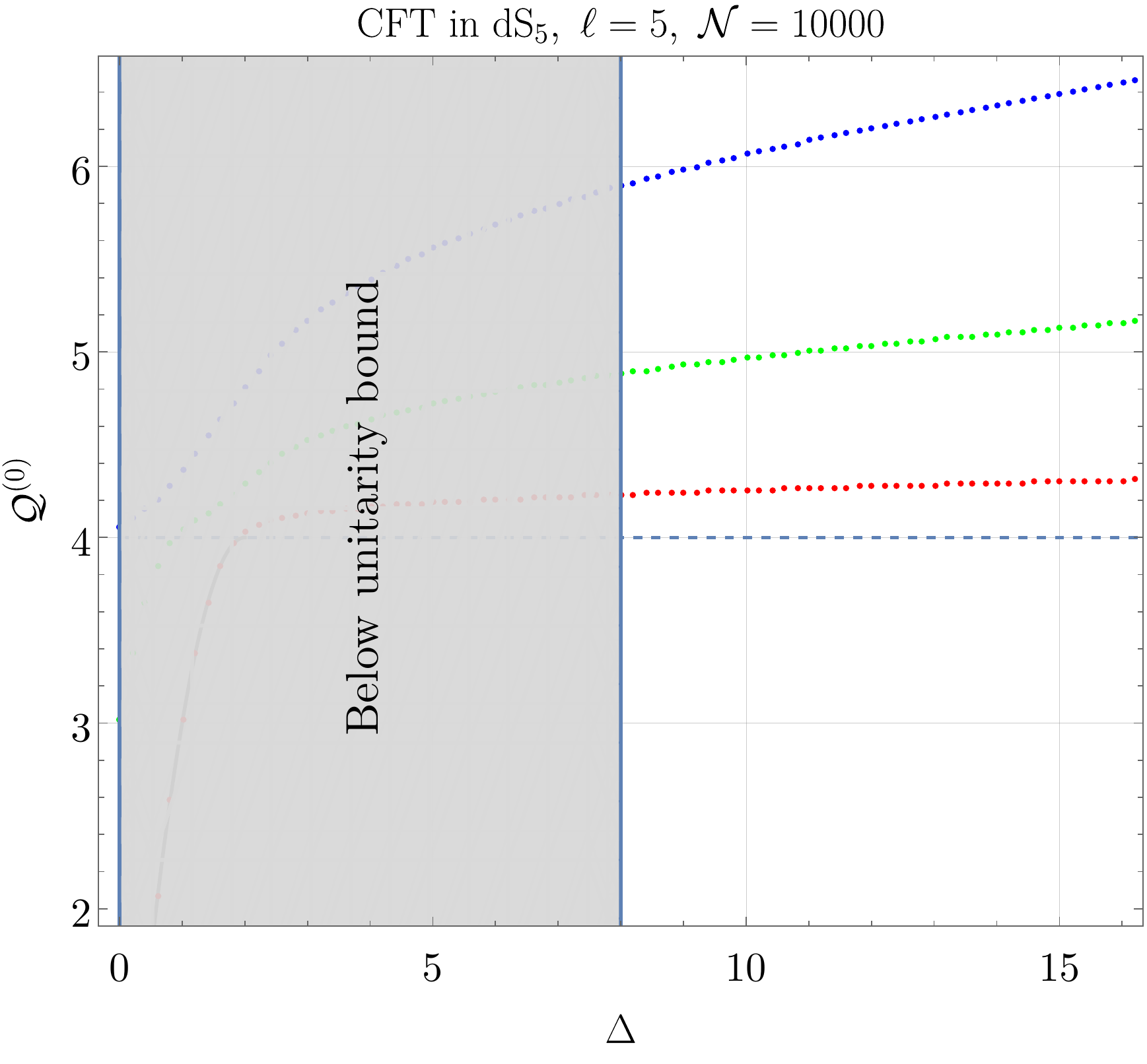}
     \end{subfigure}
          \hfill
     \begin{subfigure}[t]{0.3\textwidth}
         \centering
         \includegraphics[width=\textwidth]{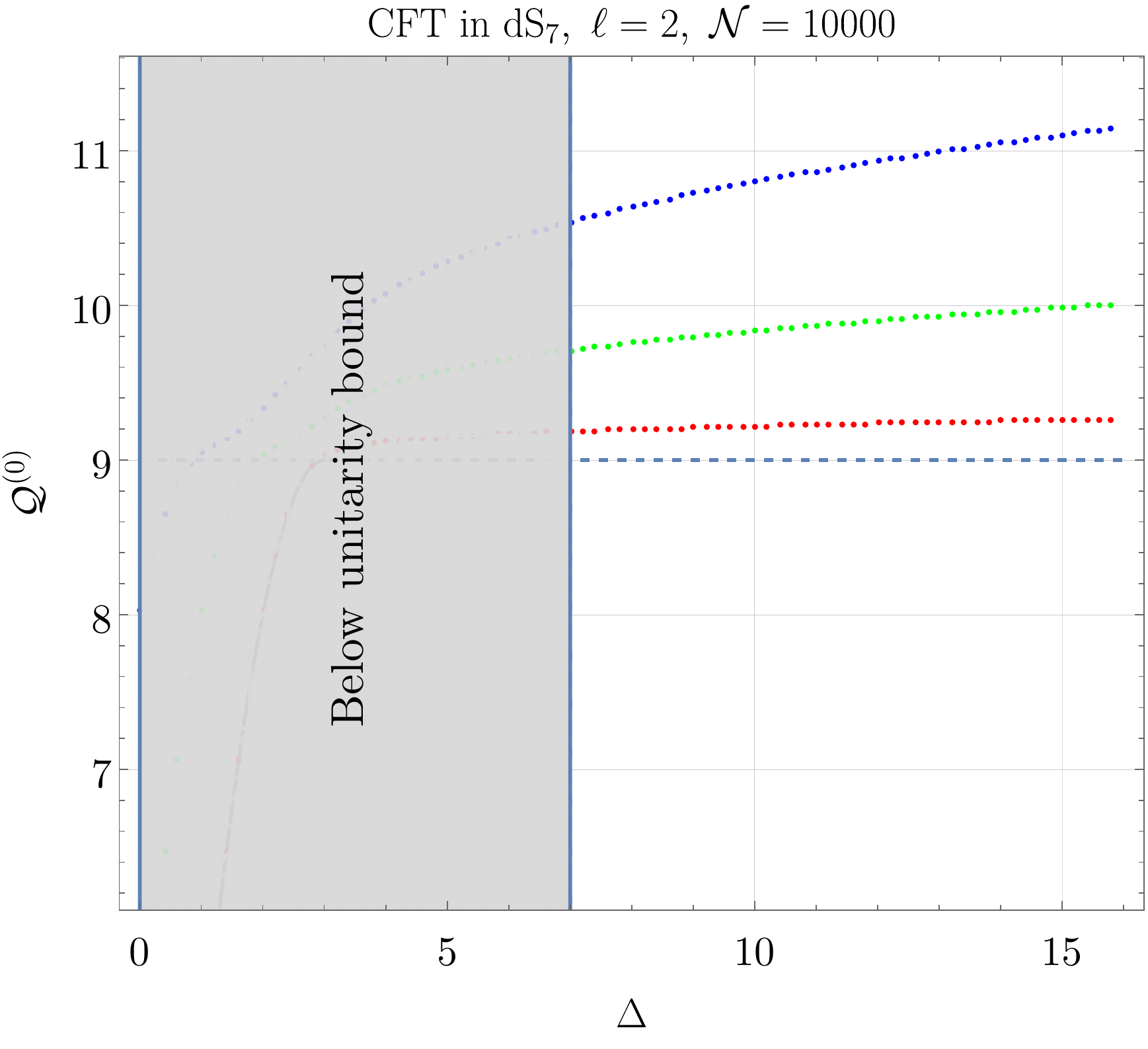}
     \end{subfigure}
             \caption{Plot of first three low-lying eigenvalues of $\CQ^{(0)}$ for various spacetime dimensions and spins. The dashed lines are $\CQ^{(0)}=\frac{d^2}{4}$, with $d=3,4,6$. }
                  \label{fig:CFThdl}
     \end{figure}

The matrix $\CQ^{(s)}$ cannot tell us anything about the type \rom{1} exceptional series representations since the latter do not  contain any $\mY_{ss}$ representation of $\SO(d+1)$. Instead, as mentioned in the footnote \ref{xxx}, we should study states of symmetry $\mY_s$. Although we have not solved this problem completely, we propose a potentially useful way to determine whether a type \rom{1} exceptional series in $\mathcal R_{\Delta,\ell}$ by using the $s=1$ case as an explicit example.  When $s=1$, the $\CV_{1, 0}$ is characterized by a spin 1 state $|\psi)_a$ satisfying 
\begin{align}\label{CVreq}
L_{a0}|\psi)_a=0,\,\,\,\,\, L_{a0}|\psi)_b-L_{b0}|\psi)_a=0
\end{align}
A generic $\SO(d+1)$ vector in $\mathcal R_{\Delta,\ell}$ is a linear combination of 
\begin{align}\label{yyy}
|\xi^-_n)_a=(L^+\cdot L^+)^n|\Delta\rangle_a, \,\,\,\,\,\,|\xi^+_n)_a=(L^+\cdot L^+)^nL^+_a|\Delta\rangle
\end{align}
where $|\Delta\rangle_{a_1}=L^+_{a_2}\cdots L^+_{a_\ell}|\Delta\rangle_{a_1\cdots a_\ell}$ and $|\Delta\rangle=L^+_{a_1}\cdots L^+_{a_\ell}|\Delta\rangle_{a_1\cdots a_\ell}$. By definition, $|\xi^\pm_n)_a$ has level $\ell+2n\pm1$. Let $|\psi)_c=\sum_{n\ge 0}\left(a_n|\xi^-_n)_c+b_n|\xi^+_n)_c\right)$ be a spin 1 state in $\CV_{1,0}$. 
After some lengthy computations, one can show that the requirement $L_{a0}|\psi)_b-L_{b0}|\psi)_a=0$ is equivalent to 
\begin{align}\label{antivan}
\frac{1}{2}a_n+2(n+1)\left(n+\Delta+\ell-\frac{d-1}{2}\right)a_{n+1}=\ell(\Delta-d)b_n
\end{align}
and   $L_{a0}|\psi)_a=0$ is equivalent to 
\small
\begin{align}\label{tracevan}
\frac{a_n\!+\!b_{n-1}}{2}\!+\!2(n\!+\!1)\left(n\!+\!\Delta\!+\!\ell\!+\!\frac{d\!-\!1}{2}\right)a_{n+1}\!+\!\left(2n\left(n\!+\!\Delta\!+\!\ell\!+\!\frac{d\!+\!1}{2}\right)\!+\!(d\!+\!1)\Delta\!+\!\ell(\Delta\!+\!1)\right)b_n=0
\end{align}
\normalsize
When $d=1$, the UIR $\CV_{1,0}$ reduces to a discrete series representation of $\SO(1,2)$, and in this case solving (\ref{antivan}) and (\ref{tracevan}) reduces to $
b_{n-1}+4(n+\Delta)(n+\ell+1)b_n=0$, which is consistent with \cite{Hogervorst:2021uvp}. In principle, one can derive  $\{a_n, b_n\}$ from  (\ref{antivan}) and (\ref{tracevan}), and then use them to check whether $|\psi)_c$ is normalizable. A normalizable $|\psi)_c$ corresponds to the existence of $\CV_{1, 0}$. Unfortunately, we could not solve the two couple recurrence relations for  $d\ge 2$. On the other hand, we want to mention that a significant simplification happens when $\Delta$ approaches the unitarity bound, i.e. $\Delta=d+\ell-1$. In this case, the primary state, which will be denoted by $|J_\ell\rangle_{a_1\cdots a_\ell}$ henceforth,  is a spin $\ell$ conserved current in the sense of $L_{a_1}^+|J_1\rangle_{a_1\cdots a_\ell}=0$. Because of this property, all $|\xi^+_n)_a$ vanish  for  $\ell\ge 0$, and all $|\xi^-_n)_a$ vanish for  $\ell\ge1$, which means that the primary representation generated by a conserved current of spin larger than 1 cannot have $\CV_{1,0}$. For $\ell=1$, $|\psi)_a$ should be  a linear combination of $|\xi^-_n)_a$ only. We can simply set all $b_n\equiv 0$ and take $\ell=1,\Delta=d$ in eq. (\ref{antivan}) or eq. (\ref{tracevan}). It gives
\begin{align}\label{ewpoi}
4n\left(n+\frac{d+1}{2}\right)a_n+a_{n-1}=0 \rightsquigarrow a_n=\frac{(-)^n}{4^n n! (\frac{d+3}{2})_n}
\end{align}
To obtain the norm of $|\xi_n^-)_a$, we need the following relation
\begin{align}\label{sdfdsa}
\left(L_-\cdot L_-\right) |\xi_n^-)_a&=4n\left(n+\frac{d-1}{2}\right) \left(2(d+1)H+L^+\cdot L^-\right)|\xi^-_{n-1})_a-4n\left[L^-_a, L^+_b\right] |\xi^-_{n-1})_b\nonumber\\
&=16 n\left(n+\frac{d-3}{2}\right)\left(n+\frac{d+1}{2}\right)(n+d-1)|\xi^-_{n-1})_a
\end{align}
It fixes  $_a(\xi^-_n|\xi^-_n)_b$ up to an overall constant  (which is suppressed here)
\begin{align}\label{qrpqoiewu}
_a(\xi^-_n|\xi^-_n)_b=4^{2n}n! \left(\frac{d-1}{2}\right)_n\left(\frac{d+3}{2}\right)_n (d)_n\delta_{ab}
\end{align}
Combining  (\ref{ewpoi}) and (\ref{qrpqoiewu}), we find that $|\psi)_a$ is nonnormalizable for $d\ge 2$:
\begin{align}
_a(\psi|\psi)_b=\delta_{ab}\sum_{n\ge 0}\frac{\left(\frac{d-1}{2}\right)_n (d)_n}{n! (\frac{d+3}{2})_n}=\infty
\end{align}
It excludes the existence of $\CV_{1, 0}$. Moreover, in this case, we can also prove that $\CU_{1, 0}$ cannot be present.
Recall that $\CU_{1,0}$ is characterized by a state $|\psi)_{a,b}=-|\psi)_{b,a}$ satisfying $L_{a0}|\psi)_{a,b}=0$, because $\CU_{1,0}$ does not contain spin 1 representations of $\SO(d+1)$. The state $|\psi)_{a,b}$ should be a linear combination  
$|\psi)_{a,b}=\sum_{n\ge 0}c_n |1,n)_{a,b}$,
where 
\begin{align}
|1,n)_{a,b}=(L^+\cdot L^+)^n\left(L^+_a|J_1\rangle_b-L^+_b|J_1\rangle_a\right)~.
\end{align} 
After a small calculation, one can show
\begin{align}
L_{a0}|1,n)_{a,b}&=\frac{i}{2}\left(L^+_a+L^-_a\right)|1,n)_{a,b}\nonumber\\
&=\frac{i}{2}\left(L^+\cdot L^+\right)^{n+1} |J_1\rangle_b+i (n+d)(2n+d-1)\left(L^+\cdot L^+\right)^{n} |J_1\rangle_b~.
\end{align}
Then  it is clear that $L_{a0}|\psi)_{a,b}$ vanishes if and only if $c_0=0$ and 
\begin{align}\label{rec?}
c_{n-1}+4(n+1)\left(n+\frac{d-1}{2}\right) c_n=0~.
\end{align}
For $d\ge 2$, the recurrence relation eq.~(\ref{rec?}) together with the initial condition $c_0=0$ only has trivial solution, i.e.  $c_n\equiv 0$. Therefore, the primary representation generated by a spin 1 conserved current does not contain any type $\CU$ exceptional series.

The reader may wonder where the exceptional series are hiding in the conformal multiplets. 
For example, free Maxwell theory, which is a CFT in dS$_4$, must contain $\CU_{1,0}$ which describes single photon states. 
We expect that $\CU_{1,0}$ is contained in the conformal multiplet of the gauge invariant  local operator that creates single photons, i.e. the field strength which has 2 \textit{anti-symmetric} indices.
We speculate that conformal multiplets with two row Young tableaus contain $\CU_{s,t}$.

\section*{Acknowledgement}

 We thank Dio Anninos, Tarek Anous, Frederik Denef, Petr Kravchuk, Manuel Loparco, Dalimil Maz\'ač,  Beatrix M\"uhlmann and David Simmons-Duffin for useful discussions. 
JP is supported by the Simons Foundation grant 488649 (Simons Collaboration on the Nonperturbative Bootstrap)  and the Swiss National Science Foundation through the project
200020\_197160 and through the National Centre of Competence in Research SwissMAP. ZS is supported by the US National Science Foundation under Grant No. PHY-2209997 and the  Gravity Initiative at Princeton University.

\appendix

\section{Fourier transform of $\psi$ functions}\label{psifour}
Let $z,w\in\mathbb C$ be two complex numbers with {\it positive} real parts. Consider the following integral 
\begin{align}\label{CIdef}
\CI(z,w)\equiv \int_0^\infty\, dt\, \frac{e^{-z t}-e^{-w t}}{1-e^{-t}}
\end{align}
Using the integral representation of $\psi$ function 
\begin{align}
\psi(z)=\int_0^\infty dt\left(\frac{e^{-t}}{t}-\frac{e^{-z t}}{1-e^{-t}}\right)
\end{align}
one can immediately find that 
\begin{align}\label{CIresult}
\CI(z,w)=\psi(w)-\psi(z),\quad \text{when}\,\, \Re(z)>0, \Re(w)>0
\end{align}
 Although the original integral definition of $\CI(z,w)$ only makes sense when $\Re(z),\Re(w)>0$, the $\psi$ function representation provides a natural analytical continuation to $\mathbb C\times\mathbb C$ minus a discrete set of points. It allows us to consider the following Fourier transformation
\begin{align}\label{CJdef}
\CJ(a,b)&\equiv\frac{1}{2\pi}\int_{\mathbb R}d\lambda\, \left(\CI(a+i\lambda,b+i\lambda)+\CI(a-i\lambda,b-i\lambda)\right)e^{i\lambda t}\nonumber\\
&=\frac{1}{2\pi}\int_{\mathbb R}d\lambda\, \left(\psi(b+i\lambda)+\psi(b-i\lambda)-\psi(a+i\lambda)-\psi(a-i\lambda)\right)e^{i\lambda t}
\end{align}
for any $a, b\in\mathbb C$. First consider $\Re(a),\Re(b)>0$. In this case, we are allowed to use the integral  representation (\ref{CIdef}) of $\CI$, which yields 
\begin{align}\label{CIplus}
\CI(a+i\lambda,b+i\lambda)+\CI(a-i\lambda,b-i\lambda)=\int_{\mathbb R}dt\, \frac{e^{-a|t|}-e^{-b|t|}}{1-e^{-|t|}} e^{-i\lambda t}
\end{align}
Combining eq. (\ref{CJdef}) and eq. (\ref{CIplus}), we find that
\begin{align}\label{CJfirst}
\CJ(a,b)= \frac{e^{-a|t|}-e^{-b|t|}}{1-e^{-|t|}},\quad   \text{when}\,\, \Re(a),\Re(b)>0.
\end{align}

Next, consider $\Re(a)<0$ and $\Re(b)>0$. Without loss of generality, assume that $\Re(a)$ is between $-(n+1)$ and $-n$ for some positive integer $n$. The strategy is to find a relation between $\CJ(a, b)$ and $\CJ(a+n,b)$, since the latter can be computed with eq. (\ref{CJfirst}). Applying the $\psi$ function recurrence relation $\psi(z+1)=\psi(z)+\frac{1}{z}$ to $\CI(a\pm i\lambda, b\pm i\lambda)$ yields
\begin{align}\label{CJrela}
\CI(a\pm i\lambda, b\pm i\lambda)&=\psi(b\pm i\lambda)-\psi(a\pm i\lambda)\nonumber\\
&=\psi(b\pm i\lambda)-\psi(a+n+1\pm i\lambda)+\sum_{k=0}^{n}\frac{1}{a+k\pm i\lambda}\nonumber\\
&=\CI(a+n+1\pm i\lambda, b\pm i\lambda)+\sum_{k=0}^{n}\frac{1}{a+k\pm i\lambda}
\end{align}
Plugging eq. (\ref{CJrela}) into the Fourier transform eq. (\ref{CJdef}), we obtain
\begin{align}\label{CJsec}
\CJ(a, b)&=\CJ(a+n+1,b)+\sum_{k=0}^{n}(a+k)\int_{\mathbb R}\frac{d\lambda}{\pi}\, \frac{e^{i\lambda t}}{(a+k)^2+\lambda^2}\nonumber\\
&= \frac{e^{-(a+n+1)|t|}-e^{-b|t|}}{1-e^{-|t|}}-\sum_{k=0}^{n} e^{(a+k) |t|}
\end{align}
where we have used eq. (\ref{CJfirst}) and $\int_{\mathbb R}dt\frac{1}{x^2+t^2}e^{it\lambda}=\frac{\pi}{|x|}e^{-|x\lambda|}$. Some further rewriting of eq. (\ref{CJsec}) leads to 
\begin{align}\label{CJthired}
\CJ(a,b)= \frac{e^{-a|t|}-e^{-b|t|}}{1-e^{-|t|}}-2\sum_{k=0}^{n}\cosh\left((a+k)t\right)
\end{align}
when $-(n+1)<\Re(a)<-n, \Re(b)>0$.

As  both $\Re (a)>0$ and $\Re(a)<0$ have been discussed, the remaining case is $\Re (a)=0$. However, for a generic imaginary $a$, the integral  (\ref{CJdef}) is not well-defined because $\psi(a\pm i\lambda)$ have poles $\lambda=\pm i a$, lying on the integration contour. The only exception is $a=0$ in that as $a$ approaches $0$, the two poles $\pm i a$ collide and annihilate each other. More explicitly, using the recurrence relation of $\psi$ functions, we get $\psi(i\lambda)+\psi(-i\lambda)=\psi(1+i\lambda)+\psi(1-i\lambda)$, which yields 
\begin{align}\label{a=0J}
\CJ(0, b)=\CJ(1, b)=\frac{1-e^{-b|t|}}{1-e^{- |t|}}-1, \quad   \text{when}\,\, \Re(b)>0.
\end{align}
where we have used eq. (\ref{CJfirst}) for $\CJ(1, b)$.

\section{Numerical analysis of $L_0\neq 0$ sector of \SO$(1,2)$}\label{sec:L0s}
In section~\ref{sec:1dPxPCon}, we numerically analyzed the decomposition of tensor product $\PtP$ by focusing on $L_0=0$ sector. The advantage of focusing on $L_0=0$ sector is that the discrete series $\CD^\pm_p$ are absent. On the other hand, the formula~\reef{eq:mother of decomposition} is not written in any particular $L_0$ eigensector. Strictly speaking, the numerical analysis through which we are finding  density of states $\bar{\rho}_\CN$ is $L_0$-variant and it is not a good candidate to match with the  results in~\ref{sec:1dPxPCon} since we do not project to any particular $L_0$ eigenspace in the character analysis. In this section, we show that indeed, the $L_0=0$ sector captures all the features of \textit{relative density} of states $\rho_{\text{rel}}$ and considering it is enough to check the character analysis result. 

Let $\CH_\Ls$ be the $L_0=\Ls$ subspace of $\PtP$. Consider the basis of $|\psi_{n,\Ls}) \equiv |n,\Ls-n),~ n\in \mathbb{Z}$ for $\CH_\Ls$\footnote{There is no obvious parity symmetry here except for when $\D_1=\D_2$ where one can define 
$$|\psi^{(\pm)}_{n,\Ls}) = |n,\Ls-n) \pm  |\Ls-n,n) $$}. Following equations~\reef{act1} and~\reef{nnorm}, one finds that the matrix elements of Casimir
\begin{align}
\CQ_{mn}^{(\Ls)}\equiv \frac{(\psi_{m,\Ls}|C^{\SO(1,2)}|\psi_{n,\Ls})}{\sqrt{(\psi_{m,\Ls}|\psi_{m,\Ls})(\psi_{n,\Ls}|\psi_{n,\Ls})}}
\end{align}
in $\CH_\Ls$ is given by
 \be
\CQ_{nn}^{(\Ls)}= 2n (n-\Ls)+ \D_1\Db_1 +  \D_2\Db_2~, \qquad \CQ_{n+1,n}^{(\Ls)} = \left(\CQ_{n,n+1}^{(\Ls)}\right)^* = \beta_{n,\Ls}~
 \ee
where 
\begin{align}\label{betacases}
\beta_{n,\Ls}=-\begin{cases}(n+\Delta_1)(n-\Ls+\Delta_2), &\CP_{\Delta_1}\otimes\CP_{\Delta_2}\\ 
(n+\Delta_1)\sqrt{(n-\Ls+\Delta_2)(n-\Ls+\bar\Delta_2)}, &\CP_{\Delta_1}\otimes\CC_{\Delta_2}\\ 
(n-\Ls+\Delta_2)\sqrt{(n+\Delta_1)(n+\bar\Delta_1)}, &\CC_{\Delta_1}\otimes\CP_{\Delta_2}\\ 
\sqrt{(n+\Delta_1)(n+\bar\Delta_1)(n-\Ls+\Delta_2)(n-\Ls+\bar\Delta_2)}, &\CC_{\Delta_1}\otimes\CC_{\Delta_2}\end{cases}
\end{align}
A similar large $n$ argument as in section~\ref{sec:1dPxPCon}, with the same $\alpha_{\pm}(\lambda)$ in~\reef{eq:alphapm}, shows that all the principal series representations appear in this tensor product as well. We may now diagonalize the $\CQ^{(j)}$ matrix  introducing a cutoff $-\CN\le n\le \CN$ and study the density of states corresponding to $L_0=\Ls$ subspace, $\bar{\rho}_\CN^{(\Ls)}$, and its relative density $\rho^{(\Ls)}_\text{rel}$.  Discrete series $\CD^+_{p}$ with $0<p\leq \Ls$ show up in the decomposition of subspace $\CH_\Ls $. This can be seen in fig.~\reef{fig:PtP1dS} where the Casimir values of discrete series $p(1-p)$ are apparent. 

 \begin{figure}[t]
     \centering
     \begin{subfigure}[t]{0.45\textwidth}
         \centering
         \includegraphics[width=\textwidth]{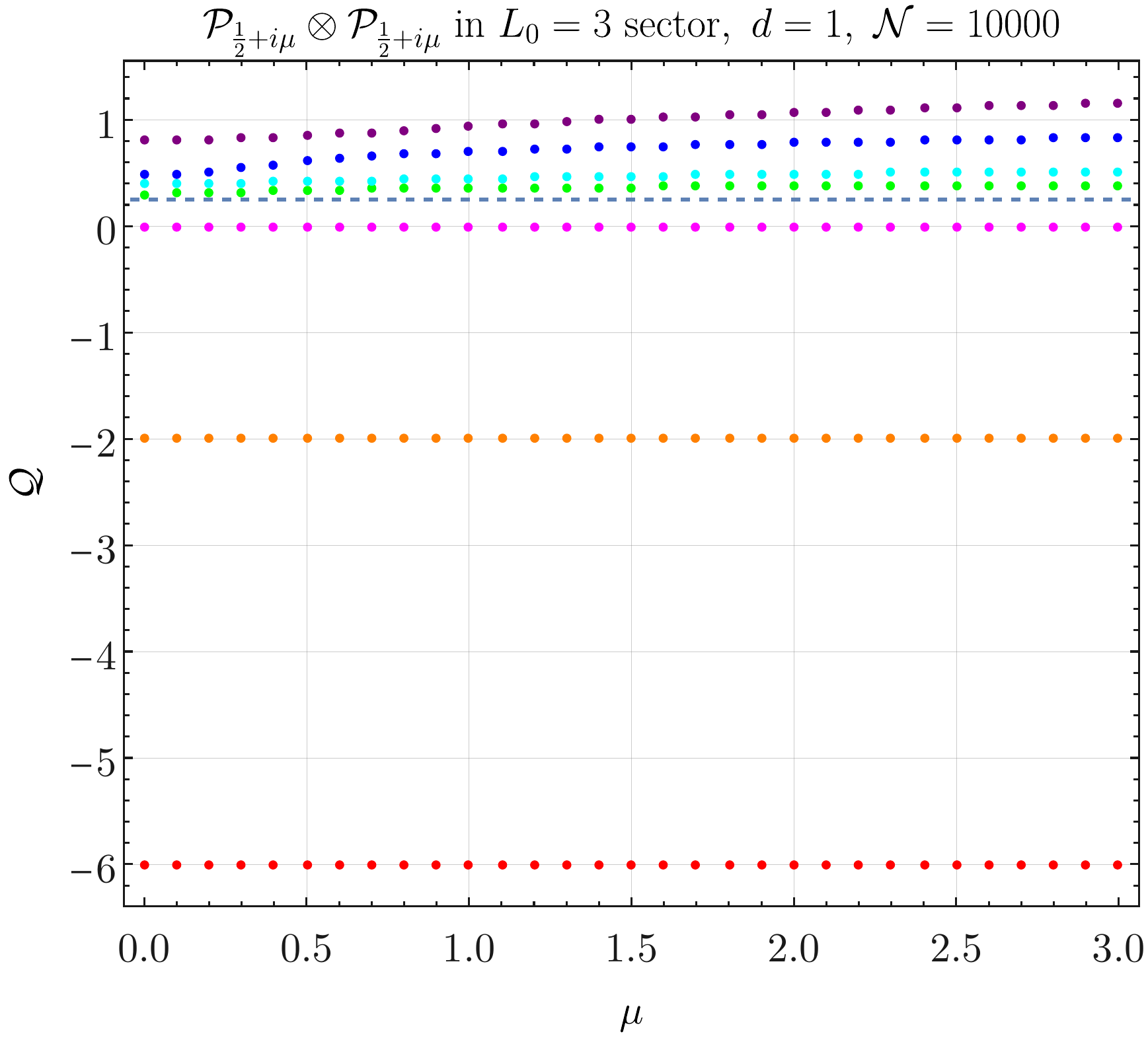}
         \label{PxPd1S}
     \end{subfigure}
     \hfill
     \begin{subfigure}[t]{0.45\textwidth}
         \centering
         \includegraphics[width=\textwidth]{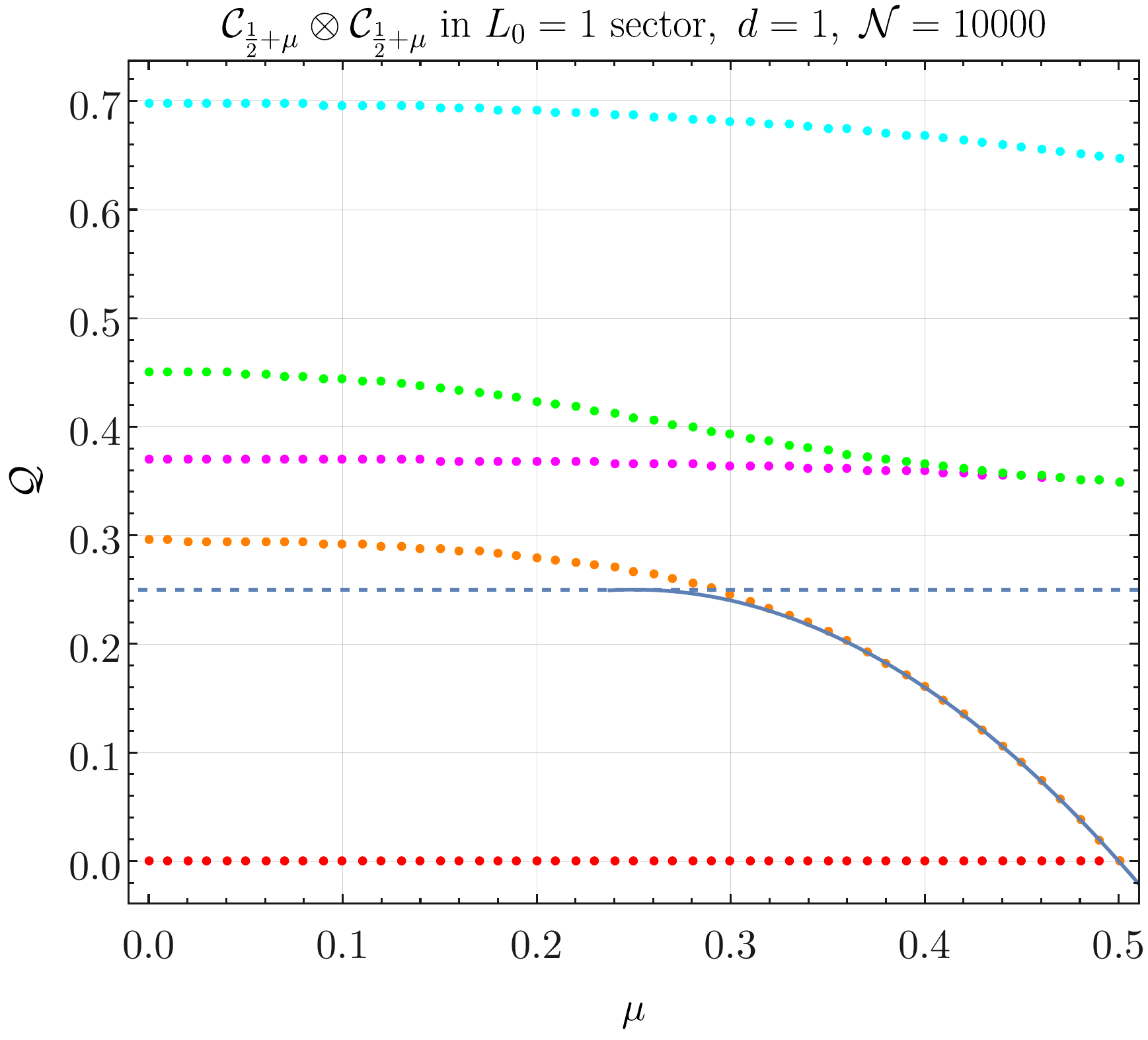}
         \label{fig:CxCd1S}
     \end{subfigure}
     \caption{ First a few eigenvalues of Casimir $\CQ^{(\Ls)}$. Left:  tensor product of two principal series  representations in the $\CH_3$ subspace. Right: tensor product of two complementary series representations in the $\CH_1$ subspace. Discrete series representations with $\CD^\pm_{p}$ with $p\in\{1,2,\cdots, \Ls\}$  also appear in the spectrum. The solid line $\CQ=2\mu(1-2\mu) $ in the right panel corresponds to  the expected single complementary series representation $\CC_{2\mu}$  in the tensor $\CC_{\frac{1}{2}+\mu} \otimes \CC_{\frac{1}{2}+\mu}$ when $\frac{1}{4}<\mu<\frac{1}{2}$. \label{fig:PtP1dS} }
\end{figure}

 \begin{figure}[t]
     \centering
     \begin{subfigure}[t]{0.45\textwidth}
         \centering
         \includegraphics[width=\textwidth]{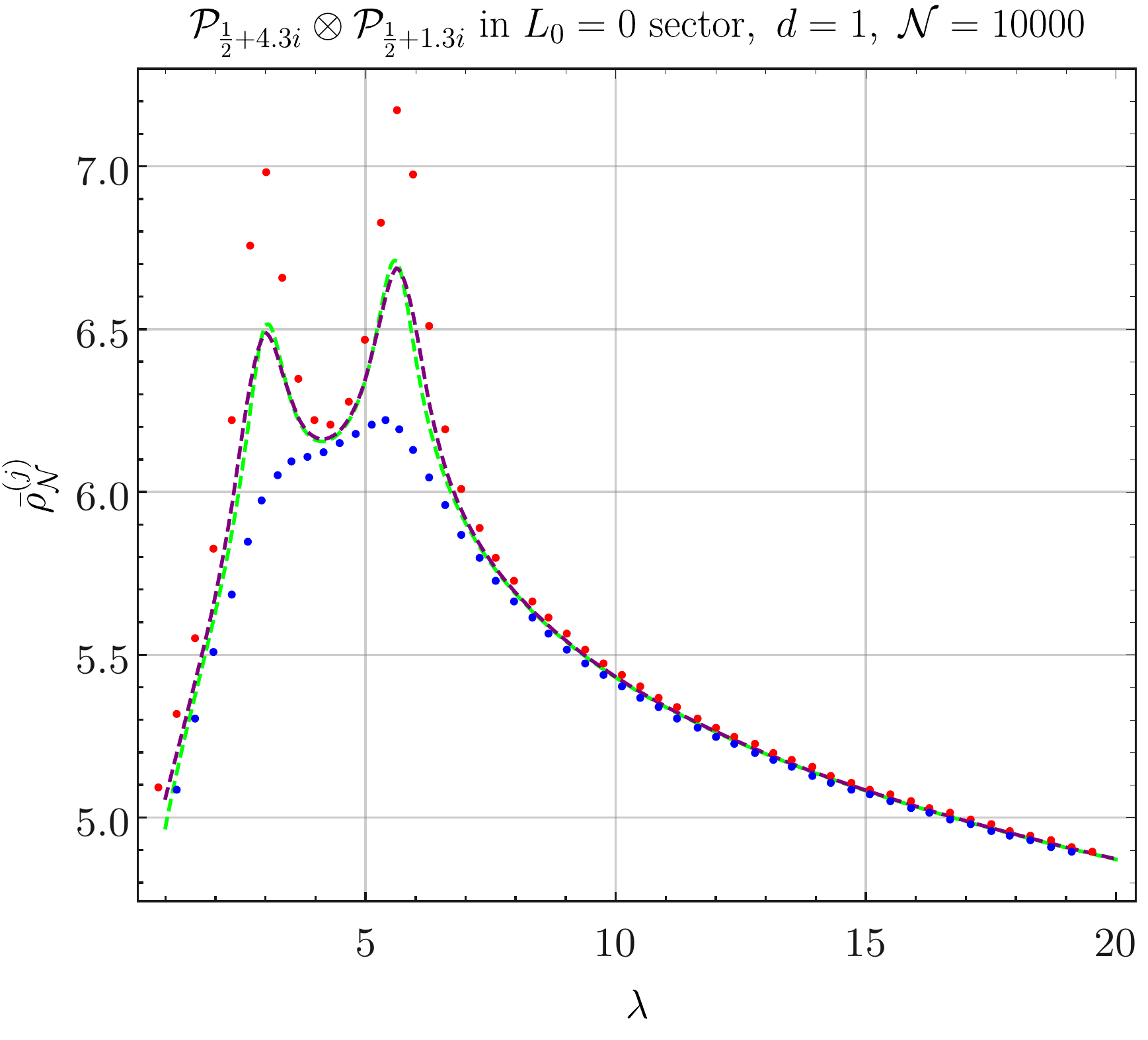}
         \label{rbarsraw}
     \end{subfigure}
     \hfill
     \begin{subfigure}[t]{0.5\textwidth}
         \centering
         \includegraphics[width=\textwidth]{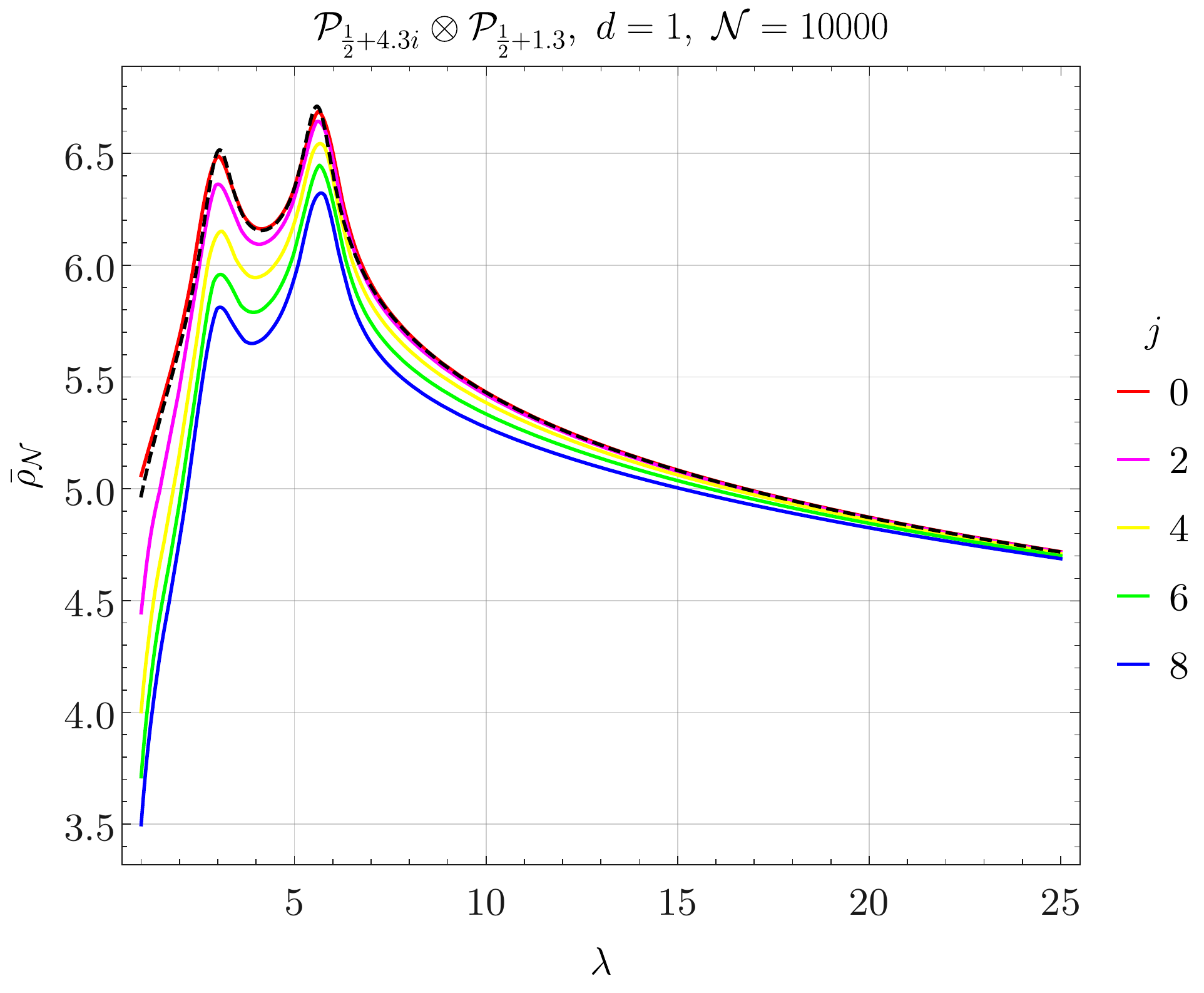}
         \label{rho1dslist}
     \end{subfigure}
     \caption{Left: Illustration of how the interpolation of $\bar{\rho}_\CN^{(\Ls)}$ is derived. The raw data of $\bar{\rho}_\CN^{(\Ls)}$ has oscillations (the red and blue dots). To derive a smooth $\bar{\rho}_\CN^{(\Ls)}$ we first interpolate each red and blue points individually and then take an average of them to find the purple dashed line. The green line is the shifted $\CK_{\text{hc}}$.
     Right: A plot of $\bar{\rho}_\CN^{(\Ls)}$  for various values of $\Ls$ in small $\lambda$ region. The dashed black line is the $\CK_{\text{hc}}$ which is shifted by~$\sim+6.745$ in y-axis. The  $\bar{\rho}_\CN^{(\Ls)}$ is $\Ls$-independent in large $\lambda$ region but in small $\lambda$ region they do not match as one varies $\Ls$. Only $\bar{\rho}_\CN^{(0)}$ agrees with $\CK_{\text{hc}}$.
     \label{fig:rbars}}
\end{figure}

In fig.~\reef{fig:rhobar1d}, we observe a great agreement between $\bar{\rho}_\CN^{(\Ls)}$ and hard-cutoff renormalized density of states $\CK_{\text{hc}}$ defined in~\reef{eq:hard cutoff} up to a shift. However, this is solely true in $L_0=0$ sector.
For $L_0\neq 0$ sectors, in the small $\lambda$ region the numerics do not agree with  $\CK_{\text{hc}}$ as shown in fig.~\reef{fig:rbars}. On the other hand, the relative density of different $L_0=\Ls$ sectors defined as
\be\label{eq:rhorels}
\rrel^{(\Ls)}  = \bar{\rho}^{(\Ls),\CP_{\Delta_1}\otimes \CP_{\Delta_2}}_{\CN} - \bar{\rho}^{(\Ls),\CP_{\Delta_3}\otimes \CP_{\Delta_4}}_{\CN} 
\ee
converges to the $\CK_\text{rel}$ in~\reef{CKrelf}. In addition to this numerical evidence for the claim that $\rho_\text{rel}$ is unique in all $L_0$ sectors and matches with $\CK_\text{rel}$, in what follows we present an analytical argument why this is the case. 
\begin{figure}[t]
\centering
         \includegraphics[width=0.8\textwidth]{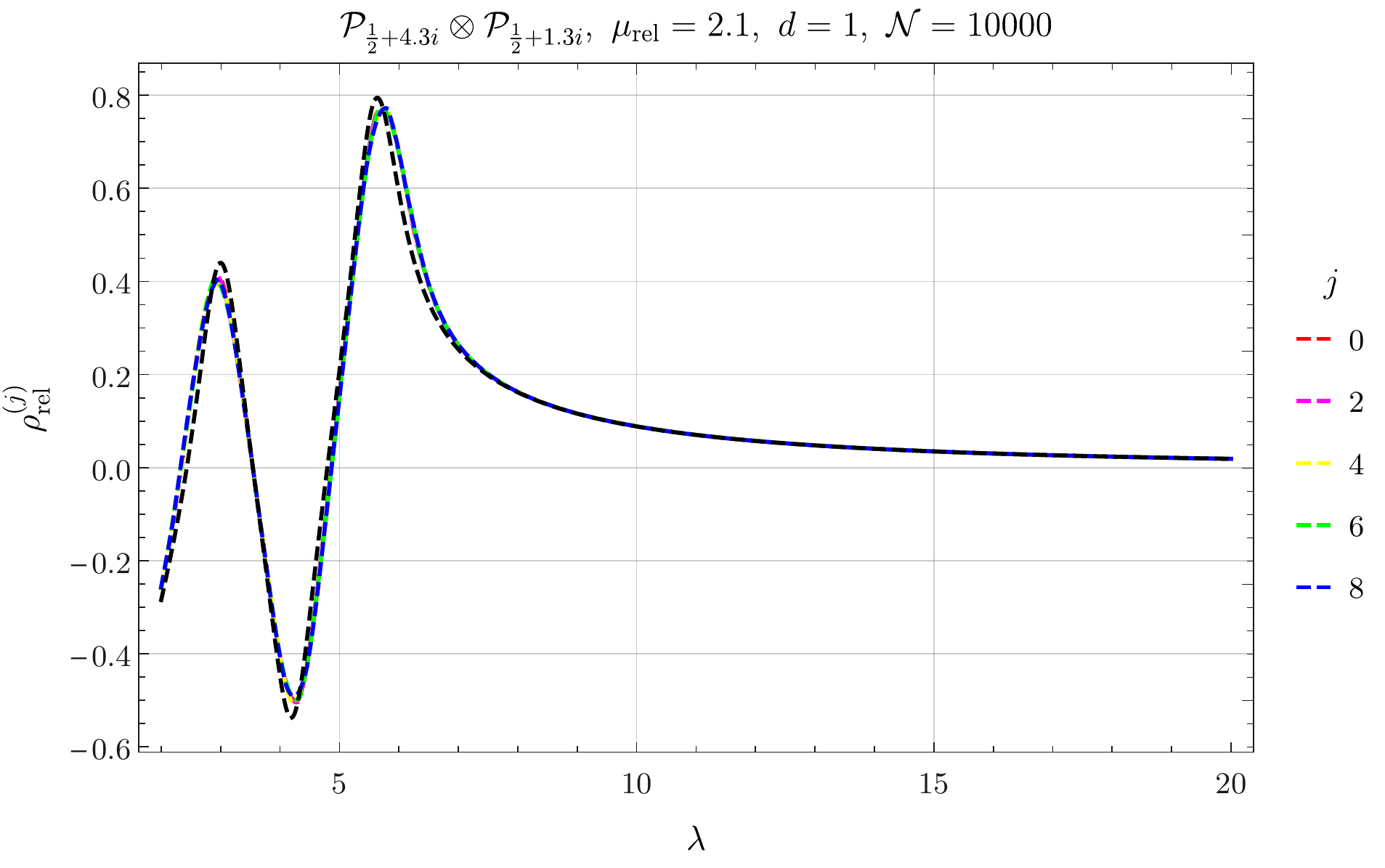}
         \caption{A perfect match of $\CK_\text{rel}$ (the dashed black line) and $\rho_\text{rel}^{(j)}$s. The $\rho_\text{rel}^{(\Ls)}$ for different $\Ls$ converges to the same function in large $\CN$. The agreement is good enough that puts them on top of each other in the plot and make them indistinguishable.}
         \label{fig:rhorelslist}
\end{figure}

The tensor product character in~\reef{khlk} can be computed in $|n,m)$ basis as
\be
\sum_{\Ls\in\mathbb{Z}} \sum_{n\in \mathbb{Z}} (n,\Ls-n|q^D|n,\Ls-n) = \sum_{\Ls\in\mathbb{Z}} \Aj
\ee
where 
\be
\Aj \equiv \sum_{n\in \mathbb{Z}} (n|q^D|n) (\Ls-n|q^D|\Ls-n)
\ee
and its dependence on $\D_1=\half+i\mu_1$ and $\D_2=\half+i\mu_2$ is implicit.  Moreover, concerning the right side of~\reef{khlk}, each character admits a similar expression:
\be
\frac{q^{\half+i\lambda}+q^{\half-i\lambda}}{1-q} = \sum_{\Ls\in\mathbb{Z}} \Bj~,\qquad  \Bj \equiv (\Ls|q^D|\Ls)~.
\ee
We would like to see whether, within each $\Ls$ sector, we can write an equation similar to~\reef{khlk}:
\be\label{eq:AjBjK}
\Aj = \int_\Real d\lambda~\CK_\Ls(\lambda)\Bj~ + \text{Discrete series}~.
\ee
To this end, let us write $\Aj$ and $\Bj$ in a $\delta$-function normalized basis $\Ket{x}$ of $\CP_{\Delta}$ mentioned in~\cite{Sun:2021thf} which satisfies the following relations:
\be
q^D\Ket{x} = q^{\bar{\D}} \ket{q x},~ \quad \Bra{x}n) = \frac{2^\D}{\sqrt{2\pi}} \left(\frac{1-ix}{1+ix}\right)^n \frac{1}{(1+x^2)^\D}~.
\ee
Plugging in the identity operator $\mathds{1} = \int_\Real dx \Ket{x}\Bra{x} $ into definition of $\Bj$, we find 
\be\label{eq:Bj def}
\Bj = \frac{1}{\pi} \int_\Real \frac{d x} {\sqrt{(1+x^2)(q^{-1}+qx^2)}}  W(q,x)^\Ls Q(q,x)^{i\lambda}
\ee
where we define
\be
Q(q,x) \equiv \frac{1+q^2x^2}{q(1+x^2)}~,\quad W(q,x)\equiv\frac{(1-ix)(1+iqx)}{(1+ix)(1-iqx)}~.
\ee
Similarly, for $\Aj$, we have
\be
\begin{split}\label{eq:Aj def}
\Aj&=\frac{1}{\pi^2} \int_{\Real^2} dx_1 dx_2 ~W(q,x)^\Ls \left(\sum_{n\in\intg}\Phi(q;x_1,x_2)^n\right)  \prod_{i=1,2} \frac{1}{(1+x^2)^{\D_i}(q^{-1}+qx_i^2)^{\bar{\D}_i}}\\
&= \frac{q}{2\pi(1-q)} \int_\Real \frac{dx}{|1-qx^2|} W(q,x)^\Ls \sum_{\pm\pm} Q(q,x)^{\pm i \mu_\pm}
\end{split}
\ee
where in the second line we perform the integral of the phase
\be
\Phi(q;x_1,x_2) \equiv W(q,x_1) W(q,-x_2) = e^{i\phi(q;x_1,x_2)}
\ee
over $x_2$ by noticing that 
\be
\sum_{n\in\intg} e^{in\phi(q;x_1,x_2)} = 2\pi\delta\left(\phi(q;x_1,x_2)\right)~,
\ee
with support at $\Phi(q;x_1,x_2)=1$.

Equation~\reef{eq:AjBjK} is formal since the integral representation for $\Aj$ is divergent because of the singularity at $x=1/\sqrt{q}$. However, if we take a derivative with respect to one of the $\mu_i$'s ($i=1,2$), then this equation becomes well defined. Of course, in this way we will lose information about the $\mu$-independent part of $\CK_{\Ls}(\lambda)$ including the discrete series.  
 Consider derivative of~\reef{eq:AjBjK}:
 \be
 \frac{\partial}{\partial \mu_i} \Aj = \int_\Real d\lambda~  \Bj \frac{\partial}{\partial \mu_i} \CK_j^{(\D_1,\D_2)} (\lambda)~
 \ee
 where we introduce the superscript to remark the $\mu_i$-dependence. 
 Plugging in the integral representations of $A_j$ and $B_j$ introduced in~\reef{eq:Aj def} and~\reef{eq:Bj def}, and using the fact that the integrals are well-defined, so we may interchange the integrals over $x$ and $\lambda$, one finds that both equations have the same $\Ls$ dependence through $W(q,x)^\Ls$ and what is left is:
 \be
 i \log Q\,\frac{Q^{i\mu_+}+Q^{i\mu_-}-Q^{-i\mu_+}-Q^{-i\mu_-}}{|Q^\half -Q^{-\half}|} = \int_\Real d\lambda~\frac{\partial}{\partial \mu_i} \CK_j^{(\D_1,\D_2)} (\lambda)~Q^{i\lambda}~,
 \ee
 where $\mu_\pm = \mu_1 \pm \mu_2$.
The left hand side of this equation is $\Ls$-independent which means that $\frac{\partial}{\partial \mu_i} \CK_j^{(\D_1,\D_2)} (\lambda)$ is also $\Ls$-independent. From this, one concludes that the difference of $\CK_j^{(\D_1,\D_2)} (\lambda) - \CK_j^{(\D_3,\D_4)} (\lambda)$ is also $\Ls$-independent. 
This completes the argument that $\CK_{\text{rel}}(\lambda)$ within each $L_0=\Ls$ sector -- which corresponds to the tensor product character $\Aj$ and single particle character $\Bj$ -- has to be $\Ls$-independent and illustrates results in fig.~\reef{fig:rhorelslist} where $\rho_\text{rel}$ is $\Ls$-independent in large $\CN$. 

\subsection*{CFT in dS}
There is a similar story for the case of CFT in dS$_2$. Given a primary representation  $\mathcal R_{\Delta, \ell}$ of $\SO(2,2)$, let $\CH^{(s)}_{\Delta, \ell}$ be the  subspace spanned by all descendants of spin $s\in\mathbb Z$ \footnote{The spin $s$ is an eigenvalue of the generator $L_0\in\SO(1,2)\subset \SO(2,2)$. So $\CH^{(s)}_{\Delta, \ell}$  is the counterpart of $\CH_j$ in the tensor product case.}. It is shown in  the appendix D of~\cite{Hogervorst:2021uvp} that the $\SO(1, 2)$ Casimir restricted in the subspace $\CH^{(s)}_{\Delta, \ell}$  can be represented by the following matrix: 
\begin{align}
&\CQ^{(s)}_{n,n}=2n(n+\ell-s)+\Delta(\ell+2n+1)-s(\ell+\D)~,\nonumber\\
&\CQ^{(s)}_{n+1,n}=\CQ^{(s)}_{n,n+1}=-\sqrt{(n+1)(n+\Delta-s)(n+\ell-s+1)(n+\Delta+\ell)}
\end{align}
with $n\geq \left|\text{Min}(0,\ell - s)\right|$. 
 \begin{figure}[t!]
     \centering
     \begin{subfigure}[t]{0.45\textwidth}
         \centering
         \includegraphics[width=\textwidth]{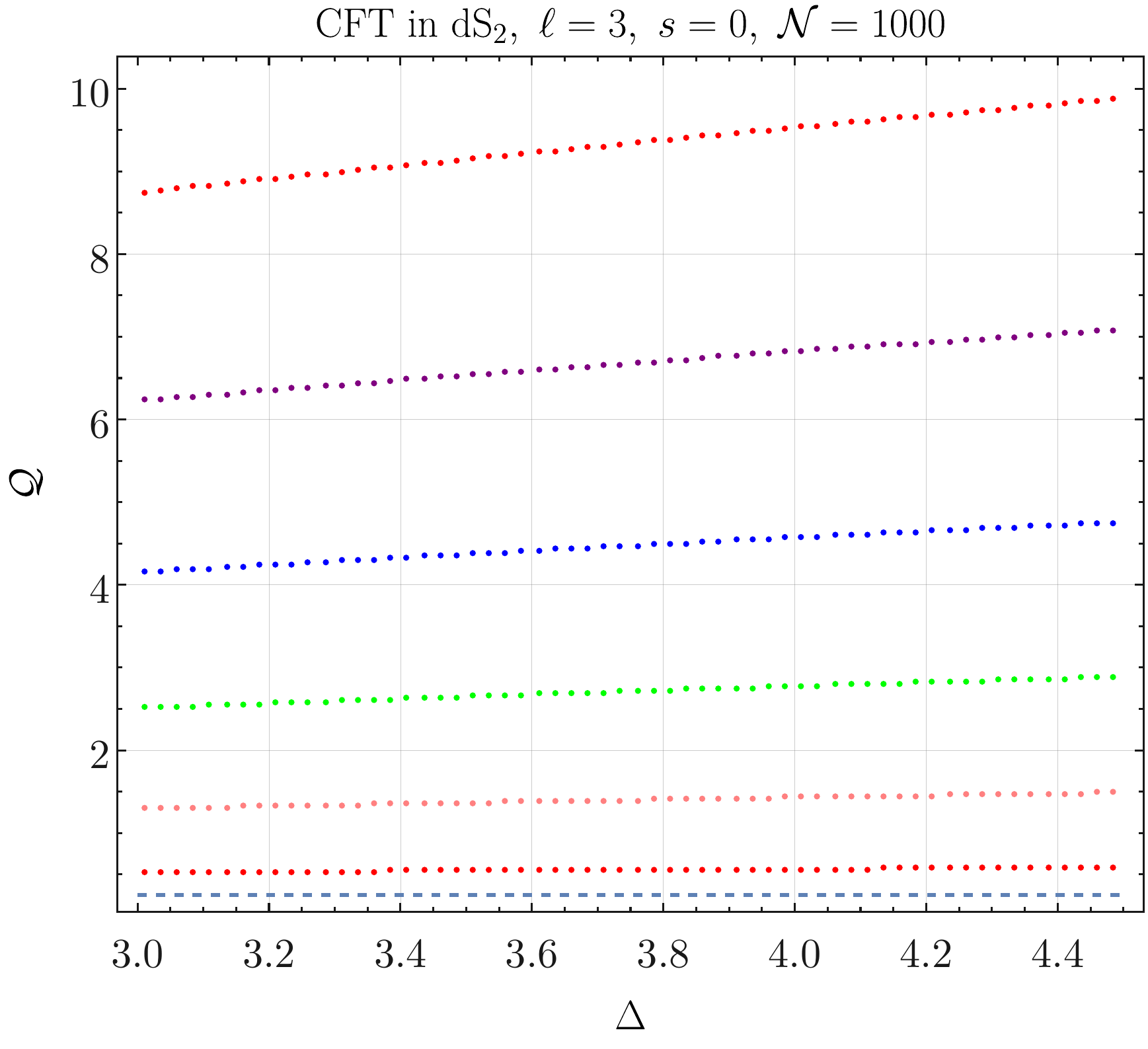}
     \end{subfigure}
     \hfill
     \begin{subfigure}[t]{0.45\textwidth}
         \centering
         \includegraphics[width=\textwidth]{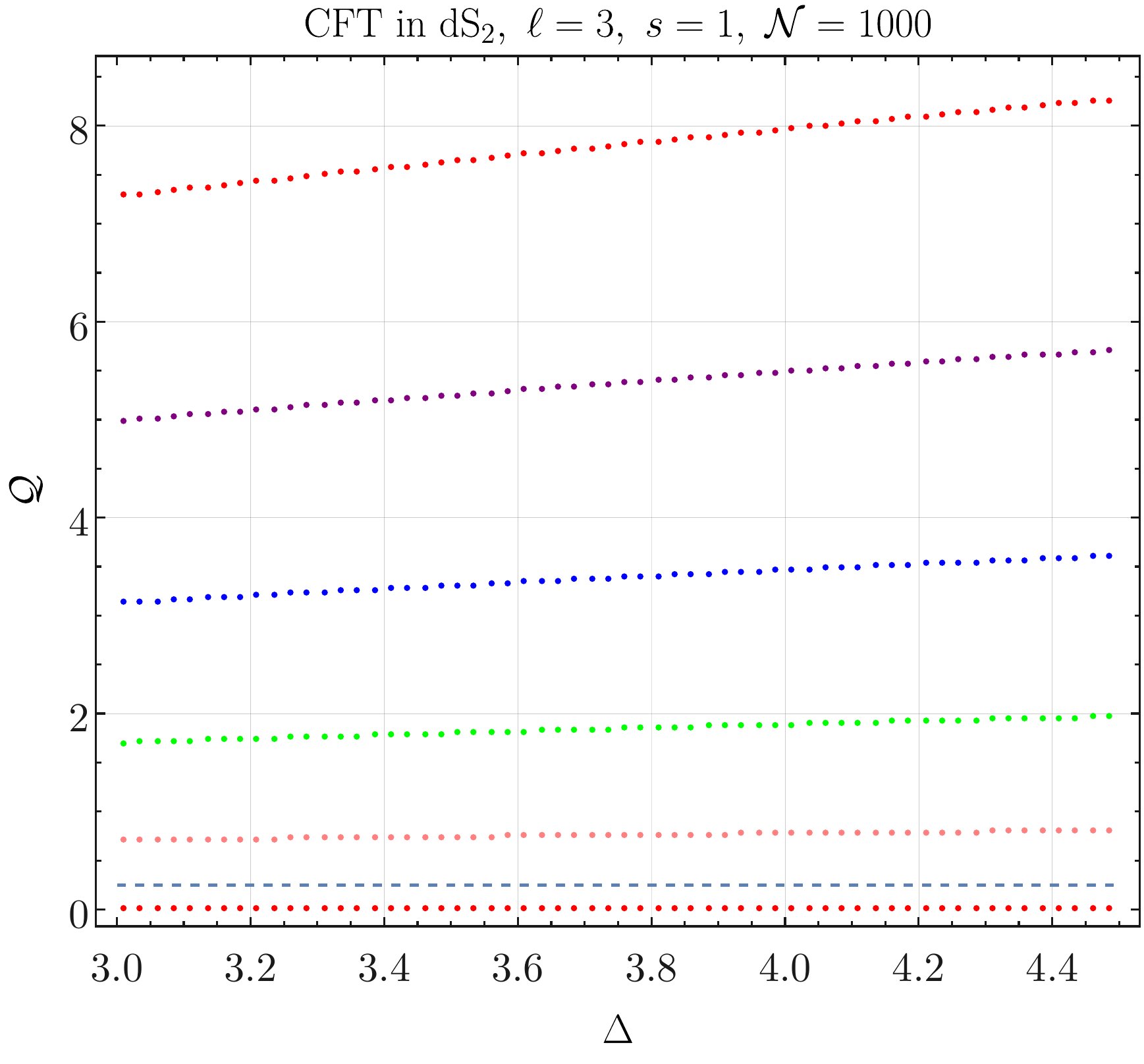}
     \end{subfigure}
          \hfill
     \begin{subfigure}[t]{0.45\textwidth}
         \centering
         \includegraphics[width=\textwidth]{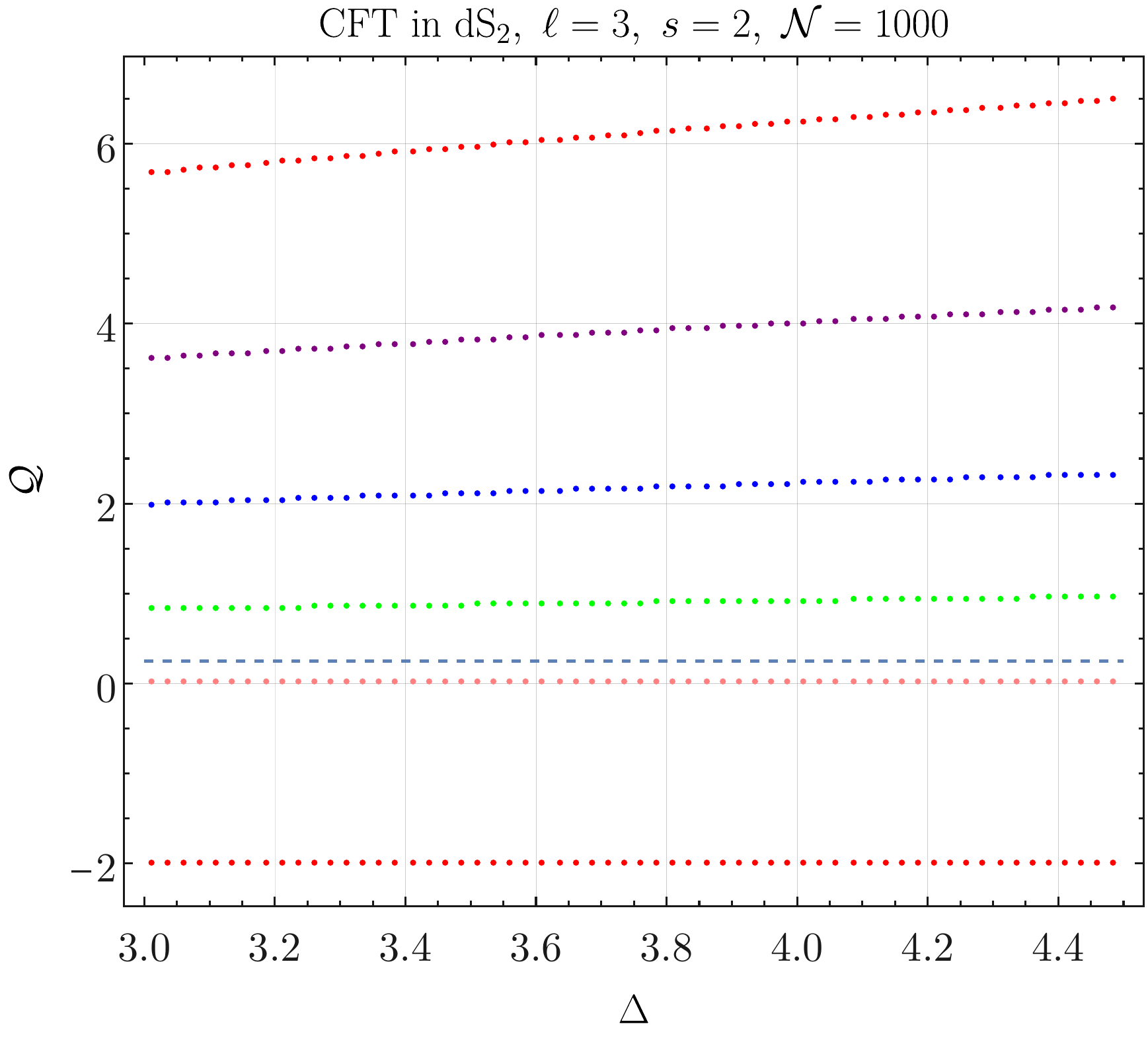}
     \end{subfigure}
          \hfill
     \begin{subfigure}[t]{0.45\textwidth}
         \centering
         \includegraphics[width=\textwidth]{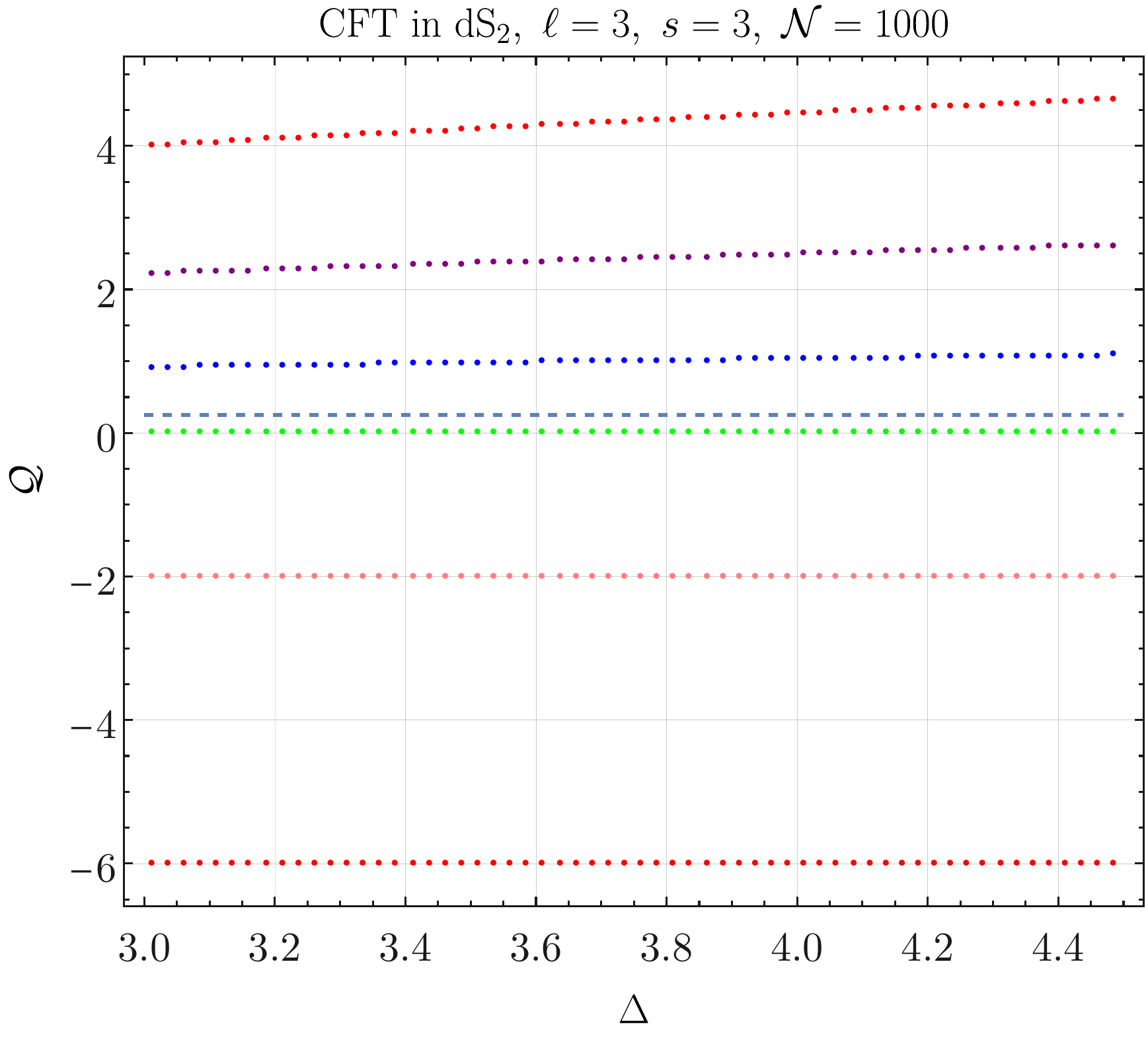}
     \end{subfigure}
     \caption{First five low-lying eigenvalues of \SO$(1,2)$ Casimir matrix restricted to the subspace $s=0,1,2,3$ for $\mathcal{R}_{\D,\ell=3}$. The points above the dashed line belong to principal series. In addition, the discrete series with Casimir $p(1-p)$ where $p\in\{1,2,\cdots,s\}$ show up as discussed in the text.  \label{fig:CFT1dCasS}}
\end{figure}

In section~\ref{sec:SO22 to SO12 Num},  
we have studied the spectrum property of $\CQ^{(0)}$ numerically.
Now we show some observations for the $s\not= 0$ case. 
As before, we  truncate  the size of  $\CQ^{(s)}$ by a large number $\mathcal{N}$ and diagonalize the truncated  matrix  numerically. As proven in~\cite{Hogervorst:2021uvp}, in decomposition of \SO$(2,2)$ primary representation $\mathcal{R}_{\D,\ell}$, discrete series representations 
$\{\CD^{\rm sign(\ell)}_1, \CD^{\rm sign(\ell)}_2,\cdots, \CD^{\rm sign(\ell)}_\ell\}$ appear. In section~\ref{sec:SO22 to SO12 Num}, we were blind to them in the numerics as we consider $\SO(1,2)$ Casimir in the $s=0$ subspace. 
In fig.~\reef{fig:CFT1dCasS}, we identify these discrete series representations as certain eigenvalues of $\CQ^{(s)}$ for $s\not=0$.

 \begin{figure}[t!]
     \centering
     \begin{subfigure}[t]{0.47\textwidth}
         \centering
         \includegraphics[width=\textwidth]{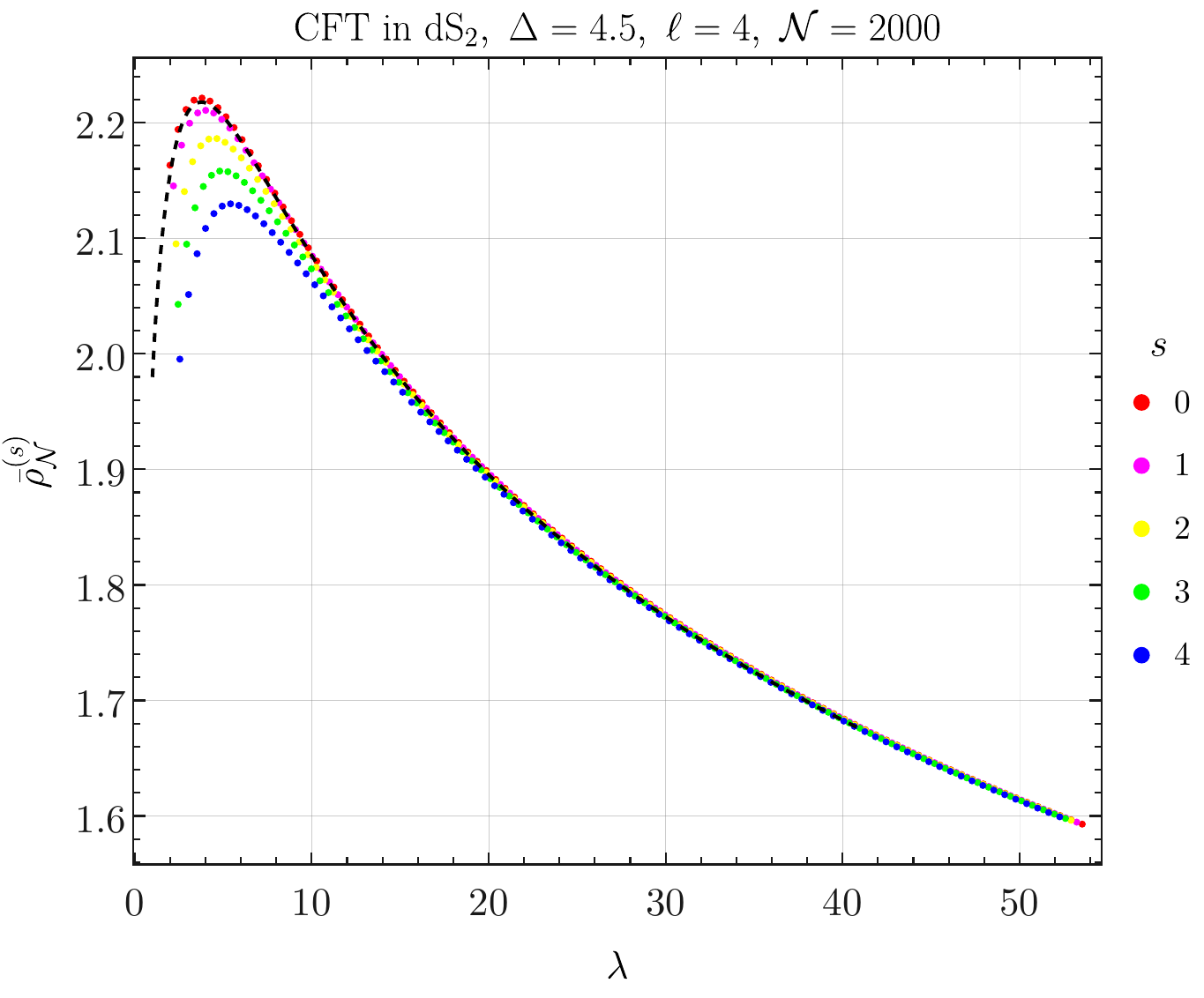}
         \label{fig:rbarCFTS}
     \end{subfigure}
     \hfill
     \begin{subfigure}[t]{0.44\textwidth}
         \centering
         \includegraphics[width=\textwidth]{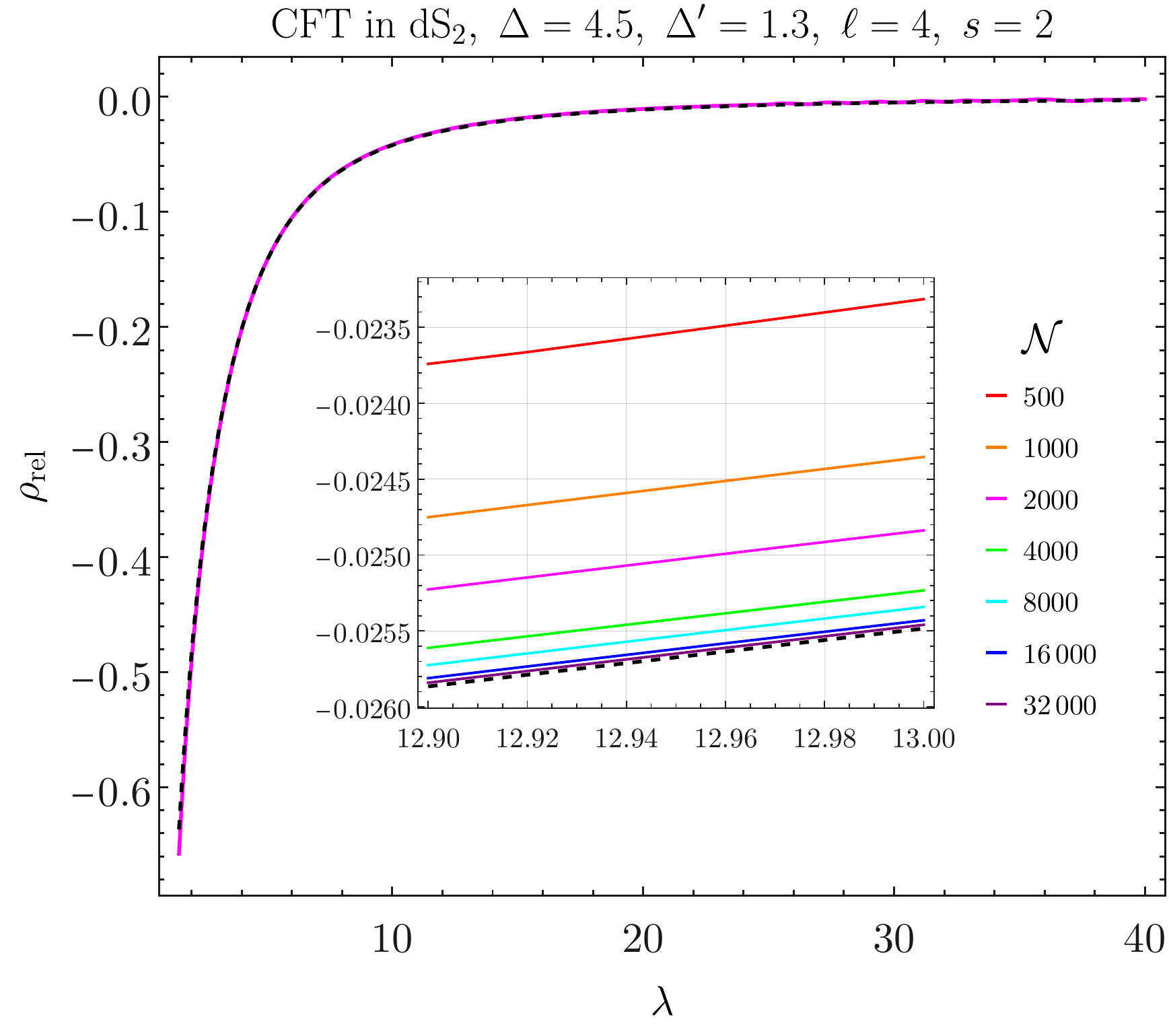}
         \label{fig:rrelCFTS}
     \end{subfigure}
     \caption{Left: Plot of $\bar{\rho}^{(s)}_\CN$ for different $s$. They differ in the small $\lambda$ region. The dashed line is the renormalized density $\CK_\text{PV}(\lambda;\ell)$ in~\reef{eq:hc CFT s} with $\Lambda\sim 8\times 10^3$.
    Right: The plot of $\rho_\text{rel}$ for $\D=4.5$ and $\D^\prime=1.3$ in $s=2$ subspace with truncation cutoff $\CN=2000$ (pink line) that matches with  $\CK_\text{rel} (\lambda,\ell)$ (black dashed line). The inset plot shows the convergence to  $\CK_\text{rel} (\lambda;\ell)$ (c.f. eq.~\reef{CKrel90}) in large $\CN$ limit by zooming in the region $\lambda \in (12.9,13)$.
     \label{fig:rhoCFTs}}
\end{figure}
Analogous to eq.~\reef{11L} and eq.~\reef{eq:CKepsilonhd}, we can define a regularized version of $\CK(\lambda;\ell)$ formally defined in~\reef{eq:theta SO22} by taking the limit $\Delta' =\Lambda \to \infty$ in eq.~\reef{CKrel90}:
\be\label{eq:hc CFT s}
\CK_\text{PV}(\lambda;\ell)=\frac{\log(\Lambda)}{\pi}-\frac{1}{2\pi} \sum_\pm \psi\left(\D-\half\pm i\lambda\right) -\frac{1}{\pi} \sum_{k=1}^{|\ell|}\frac{k-\frac{1}{2}}{(k-\frac{1}{2})^2+\lambda^2} ~.
\ee
Again, similar to what we see for the tensor products, the coarse-grained densities $\bar{\rho}^{(s)}_\CN$ extracted from $\CQ^{(s)}$ for various $s$ differ in the small $\lambda$ region and $\CK_\text{PV}(\lambda;\ell)$ matches with $\bar{\rho}^{(s=0)}_\CN$ up to fixing a value for $\Lambda$ -- see the left panel of fig.~\reef{fig:rhoCFTs} -- similar to the observasion we had for the case of tensor products illustrated in fig.~\reef{fig:rhobar1d} and fig.~\reef{fig:rbars}.
However, the relative density $\rho_\text{rel}$ has been observed to be $s$-independent in large $\CN$ limit and it matches perfectly with $\CK_\text{rel}(\lambda;\ell)$ (c.f. eq.~\reef{CKrel90}), as shown in the right panel of fig.~\reef{fig:rhoCFTs}. 

In this section, we gave numerical and analytical evidence that for $d=1$ and in the large $\CN$ limit, the \textit{relative} coarse-grained density extracted from diagonalizing the truncated Casimir matrix is the same for different subsectors labeled by $L_0=\Ls$ (for tensor product) or $L_0=s$ (for CFT). Therefore, for convenience, we pick the $L_0=0$ subspace in sections \ref{tensorlow} and \ref{CFTsec} for our numerical investigation. We conjecture this to be true in higher dimensions, although we did not explicitly check it.  The matching of the relative density from the character analysis with the numerical results in sections \ref{tensorhigh} and \ref{CFTsec}, on the other hand, is an evidence for this conjecture.

\section{Summing over $\SO(d)$ Weyl character}\label{sumW}
In this appendix, we will give a proof of the following series involving $\SO(d)$ Weyl characters
\begin{align}\label{techguess}
\sum_{s=0}^\infty \, \chi^{\SO(d)}_{\mY_s} (\bm x)\, q^s=\frac{1-q^2}{P_d(q,\bm x)}
\end{align}
where $0<q<1$ and $P_d(q,\bm x)$ is defined in eq. (\ref{Pddef}).  First we want to argue that it suffices to prove (\ref{techguess}) for even $d$. According to the branching rule, $\SO(2r+1)$ and $\SO(2r)$ characters are related by
\begin{align}
\chi^{\SO(2r+1)}_{\mY_s} (\bm x)=\sum_{m=0}^s \chi^{\SO(2r)}_{\mY_m} (\bm x)
\end{align}
which implies
\begin{align}\label{oande}
\sum_{s=0}^\infty \, \chi^{\SO(2r+1)}_{\mY_s} (\bm x)\, q^s=\sum_{m\ge 0}\chi^{\SO(2r)}_{\mY_m} (\bm x)\sum_{s\ge m}\, q^s=\frac{1}{1-q}\sum_{m\ge 0}\chi^{\SO(2r)}_{\mY_m} (\bm x) q^m
\end{align}
If (\ref{techguess}) holds for even $d=2r$, i.e. $\sum_{m\ge 0}\chi^{\SO(2r)}_{\mY_m} (\bm x) q^m=\frac{1-q^2}{P_{2r}(q,\bm x)}$,  then it is obviously correct for $d=2r+1$ since $P_{2r+1}(q,\bm x)=(1-q)P_{2r}(q,\bm x)$.

To prove (\ref{techguess}) for even $d=2r$, we need an explicit expression of  $\chi^{\SO(d)}_{\mY_s} (\bm x)$, which is given by the famous Weyl character formula \cite{fulton2013representation}
\begin{align}
 \chi^{\SO(d)}_{\mY_s} (\bm x)=\frac{|x_i^{\ell_j}+x_i^{-\ell_j}|}{|x_i^{n-j}+x_i^{-(n-j)}|}
\end{align}
Let's briefly explain the notations in this formula:
\begin{itemize}
\item $\ell_j$ is the $j$-th component of  the vector $\bm\ell=(s+r-1, r-2, r-3, \cdots,1,0)$.
\item The numerator $|x_i^{\ell_j}+x_i^{-\ell_j}|$ means the determinant of the matrix $N_\rho$ whose $(j,i)$ entry is $x_i^{\ell_j}+x_i^{-\ell_j}$.
\item The denominator $ |x_i^{n-j}+x_i^{-(n-j)}|$ means the determinant of the matrix $D_\rho$ whose $(j,i)$ entry is $x_i^{n-j}+x_i^{-(n-j)}$. It is well known that $D_\rho$ can be alternatively expressed as 
\begin{align}
|D_\rho|=2\,\Psi(x_i+x_i^{-1}), \,\,\,\,\ \Psi(\xi_i)\equiv \prod_{i<j} (\xi_i-\xi_j)
\end{align}
\end{itemize}
Since only the first row of $N_\rho$ depends  on $s$, the sum $\sum_{s\ge 0}|N_\rho| q^s$ effectively changes the first row in the following way:
\begin{align}
x_i^{s+r-1}+x_i^{1-r-s}\to \frac{x_i^{r-1}}{1-x_i q}+\frac{x_i^{1-r}}{1-x_i^{-1}q}
\end{align}
Denote this new matrix  by $N'_\rho$ and it is related to the series $\sum_{s\ge 0} \chi^{\SO(d)}_{\mY_s} (\bm x) q^s$ by
\begin{align}
\sum_{s\ge 0} \chi^{\SO(d)}_{\mY_s} (\bm x) q^s=\frac{|N'_\rho|}{|D_\rho|}
\end{align}
Then we add the rest rows to the first row of $N'_\rho$, with a weight $q^{-(j-1)}$ for the $j$-th row (except the last row, for which the weight is $\frac{1}{2}q^{1-r}$), which yields
\begin{align}
\frac{x_i^{r-1}}{1-x_i q}+\frac{x_i^{1-r}}{1-x_i^{-1}q}&\to \frac{x_i^{r-1}}{1-x_i q}+\frac{x_i^{1-r}}{1-x_i^{-1}q}+q^{-1}(x_i^{r-2}+x_i^{2-r})+\cdots +q^{-(r-1)}\nonumber\\
&=\frac{q^{2-r}\left(q^{-1}-q\right)}{1+q^2-(x_i+x_i^{-1})q}=\frac{q^{2-r}\left(q^{-1}-q\right)}{P(q,x_i)}
\end{align}
where $P(q,x)\equiv 1-(x+x^{-1})q+q^2$. 
Altogether, we can rewrite $|N_\rho'|$ as 
\begin{align}
|N_\rho'|=2q^{2-r}\left(q^{-1}-q\right)\det\begin{pmatrix} P(q,x_1)^{-1}& P(q,x_2)^{-1}&\cdots &P(q,x_r)^{-1}\\x_1^{r-2}+x_1^{2-r}& x_2^{r-2}+x_2^{2-r}&\cdots &x_r^{r-2}+x_r^{2-r} \\\cdots&\cdots&\cdots&\cdots\\ x_1+x_1^{-1}& x_2+x_2^{-1}& \cdots & x_r+x_r^{-1}\\ 1&1&\cdots&1\end{pmatrix}
\end{align}
where the factor 2 appears because we have rescaled the last row. Now a  crucial step is to replace $x_i^j+x_i^{-j}$ by $(-q)^{-j}P(q,x_i)^j$, which does not change the determinant. With this replacement, the matrix becomes essentially a Vandermonde matrix up to some reshuffling and rescaling 
\begin{align}
|N_\rho'|&=\frac{2q^{2-r}\left(q^{-1}-q\right)}{(-q)^{\frac{(r-1)(r-2)}{2}}}\det\begin{pmatrix} P(q,x_1)^{-1}& P(q,x_2)^{-1}&\cdots &P(q,x_r)^{-1}\\P(q,x_1)^{r-2}& P(q,x_2)^{r-2}&\cdots &P(q,x_r)^{r-2} \\\cdots&\cdots&\cdots&\cdots\\ P(q,x_1)&  P(q,x_2)& \cdots &P(q,x_r)\\ 1&1&\cdots&1\end{pmatrix}\nonumber\\
&=\frac{2\, (-)^{r-1}q^{2-r}\left(q^{-1}-q\right)}{(-q)^{\frac{(r-1)(r-2)}{2}}\prod_{i=1}^r P(q,x_i)} \det V(P(q,x_i))
\end{align}
where $V(\xi_i)$ is the $r\times r$ Vandermonde matrix of $\xi_1,\xi_2,\cdots, \xi_r$. Plugging in $\det V(\xi_i)=\prod_{i<j}(\xi_j-\xi_i)$ yields
\begin{align}
|N_\rho'|&=\frac{2\, (-)^{r-1}q^{2-r}\left(q^{-1}-q\right)}{(-q)^{\frac{(r-1)(r-2)}{2}}\prod_{i=1}^r P(q,x_i)} \prod_{i<j}\left(P(q,x_j)-P(q,x_i)\right)\nonumber\\
&=\frac{(-)^{r-1}(-q)^{\frac{r(r-1)}{2}}q^{2-r}\left(q^{-1}-q\right)}{(-q)^{\frac{(r-1)(r-2)}{2}}\prod_{i=1}^r P(q,x_i)} \left(2\Psi(x_i+x_i^{-1})\right)
\end{align}
Notice that the last term is exactly the denominator $|D_\rho|$ and hence we obtain
\begin{align}
\sum_{s\ge 0} \chi^{\SO(d)}_s (\bm x) q^s=\frac{(-)^{r-1}(-q)^{\frac{r(r-1)}{2}}q^{2-r}\left(q^{-1}-q\right)}{(-q)^{\frac{(r-1)(r-2)}{2}}\prod_{i=1}^r P(q,x_i)}=\frac{1-q^2}{P_d(q,\bm x)}
\end{align}
This finishes our proof of (\ref{techguess}).

\section{Matrix elements of  the noncompact generators in $\SO(1,d+1)$ }\label{matrixcomp}
The noncompact generators of $\SO(1, d+1)$ are denoted by $L_{0a}$ where $a=1,\cdots, d+1$. In particular, $L_{0, d+1}=D$ is the dilatation operator. 
The quadratic Casimir of $\SO(1, d+1)$ can be expressed as
\begin{align}\label{Casrelation}
\Cas^{\SO(1, d+1)}=-L^2_{0a}-\Cas^{\SO(d+1)}
\end{align}
where $\Cas^{\SO(d+1)}$ is the usual $\SO(d+1)$ Casimir, which equals $n(n+d-1)$ for a spin-$n$ representation.
In this appendix, we will derive explicitly how $L_{0a}$ acts on the $\SO(d+1)$ content of a continuous UIR $\CF_\Delta$.
We will first focus on the principal series case, e.g. $\Delta=\frac{d}{2}+i\mu$, and then show how the results can be generalized to complementary series. With the action of $L_{0a}$ being derived, we will use eq.~(\ref{Casrelation}) to compute the matrix elements of  $\Cas^{\SO(1, d+1)}$ in certain subsectors of the tensor product $\CF_{\Delta_1}\otimes\CF_{\Delta_2}$. 

As we have reviewed in section \ref{repreview}, the Hilbert space of $\CP_{\Delta}$ consists of all single-row representations of  $\SO(d+1)$ with multiplicity one for each. An obvious orthogonal and normalizable basis is 
\begin{align}
|n\rangle_{a_1\cdots a_n}, \,\,\,\,\, n\in\mathbb Z, \,\,\,\,\, 1\le a_i\le d+1
\end{align}
The indices $(a_1, a_2,\cdots, a_n)$ are symmetric and traceless, and hence for a fixed $n$, the different components $|n\rangle_{a_1\cdots a_n}$ furnish the spin-$n$ representation of $\SO(d+1)$. By construction, $|n\rangle_{a_1\cdots a_n}$ is an eigenstate of $L_{0a}^2$
\begin{align}\label{L2cas}
L_{0a}^2 |n\rangle_{a_1\cdots a_n}=-\lambda_n |n\rangle_{a_1\cdots a_n}, \,\,\,\,\, \lambda_n =\Delta\bar\Delta+n(n+d-1)
\end{align}
The normalization of inner product is chosen to be 
\begin{align}\label{expinner}
\text{With summation}: \,\, _{a_1\cdots a_n}\langle n|n\rangle_{a_1\cdots a_n}=D^{d+1}_{n}
\end{align}
where $D^{d+1}_n=\frac{(d+2n-1)\Gamma(n+d-1)}{\Gamma(d)\Gamma(n+1)}$ denotes the dimension of the spin-$n$ representation of $\SO(d+1)$. For example, when $n=2$, the normalization in (\ref{expinner}) yields
\begin{align}
_{a_1 a_2}\langle 2|2\rangle_{b_1 b_2}=\frac{1}{2}\left(\delta_{a_1 b_1}\delta_{a_2 b_2}+\delta_{a_1 b_2}\delta_{a_2 b_1}-\frac{2}{d+1}\delta_{a_1 a_2}\delta_{b_1b_2}\right)
\end{align}

 In the index free formalism, all $|n\rangle_{a_1\cdots a_n}$ can be encoded in 
\begin{align}
|n,z\rangle\equiv |n\rangle_{a_1\cdots a_n}z^{a_1}\cdots z^{a_n}
\end{align}
where $z^a$ is an auxiliary null vector in $\mathbb C^{d+1}$. Introducing a different null vector $w^a$, the inner product (\ref{expinner}) is equivalent to 
\begin{align}\label{nmzw}
\langle n,\bar w|m, z\rangle=\delta_{n,m}(\bar w\cdot z)^n
\end{align}
Stripping off $z^a$ is realized by the so-called interior derivative
\begin{align}\label{intD}
\CD_a=\partial_{z^a}-\frac{1}{d-1+2 \, z\cdot\partial_z} z_a\partial_z^2
\end{align}

Since $L_{0a}$ transforms as a vector under $\SO(d+1)$, $L_{0a} |n\rangle_{a_1\cdots a_n}$ has the symmetry $\mY_1\otimes \mY_{n}$, which can be decomposed into three irreducible components: $\mY_{n-1}$, $\mY_{n+1}$ and $\mY_{n,1}$. The $\mY_{n,1}$ part should vanish identically because it does not belong to the $\SO(d+1)$ content of $\CP_{\Delta}$. \footnote{When $d=2$, $\mY_{n,1}$ is the same as $\mY_{n}$ due to the totally antisymmetric tensor $\epsilon_{abc}$. However, one can still show that the $\mY_n$ component cannot enter eq. (\ref{L0an}) if $|n\rangle_{a_1\cdots a_n}$ belongs to $\CF_\Delta$. This is not true if $|n\rangle_{a_1\cdots a_n}\in \CF_{\Delta,s}$ where $s\not=0$.} Therefore, the action of $L_{0a}$ on $|n\rangle_{a_1\cdots a_n}$ should take the following form\footnote{More explicitly, $\text{trace}=\frac{n-1}{d+2n-3}\delta_{(a_1 a_2}|n-1\rangle_{a_3\cdots a_n) a}$ in this equation.}
\begin{align}\label{L0an}
L_{0a}|n\rangle_{a_1\cdots a_n}=\alpha_{n}|n+1\rangle_{aa_1\cdots a_n}+\beta_n\left(\delta_{a(a_1}|n-1\rangle_{a_2\cdots a_n)}-\text{trace}\right)
\end{align}
where the coefficients $\alpha_n$ and $\beta_n$ are to be fixed. In the index-free formalism, eq.~(\ref{L0an}) is equivalent to 
\begin{align}\label{L0anz}
L_{0a}|n,z\rangle=\frac{\alpha_n}{n+1}\CD_a |n+1,z\rangle+\beta_n  z_a |n-1,z\rangle
\end{align}
Acting $\CD_a$ on both sides of (\ref{L0anz}) and summing over $a$ yields 
\begin{align}\label{L0anc}
L_{0a_1}|n\rangle_{a_1\cdots a_n}=\beta_n\frac{(d+n-2)(d+2n-1)}{n(d+2n-3)}|n-1\rangle_{a_2\cdots a_n}
\end{align}
where we have used $\CD^2=0$ and 
\begin{align}
\CD_a (z_a|n-1, z\rangle)=\frac{(d+n-2)(d+2n-1)}{d+2n-3}|n-1,z\rangle
\end{align}
Acting $L_{0a}$ on both sides of (\ref{L0anz}), we obtain our first recurrence relation 
\begin{align}
\frac{(d+n-1)(d+2n+1)}{(n+1)(d+2n-1)}\alpha_n\beta_{n+1}+\alpha_{n-1}\beta_n=-\lambda_n
\end{align}
Imposing the initial condition $\beta_0=0$, we can completely fix the product $\alpha_n\beta_{n+1}$
\begin{align}\label{rrec1}
\alpha_n\beta_{n+1}=-\frac{(n+1)(\Delta+n)(\bar\Delta+n)}{d+2n+1}
\end{align}
To derive another recurrence relation, let's compute the norm of $L_{0a}|n,z\rangle$ using (\ref{nmzw}) and (\ref{L0anz})
\begin{align}
\langle n, \bar w|L_{0a}^\dagger L_{0a}|n,z\rangle=\lambda_n(\bar w\cdot z)^n=\frac{|\alpha_n|^2}{(n+1)^2}\CD_a^{(z)}\CD_a^{(\bar w)}(\bar w\cdot z)^{n+1}+ |\beta_n|^2 (\bar w\cdot z)^n
\end{align}
which yields 
\begin{align}\label{rrec2}
|\alpha_n|^2\frac{(d+n-1)(d+2n+1)}{(n+1)(d+2n-1)}+|\beta_n|^2=\lambda_n
\end{align}
The two recurrence relations (\ref{rrec1}) and (\ref{rrec2}) are invariant under an $n$-dependent $\text{U}(1)$ transformation:
$\alpha_n\to e^{i\theta_n}\alpha_n$ and $\beta_{n+1}\to e^{-i\theta_n}\beta_{n+1}$. The angle $\theta_n$  can always be absorbed into a redefinition of $|n\rangle_{a_1\cdots a_n}$. We will fix this $\text{U}(1)$ degree of freedom by choosing $\alpha_n\propto \Delta+n$ and $\beta_{n+1}\propto \bar\Delta+n$. In particular, $\alpha_0=\frac{\Delta}{\sqrt{d+1}}$. Then the solution of (\ref{rrec1}) and (\ref{rrec2}) is uniquely determined
\begin{align}\label{abP}
 \alpha_n=\sqrt{\frac{n+1}{d+2n+1}}(\Delta+n), \,\,\,\,\, \beta_{n}=-\sqrt{\frac{n}{d+2n-1}}(\bar\Delta+n-1)
\end{align}
The  eq.~(\ref{abP})  requires a minor modification if we replace the principal series $\CP_{\Delta}$ by complementary series $\CC_{\Delta}$ with $\Delta=\frac{d}{2}+\mu$. The recurrence relations (\ref{rrec1}) and (\ref{rrec2}) still hold in this case, but $\alpha_n\propto \Delta+n$  is not a proper choice anymore. It is easy to check that  $\alpha_0=\frac{\Delta}{\sqrt{d+1}}$ does not satisfy the $n=0$ case of eq.~(\ref{rrec2}) when $\Delta$ is real. Instead, we will fix the $\text{U}(1)$ degree of freedom differently by choosing all $\alpha_n$ to be real and positive. In particular, it means $\alpha_0=\sqrt{\frac{\Delta\bar\Delta}{d+1}}$.
The general expressions of $\alpha_n$ and $\beta_n$ in eq.~(\ref{abP}) should change accordingly as
\begin{align}\label{abP1}
\alpha_n=\sqrt{\frac{(n+1)(\Delta+n)(\bar\Delta+n)}{d+2n+1}}, \,\,\,\,\,\beta_{n}=-\sqrt{\frac{n(\Delta+n-1)(\bar\Delta+n-1)}{d+2n-1}}
\end{align}

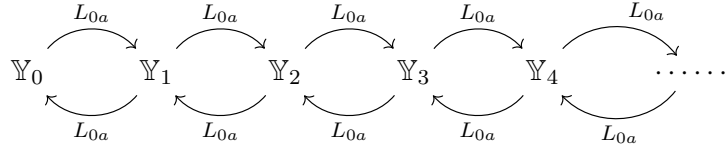
\begin{figure}[h]
 \centering
\begin{tikzcd}
\mathbb{Y}_0 \arrow[r,bend right=-50, "L_{0a}"]  \arrow[r,leftarrow,bend left=-50,swap, "L_{0a}"] 
& \mathbb{Y}_1 \arrow[r, bend right=-50, "L_{0a}"]  \arrow[r, leftarrow,bend left=-50,swap, "L_{0a}"] &\mathbb{Y}_2 \arrow[r,bend right=-50, "L_{0a}"]  \arrow[r,leftarrow,bend left=-50,swap, "L_{0a}"] & \mathbb{Y}_3 \arrow[r, bend right=-50, "L_{0a}"]  \arrow[r, leftarrow,bend left=-50,swap, "L_{0a}"] &\mathbb{Y}_4  \arrow[r,bend right=-50, "L_{0a}"]  \arrow[r,leftarrow,bend left=-50,swap, "L_{0a}"] &\cdots\cdots
\end{tikzcd}
\caption{The action of noncompact generators of $\SO(1,d+1)$ in scalar principal and complementary series representations.}
\label{f111}
\end{figure}

For principal series or complementary series, the coefficients $\alpha_n$ and $\beta_n$ given by eq.~(\ref{abP}) or eq.~(\ref{abP1}) are nonvanishing, and hence these representations are irreducible. Pictorially, this property is shown in fig.(\ref{f111}).
\noindent{}If we formally take $\Delta=d+p-1, p\in\mathbb Z_{+}$ in eq.~(\ref{abP1}), which is clearly out of the unitarity regime of complementary series, $\beta_p$ vanishes and all the rest $\alpha_n,\beta_n$ are novanishing. Using the  definition of $\alpha_n$ and $\beta_n$ given by eq.~(\ref{L0an}), the vanishing of $\beta_p$ implies that $\{|n\rangle_{a_1\cdots a_n}, n\ge p\}$ span an invariant subspace of $\CC_\Delta$. See fig.(\ref{f112}) for the $p=2$ example: 
 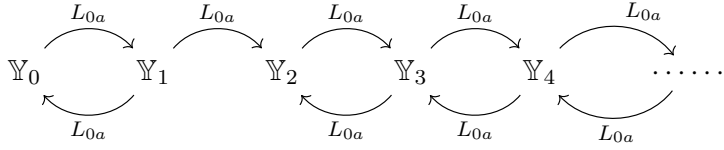
\begin{figure}[h]
 \centering
\begin{tikzcd}
\mathbb{Y}_0 \arrow[r,bend right=-50, "L_{0a}"]  \arrow[r,leftarrow,bend left=-50,swap, "L_{0a}"] 
& \mathbb{Y}_1 \arrow[r, bend right=-50, "L_{0a}"] &\mathbb{Y}_2 \arrow[r,bend right=-50, "L_{0a}"]  \arrow[r,leftarrow,bend left=-50,swap, "L_{0a}"] & \mathbb{Y}_3 \arrow[r, bend right=-50, "L_{0a}"]  \arrow[r, leftarrow,bend left=-50,swap, "L_{0a}"] &\mathbb{Y}_4  \arrow[r,bend right=-50, "L_{0a}"]  \arrow[r,leftarrow,bend left=-50,swap, "L_{0a}"] &\cdots\cdots
\end{tikzcd}
\caption{The action of noncompact generators of $\SO(1,d+1)$ in $\CV_{2,0}$.}
\label{f112}
\end{figure}

\noindent{}Indeed, this invariant subspace together with the inner product (\ref{expinner}) is the type \rom{1} exceptional series representation $\CV_{p,0}$. This construction manifests the $\SO(d+1)$ content of $\CV_{p,0}$ given by eq.~(\ref{CVcomp}).

Next, we apply the results above to the tensor product of $\CF_{\Delta_1}\otimes \CF_{\Delta_2}$. The space of $\SO(d+1)$ singlets in $\CF_{\Delta_1}\otimes \CF_{\Delta_2}$ is spanned by 
\begin{align}\label{singdef}
|\psi_n\rangle=\frac{1}{\sqrt{D^{d+1}_n}}|n,\Delta_1\rangle_{a_1\cdots a_n}|n,\Delta_2\rangle_{a_1\cdots a_n}, \,\,\,\,\, n\in\mathbb N
\end{align}
where we have introduced the factor $\frac{1}{\sqrt{D^{d+1}_n}}$ such that $|\psi_n\rangle$ is normalized to be 1. Since $|\psi_n\rangle$ is an $\SO(d+1)$ singlet, the action of $\Cas^{\SO(1, d+1)}$ on it is equivalent to $-L_{0, a}^2$ due to eq.~(\ref{Casrelation}):
\begin{align}\label{L2exp}
\Cas^{\SO(1, d+1)}|\psi_n\rangle&=-\frac{1}{\sqrt{D^{d+1}_n}}\left(L^2_{0a}|n,\Delta_1\rangle_{a_1\cdots a_n}|n,\Delta_2\rangle_{a_1\cdots a_n}+|n,\Delta_1\rangle_{a_1\cdots a_n}L^2_{0a}|n,\Delta_2\rangle_{a_1\cdots a_n}\right)\nonumber\\
&-\frac{2}{\sqrt{D^{d+1}_n}}L_{0a}|n,\Delta_1\rangle_{a_1\cdots a_n}L_{0a}|n,\Delta_2\rangle_{a_1\cdots a_n}
\end{align}
Using eq.~(\ref{L2cas}), we can easily write down the diagonal entry $\langle \psi_n|\Cas^{\SO(1, d+1)}|\psi_n\rangle$, which corresponds to the first line of (\ref{L2exp})
\begin{align}
\langle \psi_n|\Cas^{\SO(1, d+1)}|\psi_n\rangle=\Delta_1\bar\Delta_1+\Delta_2\bar\Delta_2+2n(n+d-1)
\end{align}
Combining (\ref{L0an}), (\ref{abP}) and the second line of (\ref{L2exp}), we  obtain the off-diagonal entry 
\begin{align}\label{C2off}
\langle \psi_{n+1}|\Cas^{\SO(1, d+1)}|\psi_n\rangle=-2\sqrt{\frac{D^{d+1}_{n+1}}{D^{d+1}_n}}\alpha_n(\Delta_1)\alpha_n(\Delta_2)
\end{align}
which is the complex conjugate of $\langle \psi_{n}|\Cas^{\SO(1, d+1)}|\psi_{n+1}\rangle$. When both $\CF_{\Delta_i}$ belong to principal series, plugging eq.~(\ref{abP}) into eq.~(\ref{C2off}) yields 
\begin{align}\label{PPoff}
\CP_{\Delta_1}\otimes \CP_{\Delta_2}: \quad \langle \psi_{n+1}|\Cas^{\SO(1, d+1)}|\psi_n\rangle=-\sqrt{\frac{(n+1)(n+d-1)}{(n+\frac{d-1}{2})(n+\frac{d+1}{2})}}(\Delta_1+n)(\Delta_2+n)
\end{align}
and when both $\CF_{\Delta_i}$ belong to complementary series, plugging eq.~(\ref{abP1}) into eq.~(\ref{C2off}) yields 
\small
\begin{align}\label{CCoff}
\CC_{\Delta_1}\otimes \CC_{\Delta_2}: \quad \langle \psi_{n+1}|\Cas^{\SO(1, d+1)}|\psi_n\rangle=-\sqrt{\frac{(n+1)(n+d-1)(\Delta_1+n)(\bar\Delta_1+n)(\Delta_2+n)(\bar\Delta_2+n)}{(n+\frac{d-1}{2})(n+\frac{d+1}{2})}}
\end{align}
\normalsize

\section{Absence of exceptional series in $\CF_{\Delta_1}\otimes\CF_{\Delta_2}$}\label{Noex}
In this appendix, we will show that the tensor product $\CF_{\Delta_1}\otimes\CF_{\Delta_2}$ does not contain any exceptional series, where $\CF_{\Delta_i}$ belong to either principal series or complementary series. The main computational tool involved in our proof is established in the previous appendix.

For the type \rom{1} exceptional series, we will use $\CV_{1,0}$ as an example  to illustrate the main idea of the proof, which can be easily generalized to any $\CV_{p,0}$. Its $\SO(d+1)$ content are given by 
\begin{align}\label{CVcont}
\left.\CV_{1,0}\right|_{\SO(d+1)}=\bigoplus_{n\ge 1}\mY_{n}
\end{align}
As a result of (\ref{CVcont}), any state $|\psi\rangle_a$ in the $\mY_1$ component should satisfy 
\begin{align}
L_{0a}|\psi\rangle_a=0,\,\,\,\,\, L_{0a}|\psi\rangle_b-L_{0b}|\psi\rangle_a=0
\end{align}
We will show that such states do not exists in $\CF_{\Delta_1}\otimes \CF_{\Delta_2}$. First, notice that a generic state with the $\mY_1$ symmetry takes the following form
\begin{align}
|\psi\rangle_{a_1}=\sum_{n\ge 0}\left(c_n |n+1,\Delta_1\rangle_{a_1\cdots a_{n+1}}|n,\Delta_2\rangle_{a_2\cdots a_{n+1}} +d_n |n,\Delta_1\rangle_{a_2\cdots a_{n+1}}|n+1,\Delta_2\rangle_{a_1\cdots a_{n+1}} \right)
\end{align}
Imposing the constraint $L_{0[a_1}|\psi\rangle_{b_1]}=0$ yields $c_n\alpha_n(\Delta_2)=d_n \alpha_{n}(\Delta_1)$. Imposing the other constraint $L_{a_1}|\psi\rangle_{a_1}=0$ leads to $c_0\beta_1(\Delta_1)+d_0\beta_1(\Delta_2)=0$ and 
\begin{align}
c_{n-1}\alpha_{n-1}(\Delta_2)+d_{n-1}\alpha_{n-1}(\Delta_1)+\left(c_n\beta_{n+1}(\Delta_1)+d_n\beta_{n+1}(\Delta_2)\right)\frac{(d+n-1)(d+2n+1)}{(n+1)(d+2n-1)}=0
\end{align}
Using (\ref{rrec1}), we find $\alpha_0(\Delta_1)\beta_1(\Delta_1)+\alpha_0(\Delta_2)\beta_1(\Delta_2)\not=0$ which implies  that $c_0, d_0$ vanish. Then it is easy to conclude that all $c_n$ and $d_n$ are identically zero. This excludes the exceptional series $\CV_{1,0}$.

For the type \rom{2} exceptional series, we will use $\CU_{1,0}$ as an example. As reviewed in section \ref{repreview}, the $\SO(d+1)$ content of $\CU_{1,0}$ are 
\begin{align}
\left.\CU_{1,0}\right|_{\SO(d+1)}=\bigoplus_{n\ge 1} \mY_{n, 1}
\end{align}
Unlike the other spin 1 UIRs, e.g. $\CF_{\Delta, 1}$, $\CU_{1,0}$ does not contain the vector representation of $\SO(d+1)$. It implies that  for any state $|\chi\rangle_{a, b}$ with symmetry $\mY_{1,1}$, i.e. $|\chi\rangle_{a, b}=-|\chi\rangle_{b, a}$, $L_{0b}|\chi\rangle_{a,b}$ should vanish since it carries the spin 1 representation of $\SO(d+1)$. We will show that such $|\chi\rangle_{a, b}$ does not exist in the tensor product $\CF_{\Delta_1}\otimes \CF_{\Delta_2}$. Let's first write down a basis of states in $\CF_{\Delta_1}\otimes \CF_{\Delta_2}$ that carry the two-form symmetry of $\SO(d+1)$:
\begin{align}
|\chi_n)_{a_1,b_1}=|n,\Delta_1 \rangle_{a_1 a_2\cdots a_n}|n,\Delta_2 \rangle_{b_1 a_2\cdots a_n}-(a_1\leftrightarrow b_1), \,\,\,\,\, n\ge 1
\end{align}
Then a generic state $|\chi\rangle_{a,b}$ belonging to $\mY_{1,1}$ should be a linear combination of $|\chi_n)_{a,b}$, e.g.
\begin{align}
|\chi\rangle_{a,b}=\sum_{n\ge 1} c_n |\chi_n)_{a,b}
\end{align}
Computing $L_b |\chi\rangle_{a,b}$ using eq. (\ref{L0an}) and (\ref{L0anc}) yields
\small
\begin{align}
&\sum_{n\ge 1} c_n \left(\alpha_n(\Delta_1) |n\!+\!1,\Delta_1\rangle_{a_1\cdots a_{n+1}} |n,\Delta_2\rangle_{a_2\cdots a_{n+1}} \!+\!\frac{(n\!+\!d\!-\!1)\beta_n(\Delta_2)}{n}|n,\Delta_1\rangle_{a_1\cdots a_{n}}  |n\!-\!1,\Delta_2\rangle_{a_2\cdots a_{n}} \right)\nonumber\\
&-\sum_{n\ge 1} c_n\left(\alpha_n(\Delta_2) |n,\!\Delta_1\rangle_{a_2\cdots a_{n+1}}  |n\!+\!1,\!\Delta_2\rangle_{a_1\cdots a_{n+1}} \!+\!\frac{(n\!+\!d\!-\!1)\beta_n(\Delta_1)}{n}|n\!-\!1,\!\Delta_1\rangle_{a_2\cdots a_{n}}  |n,\!\Delta_2\rangle_{a_1\cdots a_{n}} \right)
\end{align}
\normalsize
Requiring $L_b |\chi\rangle_{a,b}=0$ gives us two recurrence relations 
\begin{align}\label{crec2}
&c_n \alpha_n(\Delta_1)+c_{n+1}\beta_{n+1}(\Delta_2)\frac{n+d}{n+1}=0\nonumber\\
&c_n \alpha_n(\Delta_2)+c_{n+1}\beta_{n+1}(\Delta_1)\frac{n+d}{n+1}=0
\end{align}
and one initial condition $c_1=0$. Apparently, the only solution is  $c_n=0$ for $n\ge 1$. Therefore $\CF_{\Delta_1}\otimes \CF_{\Delta_2}$ does not contain $\CU_{1,0}$.

Although we have excluded $\CU_{1,0}$ in $\CF_{\Delta_1}\otimes\CF_{\Delta_2}$, let's still compute the norm of all $|\chi_n)_{a,b}$, since it is needed in subsection \ref{EEt} for the tensor product of $\CV_{1,0}\otimes\CV_{1,0}$. By definition, we have 
\begin{align}\label{ccnorm}
\frac{1}{2}\,_{a_1, b_1}(\chi_n|\chi_n)_{a_1, b_1}&=\,_{a_1 c_2\cdots c_n}\langle n|n\rangle_{a_1a_2\cdots a_n} \,_{b_1 c_2\cdots c_n}\langle n|n\rangle_{b_1a_2\cdots a_n}\nonumber\\
&-\,_{a_1 c_2\cdots c_n}\langle n|n\rangle_{b_1a_2\cdots a_n} \,_{b_1 c_2\cdots c_n}\langle n|n\rangle_{a_1a_2\cdots a_n}
\end{align}
where we have suppressed the label of scaling dimensions because they are irrelevant for the calculation here.
To compute each term in (\ref{ccnorm}), consider the  state
\begin{align}\label{Sst}
|S\rangle_{a_1b_1, c_1\cdots c_n}\equiv \,_{b_1a_2\cdots a_n}\langle n|n\rangle_{c_1c_2\cdots c_n} |n\rangle_{a_1a_2\cdots a_n}
\end{align}
in terms of which, eq. (\ref{ccnorm}) can be rewritten in the following form
\begin{align}\label{Sexps}
\frac{1}{2}\,_{a_1, b_1}(\chi_n|\chi_n)_{a_1, b_1}=\,_{a_1 c_2\cdots c_n}\langle n|S\rangle_{a_1b_1, b_1c_2\cdots c_n}-\,_{b_1 c_2\cdots c_n}\langle n|S\rangle_{a_1b_1, a_1c_2\cdots c_n}
\end{align}
Without doing any explicit computation, one can almost fix $|S\rangle_{a_1b_1, c_1\cdots c_n}$ up to two unkown coefficients $A, B$ 
\begin{align}
|S\rangle_{a_1b_1, c_1\cdots c_n}=A\left(\sum_{k=1}^n \delta_{b_1 c_k}|n\rangle_{a_1 c_1\cdots\hat c_k\cdots c_n}-B\sum_{1\le k<\ell\le n}\delta_{c_k c_\ell}|n\rangle_{a_1 b_1 c_1\cdots \hat c_k\cdots \hat c_\ell\cdots c_n}\right)
\end{align}
Contracting the indices $a_1$ and $b_1$, $|S\rangle$ should reduce to $|n\rangle_{c_1\cdots c_n}$ because $|n\rangle_{a_1\cdots a_n}\,_{a_1\cdots a_n}\langle n|$ is the identity operator. This condition fixes $A=\frac{1}{n}$. Next, we impose the traceless condition among the $c_j$ indices. For instance, it we contract $c_1$ and $c_2$, the first sum in eq. (\ref{Sst}) yields $2 |n\rangle_{a_1 b_1c_2\cdots c_n}$ and the second double sum yields $(d+2n-3)|n\rangle_{a_1 b_1c_2\cdots c_n}$. Therefore $B$ is equal to $\frac{2}{d+2n-3}$. With $A$ and $B$ known, the remaining task is to contract $(a_1, c_1)$ and $(b_1, c_1)$. After a simple calculation, we find 
\begin{align}\label{twocontr}
&|S\rangle_{a_1b_1, b_1c_2\cdots c_n}=\frac{(n+d-2)(2n+d-1)}{n(2n+d-3)}|n\rangle_{a_1c_2\cdots c_n}\nonumber\\
&|S\rangle_{a_1b_1, a_1c_2\cdots c_n}=\frac{d-1}{n(2n+d-3)}|n\rangle_{b_1c_2\cdots c_n}
\end{align}
Plugging (\ref{twocontr}) into (\ref{Sexps}) and using the normalization condition (\ref{expinner}), we find 
\begin{align}
\,_{a_1, b_1}(\chi_n|\chi_n)_{a_1, b_1}=\frac{2(n+d-1)}{n}D^{d+1}_n=4\left(1+\frac{d-1}{2n}\right)\binom{n+d-1}{d-1}
\end{align}
When $d=3$, it becomes 
\begin{align}\label{cc3}
d=3:\,\,\,\, \,_{a_1, b_1}(\chi_n|\chi_n)_{a_1, b_1}=\frac{2(n+2)(n+1)^2}{n}
\end{align}

\section{A review of $\SO(2,d+1)$}\label{SO2dreview}
The Lie algebra of $\SO(2,d+1)$ is generated by $L_{IJ}=-L_{IJ}, I,J=0,1,\cdots, d+2$, which are anti-hermitian operators satisfying commutation relations
\begin{align}
\left[L_{IJ}, L_{MN}\right]=\eta_{JM}L_{IN}-\eta_{IM}L_{JN}+\eta_{IN}L_{JM}-\eta_{JN}L_{IM}
\end{align} 
where $\eta_{IJ}=\text{diag}(-1,1,\cdots, 1,-1)$. A convenient basis of $\so(2,d+1)$ is
\begin{align}\label{confgen}
H=-iL_{0, d+2}, \,\,\,\,\, L^\pm_a=-i L_{a0}\mp L_{a,d+2}, \,\,\,\,\, M_{ab}=L_{ab}
\end{align}
where $1\le a,b\le d+1$ and $M_{ab}$ generate the $\SO(d+1)$ subgroup. Some nontrivial commutation relations are
\begin{align}
[H,L^\pm_a]=\pm L^\pm_a, \,\,\,\,\, [ M_{ab}, L^\pm_c]=\delta_{bc} L^\pm_a-\delta_{ac} L^\pm_b, \,\,\,\,\, [L^-_a, L^+_b]=2\delta_{ab}H-2 M_{ab}
\end{align}
The $\SO(2, d+1)$ quadratic  Casimir is given by 
\begin{align}\label{2dCas}
C^{\SO(2,d+1)}=H(H-d-1)-L^+\cdot L^--C^{\SO(d+1)}
\end{align}
Consider an $\SO(1,d+1)$ subgroup, that is generated by all $L_{AB}$ with $0\le A,B\le d+1$. Following the convention used in section \ref{repreview}, the $\SO(1,d+1)$ Casimir can be expressed as 
\small
\begin{align}\label{1dCas}
C^{\SO(1,d+1)}=-L_{0a}^2+C^{\SO(d+1)}=\frac{1}{4}L^+\cdot L^++\frac{1}{4}L^-\cdot L^-+\left(\frac{1}{2}L^+\cdot L^-+\frac{d+1}{2}H+C^{\SO(d+1)}\right)
\end{align}
\normalsize

A primary representation $\mathcal{R}_{\Delta, \ell}$ is built upon a primary state $|\Delta,\ell\rangle_{a_1\cdots a_\ell}$ satisfying
\begin{align}
&H|\Delta,\ell\rangle_{a_1\cdots a_\ell}=\Delta |\Delta,\ell\rangle_{a_1\cdots a_\ell},\,\,\,\,\, L^-_a|\Delta,\ell\rangle_{a_1\cdots a_\ell}=0\nonumber\\
& C^{\SO(d+1)}|\Delta,\ell\rangle_{a_1\cdots a_\ell}=-\ell(\ell+d-1)|\Delta,\ell\rangle_{a_1\cdots a_\ell}
\end{align}  
where the Casimir condition of $\SO(d+1)$ implies that the indices $(a_1\cdots a_\ell)$ are symmetric and traceless. We choose the inner product such that $|\Delta,\ell\rangle_{a_1\cdots a_\ell}$ is normalized as follows
\begin{align}\label{expinner1}
\text{With summation}: \,\, &_{a_1\cdots a_\ell}\langle \Delta, \ell|\Delta, \ell\rangle_{a_1\cdots a_\ell}=D^{d+1}_{\ell}
\end{align}
where $D^{d+1}_\ell$ is the dimension of the spin $\ell$ representation of $\SO(d+1)$.
The Hilbert space of $\mathcal{R}_{\Delta, \ell}$ is spanned by the primary state and its descendants of the form 
\begin{align}
L_{b_1}^+\cdots L_{b_k}^+  |\Delta,\ell\rangle_{a_1\cdots a_\ell}
\end{align}
The unitarity of $\mathcal{R}_{\Delta, \ell}$ requires $\Delta\ge \frac{d-1}{2}$ when $\ell=0$ and $\Delta\ge d+\ell-1$ when $\ell\ge 1$, known as the unitarity bound. When the unitarity bound is saturated, say $\Delta=d+\ell-1$, the primary state becomes a conserved current in the sense 
\begin{align}
L^+_{a_1} |\Delta\rangle_{a_1\cdots a_\ell}=0
\end{align}
Acting with $C^{\SO(2,d+1)}$ on the primary state $|\Delta,\ell\rangle_{a_1\cdots a_\ell}$ yields its value on any state in $\mathcal{R}_{\Delta, \ell}$
\begin{align}
C^{\SO(2,d+1)}\left(\mathcal{R}_{\Delta, \ell}\right)=\Delta(\Delta-d-1)+\ell(\ell+d-1)
\end{align}
When $d=1$, $\ell$ is an eigenvalue of the $\SO(2)$ subgroup and hence takes values in all integers, i.e. 
\begin{align}
iM_{12} |\Delta,\ell\rangle=\ell |\Delta,\ell\rangle, \,\,\,\,\, \ell\in\mathbb Z
\end{align}
A generic descendant can be written as 
\begin{align}\label{nbn}
|n,\bar n) = \left(L_1^+-i L_2^+\right)^{n}\left(L_1^++i L_2^+\right)^{\bar n} |\Delta, \ell\rangle
\end{align}
It has scaling dimension $\Delta+n+\bar n$ and spin $\ell+n-\bar n$. Fixing the normalization $\langle \Delta,\ell|\Delta,\ell\rangle=1$, then the norm of $|n,\bar n)$ becomes 
\begin{align}\label{nbnbnn}
(n,\bar n|n, \bar n)=4^{n+\bar n}n!\bar n!(\Delta+\ell)_n(\Delta-\ell)_{\bar n}
\end{align}

\section{Various properties of $|\phi^s_n)$}\label{phinsprop}
In this appendix, we compute the action of various $\SO(1,d+1)$ generators and their products on the states  $|\phi_n^s)$, defined by eq. (\ref{phinsdef}). Let's first start with the $s=0$ case as a simple exercise. Consider $L^-_a |\phi_n)$ and it has to take the following form,  by symmetry and level counting 
\begin{align}\label{L-mn}
L^-_a|\phi_n)=\alpha_n\, L^+_a |\phi_{n-1})
\end{align}
where $\alpha_n$ is an unknown coefficient. To obtain $\alpha_n$, we act with $L_a^+$ on both sides of eq. (\ref{L-mn})
\begin{align}\label{Lpmn}
\left(L^+\cdot L^-\right)|\phi_n)=\alpha_n |\phi_{n})
\end{align}
which implies that $\alpha_n$ is nothing but the eigenvalue of $|\phi_{n})$ with respect to $L^+\cdot L^-$. This eigenvalue can be easily derived by using eq. (\ref{2dCas})
\begin{align}\label{Lpmneigen}
\alpha_n=(\Delta+2n)(\Delta+2n-d-1)-\Delta(\Delta-d-1)=4n\left(n+\Delta-\frac{d+1}{2}\right)
\end{align}
Acting with $L_a^-$ on both sides of eq. (\ref{L-mn}) yields 
\begin{align}
\left(L^-\cdot L^-\right)|\phi_n)=\alpha_n \left(L^-\cdot L^+\right)|\phi_{n-1})
\end{align}
With the commutation relation $[L^-_a,L^+_b]=2\delta_{ab}H-2M_{ab}$, we are allowed to replace $L^-\cdot L^+$ by $L^+\cdot L^-+2(d+1)H$ and then we have 
\begin{align}\label{Lm2}
\left(L^-\cdot L^-\right)|\phi_n)&=\alpha_n (\alpha_{n-1}+2(d+1)(\Delta+2n-2))|\phi_{n-1})\nonumber\\
&=16\,n\,(n+\Delta-1)\left(n+\frac{d-1}{2}\right)\left(n+\Delta-\frac{d+1}{2}\right)|\phi_{n-1})
\end{align}
where we have used the eigenvalue interpretation of $\alpha_{n-1}$. The eq. (\ref{Lm2}) together with the normalization $(\phi_0|\phi_0)=\langle \Delta|\Delta)=1$, also fixes the norm of all $|\phi_n)$
\begin{align}\label{phinnorm}
(\phi_n|\phi_n)=16^n \,n!\, (\Delta)_n\left(\frac{d+1}{2}\right)_n\left(\Delta-\frac{d-1}{2}\right)_n
\end{align}

For any  $s\ge 1$, using the same symmetry argument, we can write down the most general form of $L^-_a|\phi^s_n)$ as 
\begin{align}\label{L-mns}
L^-_a|\phi^s_n)=\beta_n^s \,L^+_a\, |\phi^s_{n-1})+\gamma_n^s \,z_a\,|\phi_n^{s-1})
\end{align}
where $\beta_n^s$ and $\gamma_n^s$ are coefficients to be fixed. Acting with $L_a^+$ on both sides of eq.(\ref{L-mns}) yields that $|\phi_n^s)$ is eigenstate of $L^+\cdot L^-$ with eigenvalue $\alpha_n^s\equiv \beta_n^s+\gamma_n^s$. The latter can be easily computed via eq. (\ref{2dCas}), noticing that $|\phi_n^s)$ has energy $\Delta+s+2n$ and spin $s$
\begin{align}\label{ansdef}
\alpha_n^s=(\Delta+s+2n)(\Delta+s+2n-d-1)-\Delta(\Delta-d-1)+s(s+d-1)
\end{align}
In particular, when $n=0$, $\beta_0^s$ vanishes and then $\gamma_0^s=\alpha_0^s=2s(s+\Delta-1)$. The vanishing of $\beta_0^s$  also implies that $|\phi_0^s)$ is annihilated by $z\cdot L^-$. To obtain $ \beta_n^s $ and $\gamma_n^s$ separately, we contract both sides of  eq.(\ref{L-mns}) with $z_a$: $z\cdot L^- |\phi_n^s)=\beta_n^s |\phi^{s+1}_{n-1})$.
Then the remaining task is to compute $z\cdot L^- |\phi_n^s)$ explicitly
\small
\begin{align}
z\cdot L^- |\phi_n^s)
&=\sum_{m=0}^{n-1}\left(L^+\cdot L^+\right)^{n-m-1}\left(4\,z\cdot L^+\, H-4\, z_a \, L^+_b M_{ab}-2(d-1)z\cdot L^+\right)\left(L^+\cdot L^+\right)^{m}|\phi_0^s)\nonumber\\
&=\sum_{m=0}^{n-1}\left(4(\Delta+s+2m)-2(d-1)-4s\right)|\phi^{s+1}_{n-1})=4n\left(n+\Delta-\frac{d+1}{2}\right)|\phi^{s+1}_{n-1})
\end{align}
\normalsize
where we have used $z\cdot L^- |\phi_0^s)=0$ and $ z_a M_{ab}|\phi_0^s)= s z_b |\phi_0^s)$. Altogether, we get 
\begin{align}\label{bgexplicit}
\beta_n^s=4n\left(n+\Delta-\frac{d+1}{2}\right), \quad \gamma_n^s=\alpha_n^s-\beta_n^s
\end{align}
With $ \beta_n^s $ and $\gamma_n^s$ derived, we can proceed to compute the action of $L^-\cdot L^-$ on $|\phi_n^s)$
\begin{align}
L^-\cdot L^-|\phi_n^s)&=\beta_n^s\left(2(d+1) H+L^+\cdot L^-\right)|\phi^s_{n-1})+\gamma_n^s\,z_a \left(\beta_n^{s-1} \,L^+_a\, |\phi^{s-1}_{n-1})+\gamma_n^s \,z_a\,|\phi_n^{s-2})\right)\nonumber\\
&=\left[\beta_n^s\left(2(d+1) (\Delta+s+2n-2)+\alpha_{n-1}^s\right)+\gamma_n^s\beta_n^{s-1} \right]|\phi_{n-1}^s)\nonumber\\
&=16n(n+s+\Delta-1)\left(n+\Delta-\frac{d+1}{2}\right)\left(n+s+\frac{d-1}{2}\right)|\phi_{n-1}^s)
\end{align}
which fixes the norm of $|\phi_n^s)$ up to an $n$-independent factor:
\begin{align}\label{phinsnorm}
\frac{(\phi_n^s|\phi_n^s)}{(\phi_0^s|\phi_0^s)}=16^n n! (s+\Delta)_n\left(\Delta-\frac{d-1}{2}\right)_n\left(s+\frac{d+1}{2}\right)_n
\end{align}

\section{Various properties of $|s, n)_{a_1\cdots a_s, b_1\cdots b_s}$}\label{snprop}

In this appendix, we derive the action of $C^{\SO(1,d+1)}$ on the states $|s, n)$ (dropping spin indices for the simplicity of notation), defined in eq. (\ref{sndef}), for $s=0$ and $s=1$. For $C^{\SO(1,d+1)}$, we use eq. (\ref{1dCas})
\begin{align}
C^{\SO(1,d+1)}=\frac{1}{4}L^+\cdot L^++\frac{1}{4}L^-\cdot L^-+\left(\frac{1}{2}L^+\cdot L^-+\frac{d+1}{2}H+C^{\SO(d+1)}\right)
\end{align}
where the first term maps $|s, n)$ to $\frac{1}{4}|s, n+1)$ and the last term admits $|s, n)$ as an eigenstate.  To compute this eigenvalue, we can use the $\SO(2,d+1)$ Casimir, c.f. eq. (\ref{2dCas}), which yields the action of $L^+\cdot L^-$ on $|s, n)$
\begin{align}\label{LpLm}
L^+\cdot L^-|s,n)&=\left(H(H-d-1)-C^{\SO(d+1)}(\mY_{ss})-C^{\SO(2,d+1)}(\mathcal R_{\Delta,\ell})\right)|s,n)\nonumber\\
&=\left((2n+\ell) (2\Delta+2 n+\ell-d-1)+C^{\SO(d+1)}(\mY_{\ell})-C^{\SO(d+1)}(\mY_{ss})\right)|s,n)
\end{align}
and hence
\small
\begin{align}
\left(\frac{1}{2}L^+\!\cdot \!L^-\!+\!\frac{d\!+\!1}{2}H\!+\!C^{\SO(d\!+\!1)}\right)|s,n)\!=\!\left[2n(n\!+\!\Delta\!+\!\ell)\!+\!\frac{(d\!+\!2\ell\!+\!1)\Delta\!-\!(d\!-\!1)\ell}{2}\!+\!C^{\SO(d)}(\mY_s)\right]|s,n)
\end{align}\normalsize
where we have used $C^{\SO(d+1)}(\mY_\ell)=-\ell(\ell+d-1)$ and $\frac{1}{2}C^{\SO(d+1)}(\mY_{ss})=C^{\SO(d)}(\mY_s)=-s(s+d-2)$. So the only nontrivial task is 
$L^-\cdot L^- |s, n)$. We will compute it explicitly for $s=0$ and $s=1$. Before diving into the computation, we write down a commutation relation that will be used a lot 
\begin{align}\label{KP2}
[L^-_a, L^+\cdot L^+]&=2 (\delta_{ab}H- M_{ab}) L^+_b+2 L^+_b(\delta_{ab}H-M_{ab}) \nonumber\\
&=4L_a^+H-4L^+_b M_{ab}-2(d-1)L^+_a.
\end{align}

\subsection{$s=0$}
With eq. (\ref{KP2}), we first compute the action of $L^-_a$ on $|0, n)$
\begin{align}\label{oneLm0}
L^-_a|0,n)&=\sum_{t=0}^{n-1}(L^+\!\cdot\! L^+)^{n-t-1}\left(4L_a^+H-4L^+_b M_{ab}-2(d-1)L^+_a\right)|0,t)+(L^+\!\cdot\! L^+)^{n} L^-_a|0,0)\nonumber\\
&=\sum_{t=0}^{n-1}\left(4(\Delta+\ell+2t)-2(d-1)\right)L_a^+|0,n-1)+(L^+\cdot L^+)^{n} L^-_a|0,0)\nonumber\\
&=4n\left(n+\Delta+\ell-\frac{d+1}{2}\right)L_a^+|0,n-1)+(L^+\cdot L^+)^{n} L^-_a|0,0)
\end{align}
where we have used $M_{ab}|0,t)=0$ because  $|0,t)$ is an $\SO(d+1)$ scalar. 
Further acting with another $L^-_a$ yields 
\begin{align}\label{ss1}
L^-\cdot L^-|0,n)&=4n\left(n+\Delta+\ell-\frac{d+1}{2}\right)(L^-\cdot L^+)|0,n-1)\nonumber\\
&+4n\left(n+\Delta+\ell-1-\frac{d+1}{2}\right)(L^+\cdot L^+)^{n-1}(L^+\cdot L^-)|0,0)\nonumber\\
&-4n (L^+\cdot L^+)^{n-1} L^+_b M_{ab}L^-_a|0,0)+(L^+\cdot L^+)^{n}(L^-\cdot L^-)|0,0)
\end{align}
In the first line of (\ref{ss1}), we  write $L^-\cdot L^+$ as $2(d+1) H+L^+\cdot L^-$, and notice that $|0,n-1)$ is an eigenstate of both $H$ and $L^+\cdot L^-$. The eigenvalue of any $|s,n)$ with respect to $L^+\cdot L^-$, denoted by $\alpha_{s,n}$, is given by eq. (\ref{LpLm})
\begin{align}\label{asn}
\alpha_{s,n}=(\ell+2n)(\ell+2n+2\Delta-d-1)+2s(s+d-2)-\ell(\ell+d-1)
\end{align} 
The second line of eq. (\ref{ss1}) is a special case of (\ref{asn}) corresponding to $s=n=0$. In the third term of (\ref{ss1}), we first replace $M_{ab}L^-_a$ by the commutator $[M_{ab},L^-_a]=-dL^-_b$ and then it is becomes equivalent to the second line. The last term of (\ref{ss1}) vanishes because $(L^-\cdot L^-)|0,0)$ is at level $\ell-2$ and it is clear that such a state cannot be an $\SO(d+1)$ singlet. 
Altogether, we have 
\begin{align}\label{ss2}
L^-\cdot L^-|0,n)&=4n\left(n+\Delta+\ell-\frac{d+1}{2}\right)\left(2(d+1)(\Delta+\ell+2n-2)+\alpha_{0,n-1}\right)|0,n-1)\nonumber\\
&+4n\left(n+\Delta+\ell+\frac{d+1}{2}-2\right)\alpha_{0,0}|0,n-1)\nonumber\\
&=16 n(n+\Delta+\ell-1)\left(n+\Delta-\frac{d+1}{2}\right)\left(n+\ell+\frac{d-1}{2}\right)|0,n-1)
\end{align}

\subsection{$s=1$}
For the $s=1$ case, we also start with computing $L_c^-|1,n)_{a, b}$
\small
\begin{align}\label{ss3}
L^-_c |1,n)_{a,b}&=4n\left(n+\Delta+\ell-\frac{d+1}{2}\right) L^+_c |1,n-1)_{a,b}\nonumber\\
&-4n (L^+\cdot L^+)^{n-1} L^+_{c'} M_{cc'}|1,0)_{a,b}+(L^+\cdot L^+)^{n}L^-_c|1,0)_{a,b}
\end{align}
\normalsize
Since $|1,0)_{a,b}$ carries the 2-form representation of $\SO(d+1)$, we can get rid of the generator $M_{cc'}$
\begin{align}\label{ss4}
L^+_{c'} M_{cc'}|1,0)_{a,b}=\left(L^+_a|1,0)_{c, b}-L^+_b|1,0)_{c, a}\right)-\left(\delta_{ac} L^+_{c'}|1,0)_{c',b}-\delta_{bc} L^+_{c'}|1,0)_{c',a}\right)
\end{align}
where the first bracket is equal to $L^+_c |1,0)_{a,b}$, due to a Bianchi identity following from the identification of $ |1,0)_{a, b}=L^+_a|\Delta\rangle_b-L^+_b|\Delta\rangle_a$ as a curvature  and $L^+_a$  as a derivative. 
Combining the first term of (\ref{ss3}) and the first term of (\ref{ss4}) leads to 
\begin{align}\label{hyiu}
L^-_c |1,n)_{a, b}&=(L^+\cdot L^+)^{n}L^-_c|1,0)_{a,b}+4n\left(n+\Delta+\ell-\frac{d+3}{2}\right) L^+_c |1,n-1)_{a,b}\nonumber\\
&+4n\left(\delta_{ac} L^+_{c'}|1,n-1)_{c',b}-\delta_{bc} L^+_{c'}|1,n-1)_{c',a}\right)
\end{align}
and hence the action of $L^-\cdot L^-$  can be written as a sum of the following three pieces
\begin{align}
\rom{1}_1&=4n\left(n+\Delta+\ell-\frac{d+3}{2}\right) L^-\cdot L^+ |1,n-1)_{a,b} \nonumber\\
\rom{1}_2&=4n\left(L^-_a L^+_{c}|1,n-1)_{c,b}-L^-_b L^+_{c}|1,n-1)_{c,a}\right)\nonumber\\
\rom{1}_3&=L^-_c(L^+\cdot L^+)^{n}L^-_c|1,0)_{a,b}.
\end{align}
For the term $\rom{1}_1$, it suffices to use $L^-\cdot L^+=2(d+1)H+L^+\cdot L^-$ and eq. (\ref{asn})
\begin{align}\label{ss5.5}
(L^- \cdot L^+) |1,n-1)_{a, b}=\left(2(d+1)(\Delta+\ell+2n-2)+\alpha_{1,n-1}\right)|1,n-1)_{a, b}
\end{align}
The term $\rom{1}_2$ can be computed as follows:
\begin{align}\label{ss6}
L^-_a L^+_{c}|1,n-1)_{c,b}&=2(\Delta+\ell+2n-2)|1,n-1)_{a, b}-2M_{ac}|1,n-1)_{c,b}+L^+_{c}L^-_a|1,n-1)_{c,b}\nonumber\\
&=2(\Delta+\ell+2n-d-1)|1,n-1)_{a, b}+L^+_{c}L^-_a|1,n-1)_{c,b}
\end{align}
where 
\small
\begin{align}\label{ss7}
L^+_cL^-_a|1,n-1)_{c,b}&=(L^+\cdot L^+)^{n-1}L^+_cL^-_a|1,0)_{c,b}+4(n-1)\left(n+\Delta+\ell-\frac{d+3}{2}\right)L^+_c L^+_a |1,n-2)_{c, b}
\end{align}
\normalsize
because of eq. (\ref{hyiu}).
After antisymmetrizing $a$ and $b$, the second term of (\ref{ss7}) is proportional to $|1,n-1)_{a,b}$, again as a result of Bianchi identity, and hence we have 
\begin{align}\label{ss8}
L^-_a L^+_{c}|1,n-1)_{c,b}\!-\!(a\leftrightarrow b)&=4\left[\Delta\!+\!\ell\!+\!2n\!-\!d\!-\!1\!+\!(n\!-\!1)\left(n\!+\!\Delta\!+\!\ell\!-\!\frac{d\!+\!3}{2}\right)\right]|1,n\!-\!1)_{a,b}\nonumber\\
&+(L^+\cdot L^+)^{n-1}\left(L^+_cL^-_a|1,0)_{c,b}-L^+_cL^-_b|1,0)_{c,a}\right)
\end{align}
Using the notation $|\Delta\rangle_{a_1\cdots a_s}\equiv L^+_{a_{s+1}}\cdots L^+_{a_{\ell}}|\Delta\rangle_{a_1\cdots a_\ell}$, then $L^-_a|1,0)_{c,b}$ can be written as 
\begin{align}\label{ss9}
L^-_a|1,0)_{c,b}=2\delta_{ac}(\Delta+\ell-2)|\Delta\rangle_b+L^+_cL^-_a|\Delta\rangle_b-(b\leftrightarrow c)
\end{align}
Now a crucial step is to compute $L^-_a|\Delta\rangle_b$, which can be significantly simplified with the aid of group theory. On the one side, treating $L^-_a|\Delta\rangle_b$ as the tensor product of two fundamental representation of $\SO(d+1)$, it should has the structure $\bullet\oplus\tiny\yng(2)\oplus\yng(1,1)$ group theoretically. On the other side, according to the analysis in section \ref{CFTsec}, the $\SO(d+1)$ structures $\bullet$ and $\tiny\yng(1,1)$ exist at level $\ell$ or higher. Since $L^-_a|\Delta\rangle_b$ is of level $\ell-2$, only the $\tiny\yng(2)$ part, which is proportional to $|\Delta\rangle_{ab}$,  can be nonvanishing. With that being said, $L^-_a|\Delta\rangle_b=\kappa |\Delta\rangle_{ab}$ and the coefficient $\kappa$ is the eigenvalue of $|\Delta\rangle_a$ with respect to $L^+\cdot L^-$. We find $\kappa=2(\Delta-d-1)(\ell-1)$ by using the Casimir of $\SO(1,d+1)$ and $\SO(2,d+1)$.
Combining all the ingredients together, we obtain
\begin{align}\label{c3}
\rom{1}_2=16n\left[n\left(n+\Delta-\frac{d+1}{2}\right)+\ell\left(n+\frac{\Delta-d}{2}\right)\right]|1,n-1)_{a,b}
\end{align}

For $\rom{1}_3$, we  move $L^-_c$ rightwards by using (\ref{KP2}), and the resulting terms can be computed using (\ref{ss9}) 
\begin{align}\label{ss10}
\rom{1}_3&=4n\left(n+\Delta+\ell-\frac{d+3}{2}\right)(L^+\cdot L^+)^{n-1} \left(L^+\cdot L^-\right)|1,0)_{i,j}\nonumber\\
&-4n(L^+\cdot L^+)^{n-1} L^+_{c'}M_{cc'}L^-_{c}|1,0)_{a,b}\nonumber\\
&=4\alpha_{1,0}n\left(n+\Delta+\ell-\frac{d+3}{2}\right)|n-1,0)_{i,j}\nonumber\\
&+8n\left(d(\ell-1)(\Delta-d-1)+\Delta+\ell-2\right)|n-1,0)_{i,j}
\end{align}
Altogether, (\ref{ss5.5}), (\ref{c3}) and (\ref{ss10}) yield
\begin{align}\label{1nmm}
(L^-\cdot L^-)|1,n)_{a,b}=16 n(n+\Delta+\ell-1)\left(n+\Delta-\frac{d+1}{2}\right)\left(n+\ell+\frac{d-1}{2}\right)|1,n-1)_{a,b}
\end{align}
which takes exactly the same form as eq. (\ref{ss2}).
\subsection{Matrix elements}

For $s=0$ or $1$, the action of $C^{\SO(1, d+1)}$ on $|s, n)$ can be summarized as (dropping the spin indices again)
\begin{align}
C^{\SO(1, d+1)}|s, n)&=\left[2n(n\!+\!\Delta\!+\!\ell)\!+\!\frac{(d\!+\!2\ell\!+\!1)\Delta\!-\!(d\!-\!1)\ell}{2}\!+\!C^{\SO(d)}(\mY_s)\right]|s,n)+\frac{1}{4}|s, n+1)\nonumber\\
&+16 n(n+\Delta+\ell-1)\left(n+\Delta-\frac{d+1}{2}\right)\left(n+\ell+\frac{d-1}{2}\right)|s,n-1)
\end{align}
Using eq. (\ref{ss2}) and eq. (\ref{1nmm}), we find the $n$-dependent part of $(s,n|s,n)$ to be 
\begin{align}
(s, n|s, n)\propto 16^n n!(\Delta+\ell)_n\left(\Delta-\frac{d-1}{2}\right)_n\left(\ell+\frac{d+1}{2}\right)_n
\end{align}
and hence the action $C^{\SO(1, d+1)}$ on $|s, n\rangle = \frac{1}{\sqrt{(s,n|s,n)}}|s, n)$ is 
\begin{align}\label{needs}
C^{\SO(1, d+1)}|s, n\rangle&=\left[2n(n\!+\!\Delta\!+\!\ell)\!+\!\frac{(d\!+\!2\ell\!+\!1)\Delta\!-\!(d\!-\!1)\ell}{2}\!+\!C^{\SO(d)}(\mY_s)\right]|s,n)\nonumber\\
&+\sqrt{(n+1)(n+\Delta+\ell)\left(n+\Delta-\frac{d-1}{2}\right)\left(n+\ell+\frac{d+1}{2}\right)}|s,n+1\rangle\nonumber\\
&+\sqrt{n(n+\Delta+\ell-1)\left(n+\Delta-\frac{d+1}{2}\right)\left(n+\ell+\frac{d-1}{2}\right)}|s,n-1)
\end{align}

{
\footnotesize{
\bibliographystyle{utphys}
\bibliography{biblio}
 }
}
\end{document}